\documentclass{elsart}

\usepackage{graphicx}
\usepackage{amsmath}
\usepackage{bm}  
\usepackage[titles]{tocloft}

\journal{Physics Reports}

\newcommand{\beq}{\begin{eqnarray}}
\newcommand{\eeq}{\end{eqnarray}}
\newcommand{\bqa}{\begin{eqnarray}}
\newcommand{\eqa}{\end{eqnarray}}

\newcommand{\gsim}{\hspace*{0.2em}\raisebox{0.5ex}{$>$}
     \hspace{-0.8em}\raisebox{-0.3em}{$\sim$}\hspace*{0.2em}}
\newcommand{\lsim}{\hspace*{0.2em}\raisebox{0.5ex}{$<$}
     \hspace{-0.8em}\raisebox{-0.3em}{$\sim$}\hspace*{0.2em}}
\def\mqo2{{\!\!\!}}

\begin{document}
\begin{frontmatter}

\title{Universality in Few-body Systems with Large Scattering Length}

\author[OSU]{Eric Braaten}
\ead{braaten@mps.ohio-state.edu}
\address[OSU]{Department of Physics, The Ohio State University,
                Columbus, OH 43210, USA}
\author[INT,Bonn]{H.-W. Hammer}
\ead{hammer@itkp.uni-bonn.de}
\address[INT]{Institute for Nuclear Theory, University of Washington,
             Seattle, WA 98195, USA}
\address[Bonn]{Helmholtz-Institut f\"ur Strahlen- und Kernphysik,
 Universit\"at Bonn, 53115 Bonn, Germany}

\date{March 7, 2006}

\begin{abstract}
Particles with short-range interactions and a large scattering length 
have universal low-energy properties that do not depend on the details 
of their structure or their interactions at short distances.
In the 2-body sector, the universal properties are familiar and 
depend only on the scattering length $a$.  
In the 3-body sector for identical bosons, the universal properties 
include the existence of a sequence of shallow 3-body bound states 
called ``Efimov states" 
and log-periodic dependence of scattering observables
on the energy and the scattering length. 
The spectrum of Efimov states in the limit $a \to \pm \infty$
is characterized by an asymptotic discrete scaling symmetry 
that is the signature of renormalization group flow to a limit cycle.
In this review, we present a thorough treatment of universality for 
the system of three identical bosons and we summarize the  
universal information that is currently available for other 3-body systems.
Our basic tools are the hyperspherical
formalism to provide qualitative insights, Efimov's radial laws for deriving
the constraints from unitarity, and effective field theory for quantitative
calculations.   We also discuss topics on the frontiers of universality,
including its extension to systems with four or more particles
and the systematic calculation of deviations from universality.
\end{abstract}


\end{frontmatter}

\newpage
\tableofcontents

%
%

\newpage
\section{Introduction}
        \label{sec:intro}

The scattering of two particles with short-range interactions at
sufficiently low energy is determined by their 
S-wave {\it scattering length}, 
which is commonly denoted by $a$.
By {\it low energy}, we mean energy close to the scattering threshold 
for the two particles.  The energy is sufficiently low if the de Broglie
wavelengths of the particles are large compared to the range of the
interaction. The scattering length $a$ is important not only for
2-body systems, but also for few-body and many-body systems.
If all the constituents of a few-body system have sufficiently 
low energy, its scattering properties are determined primarily 
by $a$.  A many-body system has properties 
determined by $a$ if its constituents have
not only sufficiently low energies but also 
separations that are large compared to the range of the interaction.
A classic example is the interaction energy per particle in the 
ground state of a sufficiently dilute homogeneous 
Bose-Einstein condensate:
\begin{eqnarray}
{\mathcal E}/n \approx  {2 \pi \hbar^2 \over m} a n ,
\label{Eovern}
\end{eqnarray}
where ${\mathcal E}$ and $n$ are the energy density and number density,
respectively.
In the literature on Bose-Einstein condensates,
properties of the many-body system that are determined by the 
scattering length are called {\it universal} \cite{Andersen03}.
The expression for the energy per particle in Eq.~(\ref{Eovern})
is an example of a universal quantity.
Corrections to such a quantity from the 
effective range and other details of the interaction
are called {\it nonuniversal}. 
Universality in physics generally refers to situations 
in which systems that are very different at short distances 
have identical long-distance behavior.  
In the case of a dilute Bose-Einstein condensate,
the constituents may have completely different 
internal structure and completely different interactions,
but the many-body systems will have the same macroscopic behavior
if their scattering lengths are the same.

Generically, the scattering length $a$ is comparable in magnitude 
to the range $r_0$ of the interaction:  $|a| \sim r_0$.
Universality in the generic case is essentially a 
perturbative weak-coupling phenomenon.  The scattering length $a$
plays the role of a coupling constant.  Universal properties 
can be calculated as expansions in the dimensionless combination
$a \kappa$, where $\kappa$ is an appropriate wave number variable.
For the energy per particle in the dilute Bose-Einstein condensate,
the wave number variable is the inverse of the coherence length:
$\kappa = (16 \pi a n)^{1/2}$.
The weak-coupling expansion parameter $a \kappa$ is therefore 
proportional to the diluteness parameter $(n a^3)^{1/2}$.
The order $(n a^3)^{1/2}$ and $n a^3 \ln(n a^3)$ corrections 
to Eq.~(\ref{Eovern}) are both universal \cite{HuPi59}.
Nonuniversal effects, in the form of sensitivity to 3-body physics, 
appear first in the order $n a^3$ correction \cite{Braaten:2000eh}.

In exceptional cases, the scattering length can be much larger 
in magnitude than the range of the interaction:  $|a| \gg r_0$.
Such a scattering length necessarily requires a {\it fine-tuning}.
There is some parameter characterizing the interactions 
that if tuned to a critical value would give a 
divergent scattering length $a \to \pm \infty$.
Universality continues to be applicable in the case
of a large scattering length, but it is a much richer phenomenon.
We continue to define {\it low energy} by the condition that the 
de Broglie wavelengths of the constituents be large compared to $r_0$,
but they can be comparable to $|a|$.
Physical observables are called {\it universal} if they are insensitive 
to the range and other details of the short-range interaction.
In the 2-body sector, the consequences of universality 
are simple but nontrivial.  For example, in the case of 
identical bosons with $a>0$, there is a 2-body bound state 
near the scattering threshold with binding energy
\begin{eqnarray}
E_D  =  {\hbar^2 \over m a^2} \,.
\label{B2-eq}
\end{eqnarray}
The corrections to this formula are parametrically small: 
they are suppressed by powers of $r_0/a$.
Note the nonperturbative dependence of the binding energy on the
interaction parameter $a$.  This reflects the fact that  
universality in the case of a large scattering length is a 
nonperturbative strong-coupling phenomenon.  It should therefore  
not be a complete surprise that counterintuitive effects can 
arise when there is a large scattering length. 

A classic example of a system with a large scattering length is 
$^4$He atoms, whose scattering length is more than a factor of 10 larger
than the range of the interaction.
More examples ranging from atomic physics to nuclear and particle physics 
are discussed in detail in subsections \ref{subsec:atomexamp} and 
\ref{PartNuclargeA}.

The first strong evidence for universality in the 3-body system 
was the discovery by Vitaly Efimov in 1970 of the 
{\it Efimov effect} \cite{Efimov70},\footnote{
Some early indications of universality were already observed in 
Refs.~\cite{STM57,Dan61}.}
a remarkable feature of the 3-body spectrum 
for identical bosons with a short-range interaction 
and a large scattering length $a$.  
In the {\it resonant limit} $a \to \pm \infty$, there is a 2-body bound state 
exactly at the 2-body scattering threshold $E = 0$.  
Remarkably, there are also infinitely many,
arbitrarily-shallow 3-body bound states with binding energies 
$E_T^{(n)}$ that have an accumulation point at $E = 0$.  
As the threshold is approached, the ratio of the binding energies 
of successive states approaches a universal constant:
\begin{eqnarray}
E_T^{(n+1)}/E_T^{(n)} & \longrightarrow & 1/515.03\,, 
\qquad
{\rm as \ } n \to + \infty {\rm \ \ with \ } a = \pm \infty \,.
\label{R-resonant}
\end{eqnarray}
The universal ratio in Eq.~(\ref{R-resonant}) is independent of the mass 
or structure of the identical
particles and independent of the form of their short-range interaction. 
The Efimov effect is not unique to the system of three identical bosons.
It can also occur in other 3-body systems if at least two 
of the three pairs have a large S-wave scattering length.
If the Efimov effect occurs, there are infinitely many, arbitrarily-shallow 
3-body bound states in the resonant limit $a = \pm \infty$.  
Their spectrum is characterized 
by an asymptotic {\it discrete scaling symmetry}, 
although the numerical value of the discrete scaling factor 
may differ from the value in Eq.~(\ref{R-resonant}).  

For systems in which the Efimov effect occurs, 
it is convenient to relax the traditional definition of 
universal which allows dependence on the scattering length only.
In the resonant limit $a \to \pm \infty$, the scattering length 
no longer provides a scale.  However, the discrete Efimov spectrum
in Eq.~(\ref{R-resonant}) implies the existence of a scale.
For example, one can define a wave number $\kappa_*$ by expressing the 
asymptotic spectrum in the form
\begin{eqnarray}
E_T^{(n)} & \longrightarrow &
\left( {1 \over 515.03} \right)^{n-n_*} {\hbar^2 \kappa_*^2\over m}\,,
\qquad {\rm as \ } n \to + \infty {\rm \ \ with \ } a = \pm \infty 
\label{B3star}
\end{eqnarray}
for some integer $n_*$.
If the scattering length is large but finite,
the spectrum of Efimov states will necessarily depend on both $a$ 
and the 3-body parameter $\kappa_*$.  
Thus although the existence of the Efimov states
is universal in the traditional sense, their binding energies are not.
The dependence of few-body observables 
on $\kappa_*$ is qualitatively different from the dependence on typical
nonuniversal parameters such as the effective range.
As $a \to \pm \infty$, the dependence on typical
nonuniversal parameters decreases as positive powers of $r_0/a$,
where $r_0$ is the range of the interaction.  In contrast, 
the dependence on $\kappa_*$ 
not only does not disappear in the resonant limit,  
but it takes a particularly remarkable form.
Few-body observables are log-periodic functions of $\kappa_*$,
i.e. the dependence on $\kappa_*$ enters only through trigonometric 
functions of $\ln(\kappa_*)$.  For example, the asymptotic spectrum in 
Eq.~(\ref{B3star}) consists of the zeroes of a log-periodic function:
\begin{eqnarray}
\sin \left( \mbox{$1\over2$} s_0 \ln [m E_T/(\hbar^2 \kappa_*^2)] \right) 
= 0 \,,
\label{B3-eq}
\end{eqnarray}
where $s_0 \approx 1.00624$. Instead of regarding $\kappa_*$ 
as a nonuniversal parameter, it is more appropriate to regard it as a
parameter that labels a continous family of universality classes.
Thus for systems in which the Efimov effect occurs, 
it is convenient to relax the definition of universal
to allow dependence not only on the 
scattering length $a$ but also on the 3-body parameter
associated with the Efimov spectrum.
This definition reduces to the standard 
one in the 2-body system, because the 3-body parameter cannot 
affect 2-body observables. It also reduces to the standard definition 
for dilute systems such as the weakly-interacting Bose gas, 
because 3-body effects are suppressed by at least 
$n a^3 \ll 1$ in a dilute system.
We will refer to this extended universality simply as ``universality''
in the remainder of the paper.

If the problem of identical bosons with large scattering length 
is formulated in a {\it renormalization group} framework, 
the remarkable behavior of the system of three identical bosons in the 
resonant limit is associated with a renormalization group 
{\it limit cycle}.  The 3-body parameter associated with the Efimov 
spectrum can be regarded as
parameterizing the position along the limit cycle.
The asymptotic behavior of the spectrum in Eq.~(\ref{R-resonant}) 
reflects a {\it discrete scaling symmetry} 
that is characteristic of a renormalization group limit cycle.  
In contrast to renormalization group {\it fixed points}, 
which are ubiquitous in condensed matter physics and in high energy 
and nuclear physics, few physical applications of renormalization group 
limit cycles have been found.  Consequently, the renormalization group 
theory associated with limit cycles is largely undeveloped.
The development of such a theory could prove to be very helpful 
for extending universality into a systematically 
improvable calculational framework.

Since universality has such remarkable consequences in the 
2-body and 3-body sectors, we expect it to also have important 
implications in the $N$-body sector with $N \geq 4$.  
This is still mostly unexplored territory.  
Universality may also have important applications to the 
many-body problem.  These applications are particularly topical, 
because of the rapid pace of experimental developments in the study 
of ultracold atoms.  
By cooling an atom with a large scattering length to sufficiently 
low temperature, one can reach a regime where universality 
is applicable.  Fascinating many-body phenomena can occur 
within this regime, including Bose-Einstein condensation in the case of 
bosonic atoms and superfluidity in the case of 
fermionic atoms.  Universality has particularly exciting 
applications to these many-body phenomena, but they are 
beyond the scope of this review.

Most of this review is focused on the problem of identical bosons, 
because this is the system for which the consequences of  
universality have been most thoroughly explored.  
Identical bosons have the advantage of simplicity while still exhibiting 
the nontrivial realization of universality associated 
with the Efimov effect. 
However, we also summarize the universal results 
that are currently known for other few-body systems, 
including ones that include identical fermions.
We hope this review will stimulate the further development of 
the universality approach to such systems.

The idea of universality in systems with a large scattering length
has its roots in low-energy nuclear 
physics and has many interesting applications in 
particle and nuclear physics. 
Most of these applications are to systems with fortuitously large 
scattering lengths that arise from some accidental fine-tuning.
Universality also has many interesting applications 
in atomic and molecular physics. There are some atoms that have fortuitously 
large scattering lengths, but there are also atoms whose scattering lengths 
can be tuned experimentally to arbitrarily large values.
This makes universality particularly important
in the field of atomic and molecular physics.
We will therefore develop the ideas of universality using
the language of atomic physics: ``atom'' for  a particle, 
``dimer'' for a 2-body bound state, etc.

We begin by introducing some basic scattering concepts in 
Section~\ref{sec:scat}. 
We introduce the {\it natural low-energy length scale} and use it to define
{\it large scattering length}.  In Section~\ref{sec:RG}, we 
describe the Efimov effect and introduce some 
renormalization group concepts that are relevant to the few-body problem 
with large scattering length.  We define the {\it scaling limit} 
in which universality becomes exact and the {\it resonant limit} 
in which $a \to \pm \infty$.  We point out that these limits are 
associated with a {\it renormalization group limit cycle} 
that is characterized by a {\it discrete scaling symmetry}.  
The simple and familiar features of universality 
in the problem of two identical bosons are described in 
Section~\ref{sec:uni2}.  
We point out that there is a trivial {\it continuous scaling symmetry} 
in the scaling limit and we calculate the leading {\it scaling violations} 
which are determined by the effective range.  

In Section~\ref{sec:hyper}, we develop the hyperspherical formalism 
for three identical bosons.  We use this formalism to deduce some 
properties of Efimov states in the resonant limit
and near the atom-dimer threshold. In Section~\ref{sec:uni3}, 
we describe the most important features of universality 
for three identical bosons in the scaling limit.  
Logarithmic scaling violations reduce the continuous scaling symmetry 
to a discrete scaling symmetry.  
They also imply that low-energy 3-body observables depend not only 
on $a$ but also on an additional scaling-violation parameter.  
We then present explicit results for 3-body observables, 
including the binding energies of Efimov states, atom-dimer scattering, 
3-body recombination, and 3-atom scattering. 
We illustrate these results by applying them to the case
of helium atoms, which have a large scattering length.
We use universal scaling curves to illustrate the nontrivial realization 
of universality in the 3-body sector for identical bosons.
In Section~\ref{sec:deep}, we describe how the predictions from 
universality are modified by effects from deep 2-body bound states.

In Section~\ref{sec:EFT},
we describe a powerful method called effective field theory 
for calculating the predictions of universality.  
Using an effective field theory that describes identical bosons 
in the scaling limit, we derive a generalization of the
Skorniakov-Ter-Martirosian (STM) integral equation and use it 
to calculate the most important low-energy 3-body observables.
We also show that the renormalization 
of the effective field theory involves an ultraviolet limit cycle.

Sections~\ref{sec:uni2}--\ref{sec:EFT}
are focused exclusively on the problem of identical bosons. 
In Section~\ref{sec:beyond}, we discuss universality 
for other 3-body systems.  
We summarize what is known about the generalizations to distinguishable 
particles, fermions, unequal scattering lengths, and unequal masses.
In Section~\ref{sec:frontiers}, we discuss some of the most important 
frontiers of universality in few-body systems. They include 
power-law scaling violations
such as those associated with the effective range, 
the $N$-body problem with $N\geq 4$,
unnaturally large low-energy constants 
in other angular momentum channels such as P-waves,
and the approach to universality in scattering models.

Some of the sections of this review could be omitted by the first-time 
reader.  He or she should begin by reading Section~\ref{sec:scat} 
on {\it Scattering Concepts} and subsections 3.1 and 3.2 of 
Section~\ref{sec:RG} on {\it Renormalization Group Concepts}. 
The first-time reader should then read 
Section~\ref{sec:uni2} on {\it Universality for Two 
Identical Bosons}, subsections 5.1, 5.5, and 5.6 of Section~\ref{sec:hyper}  
on the {\it Hyperspherical Formalism}, and Section~\ref{sec:uni3} 
on {\it Universality for Three Identical Bosons}.  
The reader who is primarily interested in systems for which
there are no tightly-bound 2-body bound states
could skip Section~\ref{sec:deep} on  
{\it Effects of Deep Two-body Bound States}. 
The first-time reader could also skip Section~\ref{sec:EFT} on 
{\it Effective Field Theory}.
A first pass through the review could be completed by reading 
Section~\ref{sec:beyond} on {\it Universality in Other Three-body Systems}.  
We hope this will whet the reader's appetite for a more thorough reading
of all the sections.


%
%

\section{Scattering Concepts}
        \label{sec:scat}

In this section, we introduce the concept of the
{\it natural low-energy length scale}
and use it to define a {\it large scattering length}. 
We also give examples of 2-body systems with large scattering lengths.


\subsection{Scattering length}

In order to introduce some basic concepts
associated with universality in systems with a large scattering length, 
we begin with a brief review of scattering theory \cite{Newton,Sakurai}.
More thorough reviews with a focus on ultracold atoms can be found in 
Refs.~\cite{Dalibard} and \cite{Heinzen}.  

The most important parameter governing the interactions of
low-energy atoms is the 2-body S-wave scattering length $a$, 
which we will refer to
simply as the {\it scattering length}. It can be defined in terms of
the partial wave expansion for the scattering amplitude.
Consider atoms with mass $m$ 
that interact through a short-range potential.
The elastic scattering of two such atoms with opposite momenta 
$\pm \hbar {\bm k}$ and total kinetic energy $E = \hbar^2 k^2/m$ 
can be described by a stationary wave function $\psi({\bm r})$
that depends on the separation vector ${\bm r}$ of the two atoms.
Its asymptotic behavior as $r \to \infty$ is the sum of a 
plane wave and an outgoing spherical wave:
\begin{eqnarray}
\psi({\bm r}) = e^{i k z} + f_k(\theta) {e^{ikr} \over r} \,.
\label{psi-f}
\end{eqnarray}
This equation defines the {\it scattering amplitude} $f_k (\theta)$,
which depends on the scattering angle $\theta$ and the wave number $k$.
The {\it differential cross section} $d \sigma/d \Omega$ can be 
expressed in the form
\begin{eqnarray}
\frac{d \sigma}{d \Omega} = 
\left|f_k (\theta) \pm f_k (\pi - \theta)\right|^2 \,,
\label{dcross}
\end{eqnarray}
where the plus (minus) sign holds for identical bosons (fermions).
If the two atoms are distinguishable, 
the term $\pm f_k (\pi-\theta)$ should be omitted.
The elastic cross section $\sigma(E)$ is obtained by integrating 
over only ${1 \over 2}$ the $4 \pi$ solid angle 
if the two atoms are identical bosons or identical fermions
and over the entire $4 \pi$ solid angle if they are distinguishable.

In the low-energy limit, the scattering becomes isotropic.
The scattering length $a$ can be defined by the low-energy limit 
of the scattering amplitude:
\begin{eqnarray}
f_k(\theta) \longrightarrow - a
\qquad {\rm as\ } k \to 0 \,.
\label{fk0}
\end{eqnarray}
The absolute value of the scattering length can be determined 
by measuring the low-energy limit of the elastic cross section:
\begin{eqnarray}
\sigma(E) \longrightarrow 8 \pi a^2 
\qquad {\rm as\ } E \to 0 \,.
\label{sig0}
\end{eqnarray}
If the atoms are distinguishable, the prefactor
is $4\pi$ instead of $8\pi$.  If there are inelastic channels
at $E=0$, the scattering length is complex-valued 
and $a^2$ on the right side of Eq.~(\ref{sig0}) is replaced by $|a|^2$.
Determining the sign of $a$ is more complicated, 
because it requires measuring an interference effect.

The {\it partial-wave expansion} resolves the scattering amplitude $f_k 
(\theta)$ into contributions from definite angular momentum quantum number 
$L$ by expanding it in terms of Legendre polynomials of $\cos \theta$:
\begin{eqnarray}
f_k (\theta) = {1 \over k}
\sum_{L=0}^\infty (2L+1) c_L(k) P_L (\cos \theta) \,.
\label{fk-theta}
\end{eqnarray}
If the atoms are identical bosons (fermions), 
only even (odd) values of $L$ contribute to the 
differential cross section in Eq.~(\ref{dcross}).
The coefficients in Eq.~(\ref{fk-theta}) are constrained
by unitarity to satisfy $|c_L(k)| \le 1$. The unitarity constraints
can be taken into account automatically by expressing the coefficients 
in terms of {\it phase shifts} $\delta_L (k)$:
\begin{eqnarray}
c_L(k) = e^{i \delta_L(k)} \sin \delta_L(k) \,.
\end{eqnarray}
The expression for the scattering amplitude can be written
\begin{eqnarray}
f_k (\theta) = \sum_{L=0}^\infty
{ 2L+1 \over k \cot \delta_L(k) - ik} P_L (\cos \theta) \,.
\label{pwe}
\end{eqnarray}
If there are inelastic channels, the phase shifts can be
complex-valued with positive imaginary parts.
If there are no inelastic 2-body channels, 
the phase shifts $\delta_L(k)$ are real-valued.
The expression for the cross section integrated 
over the scattering angle is then
\begin{eqnarray}
\sigma(E)  = \frac{8\pi}{k^2} \sum_{L=0}^\infty (2 L+1) 
\sin^2 \delta_L(k) \,,
\label{sigtot2bdy}
\end{eqnarray}
where $E = \hbar^2 k^2/m$ is the total kinetic energy
of the two atoms.
If the atoms are identical bosons (fermions), the sum is only over
even (odd) values of $L$. If the atoms are distinguishable, 
the prefactor in Eq.~(\ref{sigtot2bdy}) is $4\pi$ instead of $8\pi$.

The {\it optical theorem} relates the total cross section
to the forward-scattering limit ($\theta \to 0$)
of the scattering amplitude:
\begin{eqnarray}
\sigma^{\rm (total)}(E)  
= \frac{8\pi}{k} \, {\rm Im} \, f_k(\theta = 0) \,.
\label{optical}
\end{eqnarray}
If the atoms are distinguishable, the prefactor
is $4\pi$ instead of $8\pi$.  
If there are no inelastic 2-body channels, the total cross section
on the left side of Eq.~(\ref{optical}) is the elastic cross section 
in Eq.~(\ref{sigtot2bdy}). If there are inelastic 2-body channels, 
the total cross section is the sum of the elastic and inelastic 
cross sections.

If the atoms interact through a short-range 2-body potential, then
the phase shift $\delta_L (k)$ approaches zero like $k^{2L+1}$
in the low-energy limit $k \to 0$.
Thus S-wave ($L=0)$ scattering dominates in the low-energy limit
unless the atoms are identical fermions, 
in which case P-wave ($L=1)$ scattering dominates.
At sufficiently low energies, the S-wave phase shift
$\delta_0 (k)$ can be expanded in powers of $k^2$
\cite{Schwinger47}. The expansion is
called the {\it effective-range expansion} and is
conventionally expressed in the form
\begin{eqnarray}
k \cot \delta_0 (k) = - 1/a + \mbox{$1\over 2$} r_s k^2 
                        - \mbox{$1\over 4$} P_s k^4 + \ldots \,.
\label{kcot}
\end{eqnarray}
The first few terms define the scattering length\footnote{
We caution the reader that the opposite sign convention for the
scattering length is used in some of the literature.} 
$a$, the S-wave {\it effective range} $r_s$,
and the S-wave {\it shape parameter} $P_s$.

If the 2-body potential $V(r)$ has a long-range tail that falls off like
$1/r^n$ with $n > 1$, the higher partial waves are not as strongly 
suppressed at low energy.  For $L>(n-2) /2$, 
the phase shifts $\delta_L(k)$ approach 0 like $k^{n-1}$.
Thus in the low-energy expansion of the scattering amplitude, 
all partial waves contribute at order $k^{n-2}$ and beyond.
For example, the interatomic potential between atoms in their ground state 
has a van der Waals tail that falls off like $1/r^6$.
The phase shifts for $L=0$, 1, and 2 approach 0 like $k$, $k^3$,
and $k^5$, respectively, just as in the case of a short-range potential.
However, all the phase shifts $\delta_L(k)$ for $L \ge 2$ approach 0 like 
$k^5$.  Thus the low-energy expansion of the scattering amplitude receives
contributions from all partial waves beginning at order $k^4$ \cite{LanLifQM}.

If the potential has an attractive region that is sufficiently deep, 
the two atoms can form bound states.
The binding energies $E_2$ for 2-body bound states are determined 
by the poles of the scattering amplitude in the 
upper half-plane of the complex variable $k$.
It is convenient to define the binding wave number $\kappa$  by
\begin{eqnarray}
E_2 = \hbar^2 \kappa^2/m \,.
\label{B2-kappa}
\end{eqnarray}
For S-wave bound states, the binding wave number $\kappa$ is a
positive real-valued solution to the equation
\begin{eqnarray}
i\kappa\cot\delta_0(i\kappa)+\kappa=0\,.
\label{BE-eq}
\end{eqnarray}

If the potential is sufficiently weak,
the scattering amplitude is given by the Born approximation:
\begin{eqnarray}
f_k(\theta) \approx - {m \over 4 \pi \hbar^2}
\int d^3 r \;  e^{-i {\bm q} \cdot {\bm r} } \; V(r) \,.
\label{F-Born}
\end{eqnarray}
The integral is a function of $|{\bf q}| = 2 k \sin(\theta/2)$.
The scattering length is linear in the potential:
\begin{eqnarray}
a  \approx {m \over 4 \pi \hbar^2} 
\int d^3 r   \; V(r) \,.
\label{a-Born}
\end{eqnarray}
A simple condition for the applicability 
of the Born approximation to scattering with wave number $k$ is
\begin{eqnarray}
{m \over 4 \pi \hbar^2}
\left| \int d^3 r \;  \frac{e^{-i kr }}{r} \; V(r)\right| \ll 1 \,.
\label{Born-cond}
\end{eqnarray}
In the case of a large scattering length, this condition is not
satisfied and the dependence of $a$ on $V(r)$ is highly 
nonlinear.


\subsection{Natural low-energy length scale}
        \label{subsec:natural}

At sufficiently low energies, atoms behave like point particles with 
short-range interactions.
The length scale that governs the quantum behavior of the center-of-mass 
coordinate of an atom is the de Broglie wavelength $\lambda = 2 \pi \hbar/p$,
where $p$ is the momentum of the atom.
If the relative momentum $p$ of two atoms is sufficiently small,
their de Broglie wavelengths are larger than the spacial extent of the
atoms and they are unable to resolve each other's 
internal structure, which is provided by their electron clouds.  
Their interactions will therefore be indistinguishable 
from those of point particles.
 
If the atoms interact through a short-range potential
with range $r_0$ and if the relative momentum of the two atoms
satisfies $p \ll  \hbar/ r_0$,
then their de Broglie wavelengths prevent
them from resolving the structure of the potential. 
In this case, the effects of the interactions would be indistinguishable 
from those of a local potential consisting only of contact terms proportional 
to the delta function $\delta^3({\bm r} - {\bm r}')$
and derivatives of the delta function.  The effects of the interaction
could be reproduced with higher and higher accuracy by including
higher and higher derivatives of the delta function.
Equivalently, the scattering amplitude could be approximated 
by a truncated expansion in powers of the relative momentum.
Approximations with higher and higher accuracy could be obtained by including
more and more terms in the expansion.

For real atoms, the potential is not quite short-range.
The interatomic potential $V(r)$ between two neutral atoms 
in their ground states consists of a short range potential and
a long-range tail provided by the van der Waals interaction.
The short-range part of the potential $V(r)$ is essentially the 
Born-Oppenheimer potential,
which is the ground-state energy of the electron clouds 
when the nuclei of the atoms have a fixed separation $r$. 
For separations much larger than the size of the electron cloud
of an individual atom, the interaction energy is dominated by 
the van der Waals interaction, which
arises from the polarizability of the electron clouds.
The Coulomb interaction between the polarized electron clouds
produces a power-law potential  that falls off asymptotically like $1/r^6$:%
\footnote{Retardation effects ultimately change the asymptotic power-law 
behavior 
to $1/r^7$, but this complication is not essential for our purposes.}
\begin{eqnarray}
V(r) \longrightarrow - {C_6 \over r^6}
\qquad \mbox{ as\ }  r \to \infty \,.
\label{V-vdW}
\end{eqnarray}
At sufficiently low energy, the interactions between atoms 
are dominated by the van der Waals interaction.
The natural  length scale  associated with these interactions
can be determined by assuming that the wave function 
has only one important length scale $\ell_{\rm vdW}$ and 
requiring a balance between the typical kinetic and potential energies:
\begin{eqnarray}
{\hbar^2 \over m r^2} \sim {C_6 \over r^6}
\qquad {\rm at} \; r \sim \ell_{\rm vdW} \,.
\end{eqnarray}
The resulting length scale is called the van der Waals length:
\begin{eqnarray}
\ell_{\rm vdW} = \left(m C_6/\hbar^2 \right)^{1/4} \,.
\label{ellvdW}
\end{eqnarray}

For real atoms whose potential has a long-range van der Waals tail,
the interactions can be described accurately by a local potential 
if their relative momentum $p$ satisfies 
$p \ll  \hbar/  \ell_{\rm vdW}$, where $\ell_{\rm vdW}$ 
is the van der Waals length scale given in Eq.~(\ref{ellvdW}).
However, there is a limit to the accuracy of such a description.
The scattering amplitude can be expanded in powers of the 
relative momentum ${\bm p}$, but at 4$^{\rm th}$ order the dependence 
on ${\bm p}$ becomes nonpolynomial.
The analogous statement in coordinate space is that for 
$p \ll \hbar/\ell_{\rm vdW}$, the interactions can be approximated by a 
local potential proportional to $\delta^3({\bm r} - {\bm r}')$
and more accurately by a potential with two terms proportional to 
$\delta^3({\bm r} - {\bm r}')$ and $\nabla^2\delta^3({\bm r} - {\bm r}')$.
However, any further improvement in accuracy must take into account 
the nonlocal behavior of the potential.  
The relative errors cannot be decreased below $(p \ell_{\rm vdW}/\hbar)^4$
without taking into account the van der Waals tail of the potential 
explicitly.
This limit will not be of much concern to us,
because our primary goal will be calculations of the universal properties of
low-energy atoms up to relative errors of order $p \ell_{\rm vdW}/\hbar$.

In order to define {\it large scattering length}, 
we introduce the concept of the {\it natural low-energy length scale} 
$\ell$ associated with an interaction potential. 
It is sometimes referred to as the 
{\it characteristic radius of interaction} and often denoted $r_0$. 
The natural low-energy length scale sets the natural scale for the
coefficients in the low-energy expansion 
of the scattering amplitude $f_k(\theta)$.
By dimensional analysis, the coefficient of a term proportional to $k^n$ 
can be expressed as $\ell^{n+1}$ with a dimensionless coefficient.
There is no general constraint on the magnitude of these coefficients.
However, for a generic potential, one usually finds that 
there is a length scale $\ell$ such that the coefficients 
all have magnitudes of order 1.  The absolute value of any specific 
coefficient can be orders of magnitude larger than 1, but this typically 
requires the {\it fine-tuning} of  parameters in the potential,
such as its depth or its range. 
We define the natural low-energy length scale $\ell$ 
by the condition that most of the coefficients of $\ell^{n+1} k^n$
in the low-momentum expansion of the scattering amplitude 
have magnitudes close to 1. 
If the magnitude $|a|$ of the scattering length is comparable to $\ell$, 
we say that $a$ has a {\it natural size}.
If $|a| \gg \ell$, we call the scattering length {\it unnaturally large},
or just {\it large} to be concise.
As mentioned above, this case typically requires 
the fine-tuning of some parameter in the potential $V(r)$.
The natural low-energy length scale sets the natural scale for the other
coefficients in the low-energy expansion of the scattering amplitude,
such as the effective range $r_s$ defined by Eq.~(\ref{kcot}). 
Even if $a$ is large, we should expect $r_s$ to have a natural
magnitude of order $\ell$.  For $a$ and $r_s$ to both be unnaturally large
would require the simultaneous fine-tuning of two parameters in the potential.

The natural low-energy length scale $\ell$ also sets the natural scale for the
binding energies of the 2-body bound states closest to threshold. The
binding energy of the shallowest bound state is expected to be
proportional to $\hbar^2 /m \ell^2$, with a coefficient whose magnitude
is roughly one. The coefficient can be orders of magnitude smaller, but
this again requires the fine-tuning of a parameter in the potential.
This is precisely the same fine-tuning required to get a large positive
scattering length $a \gg \ell$.  This can be seen by inserting the effective
range expansion in Eq.~(\ref{kcot}) into the bound-state equation 
in Eq.~(\ref{BE-eq}).   If $a \gg \ell$ and if 
all higher coefficients in the effective-range expansion
have natural sizes set by $\ell$, then the bound-state equation 
has the approximate solution in Eq.~(\ref{B2-eq}).
Thus a large positive scattering length can be obtained by tuning 
the binding energy to a value much smaller than $\hbar^2/m \ell^2$.

If a short-range potential $V(r)$ is sufficiently weak, 
the Born approximation in Eq.~(\ref{a-Born}) is applicable 
and the scattering length  scales like $(mV_0/\hbar^2) \ell^3$,
where $V_0$ is the depth of the potential.
On the other hand, if the potential is very strong
and if there is a well-behaved limit as $V_0 \to \infty$, then
dimensional analysis implies that $a$ should  scale like  $\ell$.
If $V(r)$ is completely repulsive, the scattering length $a$ 
is necessarily positive and always comparable to $\ell$.  If $V(r)$ 
has regions that are attractive, the scattering length $a$ 
can be positive or negative.  In general, its value can be anywhere
between $-\infty$ and $+\infty$, but
its absolute value $|a|$ is most likely to be of order $\ell$.
This vague statement can be made into a more precise probabilistic
statement by considering a
1-parameter deformation $V_\lambda (r)$ that allows the number of bound
states in the potential to be changed, such as $V_\lambda (r)=\lambda
V(r)$. The corresponding scattering length $a_\lambda$ depends on
$\lambda$. As $\lambda$ increases from its initial value 1, the depth of
the potential increases and $a_\lambda$ decreases.  It eventually
decreases to $-\infty$ and then jumps 
discontinuously to $+\infty$ at the critical value
$\lambda_c$ at which a virtual state drops below threshold to become an
additional bound state.  If we continue to increase $\lambda$, 
$a_\lambda$ will ultimately return to its original
value $a$ at some value $\lambda_1$.  A uniform probability distribution
for $\lambda$ in the range $1<\lambda <\lambda_1$ 
defines a probability distribution for the scattering
length $a_\lambda$. Most of the probability is concentrated in regions
where $|a|$ is comparable to $\ell$. 
The scattering length is more than an order of magnitude larger
only if $\lambda$ is in a narrow interval around $\lambda_c$ that has a
very small probability. Thus a large scattering length 
requires a fine-tuning of the parameter
$\lambda$ to the region near $\lambda_c$. In the absence of any
fine-tuning, one should expect $|a|$ to be comparable to $\ell$.

\begin{figure}[htb]
\bigskip
\centerline{\includegraphics*[width=5cm,angle=0]{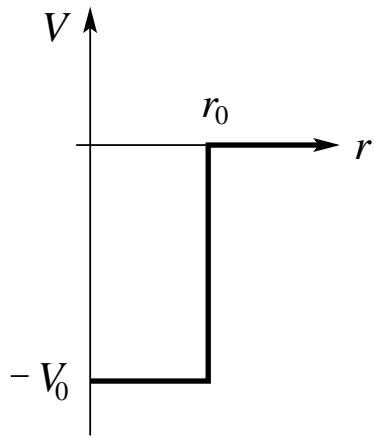}}
\medskip
\caption{Attractive square well potential with range $r_0$ and depth 
 $V_0$.}
\label{fig:pot_sqw}
\end{figure}

\begin{figure}[htb]
\bigskip
\centerline{\includegraphics*[width=8cm,angle=0]{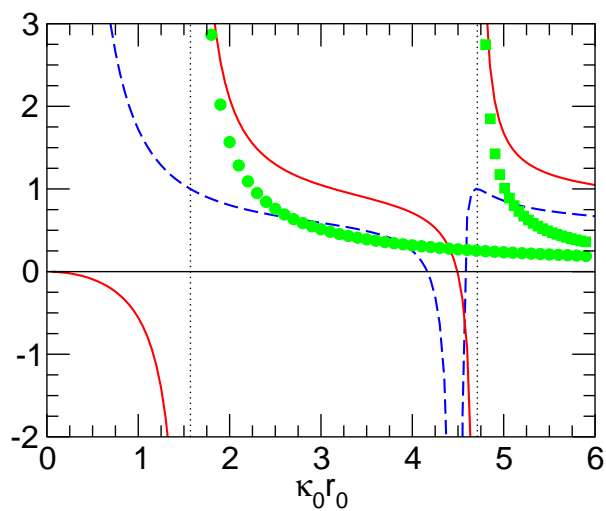}}
\medskip
\caption{Two-body observables for the attractive square-well potential.
The scattering length $a$ (solid curve), 
the effective range $r_s$ (dashed curve), 
and the inverse binding wave numbers $(m E_2 /\hbar^2)^{-1/2}$
for the first two bound states (dots and squares) in units of
$r_0$ are shown as functions of $\kappa_0 r_0$.
The vertical dotted lines are the critical values where $a$ diverges.}
\label{fig:sqwell}
\end{figure}

For an attractive short-range potential, 
the natural low-energy length scale $\ell$ is simply the {\it range} itself, 
i.e., the length beyond which the potential falls rapidly to zero.  
To illustrate the point that the range is the natural low-energy length scale,
we consider an attractive square well with range $r_0$ and depth $V_0$
as shown in Fig.~\ref{fig:pot_sqw}:
\begin{subequations}
\begin{eqnarray}
V(r) & = & - V_0, \qquad   r < r_0 \,,
\\
      & = & 0,   \qquad \hspace{0.5cm} r > r_0 \,.
\end{eqnarray}
\end{subequations}
The natural low-energy length scale is $\ell \approx r_0$. We will
treat
the depth $V_0$ as a parameter that can be varied to adjust the scattering
length $a$. The scattering length and effective range are
\begin{subequations}
\begin{eqnarray}
a &=& r_0 \left[1  - {\tan(\kappa_0 r_0) \over \kappa_0 r_0} \right] \,,
\label{a:sqwell}
\\
r_s &=& r_0\left[ 1- \frac{r_0^2}{3 a^2}-\frac{1}{\kappa_0^2ar_0} \right] \,,
\end{eqnarray}
\end{subequations}
where $\kappa_0=(m V_0/\hbar^2)^{1/2}$.
The binding energies $E_2>0$ satisfy the transcendental equation
\begin{eqnarray}
\left(\kappa_0^2 - \kappa^2 \right)^{1/2} 
\cot\left[ \left(\kappa_0^2 - \kappa^2 \right)^{1/2} \right] = - \kappa \,,
\end{eqnarray}
where $\kappa=(m E_2/\hbar^2)^{1/2}$ is the binding wave number.
In Fig.~\ref{fig:sqwell}, we show the scattering length $a$,
the effective range $r_s$, and the inverse binding wave number
$1/\kappa$ for the first two bound states 
as functions of the dimensionless variable $\kappa_0 r_0$.
For most values of $\kappa_0 r_0$, the variables $a$,
$r_s$, and $1/\kappa$ all have magnitudes
of order $r_0$. The scattering length is unnaturally large
only in narrow intervals of $\kappa_0 r_0$ near the
critical values ${1 \over 2} \pi$, ${3 \over 2} \pi$, ${5 \over 2} \pi$,
$\dots$, which are shown as vertical dotted
lines in Fig.~\ref{fig:sqwell}. 
The critical values can be reached by tuning either the depth $V_0$ 
or the range $r_0$ of the potential.
Wherever $a$ is unnaturally large and
positive, there is a bound state with unnaturally small binding energy 
given approximately by Eq.~(\ref{B2-eq}). 
Note that the effective range has the natural value
$r_s = r_0$ at the critical values of $\kappa_0 r_0$ where $a$ diverges. 
The effective range $r_s$ is unnaturally large only near those values of
$\kappa_0 r_0$ where $a$ vanishes, 
but $a^2 r_s$ has a natural value at those points.

\begin{figure}[htb]
\bigskip
\centerline{\includegraphics*[width=8.5cm,angle=0]{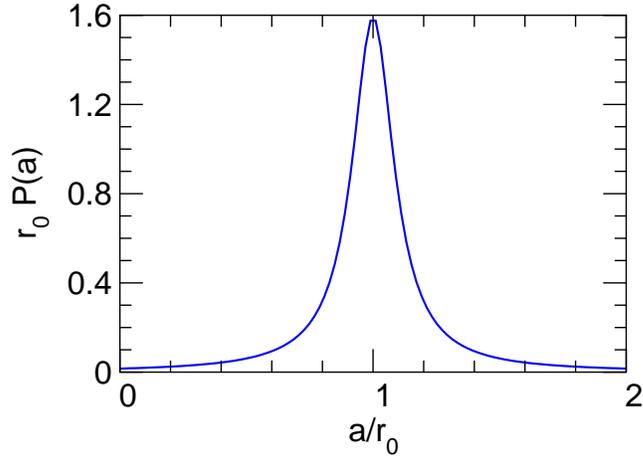}}
\medskip
\caption{Probability distribution for the scattering length $a$ for
the attractive square-well potential with $\kappa_0 r_0 =10$.}
\label{fig:probdis_sqw}
\end{figure}

If $\kappa_0 r_0 \gg 1$, we can use the expression in Eq.~(\ref{a:sqwell}) 
to make a simple probabilistic statement about the scattering length.  
A probability distribution for $V_0$ or
$r_0$ will generate a probability distribution for $a$.
If $\kappa_0 r_0 \gg \pi$, a small
fractional variation in $V_0$ or $r_0$ can generate a variation in the
argument of $\tan(\kappa_0 r_0)$  that extends over several periods.
Any probability distribution for $V_0$ or $r_0$ that is approximately
constant over intervals of $\kappa_0 r_0$ of length $\pi$
will give an approximately uniform distribution for $\kappa_0 r_0$ mod $\pi$. 
The resulting probability distribution for $a$ is
\begin{eqnarray}
P(a) da = {1 \over (a - r_0)^2 + 1/\kappa_0^2} {da \over \pi\kappa_0} \,.
\label{P:sqwell}
\end{eqnarray}
The distribution is shown in Fig.~\ref{fig:probdis_sqw}.
It peaks at $a = r_0$ 
and its full width at half maximum is $2/\kappa_0$.
Thus the probability is concentrated near $a = r_0$
and it is sharply peaked if $\kappa_0 r_0 \gg 1$.

\begin{figure}[htb]
\bigskip
\centerline{\includegraphics*[width=5.5cm,angle=0]{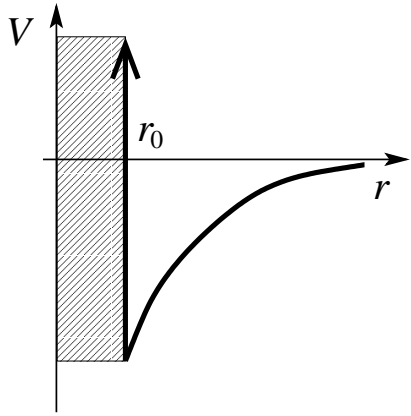}}
\medskip
\caption{Van der Waals potential with a hard-core radius $r_0$.}
\label{fig:hcvdW}
\end{figure}

\begin{figure}[htb]
\bigskip
\centerline{\includegraphics*[width=8cm,angle=0]{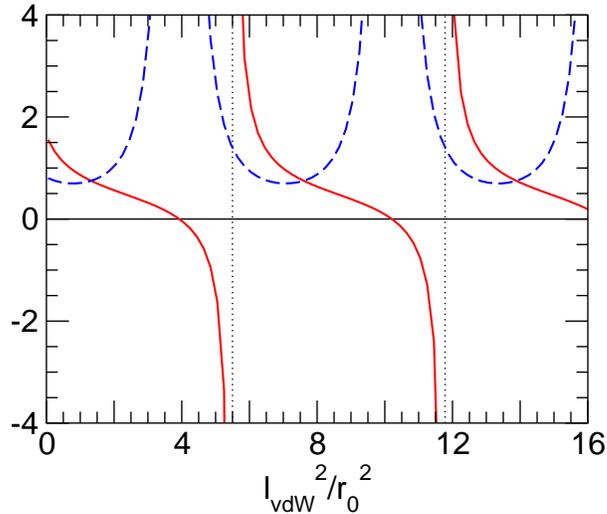}}
\medskip
\caption{Two-body scattering observables for the hard-core van der Waals 
potential. The scattering length $a$ (solid curve) and the effective range 
$r_s$ (dashed curve) in units
of $\ell_{\rm vdW}$ are shown as functions of  $\ell_{\rm vdW}^2/r_0^2$.
The vertical dotted lines are the critical values where $a$ diverges.}
\label{fig:vdwaals}
\end{figure}

For atoms interacting through a short-range potential 
with a $1/r^6$ van der Waals tail as in Eq.~(\ref{V-vdW}),
the natural low-energy length scale  is 
the van der Waals length $\ell_{\rm vdW}$
given in Eq.~(\ref{ellvdW}).
To illustrate the point that $\ell_{\rm vdW}$ 
is the natural low-energy length scale,
we consider a potential that has a hard core of
radius $r_0$ and decreases like $-C_6 / r^6$ for $r > r_0$
as illustrated in Fig.~\ref{fig:hcvdW}:
\begin{subequations}
\begin{eqnarray}
V(r) & = & +\infty, \qquad \hspace{0.1cm}   r < r_0 \,,
\\
      & = & -\frac{C_6}{r^6},   \qquad r > r_0 \,.
\label{vdWpot}
\end{eqnarray}
\end{subequations}
The scattering length and the effective range
can be calculated analytically \cite{GrF93,FGH99}:
\begin{subequations}
\begin{eqnarray}
a=\frac{\Gamma^2({3 \over 4})}{\pi}
\left(1-\tan \Phi\right)
\ell_{\rm vdW} \,,
\label{avdW}
\\
r_s=\frac{2 \pi}{3 \Gamma^2({3\over 4})}
\frac{1+\tan^2 \Phi}{(1-\tan \Phi)^2}  \ell_{\rm vdW} \,,
\label{rvdW}
\end{eqnarray}
\end{subequations}
where the angle $\Phi$ is
\begin{eqnarray}
\Phi = \frac{\ell_{\rm vdW}^2}{2r_0^2} - \frac{3\pi}{8} \,.
\end{eqnarray}
In Fig.~\ref{fig:vdwaals}, we show the scattering length
$a$ and the effective range $r_s$ as functions of the dimensionless
variable $\ell_{\rm vdW}/r_0$.
For most values of $\ell_{\rm vdW}^2/r_0^2$, 
the scattering variables $a$ and $r_s$
have natural magnitudes of order $\ell_{\rm vdW}$. 
The scattering length is unnaturally large only in narrow intervals
of $\ell_{\rm vdW}^2/r_0^2$ near the critical values
$\frac{7}{4}\pi$, $\frac{15}{4}\pi$,
$\frac{23}{4}\pi$, $\ldots$, which are shown as vertical dotted lines
in Fig.~\ref{fig:vdwaals}. The critical values can be reached
by fine-tuning either the strength $C_6$ of the long-distance potential or
the hard-core radius $r_0$. 
Note that the effective range has 
the natural value $r_s = 1.39 \;  \ell_{\rm vdW}$ at
the critical values of $\ell_{\rm vdW}/r_0$ where $a$ diverges.
The effective range $r_s$ is unnaturally large only near the values of 
$\ell_{\rm vdW}/r_0$ at which $a$ vanishes, 
but $a^2 r_s$ has a natural value at those points.

\begin{figure}[htb]
\bigskip
\centerline{\includegraphics*[width=8.5cm,angle=0]{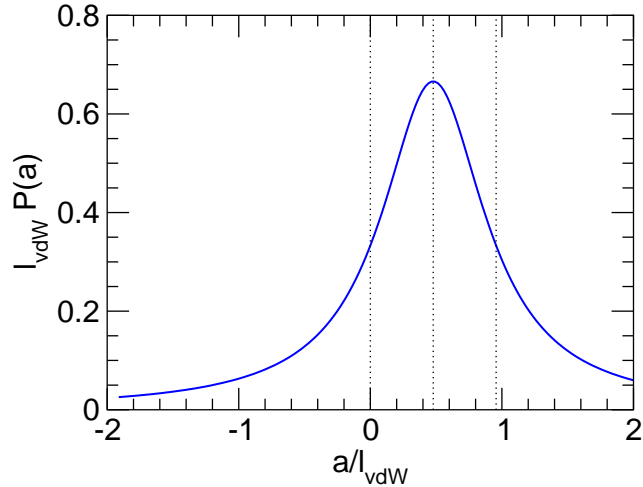}}
\medskip
\caption{Probability distribution for the scattering length $a$
for potentials with a van der Waals tail. The vertical dotted 
lines separate the horizontal axis into four regions, 
each with probability 1/4.}
\label{fig:probdis_vdW}
\end{figure}

If $\ell_{\rm vdW}^2/r_0^2 \gg 2 \pi$, we
can use the expression in Eq.~(\ref{avdW}) to make a simple probabilistic
statement about the scattering length. A probability distribution for $C_6$ or
$r_0$ will generate a probability distribution for $a$.
If $\ell_{\rm vdW}^2/r_0^2 \gg 2 \pi$, a small
fractional variation in $C_6$ or $r_0$ will generate a variation in the
argument of $\tan \Phi$ that extends over several periods.
Any probability distribution for $C_6$ or $r_0$ that is approximately
constant over intervals of $\ell_{\rm vdW}^2/r_0^2$ of length $2\pi$
will give an approximately uniform distribution for $\Phi$ mod $\pi$. 
The resulting probability distribution for $a$ is
\begin{eqnarray}
P(a) da = {1 \over (a - c \, \ell_{\rm vdW})^2 + c^2 \ell_{\rm vdW}^2} 
        {c \, \ell_{\rm vdW} da \over \pi} \,,
\label{P-vdW}
\end{eqnarray}
where $c = \Gamma^2({3\over 4})/\pi \approx 0.478$.
The distribution is shown in Fig.~\ref{fig:probdis_vdW}.
It peaks at $a = c \, \ell_{\rm vdW}$ 
and its full width at half maximum is $2 c \, \ell_{\rm vdW}$. 
Each of the 4 intervals $(- \infty, 0)$, $(0,c \, \ell_{\rm vdW})$, 
$(c \, \ell_{\rm vdW}, 2 c \, \ell_{\rm vdW})$, 
and $(2 c \, \ell_{\rm vdW}, + \infty)$ has probability 1/4.
Note that there is a significant probability for the scattering length 
to be negative, but it is 3 times more likely to be positive.


\subsection{Atoms with large scattering length}
\label{subsec:atomexamp}

We have defined {\it large scattering length} by the condition
$|a| \gg \ell$, where $\ell$ is the natural low-energy length scale.
As illustrated in Section~\ref{subsec:natural},
a large scattering length requires the {\it fine-tuning} 
of some interaction parameter. This fine-tuning can be 
due to fortuitous values of the fundamental constants of nature,
in which case we call it {\it accidental fine-tuning},
or it can be due to the adjustment of parameters 
that are under experimental control,
in which case we call it {\it experimental fine-tuning}.  
We will give examples of atoms with both kinds of fine-tunings.

The simplest example of an atom with a large positive scattering length 
is the helium atom $^4$He.  
The coefficient $C_6$ in the van der Waals potential 
for He is calculated in Ref.~\cite{YBDD96}.
The van der Waals length defined by Eq.~(\ref{ellvdW}) is 
$\ell_{\rm v dW} \approx 10.2 \, a_0$, where $a_0$ is the Bohr radius:
\begin{eqnarray}
a_0     &=& 5.29177  \times 10^{-11} \;  {\rm m} \,.
\label{convert:a0}
\end{eqnarray}
The equilibrium radius defined by the minimum of the 
interatomic potential is $r_{\rm eq} = 5.6 \, a_0$ \cite{JA95}. 
Since $\ell_{\rm v dW}$ is much larger than $r_{\rm eq}$,
the natural low-energy length scale is $\ell_{\rm v dW}$.
A pair of $^4$He atoms has a single 2-body bound state or {\it dimer},
and it is very weakly bound. 
From a measurement of the size of the $^4$He dimer,
the scattering length has been determined to be 
$a = (197^{+15}_{-34}) \, a_0$ \cite{Grisenti}. 
This is much larger than the van der Waals length.
More precise values of the scattering length can be calculated 
from model potentials for helium atoms.  
For example, the LM2M2 \cite{AS91} and TTY \cite{TTY95} 
potentials have a large scattering length 
$a = 189 \, a_0$ but a natural effective range $r_s = 14 \, a_0$. 
The binding energy of the dimer is $E_2= 1.31$ mK,
which is much smaller than the natural low-energy scale 
$\hbar^2/m \ell_{\rm vdW}^2 \approx 400$ mK.
We have expressed these energies in terms of the temperature unit mK.
The conversion factors to electron volts 
and to the natural atomic energy unit $\hbar^2 /m_e a_0^2$ are
\begin{eqnarray}
1\; {\rm mK}   = 8.61734 \times 10^{-8} \; {\rm eV}  
= 3.16682 \times 10^{-9} \; \hbar^2 /m_e a_0^2 \,.
\label{convert}
\end{eqnarray}
The scattering length of $^4$He atoms is large because of an 
{\it accidental fine-tuning}.  The mass of the $^4$He nucleus, 
the electron mass, and the fine structure constant $\alpha$ of QED 
have fortuitous values that make the potential between two $^4$He atoms 
just deep enough to have a bound state very close to threshold, 
and therefore a large scattering length.  If one of the $^4$He atoms 
is replaced by a $^3$He atom, which decreases the reduced mass by 14\% 
without changing the interaction potential, 
the scattering length has the more natural value $-33 \, a_0$.

The simplest example of an atom with a large negative scattering length 
is the polarized tritium atom $^3$H \cite{BEGKH02}.  
The van der Waals length for $^3$H is $\ell_{\rm vdW} = 13.7 \, a_0$. 
The equilibrium radius defined by the minimum of the spin-triplet
potential for H atoms is $r_{\rm eq} = 1.4 \, a_0$. 
Since $\ell_{\rm v dW}$ is much larger than $r_{\rm eq}$,
the natural low-energy length scale is $\ell_{\rm v dW}$. 
The scattering length for polarized $^3$H atoms is the
spin-triplet scattering length $a_t = -82.1 \, a_0$ \cite{BEGKH02},
which is much larger than the van der Waals length. 
Polarized tritium atoms have no 2-body bound states,
but they have a single 3-body bound state 
with a shallow binding energy of about 4.59 mK \cite{BEGKH02}. 

Other examples of atoms with large scattering lengths can be found among 
the alkali atoms.  The total spin of an alkali atom  is called the 
{\it hyperfine} spin, and its quantum number is usually denoted by $f$.
The hyperfine spin is the sum of the electronic spin, 
whose quantum number is $s = {1 \over 2}$,
and the nuclear spin, whose quantum number is denoted by $i$.
The hyperfine interaction between the electronic and nuclear spins 
splits the ground state of the alkali atom into multiplets with 
hyperfine spin quantum number $f = i+{1 \over 2}$ or $f = i-{1 \over 2}$. 
We denote the individual hyperfine states by $|f, m_f \rangle$.
Each of these hyperfine states has its own scattering length $a_{f,m_f}$,
but they are all related to the spin-singlet and spin-triplet scattering 
lengths $a_s$ and $a_t$.  These scattering lengths are associated with 
different Born-Oppenheimer potentials with the same van der Waals tail.
A compilation of the scattering lengths  $a_s$ and $a_t$ for alkali atoms 
is given in Table~\ref{tab:scattl}.
The scattering lengths for the isotopes of hydrogen were calculated in
Refs.~\cite{WJ93,BEGKH02}.  The scattering lengths for some of the 
heavier alkali were determined in 
Refs.~\cite{LRTBJ02,KKHV02,LWJ00}.  The remaining
scattering lengths are the central values 
of the most precise measurements tabulated in Ref.~\cite{Heinzen}.
Note that $1/4$ of the scattering lengths are negative,
which is exactly what is expected
for a random sample of potentials with van der Waals tails.

\begin{table}[htb]
\begin{tabular}{l|cc|llc}
atom & $r_{\rm eq}$ & $\ell_{\rm vdW}$ & $a_s$ & $a_t$ & Ref. \cr
\hline
$^{1}$H    & 1.4 & 10.5 & $+0.3$    & $+1.3$  & \cite{WJ93}    \cr
$^{2}$H    & 1.4 & 12.4 & $+13$     & $-6.9$  & \cite{WJ93}    \cr
$^{3}$H    & 1.4 & 13.7 & $+35$     & $-82$   & \cite{BEGKH02} \cr
$^{6}$Li   & 5.0 & 62.5 & $+45$     & $-2160$ & \cr
$^{7}$Li   & 5.0 & 65.0 & $+34$     & $-27.6$ & \cr
$^{23}$Na  & 5.8 & 89.9 & $+19.1$   & $+65.3$ & \cr
$^{39}$K   & 7.4 & 129  & $+139.4$  & $-37$   & \cite{LRTBJ02} \cr
$^{40}$K   & 7.4 & 130  & $+105$    & $+194$  & \cr
$^{41}$K   & 7.4 & 131  & $+85$     & $+65$   & \cr
$^{85}$Rb  & 8.0 & 164  & $+2800$   & $-388$  & \cite{KKHV02} \cr
$^{87}$Rb  & 8.0 & 165  & $+90.4$   & $+99.0$ & \cite{KKHV02} \cr
$^{133}$Cs & 12  & 202  & $+280.3$  & $+2405$ & \cite{LWJ00}   \cr
\end{tabular}
\vspace*{0.3cm}
\caption{Scattering lengths and length scales for alkali atoms 
in units of $a_0$: the equilibrium radius $r_{\rm eq}$, 
the van der Waals length $\ell_{\rm vdW}$,
the spin-singlet scattering length $a_s$, and 
the spin-triplet scattering length $a_t$.  
The scattering lengths with no reference are the central values of 
the most precise results tabulated in Ref.~\cite{Heinzen}.
}
\label{tab:scattl}
\end{table}

The coefficients $C_6$ in the van der Waals potentials
for alkali atoms have been calculated in 
Ref.~\cite{YBDD96} for H and Li, and in Ref.~\cite{DJSB99} for the heavier
alkali atoms. In Table~\ref{tab:scattl}, we list the corresponding van der
Waals length $\ell_{\rm vdW}$ for each of the alkali atoms. 
We also give the equilibrium radius $r_{\rm eq}$,
which is the radius of the minimum in the potential
between two atoms in the spin-triplet channel. It provides an estimate
of the range of the short-distance part of the potential. 
We see in Table~\ref{tab:scattl} that 
the van der Waals length $\ell_{\rm vdW}$ is
much larger than $r_{\rm eq}$ for all the alkali atoms. 
Thus the van der Waals length defined by Eq.~(\ref{ellvdW})
is the natural low-energy length scale for the alkali atoms.
The alkali atoms in Table~\ref{tab:scattl} provide several 
examples of atoms with large scattering lengths.
The spin-triplet scattering lengths $a_t$ for $^6$Li and for $^{133}$Cs
and the spin-singlet scattering length $a_s$ for $^{85}$Rb
are all more than an order of magnitude larger than the 
corresponding van der Waals scales $\ell_{\rm vdW}$. 
For these atoms, nature has provided a fortuitous fine-tuning 
of the potential and the mass of the atoms to give a large scattering length.
The fine-tuning is illustrated by the facts that $^7$Li,
whose mass is 17\% larger than  that of $^6$Li, has a natural value for $a_t$
and that $^{87}$Rb, whose mass is 2.3\% larger than  
that of $^{85}$Rb, has a natural value for $a_s$. 

\begin{figure}[htb]
\bigskip
\centerline{\includegraphics*[width=10cm,angle=0]{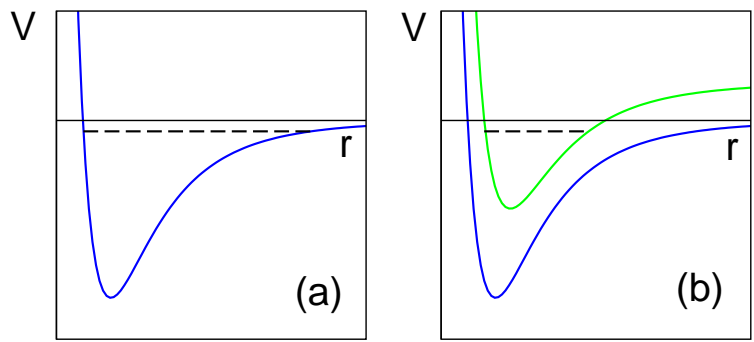}}
\medskip
\caption{
Mechanisms for generating a large scattering length by tuning 
a bound state (dashed line) to the scattering threshold for the open channel.
At a shape resonance (a), the bound state is in the potential 
for the open channel.
At a Feshbach resonance (b), the bound state is in the potential for 
a weakly-coupled closed channel.
}
\label{fig:mechanisms}
\end{figure}

The mechanism for generating a large scattering length that involves 
tuning the depth or range of the potential is called a 
{\it shape resonance} and is illustrated in Fig.~\ref{fig:mechanisms}.
With this mechanism, only the {\it open channel} defined by the 
scattering particles plays an important role.
Another mechanism for generating a  large scattering length is a
{\it Feshbach resonance}  \cite{Feshbach62}.
This requires a second {\it closed channel} in which
scattering states are energetically forbidden
that is weakly coupled to the open channel.
The closed channel might consist of particles 
in different spin states from those in the open channel.
If the interaction potential $V(r)$ for the open channel
asymptotes to 0 as $r \to \infty$,
the interaction potential for the closed channel asymptotes to 
a  positive value that is large compared to the energy scale 
of the particles in the open channel.  Thus the only states 
in the closed channel that are energetically accessible are bound states.
The weak coupling between the channels allows transitions 
between pairs of particles in the two channels.  
A large scattering length for particles in the open channel
can be generated by tuning the depth of the potential for the closed 
channel to bring one of its bound states close to the threshold for the 
open channel, as illustrated in Fig.~\ref{fig:mechanisms}. 
The resulting enhancement of the scattering of particles 
in the open channel is a Feshbach resonance.

\begin{figure}[htb]
\bigskip
\centerline{\includegraphics*[width=8cm,angle=0]{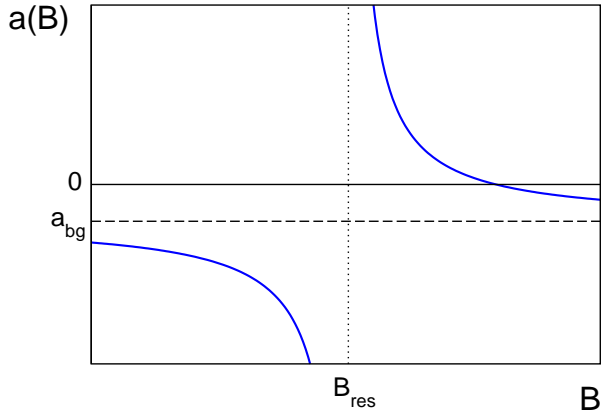}}
\medskip
\caption{The scattering length $a$ as a function of the magnetic
field $B$ near a Feshbach resonance with $a_{bg}<0$ and $c_{res}>0$.}
\label{fig:feshbach}
\end{figure}

Feshbach resonances in alkali atoms can be created by tuning the 
magnetic field \cite{TMVS92,TVS93}.
In this case, the open channel consists of a pair of atoms in a specific 
hyperfine state $|f,m_f \rangle$.  The  closed channel consists of a 
pair of atoms in different hyperfine states with a higher scattering threshold.
The weak coupling between the channels is provided by the hyperfine 
interaction.  Since different hyperfine states have different magnetic 
moments, a magnetic field can be used to vary the energy gap 
between the scattering thresholds and bring a bound state in the 
closed channel into resonance with the threshold of the open channel.
The resulting enhancement of the scattering of particles 
in the open channel is a Feshbach resonance.
The scattering lengths in Table~\ref{tab:scattl}
are in the absence of a magnetic field.  The scattering lengths generally 
vary slowly with the magnetic field $B$.  However, if
$B$ is varied through a Feshbach resonance for a particular 
hyperfine state, the scattering length changes dramatically
as illustrated in Fig.~\ref{fig:feshbach}.
It increases or decreases  to $\pm \infty$,
jumps discontinuously to $\mp \infty$, and then 
returns to a value close to its original off-resonant value.
If the Feshbach resonance is narrow, the scattering length 
near the resonance has the approximate form
\begin{eqnarray}
a(B) \approx 
a_{\rm bg} +  {c_{\rm res} \over B - B_{\rm res}} \,,
\label{Feshbach}
\end{eqnarray}
where $a_{\rm bg}$ is the off-resonant scattering length, $B_{\rm res}$ is the
location of the Feshbach resonance, and $c_{\rm res}$ controls
the width of the resonance.
In this case, the magnetic field provides an experimental 
fine-tuning parameter that can be used to make $|a|$ arbitrarily large.
In particular, the scattering length can be made
larger than the natural low-energy scale $\ell_{\rm vdW}$.

The use of a Feshbach resonance to produce a large scattering length 
in alkali atoms was first demonstrated in experiments 
with Bose-Einstein condensates of $^{23}$Na atoms \cite{Inouye98}
and with cold gases of $^{85}$Rb atoms \cite{Courteille98,Roberts98}.  
For $^{23}$Na atoms, the spin-singlet and spin-triplet scattering lengths 
$a_s$ and $a_t$ given in Table I have natural values, 
so all the hyperfine spin states $|f, m_f\rangle$ 
have natural scattering lengths $a_{f, m_f}$ at $B=0$.  
However, they diverge at values of $B$ that depend on the hyperfine state
$| f, m_f \rangle$.
For example, $a_{1, -1}$ has Feshbach resonances near $B_{\rm res} = 853$ G 
and near $B_{\rm res} = 907$ G.  
For $^{85}$Rb atoms, the spin-triplet scattering length $a_t$ 
is a factor 2.4 larger than the natural scale $\ell_{\rm vdw}$,
and the spin-singlet scattering length $a_s$ is very large.  
Thus most of the hyperfine spin states have relatively large scattering 
lengths $a_{f, m_f}$ at $B=0$.  
However, $a_{f, m_f}$ can be made arbitrarily large by tuning the
magnetic field to a Feshbach resonance.  For example, $a_{2, -2}$ has a
Feshbach resonance near $B_{\rm res} = 155$ G.


\subsection{Particles and nuclei with large scattering length}
\label{PartNuclargeA}

Systems with large scattering length also arise in particle and 
nuclear physics.  In all the subatomic systems described below,
the large scattering length arises from an
accidental fine-tuning of the parameters in the underlying theory. 

The simplest example of a particle with a large scattering length 
is the neutron.
The neutron is a spin-${1 \over 2}$ fermion.  Neutrons  with opposite spins 
can scatter in the S-wave channel.  The scattering length and 
the effective range are $a = -18.5$ fm and $r_s = 2.8$ fm.
The low-energy interactions between two neutrons 
can be described by a short-range potential that is generated by the exchange 
of pions.  The natural low-energy length scale is the range of the
one-pion-exchange potential:  $\ell_\pi \approx  \hbar/m_\pi c = 1.4$ fm.
The effective range for neutron-neutron scattering is comparable to this
natural low-energy length scale, but the absolute 
value of the scattering length is larger by more than an order of magnitude.

The best known example of a system in nuclear physics with a large scattering
length is the proton-neutron system. 
The proton ($p$), like the neutron ($n$), is a spin-${1 \over 2}$ fermion.  
Nuclear forces respect an approximate $SU(2)$ isospin 
symmetry that mixes protons and neutrons.  
It is therefore  useful to regard protons and neutrons 
as distinct isospin states of a single particle called the 
{\it nucleon} (denoted by $N$).
Isospin symmetry is broken by electromagnetic effects, 
which generate the Coulomb force between protons, 
and by small effects associated with the difference
between the masses of the up and down quarks.
Because of isospin symmetry, there are two independent S-wave
scattering lengths that govern the low-energy scattering 
of nucleons. We can take them to be the spin-singlet 
and spin-triplet $np$ scattering lengths $a_s$ and $a_t$,
which correspond to $NN$ scattering in the isospin-triplet $^1S_0$
and isospin-singlet $^3S_1$ channel, respectively.
The spectroscopic notation $^{2s+1}L_j$ encodes the angular momentum 
quantum numbers for total spin ($s$), orbital angular momentum 
($L=S,P,D, \ldots$ for $L=0,1,2,\ldots$), 
and total angular momentum ($j$).
The scattering lengths and effective ranges are
$a_s = -23.76$ fm and $r_s=2.75$ fm  in the spin-singlet channel
and $a_t = 5.42$ fm and $r_t=1.76$ fm in the spin-triplet channel.  
The effective ranges are both comparable to the 
natural low-energy length scale  $\ell_\pi \approx 1.4$ fm.
However, the scattering length $a_s$ is much larger than $\ell_\pi$
and $a_t$ is at least  significantly larger.
The {\it deuteron} is an isospin-singlet  $^3S_1$ $pn$ bound state
with binding energy $E_d = 2.225$ MeV.  This is significantly 
smaller than the natural low-energy scale
$m_\pi^2 c^2/m_N  = 21$ MeV, and fairly close 
to the universal prediction $\hbar^2/m a_t^2 = 1.4$ MeV of Eq.~(\ref{B2-eq}). 
Thus we can identify the deuteron as the shallow bound state 
associated with the large positive spin-triplet scattering length.

Two low-energy protons interact through both the nuclear force 
and the Coulomb force.  Isospin symmetry implies that, 
in the absence of the Coulomb force, the scattering length  
and effective range would have the same values 
$a_s$ and $r_s$ as for the spin-singlet $np$ system.
Thus the $pp$ scattering length is large.  Unfortunately, 
universal effects in the $pp$ system are complicated by the 
long-range Coulomb potential.

Another example of a large scattering length in nuclear physics 
is the $\alpha \alpha$ system \cite{Efimov90}, where  $\alpha$ 
stands for the $^4$He nucleus.
The scattering length and the effective range 
for $\alpha\alpha$ scattering are estimated to be
$a\approx 5$ fm and $r_s \approx 2.5$ fm.
The scattering length is significantly larger than the natural low-energy
length scale  $\ell_\pi \approx 1.4$ fm set by one-pion exchange.
The unstable $^8$Be nucleus is 
known to be clustered into two $\alpha$ particles, and
the ground state energy of $^8$Be lies only about 0.1 MeV 
above the $\alpha\alpha$ threshold. This energy is much 
smaller than the natural energy scale $m_\pi^2 c^2/m_\alpha \approx 5.3$ MeV,
and comparable in magnitude  to the universal prediction 
$-\hbar^2/m_\alpha a^2 \approx -0.4$ MeV  of Eq.~(\ref{B2-eq}). 
The difference can be partly attributed to the Coulomb repulsion of 
the doubly-charged $\alpha$ particles.
As a consequence, the ground state of $^8$Be
can be interpreted as a shallow $\alpha\alpha$ resonance resulting from 
the large scattering length.
Universal effects in the $\alpha \alpha$ system are complicated not only
by the long-range  Coulomb force between the $\alpha$ particles, 
but also by the fact that the scattering length is not terribly large.

A new example of a system with large scattering length
has recently emerged in particle physics.
In 2003, the Belle collaboration discovered a new hadronic 
resonance that decays into  $J/\psi \; \pi^+ \pi^-$, 
where $J/\psi$ is the lowest spin-triplet charmonium state \cite{Choi:2003ue}.
Its mass is $3872.0 \pm 0.6 (\mbox{stat})\pm 0.5 (\mbox{syst})$ MeV$/c^2$
and its total width is less than  2.3 MeV$/c^2$ at the 90\% confidence level.
The nature of this state, which was tentatively named $X(3872)$, 
has not yet been determined.  However, its mass is extremely close 
to the threshold $3871.2 \pm 0.7$ MeV/$c^2$ for decay into the charm mesons
$D^0$ and $\bar{D}^{*0}$ or $\bar{D}^0$ and $D^{*0}$.
This suggests that it might be a $D^0 \bar{D}^{*0}$/$\bar{D}^0 D^{*0}$
molecule \cite{Tornqvist:1991ks}.  
The natural scale for the binding energy of such a molecule
is $m_\pi^2 c^2/(2 \mu) \approx 10$ MeV/$c^2$, 
where $\mu$ is the reduced mass of the $D^0$ and $\bar{D}^{*0}$.  
The mass measurement indicates that its binding energy 
is $-0.8 \pm 1.1$ MeV.  Since this is much smaller than the natural scale,
the $D^0 \bar{D}^{*0}$ scattering length must be much larger than the 
natural length scale $\ell_\pi \approx 1.4$ fm.
If the $X(3872)$ is indeed a  $D^0 \bar{D}^{*0}$/$\bar{D}^0 D^{*0}$
molecule, then it has universal properties that are determined by the 
unnaturally large scattering length \cite{Voloshin:2003nt,Braaten:2003he}.


%
%

\section{Renormalization Group Concepts}
        \label{sec:RG}

In this section, we introduce some concepts that arise naturally 
if the problem of atoms with large scattering length 
is formulated within a renormalization group framework. 
We introduce the {\it resonant} and {\it scaling}  limits,
and explain how the nontrivial realization of universality 
in the 3-body sector is related to
{\it renormalization group limit cycles}.


\subsection{Efimov effect}
\label{Efimoveff}

The {\it Efimov effect} is a remarkable phenomenon that can occur 
in the 3-body sector for nonrelativistic particles if at least two of the 
three pairs of particles has a large scattering length.   
It was discovered by Efimov in 1970 \cite{Efimov70}.  
The Efimov effect is very well-established theoretically, but 
there is as yet no convincing experimental evidence 
for this effect.

The {\it Thomas effect} is closely related to the Efimov effect,
but it was discovered much earlier in 1935 \cite{Tho35}.
Thomas considered particles interacting through a 2-body potential 
with depth $V_0$ and range $r_0$ that supported 
a single bound state with binding energy $E_2$.
Thomas studied the {\it zero-range limit} 
defined by $r_0 \to 0$ and $V_0 \to \infty$ with $E_2$ fixed.
Using a simple variational argument, he showed that
the binding energy $E_T$ of the deepest 3-body bound state 
diverges to $\infty$ in the zero-range limit.
Thus the spectrum of 3-body bound states is unbounded from below.
The counterintuitive conclusion is that a 2-body potential 
that is only attractive enough to support a single 2-body bound state
can nevertheless produce 3-body bound states with arbitrarily 
large binding energies.  

The zero-range limit considered by Thomas produces a 
large scattering length $a \gg r_0$.
The binding energy of the 2-body bound state in this limit 
is given by Eq.~(\ref{B2-eq}).  The binding energy of the deepest 
3-body bound state produced by the variational argument
scales like $\hbar^2/m r_0^2$, 
and thus diverges as $r_0 \to 0$.  The importance of the Thomas effect 
is limited by the fact that the binding energy and other properties 
of this deepest 3-body bound state may
depend on the details of the interaction potential.

In 1970, Efimov pointed out that when $|a|$
is sufficiently large compared to the range $r_0$ of the potential, 
there is also a sequence of 3-body bound states
whose binding energies are spaced roughly geometrically
in the interval between $\hbar^2/m  r_0^2$ and $\hbar^2/m a^2$.
As $|a|$ is increased, new bound states appear 
in the spectrum at critical values of  $a$ that differ by 
multiplicative factors of $e^{\pi/s_0}$,  where $s_0$ 
depends on the statistics and the mass ratios of the particles.
In the case of identical bosons, $s_0$ is the solution to the 
transcendental equation 
\begin{eqnarray}
s_0 \cosh {\pi s_0 \over 2} = {8 \over \sqrt{3}} \sinh {\pi s_0 \over 6} \,.
\label{s0}
\end{eqnarray}
Its numerical value is $s_0\approx 1.00624$, so $e^{\pi/s_0} \approx 22.7$.
As $|a|/r_0 \to \infty$, 
the asymptotic number of 3-body bound states is 
\begin{eqnarray}
N \longrightarrow {s_0 \over \pi} \ln {|a| \over r_0} \,.
\label{N-Efimov}
\end{eqnarray}
In the limit $a \to \pm \infty$, there are infinitely 
many 3-body bound states with an accumulation point at the 
3-body scattering threshold.
A formal proof of the Efimov effect was subsequently given
by Amado and Nobel \cite{AN71,AN72}.
The Thomas and Efimov effects are closely related.
The deepest 3-body bound states found by Thomas's variational 
calculation can be identified with the deepest Efimov states
\cite{ADFGT88}.

The importance of the Efimov effect is that the sequence of 
3-body bound states he discovered have universal properties 
that are insensitive to the details of the 2-body potential 
at short distances.  The simplest such property is that 
in the resonant limit in which there are infinitely many 
arbitrarily-shallow 3-body bound states, the ratio of the 
binding energies of the successive bound states approaches a
universal number as the threshold is approached:
\begin{eqnarray}
E_T^{(n+1)} /E_T^{(n)}  & \longrightarrow & e^{- 2 \pi/s_0} \,,
\qquad {\rm as \ } n \to + \infty {\rm \ \ with \ } a = \pm \infty \,.
\end{eqnarray}
In the case of identical bosons, the universal number has 
the value 1/515.03.  This implies that the asymptotic behavior 
of the spectrum in the resonant limit has the form
\begin{eqnarray}
E_T^{(n)} & \longrightarrow &
\left( e^{-2 \pi/s_0} \right)^{n-n_*} \hbar^2 \kappa_*^2/m\,,
\qquad {\rm as \ } n \to + \infty {\rm \ \ with \ } a = \pm \infty
\label{B3-resonant}
\end{eqnarray}
for some integer $n_*$ and some parameter $\kappa_*$ 
with dimensions of wave number.
If we chose a different integer $n_*$, the  value of $\kappa_*$ 
would change by some power of $e^{\pi/s_0}$. 
Thus $\kappa_*$ is defined by Eq.~(\ref{B3-resonant})
only up multiplicative factors of $e^{\pi/s_0}$.

In subsequent papers, Efimov showed that universality 
is a general feature of the 3-body problem in the scaling limit.
It does not require the resonant limit $a = \pm \infty$,
but occurs whenever the scattering length is large.
In two brilliant papers in 1971 and 1979 \cite{Efimov71,Efimov79}, 
Efimov derived a number of universal results on  
low-energy 3-body observables for three identical bosons. 
A remarkable example is a universal formula for the
atom-dimer scattering length  \cite{Efimov79}:
\begin{eqnarray}
a_{AD} = 
\big( b_1 - b_0 \tan [s_0 \ln (a \kappa_*) + \beta]   \big) \; a \,,
\label{a12-Efimov}
\end{eqnarray}
where $b_0$, $b_1$, and $\beta$  are universal numbers
and $\kappa_*$ is the 3-body parameter 
defined by the asymptotic behavior of the spectrum of Efimov states 
in the resonant limit given in Eq.~(\ref{B3-resonant}).
The universal numbers $b_0$ and $b_1$ were first calculated
by Simenog and Sinitchenko \cite{Sim81}.
These and other universal results for the 3-body system consisting 
of three identical bosons are presented in Section~\ref{sec:uni3}.
The expression (\ref{a12-Efimov}) for $a_{AD}$ in Eq.~(\ref{a12-Efimov}) 
is universal in the sense that it holds for all 
identical bosons, independent of the short-range interactions that 
generate the large scattering length, provided that 
the shallow dimer is the only 2-body bound state.
If there are additional deep  2-body bound states,  
the universal expression for $a_{AD}$ 
is more complicated and is given in Section~\ref{sec:deep}.


\subsection{The resonant and scaling limits}
        \label{sec:scaling}
 
We have defined a large scattering length to be one that satisfies 
$|a| \gg \ell$, where $\ell$ is the natural low-energy length scale. 
The corrections to the universal behavior 
are suppressed by powers of $\ell/|a|$.
There are two obvious limits in which the size 
of these corrections decreases to zero:
\begin{itemize}

\item the {\it resonant limit}: $a \to \pm \infty$ with $\ell$ fixed,

\item the {\it scaling limit}: $\ell \to 0$ with $a$ fixed.

\end{itemize}
In either limit, the leading corrections to the universal behavior 
are suppressed by powers of $\ell/|a|$.
It will sometimes also be useful to consider systems in which
the resonant and scaling limits are achieved simultaneously: 
$a = \pm \infty$ and $\ell = 0$.
 
\begin{figure}[htb]
\bigskip
\centerline{\includegraphics*[width=10cm,angle=0]{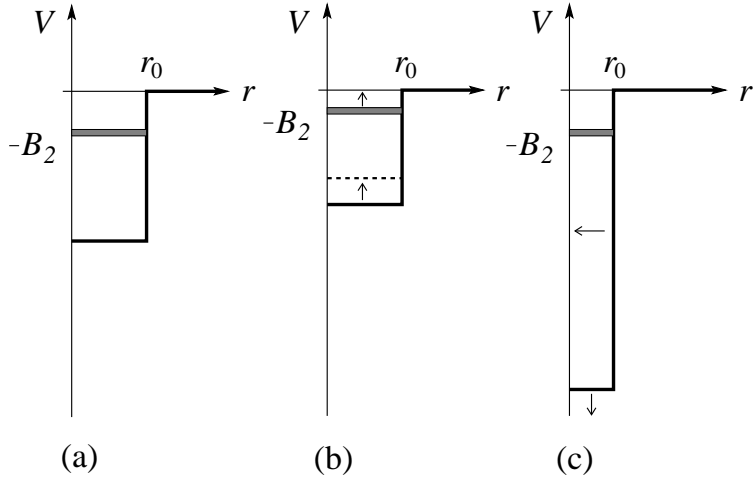}}
\medskip
\caption{
 (a) A square well potential with a single shallow bound state,
 (b) a potential with the same range $r_0$ that is approaching the 
     resonant limit $E_2 \to 0$, and
 (c) a potential with the same binding energy $E_2$ as in (a) 
     that is approaching 
     the scaling limit $r_0 \to 0$.}
\label{fig:potrange}
\end{figure}

The  {\it resonant limit} is also sometimes called the {\it unitary limit},
because in this limit the S-wave contribution to the cross section 
at low energy saturates its unitarity limit $\sigma^{(L=0)} \le 8 \pi/k^2$.
The resonant limit can often be approached by tuning a single parameter. 
This parameter could be the depth of the interatomic potential,
or an overall rescaling of the potential: $V(r) \to \lambda V(r)$.
As illustrated in Fig.~\ref{fig:potrange}(b), the parameter must be tuned to a
critical value for which there is a 2-body bound state exactly 
at the 2-body threshold.
In the case of a Feshbach resonance, the resonant limit can be approached 
by tuning the magnetic field. 
Since $a = \pm \infty$ in the {\it resonant limit}, 
one might expect that the natural low-energy length scale $\ell$ 
is the only important length scale at low energies.
This is true in the 2-body sector.
However, the Efimov effect reveals that there is another length scale
in the 3-body sector.
In the resonant limit, there are infinitely many, arbitrarily-shallow
3-body bound states with a spectrum of the form in Eq.~(\ref{B3-resonant}).
The parameter $\kappa_*$ defined by Eq.~(\ref{B3-resonant})  
can be interpreted as the approximate binding wave number
of the Efimov state labelled by the integer $n_*$.  

The {\it scaling limit} is also sometimes called the {\it zero-range limit}.
We prefer ``scaling limit'' because it emphasizes the renormalization 
aspects of the problem.  This terminology seems to have been first used 
in Ref.~\cite{FTDA99}.
The scaling limit may at first seem a little contrived,
but it has proved to be a powerful concept.
It can be defined by specifying the phase shifts for 2-body scattering.
In the scaling limit, the S-wave phase shift $\delta_0(k)$ has
the simple form 
\begin{eqnarray}
k \cot \delta_0(k) = -1/a \,,
\label{kcot-scaling}
\end{eqnarray}
and the phase shifts $\delta_L(k)$ for all higher partial waves vanish.
To approach the scaling limit typically requires tuning multiple
parameters in the interatomic potential. 
For example, as illustrated in Fig.~\ref{fig:potrange}(c), 
it can be reached by simultaneously tuning the range of the potential
to zero and its depth to $\infty$ in such a way that the binding energy of 
the shallowest 2-body bound state remains fixed.  
In the scaling limit, the scattering length $a$ sets the scale 
for most low-energy observables.
It is the only length scale in the 2-body sector.
However, as we shall see, in the 3-body sector, observables can also have 
logarithmic dependence on a second scale.  In the scaling limit,
there are infinitely many arbitrarily-deep
3-body bound states with a spectrum of the form \cite{MiF62,MiF62b}
\begin{eqnarray}
E_T^{(n)} & \longrightarrow &
\left( e^{-2 \pi/s_0} \right)^{n-n_*} {\hbar^2 \kappa_*^2 \over m}\,,
\qquad {\rm as \ } n \to - \infty {\rm \ \ with \ } \ell = 0 \,.
\label{B3-scaling}
\end{eqnarray}
Thus the spectrum is characterized by
a parameter $\kappa_*$ with dimensions of wave number.

The scaling limit may appear to be pathological, because
the spectrum of 3-body bound states in Eq.~(\ref{B3-scaling})
is unbounded from below.
However, the deep 3-body bound states have a negligible effect 
on the low-energy physics of interest.
The pathologies of the scaling limit
can be avoided simply by keeping in mind 
that the original physical problem
before taking the scaling limit had a natural low-energy length scale $\ell$.
Associated with this length scale is an energy scale $\hbar^2 /m \ell ^2$
that we will refer to as the {\it natural ultraviolet cutoff}.
Any predictions involving energies comparable to or larger than 
the natural ultraviolet cutoff are artifacts of the scaling limit.
Thus when we use the scaling limit to describe a physical system, 
any predictions involving energies $|E| \gsim \hbar^2 /m \ell ^2$ 
should be ignored.  

In spite of its pathologies, we shall take the scaling limit 
as a starting point for describing atoms with large scattering length.
We will treat the deviations from the scaling limit as perturbations.
Our motivation is that when the scattering length is large,
there are intricate 
correlations between 3-body observables associated with the Efimov effect 
that can be easily lost by numerical approximations.
By taking the scaling limit, we can build in these intricate correlations
exactly at high energy.  Although these correlations are unphysical 
at high energy, this does not prevent us from describing
low-energy physics accurately.  It does, however, guarantee that the 
intricate 3-body correlations are recovered automatically
in the resonant limit $a \to \pm \infty$.

In the 2-body sector, the scaling limit is associated with a 
{\it continuous scaling symmetry} that
consists of rescaling the scattering length $a$, 
the coordinate ${\bm r}$, and the time $t$ by appropriate powers 
of a positive number $\lambda$:
\begin{eqnarray}
a & \longrightarrow & \lambda a \,,
\qquad
{\bm r} \longrightarrow  \lambda {\bm r} \,,
\qquad
t  \longrightarrow  \lambda^2 t \,.
\label{csi}
\end{eqnarray}
The scaling of the time by the square of the scaling factor 
for lengths is natural in a nonrelativistic system.
Under the continuous scaling symmetry, 2-body observables,
such as binding energies and cross sections, 
scale with the powers of $\lambda$ implied by dimensional analysis.
This continuous scaling symmetry is a trivial consequence of the fact 
that $a$ is the only length scale that remains nonzero in the scaling limit.
For real atoms, the scaling limit can only be an approximation.
There are {\it scaling violations} that break the scaling symmetry.
In the 2-body sector, the most important scaling violations 
come from the S-wave effective range $r_s$, 
which give corrections to the scaling limit that scale as $r_s/|a|$.
All other scaling violations in the 2-body sector 
give corrections that scale as higher powers of $\ell/|a|$.
Scaling violations that give corrections that scale as powers of $\ell/|a|$
can be treated as perturbations to the scaling limit.

We will see that in the 3-body sector, 
there are {\it logarithmic scaling violations} that give corrections 
that scale as $\ln(|a|/\ell)$.
Logarithmic scaling violations do not become less important 
as one approaches the scaling limit,
and therefore cannot be treated as perturbations.
The origin of the logarithmic scaling violations is that the 
3-body problem in the scaling limit is singular at short distances.
In the 3-body sector, the scaling limit is characterized not only by the
scattering length $a$, but also by a second parameter 
that sets the scale for the logarithmic corrections. 
A convenient choice for this parameter is the wave number $\kappa_*$
defined by the spectrum of Efimov states in the resonant limit, 
which is given in Eq.~(\ref{B3-resonant}).
Low-energy observables are log-periodic functions of $\kappa_*$.
Simple observables that do not depend on any kinematic variables 
must have the form $a$ raised to a power determined by dimensional analysis
multiplied by a periodic function of $\ln(|a|\kappa_*)$.

The 3-body sector in the scaling limit has a trivial continuous 
scaling symmetry defined by Eqs.~(\ref{csi}) together with 
$\kappa_* \to \lambda^{-1} \kappa_*$. However, 
because of the log-periodic form of the logarithmic scaling violations,
it also has a nontrivial {\it discrete scaling symmetry}.
There is a discrete subgroup of the continuous scaling symmetry
that remains an exact symmetry in the scaling limit.  
The discrete scaling symmetry is 
\begin{eqnarray}
\kappa_* & \longrightarrow & \kappa_* \,,
\qquad
a \longrightarrow \lambda_0^n a \,,
\qquad
{\bm r} \longrightarrow \lambda_0^n {\bm r} \,,
\qquad
t \longrightarrow \lambda_0^{2n} t \,,
\label{dsi}
\end{eqnarray}
where $n$ is an integer, $\lambda_0 = e^{\pi/s_0}$, 
and $s_0 = 1.00624$ is the solution to the transcendental equation 
in Eq.~(\ref{s0}).
Under this symmetry, 3-body observables
scale with the powers of $\lambda_0^n$ implied by dimensional analysis.
This discrete scaling symmetry will be discussed in 
Section~\ref{sec:discrete}. 
Note that the discrete scaling symmetry transformation leaves 
the 3-body parameter $\kappa_*$ fixed.
By combining the trivial continuous scaling symmetry
with the discrete scaling symmetry given by Eqs.~(\ref{dsi}), we can see that
$\kappa_*$ is only defined modulo multiplicative factors 
of $\lambda_0$.

When we approximate a physical system by the idealized scaling limit,
it is important to have an estimate for the size of the corrections.
The logarithmic scaling violations must be taken into account without
approximation in the scaling limit.  The next most important 
scaling violations come from the effective range $r_s$,
which we expect to have a natural value of order $\ell$.
The leading correction is linear in $r_s$.
For an observable that does not involve any kinematic variables,
the only other scales are the scattering length $a$ and $\kappa_*$.
Since the dependence on $\kappa_*$ can only be logarithmic,
our estimate of the fractional corrections 
to the scaling limit is $r_s/|a|$.
For an observable involving energy $E$, the energy provides an 
alternative scale, so there can also be corrections 
that scale as $(m|E|/\hbar^2)^{1/2} r_s$.
For observables involving energy $|E| \sim \hbar^2/m a^2$, 
this reduces to our previous estimate $r_s/|a|$.
For observables involving energies that are comparable to the
natural ultraviolet cutoff $\hbar^2/m \ell^2$, 
our estimate of the fractional correction is 100\%.


\subsection{Universality in critical phenomena}

{\it Universality} refers to the fact that physical systems 
that are completely different at short distances can 
in certain limits exhibit identical behavior at long distances.
The classic examples of universality are condensed matter systems near
the critical point where a line of first order phase transitions 
end \cite{Fisher-98}.  
Low-energy atoms with large scattering length $a$ also exhibit
universal behavior that is insensitive to the details of their interactions
at separations small compared to $a$.  
Their universal behavior is more complex than the more familiar
examples provided by critical phenomena in condensed matter physics.
Before considering atoms with large scattering lengths, 
we first recall some simple examples 
of universality in critical phenomena.

\begin{figure}[htb]
\bigskip
\centerline{\includegraphics*[width=6cm,angle=0]{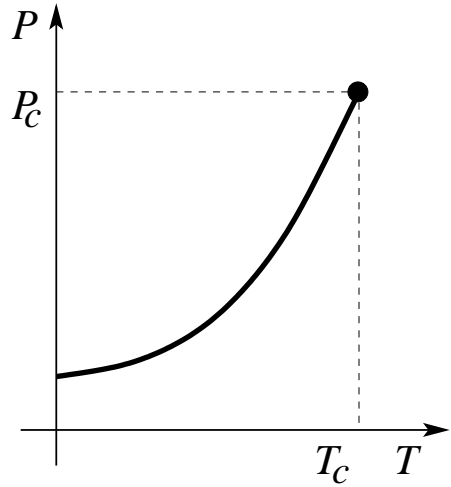}}
\medskip
\caption{The pressure $P$ versus the temperature $T$ for a liquid-gas system
with a critical point at ($T_c,P_c$).}
\label{fig:liquidgas}
\end{figure}

One of the most familiar systems in which critical phenomena can occur
is a substance with liquid and gas phases.  The thermodynamic state of 
the system is determined by the temperature $T$ and the pressure $P$.
As illustrated in Fig.~\ref{fig:liquidgas}, 
the liquid and gas phases are separated by a line 
in the $T$--$P$ plane.  As this line is crossed, the substance
undergoes a {\it phase transition}: 
a discontinuous change from a liquid with density 
$\rho_{\rm liq}(T)$ to a gas with density $\rho_{\rm gas}(T)$.
The phase transition line typically ends at some point $(T_c, P_c)$,
and this endpoint is called the {\it critical point}.
At higher temperatures and pressures, the transition between liquid and gas 
is smooth.  Just beyond the critical point, there is a rapid cross-over
between liquid and gas.  At much higher temperatures,
the transition is gradual.

Near the critical point $(T_c, P_c)$, the liquid--gas system exhibits 
{\it universal behavior}.  An example is the behavior of the coexistence curves
$\rho_{\rm liq}(T)$ and $\rho_{\rm gas}(T)$ near the critical point.
At $T=T_c$, they are both equal to the critical density $\rho_c$.
As $T \to T_c$ from below, 
the deviations of $\rho_{\rm liq}(T)$ and $\rho_{\rm gas}(T)$ 
from $\rho_c$ have a power-law behavior:
\begin{subequations}
\begin{eqnarray}
\rho_{\rm liq}(T) - \rho_c &\longrightarrow & + A (T_c - T)^\beta \,,
\\
\rho_{\rm gas}(T) - \rho_c &\longrightarrow & - A (T_c - T)^\beta \,.
\end{eqnarray}
\label{drho+-}
\end{subequations}
The coefficient $A$ varies widely from substance to substance,
but the {\it critical exponent} $\beta$ is a universal number.
It has the same value $\beta = 0.325$ for all liquid--gas systems.
Thus substances that are completely different on the atomic level
have the same universal behavior near their critical points.

\begin{figure}[htb]
\bigskip
\centerline{\includegraphics*[width=7cm,angle=0]{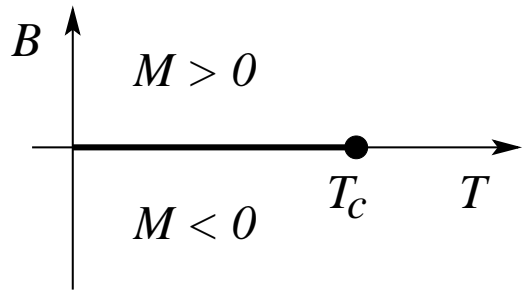}}
\medskip
\caption{The magnetic field $B$ versus the temperature $T$ for a 
ferromagnetic material with one easy axis of magnetization and
a critical point at ($T_c,B=0$).}
\label{fig:ferromagnet}
\end{figure}

Another familiar example of a system in which critical phenomena can occur 
is a ferromagnetic material with one easy axis of magnetization.
The thermodynamic state of the system is determined
by its temperature $T$ and the magnetic field $B$.
When $B=0$, the magnetization $M$ can be positive or negative.  
As illustrated in Fig.~\ref{fig:ferromagnet}, 
the $M<0$ and $M>0$ regions are separated by a phase transition line 
on the $B=0$ axis.  As this line is crossed, the magnetization changes
discontinuously from $-M_0(T)$ to $+M_0(T)$.  The phase transition line 
ends at a critical temperature $T_c$.  Above that temperature,
the transition from negative to positive magnetization is smooth.
Near the critical point $(T=T_c, B=0)$, the system exhibits universal behavior.
An example is the behavior of the discontinuity $M_0(T)$ near the critical 
point. As $T \to T_c$ from below, $M_0$ has a power-law behavior:
\begin{eqnarray}
M_0(T) &\longrightarrow & A' (T_c - T)^\beta \,.
\label{M0}
\end{eqnarray}
The coefficient $A'$ varies widely from substance to substance,
but the critical exponent $\beta$ is a universal number.
Remarkably, it has the same value $\beta = 0.325$ as for liquid--gas systems.
Thus even systems that bear as little resemblance to each other 
as a liquid--gas system and a ferromagnetic
can exhibit the same universal behavior near a critical point,
provided we make an appropriate mapping between the thermodynamic variables.

An important feature of critical phenomena is that these systems
exhibit {\it scaling behavior} in the {\it critical region} \cite{Fisher-98}.
There is some correlation length $\xi$ that diverges at the critical point.
In the liquid-gas system, $\xi$ is the correlation length for density
fluctuations.  In the ferromagnet, it is the correlation length for spin
fluctuations.  The critical region is defined by $\xi \gg \ell$, 
where $\ell$ is the natural length scale for these correlations,
which is the typical separation of the atoms. 
The rate at which the correlation length diverges 
as the critical point is approached is characterized by a critical exponent.
For example, its dependence on the temperature has the form
\begin{eqnarray}
\xi(T) &\longrightarrow & C |T-T_c|^{-\nu} \,,
\label{xi-scaling}
\end{eqnarray}
where $\nu$ is a critical exponent. In the liquid-gas system or in the 
ferromagnet, this critical exponent is $\nu= 0.63$. 
In the critical region, the correlation length $\xi$ is the only important 
length scale at long distances.
The deviations of other observables from  their critical values 
scale like powers of $\xi$.  These powers typically differ from the
values suggested by dimensional analysis, so they are called 
{\it anomalous dimensions}.  For example, the deviations in
Eqs.~(\ref{drho+-}) and (\ref{M0}) have anomalous dimensions $\beta/\nu$.

The modern understanding of universality in critical phenomena 
is based on the {\it renormalization group}.
The basic ideas of renormalization group theory are described 
in Ken Wilson's Nobel lectures \cite{Wilson-83}.
An excellent overview of this subject has been given 
more recently by Michael Fisher \cite{Fisher-98}.
The crucial concept is that of a 
{\it renormalization group (RG) transformation} 
on a Hamiltonian ${\mathcal H}$ that eliminates short-distance degrees 
of freedom while keeping long-distance observables invariant.
If all degrees of freedom with length scale shorter than $1/\Lambda$ 
have been eliminated, we refer to $\Lambda$ as the ultraviolet cutoff.
It is convenient to expand the Hamiltonian
in terms of a suitable basis of operators ${\mathcal O}_n$:
\begin{eqnarray}
{\mathcal H} = \sum_n g_n {\mathcal O}_n \,. 
\end{eqnarray}
This allows the Hamiltonian to be represented by a point 
${\bm g} = (g_1, g_2, \ldots)$ in the infinite-dimensional space 
of coupling constants. 
If the ultraviolet cutoff $\Lambda$ can be treated as a continuous variable,
the RG transformation defines a flow in the space of 
coupling constants that can be expressed as a differential equation:
\begin{eqnarray}
\Lambda {d \ \over d \Lambda} {\bm g} = {\bm \beta}({\bm g}) \,,
\label{RG-flow}
\end{eqnarray}
The {\it beta function} ${\bm \beta}({\bm g})$ is in general 
a complicated nonlinear function of the coupling constants.
If we start from a generic Hamiltonian ${\mathcal H}_0$
with coupling constants ${\bm g}_0$, the RG flow will carry 
it along a path called an {\it RG trajectory} that leads 
in the {\it infrared limit} $\Lambda \to 0$ to a
very complicated Hamiltonian that describes the long-distance physics 
in terms of the long-distance degrees of freedom only.

Wilson pointed out that the scale-invariant behavior at long distances 
that is characteristic of critical phenomena can arise from RG flow 
to a {\it fixed point} in the infrared limit $\Lambda \to 0$.
A fixed-point Hamiltonian ${\mathcal H}_*$ is one that does not change 
with $\Lambda$. Its coupling constant ${\bm g}(\Lambda)$ satisfies
\begin{eqnarray}
{\bm g}(\Lambda) = {\bm g}_* \,,
\end{eqnarray}
where ${\bm g}_*$ is a solution to ${\bm \beta}({\bm g}_*)=0$.
Since the Hamiltonian ${\mathcal H}_*$ remains invariant as one varies the 
ultraviolet cutoff $\Lambda$ that specifies the length scale of
the short-distance degrees of freedom that have been eliminated,
the Hamiltonian ${\mathcal H}_*$ must describe a scale-invariant system.
Associated with a fixed point ${\bm g}_*$, 
there is typically a {\it critical subspace} of points that flow 
asymptotically to ${\bm g}_*$ as $\Lambda \to 0$.
If the initial Hamiltonian ${\mathcal H}_0$ is carefully tuned to
a point in this subspace, it will flow 
along a {\it critical trajectory} that asymptotically approaches 
the fixed point ${\mathcal H}_*$ in the infrared limit $\Lambda \to 0$.
The tuning of macroscopic variables to a critical point corresponds 
at the microscopic level to the tuning of the coupling constants 
to the critical trajectory. 
The universality of critical phenomena
can be explained by the fact that critical trajectories can flow to the same
infrared fixed point from widely separated regions of the space of 
Hamiltonians.  Thus systems that are completely different at 
the microscopic level can have the same long-distance behavior.

Since a Hamiltonian ${\mathcal H}$ on the critical trajectory
flows to the fixed point ${\mathcal H}_*$ in the infrared limit 
$\Lambda \to 0$,  that Hamiltonian must be characterized by asymptotic 
scale invariance in the infrared limit.
Critical phenomena are associated with fixed points of the RG flow.
In the neighborhood of a fixed point, the RG flow given by Eq.~(\ref{RG-flow}) 
can be linearized:
\begin{eqnarray}
\Lambda {d \ \over d \Lambda} {\bm g} \approx B ({\bm g}-{\bm g}_*) \,,
\label{RG-lin}
\end{eqnarray}
where $B$ is a linear operator.
The eigenvalues of the operator $B$ are called the 
{\it critical exponents} of the operators associated 
with the corresponding eigenvectors of $B$.  Operators with
positive, zero, and negative critical exponents are called
{\it relevant}, {\it marginal}, and {\it irrelevant}, respectively.
As the ultraviolet cutoff $\Lambda$ increases, the coupling constants  
of relevant operators increase and the system 
flows away from the fixed point.  Thus, in order for the system 
to flow to the fixed point in the ultraviolet limit, 
the coupling constants of the relevant operators 
must be tuned to their critical values. 
The critical exponent $\beta$ in Eqs.~(\ref{drho+-})
and (\ref{M0}) and the critical exponent $\nu$ in 
Eq.~(\ref{xi-scaling}) are related in a simple way to the 
critical exponents of appropriate operators in the Hamiltonian.

RG fixed points are central to the modern understanding of critical 
phenomena and they play an important role in many other problems 
in condensed matter physics.
They also play a central role in the Standard Model 
of elementary particle physics,
and in many of the proposed extensions of the Standard Model, 
such as grand unified theories.
The many applications in condensed matter physics, high energy physics,
and nuclear physics 
have provided a strong driving force for the development of 
the renormalization group theory associated with RG fixed points.


\subsection{Renormalization group limit cycles}
\label{sec:RGlimcyc}

The RG flow defined by Eq.~(\ref{RG-flow}) can in general have a very 
complicated topology. A fixed point is only the
simplest possible topological feature.  
A more complicated possibility is a {\it limit cycle},
in which the RG trajectory flows forever around a closed loop.
A limit cycle is a family ${\mathcal H}_*(\theta)$ of Hamiltonians  
that is closed under the RG flow and can be parameterized 
by an angle $\theta$ that runs from 0 to $2\pi$.
The Hamiltonian makes a complete circuit around the limit cycle
every time the ultraviolet cutoff $\Lambda$ 
changes by some multiplicative factor $\lambda_0$.
The parameter $\theta$ can be chosen so that the RG flow 
on the limit cycle ${\bm g}_*(\theta)$ 
is just a linear increase in $\theta$ with $\ln \Lambda$.  
Thus if the coupling constant at some initial cutoff $\Lambda_0$ is 
${\bm g}(\Lambda_0) = {\bm g}_*(\theta)$, its value for a general cutoff is
\begin{eqnarray}
{\bm g}(\Lambda) =  
{\bm g}_* \big(\theta + 2 \pi \ln(\Lambda/\Lambda_0)/\ln(\lambda_0) \big) \,.
\label{RG-lc}
\end{eqnarray}
Note that the coupling constant is invariant under a discrete scale 
transformation of the ultraviolet cutoff $\Lambda$ with
discrete scaling factor $\lambda_0$.

The possibility of RG limit cycles was first discussed by Wilson
in a pioneering paper that proposed applying the renormalization group 
to the strong interactions of elementary particle physics 
\cite{Wilson:1970ag}.  
The fundamental theory of the strong interactions was not yet known 
at that time, but Wilson suggested that it ought to be a relativistic
quantum field theory whose coupling constants are governed by the 
renormalization group.
Experiments on deeply-inelastic lepton-nucleon scattering had revealed 
evidence for scaling behavior in the strong interactions at high energy.
Wilson suggested that such simple high-energy behavior could be 
explained by simple behavior of the RG flow in the
{\it ultraviolet limit} $\Lambda \to \infty$.
The simplest possibility is RG flow to an ultraviolet fixed point.
The next simplest possibility is RG flow to an {\it ultraviolet limit cycle}. 
The fundamental field theory of the strong interactions
called Quantum Chromodynamics (QCD) was subsequently developed.
QCD has a single coupling constant $\alpha_s(\Lambda)$ with an 
{\it asymptotically free} ultraviolet fixed point:
$\alpha_s(\Lambda) \to 0$ as $\Lambda \to \infty$
\cite{Gross-Wilczek,Politzer}. 
Thus ultraviolet limit cycles are not relevant to QCD.\footnote{
There is, however, a possibility that QCD in the few-nucleon sector
has an infrared limit cycle at critical values of the quark masses 
that are not far from their physical values \cite{BH03}.}

As pointed out by Wilson, one of the signatures of an RG limit cycle 
is {\it discrete scale invariance}.
If the ultraviolet cutoff is decreased by the factor $\lambda_0$,
the Hamiltonian ${\mathcal H}_*(\theta_0)$ flows around the limit cycle 
and returns to ${\mathcal H}_*(\theta_0)$.  This implies that if 
degrees of freedom with length scales less than $1/\Lambda$
have been eliminated and if one further
eliminates all degrees of freedom with length scales between $1/\Lambda$ 
and $\lambda_0/\Lambda$, the dynamics of the system is unchanged.  
This is possible only if the system has a discrete scaling symmetry under
${\bm r} \to \lambda_0^n {\bm r}$ for all $n$.
The discrete scale invariance implies that long-distance observables 
can have a periodic dependence on the logarithm of the variable that 
characterizes the critical region \cite{Wilson:1970ag}.  
An example of such a periodic dependence on the logarithm of $a$ 
is provided by Efimov's expression
for the atom-dimer scattering length in Eq.~(\ref{a12-Efimov}).  
Discrete scale invariance may also arise in other contexts 
that are as varied as turbulence, sandpiles, earthquakes, 
and financial crashes \cite{Sor97}.

In contrast to RG fixed points, the development of 
the renormalization group theory associated with RG limit cycles
is still in its infancy.  This situation is partly due to the lack of 
compelling examples. However, new examples of RG limit cycles 
have recently begun to emerge.
A very simple example of a renormalization group limit cycle
occurs in the quantum mechanics of a  particle in a $1/r^2$ potential.
This example is described in Section~\ref{sec:EFT-qm}.
Glazek and Wilson have presented a discrete Hamiltonian system 
whose renormalization involves a limit cycle \cite{Glazek:2002hq,GW03b}.
The spectrum of this model has some features in common with 
the 3-body spectrum of identical bosons, so it will be discussed 
in detail later in this section.
LeClair, Roman, and Sierra have constructed a generalization of a reduced
Hamiltonian for Cooper pairs in a superconductor whose renormalization 
involves a limit cycle \cite{LeClair:2002ux}.  
The gap equation for this model has multiple solutions related by 
a discrete scaling symmetry.

LeClair, Roman, and Sierra have also discovered a  (1+1)-dimensional 
quantum field theory whose renormalization involves a limit cycle
\cite{LeClair:2003xj,LeClair:2003hj,LeClair:2004ps}.
The model is a perturbation of a conformal field theory with an $SU(2)$ 
current algebra by a current-current interaction.  
The current-current interaction has three independent coupling constants.
For generic values of the coupling constants, the S-matrix  is a log-periodic 
function of the energy.  There are infinitely many resonances, 
with masses and widths related by a discrete scaling symmetry.
The discrete scaling symmetry is the signature of an RG limit cycle.
The conformal field theory is an interacting quantum field theory 
that corresponds to a nontrivial RG fixed point.
At a generic point in the 3-dimensional space of coupling constants
for the current-current interaction, the RG flow is log-periodic.
As the ultraviolet cutoff is decreased by the discrete scaling factor,
the coupling constants flow out to infinity in one direction, 
flow back from infinity from another direction, and return to their 
original values.

\begin{figure}[htb]
\bigskip
\centerline{\includegraphics*[width=7cm,angle=0]{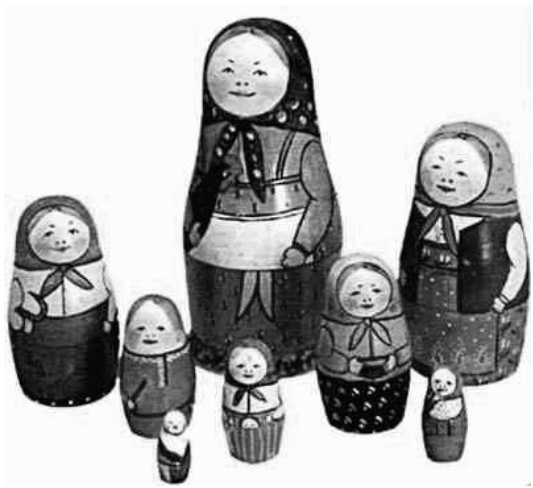}}
\medskip
\caption
{Nesting Russian dolls with a discrete scaling factor of about  1.3.}
\label{fig:russdoll}
\end{figure}

Leclair, Roman, and Sierra have used the colorful phrase 
{\it Russian doll renormalization group} to describe RG limit cycles
\cite{LeClair:2002ux,LeClair:2003xj}.
The name refers to a traditional souvenir from Russia consisting 
of a set of hollow wooden dolls that can be nested inside each other.
An example is shown in Fig.~\ref{fig:russdoll}.   
The scaling factors between each doll and the 
next smaller one are all approximately equal.

We now discuss the discrete Hamiltonian model of Glazek and Wilson
\cite{Glazek:2002hq,GW03b}  in some detail. 
The Hamiltonian $H_N$ for the $(N + 1)$-state model 
is an $(N + 1) \times (N + 1)$ matrix with entries
\begin{eqnarray}
\langle m | H_N | n \rangle 
&=& b^{(m+n)/2} \epsilon (I_{mn} - g_N S_{mn} - ih_N A_{mn}) \,,
\quad
m,n=0, 1,.., N \,,
\label{H-GW}
\end{eqnarray}
where $b > 1$ is dimensionless, $\epsilon > 0$ has units of energy, 
$I$ is the identity matrix, $S$ is the symmetric
matrix all of whose entries are $+1$, and $A$ is the antisymmetric
matrix all of whose entries above the diagonal are $+1$. 
We can interpret $\epsilon$ as an infrared cutoff 
and $\Lambda = b^N \epsilon$ as an ultraviolet cutoff.

\begin{figure}[htb]
\bigskip
\centerline{\includegraphics*[width=7.cm,angle=0]{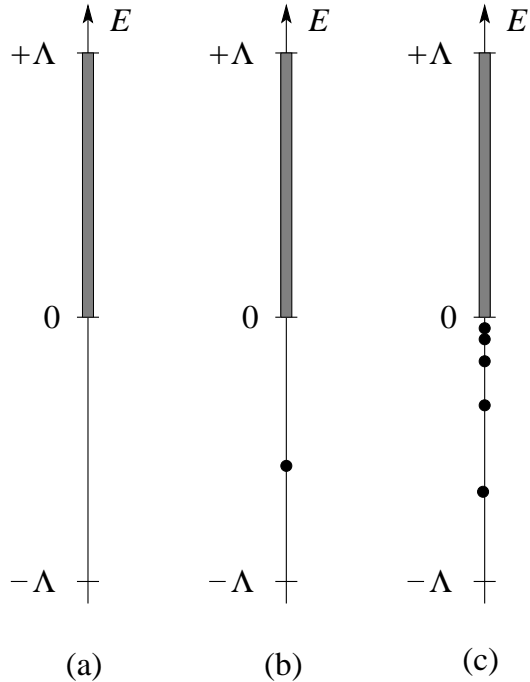}}
\medskip
\caption{Energy spectrum for the continuum limit of the Glazek-Wilson
 model in three cases: 
 (a) the noninteracting model with $g=h=0$,
 (b) the interacting model with $h=0$, and 
 (c) the interacting model with discrete scaling
     factor $\lambda_0 =2$. A shaded band represents a continuum of
     positive energy states, while a dot represents a discrete 
     negative-energy state. In case (c), there are infinitely many negative
     energy states with an accumulation point at $E=0$.}
\label{fig:wilsonmod}
\end{figure}

The Glazek-Wilson model has two coupling constants: $g_N$ and $h_N$.  
In the noninteracting case $g_N = h_N = 0$, 
the model has a geometric spectrum 
of energy eigenvalues:  $E_n = b^n \epsilon$, $n = 0, 1, \dots, N$.  
We can define a continuum limit by taking $N \to \infty$, 
$b \to 1$ and $\epsilon \to 0$ with $\Lambda = b^N \epsilon$ fixed.  
In this limit, the energy spectrum of the noninteracting model is a continuum
of positive-energy states with energies $0 < E < \Lambda$ as illustrated in
Fig.~\ref{fig:wilsonmod}(a).

In the interacting case in which $g_N$ and $h_N$ are nonzero, the energy
spectrum is more interesting \cite{Glazek:2002hq}.  In addition to positive
eigenvalues, the Hamiltonian $H_N$ has negative eigenvalues that are 
exponential in $1 /g_N$.  Thus the interaction terms are nonperturbative.
In the continuum limit, there is a continuum
of positive-energy states with energies $0 < E < \Lambda$.
If $h_N=0$, there is a single negative eigenvalue in the continuum limit
as illustrated in Fig.~\ref{fig:wilsonmod}(b).
If $h_N \ne 0$, there are multiple negative eigenvalues
as illustrated in Fig.~\ref{fig:wilsonmod}(c).

One way to study the model is using the renormalization group.  An RG
transformation that integrates out the state $| N \rangle$ is obtained by
constructing a Hamiltonian $H_{N-1}$ for the states $| n \rangle$, 
$n = 0, 1, \dots, N-1$, that reproduces the low-energy eigenvalues $E_n$
of $H_N$ that satisfy $|E_n | \ll \Lambda$.
Remarkably the Hamiltonian $H_{N-1}$ has the same form as $H_N$ 
in Eq.~(\ref{H-GW}), except that $I$, $S$, and $A$ are $N \times N$ matrices.  
The coupling constants $g_{N-1}$ and $h_{N-1}$ in $H_{N-1}$ are
\begin{eqnarray}
g_{N-1} & = & \frac{g_N + h_N^2}{1-g_N} \,,
\qquad
h_{N-1} = h_N \,.
\end{eqnarray}
Since $h=h_N$ remains fixed under the RG transformation, its
subscript is unnecessary.  
After $p$ iterations of the RG transformation,
the coupling constant $g_{N-p}$ is given by 
\begin{eqnarray}
\arctan(g_{N-p} / h) = \arctan (g_N / h) + p \arctan h \,.
\label{RGdisc}
\end{eqnarray}
This implies that if the cutoff is decreased by a factor of
$b^p$ such that $p \arctan h = n \pi$, where $n$ is an integer,
the coupling constant $g_{N-p}$ returns to its original value $g_N$.
This implies that the low-energy behavior of the system is 
characterized by as asymptotic discrete scaling symmetry 
with discrete scaling factor 
\begin{eqnarray}
\lambda_0 = b^{\pi/ \arctan (h)} \,. 
\label{lambda0-GW}
\end{eqnarray}

\begin{figure}[htb]
\bigskip
\centerline{\includegraphics*[width=8.5cm,angle=0]{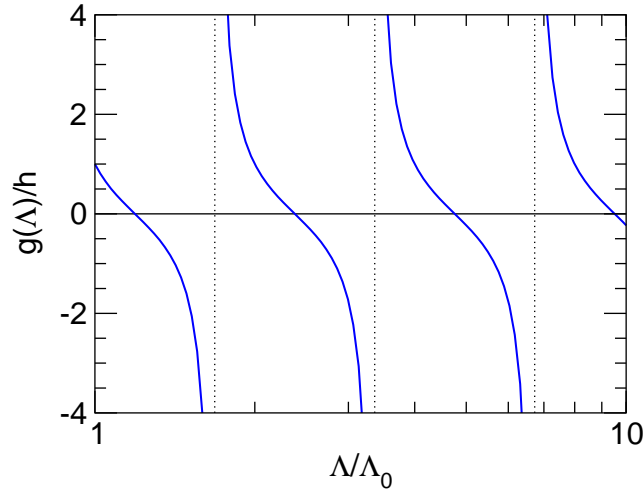}}
\medskip
\caption
{The behavior of $g(\Lambda)/h$ as a function of $\Lambda/\Lambda_0$
for the case $g (\Lambda_0) / h = 1$ and $\lambda_0 = 2$.}
\label{fig:g-GW}
\end{figure}

When the continuum limit is taken, it is convenient to take $h \to 0$ 
along with $b \to 1$ in such a way that $\lambda_0$ remains fixed.
In this case, the positive-energy spectrum approaches a continuum
while the negative-energy spectrum remains discrete 
with an accumulation point at $E = 0$ as illustrated
in Fig.~\ref{fig:wilsonmod}(c).
As the accumulation point is approached, the ratio of adjacent 
negative energy levels approaches $\lambda_0$.
The spectrum is characterized by an asymptotic discrete scaling symmetry 
with discrete scaling factor $\lambda_0$.
In the continuum limit, $g$ becomes a function 
of the ultraviolet cutoff $\Lambda$, and the discrete RG transformation
in Eq.~(\ref{RGdisc}) becomes a continuous RG flow:
\begin{eqnarray}
\arctan {g (\Lambda )\over h} = \arctan {g (\Lambda_0) \over h} 
- {\pi \over \ln \lambda_0} \ln {\Lambda \over \Lambda_0} \,.
\end{eqnarray}
The coupling constant $g (\Lambda)$ is a periodic function of $\ln (\Lambda)$. 
Thus the renormalization of $g (\Lambda)$ is governed by a limit cycle with a
discrete scaling factor $\lambda_0$.  The behavior of $g (\Lambda)$ is
illustrated in Fig.~\ref{fig:g-GW}
for the case $g (\Lambda_0) / h = 1$ and $\lambda_0 = 2$.
As $\Lambda$ decreases, $g (\Lambda)$ increases eventually to
$+ \infty$, jumps discontinuously to $- \infty$, and then continues
increasing.
The discontinuous behavior of the coupling constant is easy to understand.
When the ultraviolet cutoff is decreased to a value $\Lambda$, 
not only are positive energy states with energies 
$E > \Lambda$ removed from the spectrum, but negative energy eigenvalues 
with $E_n < - \Lambda$ are also removed.  The discontinuous change 
in $g (\Lambda)$ from $+ \infty$ to $- \infty$ occurs when $\Lambda$ 
is decreased through $| E_n |$, where $E_n$ is a negative eigenvalue.  
The discontinuous change in $g(\Lambda)$ is necessary to
compensate for the effects at low energy of the discrete state with negative 
energy $E_n$ that has been removed from the spectrum.


\subsection{Universality for large scattering length}

The universal low-energy behavior of atoms with large scattering 
length has many features in common with critical phenomena.  
The scattering length $a$ plays a role 
analogous to the correlation length $\xi$.  The region of large scattering
length $|a| \gg \ell$ is analogous to the critical region, 
and the resonant limit 
where $a$ diverges is analogous to the critical point.
In the critical region, $|a|$ is the most important length scale 
for low-energy observables.  It would be the only important length scale
if it were not for the logarithmic scaling violations 
associated with short-distance singularities in the 3-body sector
in the scaling limit.  These logarithmic scaling violations 
make the universal behavior of atoms 
with large scattering length richer and more complex 
in some ways than standard critical phenomena.

We now give a few specific examples of universality in atoms with
large scattering length.  In the 2-body sector, the universal predictions
are very simple. They are described in detail in Section~\ref{sec:uni2}.
The simplest example is that for
positive $a$, there is a shallow 2-body bound state 
that we call the {\it dimer} whose binding energy in the scaling limit is
given by Eq.~(\ref{B2-eq}).
In the 3-body sector, the universal predictions are more complex.
They are described in detail in Section~\ref{sec:uni3}.
The simplest example comes from the asymptotic behavior
of the spectrum of Efimov states in the resonant limit
given in Eq.~(\ref{R-resonant}).  This spectrum has an asymptotic 
discrete scaling symmetry with discrete scaling factor $e^{\pi/s_0}$.
Another example is the universal expression for the atom-dimer 
scattering length in Eq.~(\ref{a12-Efimov}).
The scaling behavior $a_{AD} \sim a$ expected from dimensional analysis
is violated by the logarithmic dependence of the coefficient on $a$.
Thus there are logarithmic scaling violations.
A remarkable feature of the coefficient is that it is a log-periodic 
function of $a$ that is invariant under the discrete scaling symmetry
in Eq.~(\ref{dsi}).  This discrete scaling symmetry is a signature 
of a RG limit cycle.

The connection between the Efimov effect and RG limit cycles 
was first realized in Ref.~\cite{Albe-81}.  
The connection was made manifest by Bedaque, Hammer, and van Kolck
in their formulation of the problem of identical bosons
in terms of effective field theory \cite{BHK99,BHK99b},
although they did not use the phrase ``limit cycle'' initially.
Their effective-field-theory formulation will be described in
Section~\ref{sec:EFT}. Some renormalization group aspects of the 
limit cycle in the 3-body problem with large scattering length
were discussed in 
Refs.~\cite{Barford:2004jm,Barford:2004fz,Griesshammer:2005ga,Mohr:2005pv}.

We will sketch briefly how one could 
approach the problem of low energy atoms with large scattering length
from a renormalization group perspective.
Our starting point is a Hamiltonian ${\mathcal H}_0$ 
that describes identical bosons
interacting through a short-range 2-body potential $V_0(r)$.  
We imagine constructing a renormalization group transformation  
that eliminates the interaction energy for configurations 
in which the atoms approach to within a short distance $1/\Lambda$ 
while leaving low-energy observables invariant.
The low-energy observables include the binding energies and scattering
amplitudes for states for which the total energy per atom in any 
isolated cluster of $N$ atoms satisfies $|E|/N \ll \hbar^2 \Lambda^2/m$, 
where $E=0$ is the energy of atoms at rest with infinite separations.
In the 2-body sector, the resulting interaction between atoms 
can be described by a potential $V(r; \Lambda)$ that agrees with 
$V_0(r)$ at long distances $r \gg 1/\Lambda$ 
but vanishes at short distances $r \ll 1 / \Lambda$.
At intermediate distances $r \sim 1/\Lambda$, the potential 
$V(r; \Lambda)$ must be adjusted so that it gives the same spectrum 
as $V_0(r)$ for 2-body bound states with binding energies 
$E_2 \ll \hbar^2 \Lambda^2 / m$ and the same scattering phase shifts 
$\delta_\ell(k)$ as $V_0(r)$ for wave numbers $k \ll \Lambda$.  
In the 3-body sector, the potential $V (r; \Lambda)$ will
compensate for the absence of interaction energy for configurations 
in which one pair of atoms has a small separation 
$r_{12} \ll 1 / \Lambda$ and the third atom is well-separated from the pair.  
However, it will not completely compensate for the absence 
of interaction energy for configurations in which all three atoms 
have small separations $r_{ij} \ll 1 / \Lambda$.  
To compensate for these configurations, it will be necessary to include 
a 3-body potential $V({\bm r}_1, {\bm r}_2, {\bm r}_3; \Lambda)$ 
that vanishes when any pair of atoms has a small separation 
$r_{ij} \ll 1 / \Lambda$.  In the region where all three atoms 
have intermediate separation $r_{ij} \sim 1 / \Lambda$, 
the potential must be adjusted so that 3-body observables 
with low energies satisfying $|E| \ll \hbar^2 \Lambda^2 / m$ 
agree with those for the original Hamiltonian 
with the 2-body potential $V_0 (r)$ only.  Similarly, 
to compensate for the absence of interaction energy for 
$N$-body configurations in which all atoms have separations 
$r_{ij} \ll 1 / \Lambda$, it is necessary to include an $N$-body potential 
$V ({\bm r}_1, {\bm r}_2, \dots , {\bm r}_N; \Lambda)$ 
that is adjusted to make low-energy $N$-body observables agree 
with those of the original Hamiltonian.

Thus the RG flow corresponding to decreasing $\Lambda$
carries the original Hamiltonian defined by the 2-body potential $V_0(r)$ 
into a larger space of Hamiltonians defined by $N$-body potentials
for all $N$.  We suggest that the RG flow must have a limit cycle
in this larger space of Hamiltonians. In the resonant limit,
the spectrum of Efimov states satisfies Eq.~(\ref{B3-resonant}).
It has an asymptotic scaling symmetry with scaling factor
$\lambda_0 = e ^{\pi/s_0}$.
It is therefore plausible that tuning to the resonant limit
corresponds to tuning the 2-body potential $V_0(r)$ 
to the critical trajectory for an infrared limit cycle 
in the space of Hamiltonians defined by short-range $N$-body 
potentials for all $N$.  As the momentum scale $\Lambda$ is decreased, 
3-body bound states with binding energies $E_T > \hbar^2 \Lambda^2/m$
are eliminated from the spectrum.  The spectrum of the remaining 
3-body bound states approaches closer and closer to the corresponding
state in a spectrum 
with an exact discrete scaling symmetry:
\begin{eqnarray}
E_T^{(n)} & = &
\left( e^{-2 \pi/s_0} \right)^{n-n_*} \hbar^2 \kappa_*^2/m\,,
\qquad {\rm for\ all\ } n {\rm \ \ with \ } a = \pm \infty, \ \ell = 0\,.
\label{B3-limcyc}
\end{eqnarray}
Such a spectrum is characteristic of an RG limit cycle. 

The resonant ($a \to \pm \infty$) and scaling ($\ell \to 0$) limits 
have simple interpretations in terms of the
renormalization group limit cycle.
In the subspace of Hamiltonians defined by 2-body potentials $V_0(r)$,
there is a subspace of potentials that lie on critical trajectories
that flow asymptotically to the limit cycle in the infrared limit.
The resonant limit corresponds to tuning the potential 
to one of these critical trajectories.
The RG flow will then carry the Hamiltonian asymptotically to the 
limit cycle as $\Lambda \to 0$.
One of the signatures of the infrared limit cycle is that there are
infinitely many arbitrarily-shallow 3-body bound states 
with an asymptotic
discrete scaling symmetry as in Eq.~(\ref{B3-resonant}).
The scaling limit corresponds to tuning the potential $V_0(r)$
to a critical trajectory that flows asymptotically to the limit cycle 
in the ultraviolet limit.  One of the signatures of the ultraviolet
limit cycle is that there are infinitely many, arbitrarily-deep 
3-body bound states with an asymptotic
discrete scaling symmetry as in Eq.~(\ref{B3-scaling}).
Taking the resonant limit and the scaling limit simultaneously
corresponds to tuning the Hamiltonian to the limit cycle itself.
A signature of the limit cycle is the exact
discrete scaling symmetry in Eq.~(\ref{B3-limcyc}).


%
%

\section{Universality for Two Identical Bosons}
        \label{sec:uni2}

In this section, we describe the universal aspects of the
2-body problem for identical bosons with large scattering length. 
We exhibit a trivial scaling symmetry that relates the 
2-body observables for different values of the scattering length.
We also discuss the leading scaling violations associated 
with the effective range.


\subsection{Atom-atom scattering}
\label{sec:uni2AA}

\begin{figure}[htb]
\bigskip
\centerline{\includegraphics*[width=8.5cm,angle=0]{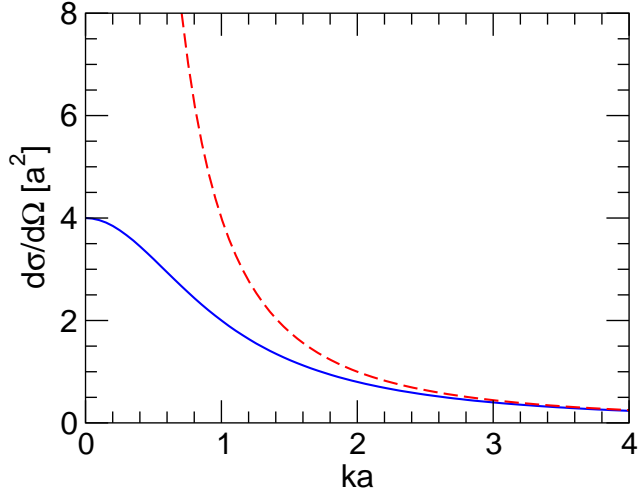}}
\medskip
\caption{ 
Differential cross section for atom-atom scattering  as a function of $k a$
(solid line).  The dashed line is the unitarity bound $4/k^2$.}
\label{fig:dsigAA}
\end{figure}

One of the universal 2-body observables is the cross section for
low-energy atom-atom scattering. 
By low energy, we mean energies $E=\hbar^2k^2/m$ 
much smaller than the natural ultraviolet cutoff $\hbar^2/m \ell^2$,
which means that the wave number satisfies $k \ll 1/ \ell$. 
The partial wave expansion in Eq.~(\ref{pwe})
expresses the scattering amplitude in terms of phase shifts
$\delta_L (k)$. The natural magnitude for the coefficients in the
low-energy expansion of $k\cot\delta_L(k)$ is $\ell$ raised to the
power required by dimensional analysis. 
In the scaling limit, all these coefficients vanish with the exception of the
scattering length.
The S-wave phase shift $\delta_0(k)$ is
given in Eq.~(\ref{kcot-scaling}).
The scattering amplitude in Eq.~(\ref{pwe}) reduces to
\begin{eqnarray}
f_k(\theta) = {1 \over - 1/a -ik} \,,
\label{f-2}
\end{eqnarray}
and the differential cross section in Eq.~(\ref{dcross}) is
\begin{eqnarray}
{d\sigma_{AA} \over d\Omega} = {4a^2 \over 1 + a^2 k^2} \,.
\label{sig-2}
\end{eqnarray}
The cross section is obtained by integrating 
over the solid angle $2 \pi$.
The differential cross section in Eq.~(\ref{sig-2}) is shown in 
Fig.~\ref{fig:dsigAA}.
For very low wave numbers $k \ll 1/|a|$,
it reduces to the constant $4a^2$.
For wave numbers $k \gg 1/|a|$, it has the
scale-invariant form $4/k^2$, which saturates the upper bound from
partial-wave unitarity in the $L=0$ channel.

The wave function for atom-atom scattering states at
separations $r \gg \ell$ is also universal.  The stationary wave function 
in the center-of-mass frame for two atoms in an $L=0$ state
with energy $E = \hbar^2k^2/m$ is
\begin{eqnarray}
\psi_{AA}({\bm r})= {1 \over kr} \sin \left[ kr + \delta_0(k) \right] \,.
\label{psi-AAsin}
\end{eqnarray}
For all other angular momentum quantum numbers $L$, 
the phase shifts vanish in the scaling limit.
The wave function associated with the scattering of two identical bosons 
with wave numbers $\pm {\bm k}$ is
\begin{eqnarray}
\psi_{AA}({\bm r})= \cos({\bm k} \cdot {\bm r}) 
- {1 \over 1/a + i k} {e^{i k r} \over r} \,.
\label{psi-AA}
\end{eqnarray}
Projecting onto $L=0$ by averaging over the angles of ${\bm r}$,
we recover the wave function in Eq.~(\ref{psi-AAsin}) up to a phase.


\subsection{The shallow dimer}
\label{sec:uni2D}

Another universal observable is the spectrum of 
shallow 2-body bound states.  By a {\it shallow bound state},
we mean one with binding energy $E_D$ 
much smaller than the natural ultraviolet cutoff $\hbar^2/m \ell^2$.
The spectrum of shallow 2-body bound states is very simple.
For $a<0$, there are no shallow bound states.  
For $a>0$, there is a single shallow bound state, 
which we will refer to as the {\it shallow dimer},
or simply as the {\it dimer} for brevity. 
The binding energy $E_D$ of the dimer can be deduced 
by inserting the expression for the phase shift in Eq.~(\ref{kcot-scaling})
into the bound-state equation (\ref{BE-eq}):
\begin{eqnarray}
E_D={\hbar^2 \over ma^2}  \,.
\label {B2-uni}
\end{eqnarray}

\begin{figure}[htb]
\bigskip
\centerline{\includegraphics*[width=8.5cm,angle=0]{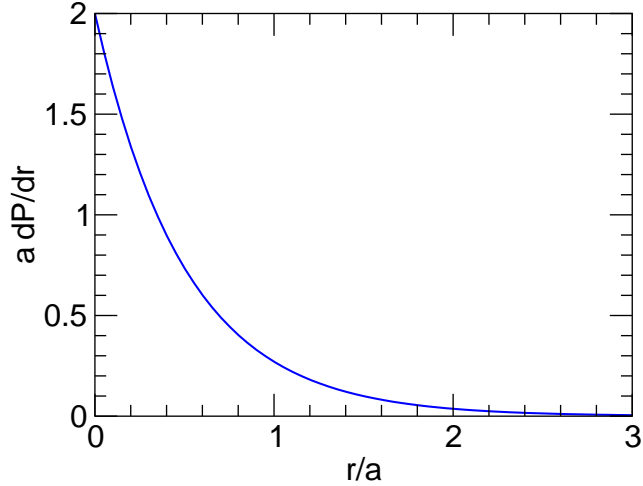}}
\medskip
\caption
{Normalized probability distribution $dP/dr$
for the separation $r$ of the atoms in the shallow dimer  
as a function of $r/a$.}
\label{fig:Pdimer}
\end{figure}

The wave function of the dimer at separations $r \gg \ell$ is also universal. 
The unnormalized coordinate-space wave function is
\begin{eqnarray}
\psi_D({\bm r})=\frac{1}{r} e^{-r/a} \,,
\label{psi-D}
\end{eqnarray}
where $r=|{\bm r}|$ is the separation of the two atoms. 
The normalized probability distribution 
for the separation $r$ of the atoms in the shallow dimer  
is shown in Fig.~\ref{fig:Pdimer}.
The size of the dimer is roughly $a$. 
A quantitative measure of the size is the mean-square separation 
of the atoms:
\begin{eqnarray}
\langle r^2 \rangle = a^2/2 \,.
\end{eqnarray}
The momentum-space wave function for the dimer is
\begin{eqnarray}
\psi_D({\bm k})= {4 \pi \over k^2 + 1/a^2} \,,
\end{eqnarray}
where ${\bm k}$ is the relative wave number of the two atoms. 
Thus the typical scale of the relative wave number is $1/a$.


\subsection{Continuous scaling symmetry}

The simple expressions for the cross-section in Eq.~(\ref{sig-2}) 
and the binding energy in Eq.~(\ref{B2-uni}) depend only on the scattering 
length. The fact that low-energy observables depend only on a single 
parameter $a$ with dimensions of length can be expressed 
formally in terms of a continuous scaling symmetry. 
Under this symmetry, the scattering length $a$ and kinematic variables
such as the energy $E$ are scaled by appropriate powers 
of a positive real number $\lambda$:
\begin{eqnarray}
a &\longrightarrow& \lambda a \,,
\qquad
E \longrightarrow \lambda ^{-2} E \,.
\label{scaling-1}
\end{eqnarray}
Under this symmetry, observables, such as the dimer binding energy 
$E_D$ or the atom-atom cross section $\sigma_{AA}$, 
scale with the powers of $\lambda$ suggested by dimensional analysis.

The scaling symmetry strongly constrains the dependence of the
observables on the scattering length and on kinematic variables.
As a simple example, consider the dimer binding energy, which scales as 
$E_D \to \lambda^{-2} E_D$.  The scaling symmetry constrains its dependence
on the scattering length:
\begin{eqnarray}
E_D(\lambda  a) =  \lambda ^{-2} E_D(a) \,.
\end{eqnarray}
This implies that $E_D$ is proportional to $1/a^2$, 
in agreement with the explicit formula in Eq.~(\ref{B2-uni}).
As another example, consider  the atom-atom cross section, 
which scales as $\sigma_{AA} \to \lambda^{2} \sigma_{AA}$.
The scaling symmetry constrains its dependence
on the scattering length and the energy:
\begin{eqnarray}
\sigma_{AA}(\lambda^{-2}E; \lambda  a) = \lambda^{2} \sigma_{AA}(E; a) \,.
\end{eqnarray}
The explicit expression for the differential cross section in
Fig.~(\ref{sig-2}) is consistent with this constraint.

The set of all possible low-energy 2-body states in the scaling limit 
can be represented as points $(a^{-1}, K)$ on the plane 
whose horizontal axis is $1/a$ 
and whose vertical axis is the wave number variable
\begin{eqnarray}
K = {\rm sign} (E) (m|E|/\hbar^2)^{1/2} \,.
\label{K-def}
\end{eqnarray}
It is convenient to also introduce polar coordinates consisting of
a radial variable $H$ and an angular variable $\xi$ defined by
\begin{eqnarray}
1/a &=& H \cos \xi \,,
\qquad
K   = H \sin \xi \,.
\label{Hxi-def}
\end{eqnarray}
We choose $\xi$ to be 0 on the positive $a$ axis and to have a
discontinuity with values $\pm \pi$ on the negative
$a^{-1}$ axis.
In terms of these polar coordinates,
the scaling symmetry given by Eqs.~(\ref{scaling-1}) is simply a rescaling 
of the radial variable: $H \to \lambda ^{-1} H$.

\begin{figure}[htb]
\bigskip
\centerline{\includegraphics*[width=8cm,angle=0]{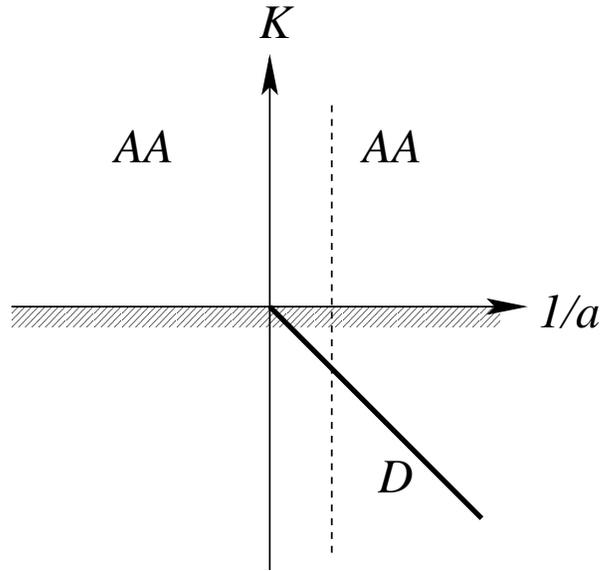}}
\medskip
\caption
{The $a^{-1}$--$K$ plane for the 2-body problem. The allowed region for
atom-atom scattering states are the two quadrants labelled $AA$. 
The heavy line labelled $D$ is the shallow dimer.
The cross-hatching indicates the 2-atom threshold.}
\label{fig:2body}
\end{figure}

The $a^{-1}$--$K$ plane for the 2-body system in the scaling limit
is shown in Fig.~\ref{fig:2body}.
The possible states are atom-atom scattering states and the shallow dimer. 
The quadrants in which there are atom-atom scattering states 
are labelled $AA$.
The threshold for atom-atom scattering states is indicated in 
Fig.~\ref{fig:2body} by the hatched area.
The shallow dimer lies along the ray $\xi = -{1\over 4}\pi$,
which is indicated by the heavy line labelled $D$.
A given physical system has a specific value of the scattering length, 
and so is represented by a vertical line.
Changing $a$ corresponds to sweeping the line horizontally across the page. 
The continuum of atom-atom scattering states is represented by the 
points on the vertical line that lie in the upper-half plane.
If $a>0$, there is also a discrete bound state (the shallow dimer)
lying on the intersection of the vertical line with the ray 
$\xi = -{1\over4}\pi$.
The resonant limit corresponds to tuning the vertical line to the $K$ axis.


\subsection{Scaling violations}
\label{sec:2bodyscalviol}

The scaling symmetry given by Eqs.~(\ref{scaling-1}) is reflected in 
the scaling behavior of the universal expressions for 2-body observables.  
The differential cross section in Eq.~(\ref{sig-2}) scales like $a^2$ with a 
coefficient that is a function of $ka$. The dimer binding energy 
in Eq.~(\ref{B2-uni}) scales like $a^{-2}$.  If the natural
low-energy length scale $\ell$ is nonzero, there are scaling violations that
give corrections to the universal expressions that decrease as powers of
$\ell / |a|$ when $a \to \pm \infty$.

The leading scaling violations decrease as a single power of $\ell / |a|$.
They come from the S-wave effective range $r_s$ defined by the 
effective-range expansion in Eq.~(\ref{kcot}). 
We can deduce the leading scaling violations to the differential cross 
section by truncating the effective-range expansion in Eq.~(\ref{kcot}) 
after the $k^2$ term:
\begin{eqnarray}
k \cot \delta_0 (k) = - 1/a + \mbox{$1 \over 2$} r_s k^2 \,.
\label{kcotdelta:rs}
\end{eqnarray}
This approximation to the 2-body problem 
is called the {\it effective-range theory} 
\cite{Schwinger47,Blatt48,Bethe49}. 
The expression for the differential cross 
section in Eq.~(\ref{dcross}) then becomes
\begin{eqnarray}
{d \sigma \over d \Omega} = 
{4 a^2 \over (1- {1 \over 2} r_s a k^2)^2 + a^2 k^2} \,.
\end{eqnarray}
If $k \ll 1/|r_s|$, this can be expanded in powers of $r_s$:
\begin{eqnarray}
{d \sigma \over d \Omega} &=& 
{4 a^2 \over 1 + a^2 k^2}
\left( 1 + {r_s \over a} \, {a^2 k^2 \over 1 + a^2 k^2} 
  + {r_s^2 \over a^2} \, {a^4 k^4(3-a^2k^2) \over 4 (1 + a^2 k^2)^2} 
        + \ldots \right) \,.
\label{dsig:exp}
\end{eqnarray}
The leading term is the universal expression in Eq.~(\ref{sig-2}).
For $k\sim 1/|a|$, the next-to-leading term is suppressed by $r_s / |a|$. 
For small wave numbers $k \ll 1/|a|$, 
there is an additional suppression factor of $(k a)^2$.  

We can study the leading scaling violations to the binding energy of the
shallow dimer by inserting the truncated effective-range expansion 
in Eq.~(\ref{kcotdelta:rs}) into  Eq.~(\ref{BE-eq}). 
The binding energy equation then reduces to a quadratic equation:
\begin{eqnarray}
- 1/a - \mbox{$1 \over 2$} r_s \, \kappa^2 + \kappa = 0 \,.
\label{kappa:rs}
\end{eqnarray}
A positive real-valued solution $\kappa$ to the binding energy equation
corresponds to a bound state with binding energy $E_2 = \hbar^2 \kappa^2 / m$. 
The two solutions to the quadratic equation (\ref{kappa:rs}) are 
\begin{eqnarray}
\kappa^{(\pm)} = \left ( 1 \pm \sqrt{1 - 2r_s /a} \right ) {1 \over r_s} \,.
\label{kappapm}
\end{eqnarray}
Assuming that the scattering length is large, the two solutions are both
real-valued.  The asymptotic solutions in the limit $|r_s| \ll |a|$ are 
\begin{subequations}
\begin{eqnarray}
\kappa^{(+)} & \longrightarrow & 2/r_s \,,
\\
\kappa^{(-)} & \longrightarrow & 1/a \,.
\end{eqnarray}
\end{subequations}
The solution $\kappa^{(+)}$ is positive and corresponds to a deep bound state 
if $r_s> 0$, while $\kappa^{(-)}$ is positive and corresponds to 
a shallow bound state if $a > 0$.

If $a > 0$, the binding energy for the shallow bound state is
obtained by inserting the solution $\kappa^{(-)}$ in Eq.~(\ref{kappapm})
into Eq.~(\ref{B2-kappa}):
\begin{eqnarray}
E_2^{(-)} = {\hbar^2 \over m r_s^2} \left ( 1 - \sqrt{1 - 2r_s / a} 
\right)^2  \,.
\label{B2minus}
\end{eqnarray}
In the limit $|r_s| \gg a$, this reduces to the universal expression
in Eq.~(\ref{B2-uni}).  The expansion of the binding energy in powers 
of the effective range is
\begin{eqnarray}
E_2^{(-)} \approx {\hbar^2 \over m a^2} \left ( 1 + {r_s \over a} + {5 r_s^2
\over 4 a^2} + \ldots \right ) \,.
\label{B2-exp}
\end{eqnarray}
Thus the leading scaling violation is linear in $r_s / a$.

One can study higher-order scaling violations by considering the effects of
higher-order terms in the low-momentum expansion of the scattering amplitudes. 
If the $k^4$ term in the effective-range expansion in Eq.~(\ref{kcot})
has a natural coefficient $P_s \sim \ell^3$, the corresponding 
correction to the
universal differential cross section in Eq.~(\ref{sig-2}) is 
suppressed by a factor
$k^4 a \ell^3$, which is of order $(\ell / a)^3$ if $k \sim 1/|a|$.  
For identical bosons, there is no P-wave
($L=1$) term in the partial wave expansion.  The leading contribution to the
differential cross section from higher partial waves $L \ge 2$ 
comes from interference with the $L=0$ term and is 
suppressed by a factor $(k\ell)^{2L} \ell/|a|$,
which is of order $(\ell/a)^{2L+1}$ for $k \sim 1/|a|$.
Thus the scaling violations of order $\ell / a$ and $\ell^2/ a^2$ 
in the differential cross section
are completely determined by the effective range $r_s$ 
and are given in Eq.~(\ref{dsig:exp}).
The correction to the dimer binding energy in Eq.~(\ref{B2-uni}) 
from a $k^4$ term in the effective-range expansion in Eq.~(\ref{kcot}) 
is suppressed
by $\ell^3 / a^3$.  
Thus the scaling violations of order $\ell / a$ and $\ell^2/ a^2$ 
in the dimer binding energy are completely determined 
by the effective range $r_s$ and are given in Eq.~(\ref{B2-exp}).

If $r_s > 0$, the solution $\kappa^{(+)}$ in Eq.~(\ref{kappapm}) to the binding
energy equation (\ref{kappa:rs}) corresponds to a deep (tightly-bound) 
diatomic molecule with binding energy
\begin{eqnarray}
E_2^{(+)} \approx 4 \hbar^2 /(m r_s^2) \,.
\label{B2plus}
\end{eqnarray}
Its binding energy is of order $\hbar^2 / m \ell^2$.  By considering higher
orders terms in the low-momentum expansion of $k \cot \delta_0(k)$, such as
the $k^4$ term in Eq.~(\ref{kcot}), one can easily show that there can be other
contributions to the binding energy of order $\hbar^2 / m \ell^2$.  Thus there
is nothing universal about the expression for the binding energy $E_2^{(+)}$,
or even the existence of that deep bound state.  It is simply an artifact of
the model defined by truncating the effective-range expansion after the $k^2$
term in Eq.~(\ref{kcot}).


\subsection{Theoretical approaches}

Historically, the idea of universality has its roots in nuclear 
physics. Early on, it was realized that the deuteron is large compared to
the range of the nuclear force. Similarly, the spin-singlet S-wave
scattering length of two nucleons was found to be large compared to the 
range. Starting in the 1930's, these observations lead to the development 
of various theoretical approaches to exploit this separation of length scales.
In the following, we will give a brief overview of these techniques.
For a more detailed discussion, see Ref.~\cite{vanKolck:1998bw}.

The {\it boundary condition method} was first used by Bethe and Peierls
\cite{Bethe35a,Bethe35b,Breit47}. For short-range interactions, 
the two-particle wave function is governed by the free Schr\"odinger equation, 
except for particle separations of the order of the range or less. 
The effect of the interactions at short distances can be 
taken into account through a boundary condition on the wave function 
as the separation vector ${\bm r}$ goes to zero:
\beq
\psi({\bm r}) \longrightarrow C 
\left( {1\over r} - {1\over a} \right) 
\qquad \mbox{as\ } r \to 0 \,,
\label{eq:boundcon}
\eeq
where $C$ is a constant.
Together with the free Schr\"odinger equation for large $r$,
Eq.~(\ref{eq:boundcon}) determines the long-distance wave function completely.
Note that the scattering wave function in Eq.~(\ref{psi-AAsin})
and the dimer wave function in Eq.~(\ref{psi-D}) satisfy the 
boundary condition in Eq.~(\ref{eq:boundcon}) with 
$C = -a/(1 + a^2 k^2)^{1/2}$ and $C=1$, respectively.

An alternative way to introduce a boundary condition on the wave function
is the {\it pseudopotential method} \cite{Breit47,Huang57}.
It involves replacing the potential $V(r)$ in the 
Schr\"odinger equation by a {\it pseudopotential} that
acts on the wave function $\psi({\bm r})$ for the separation 
vector of the two particles as
\beq
V(r) \psi({\bm r}) = \frac{4\pi \hbar^2 a}{m} 
\delta^3({\bm r}) \frac{\partial \ }{\partial r} \big( r \psi({\bm r}) 
\big) \,.
\label{eq:pseudo}
\eeq
The Schr\"odinger equation with this pseudopotential
is equivalent to the free Schr\"odinger equation
for ${\bm r} \ne 0$ supplemented by the boundary condition
\beq
\lim_{r \to 0} r^2 \frac{\partial \ }{\partial r} \psi({\bm r})
= a \, \lim_{r \to 0}  
\frac{\partial \ }{\partial r} \big( r \psi({\bm r}) \big) \,.
\eeq
This boundary condition is equivalent to that in 
Eq.~(\ref{eq:boundcon}).

A third method to exploit universality in the two-body problem is the 
{\it effective-range expansion} \cite{Schwinger47,Blatt48,Bethe49}, 
which was already discussed in Section~\ref{sec:2bodyscalviol}.
For short-range potentials, $k\cot\delta_0(k)$ is an analytic function of 
the energy $E$ and it can therefore be expanded as a power series in 
$k^2$ as in Eq.~(\ref{kcotdelta:rs}). 
Truncating this expansion after the energy-independent scattering length
term $-1/a$ is equivalent to the boundary condition and pseudopotential 
methods discussed above. The description of the phase shifts can be 
improved by including higher order terms in the effective-range expansion.

The {\it effective field theory} method, which is discussed in detail 
in Section~\ref{sec:EFT}, is another method that can be used to exploit 
universality \cite{Beane:2000fx,Bedaque:2002mn}.
All three methods discussed above can be related within an effective field 
theory for short-range forces \cite{vanKolck:1998bw}. In particular, the 
effective-range expansion is reproduced order by order in $k^2$
within an effective field theory with contact interactions. 
(See Section~\ref{sec:EFT}).
Effective field theory is particularly convenient for calculating
corrections to universality systematically and for calculating
the effects of electromagnetic and weak interactions
for charged particles.


%
%

\section{Hyperspherical Formalism}
        \label{sec:hyper}

In this section, we introduce hyperspherical coordinates,
develop the hyperspherical formalism for the low-energy 3-body problem, 
and use it to derive the Efimov effect. We will start out with
the general formalism, but later focus on the sector with 
zero total angular momentum.


\subsection{Hyperspherical coordinates}
        \label{subsec:hyper}

The universal aspects of the 3-body problem can be understood most easily by
formulating it in terms of {\it hyperspherical coordinates}.
A good introduction to hyperspherical coordinates
and a thorough review of the hyperspherical
formalism is given in a recent review article 
by Nielsen, Fedorov, Jensen, and Garrido \cite{NFJG01}.

\begin{figure}[htb]
\bigskip
\centerline{\includegraphics*[width=6cm,angle=0]{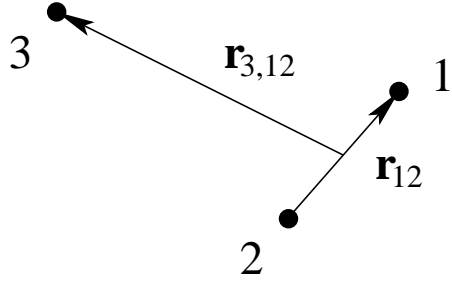}}
\medskip
\caption
{One of the three possible sets of Jacobi coordinates defined in 
Eqs.~(\ref{jacobi}).}
\label{fig:Jacobi}
\end{figure}

\begin{figure}[htb]
\bigskip
\centerline{\includegraphics*[width=8.5cm,angle=0]{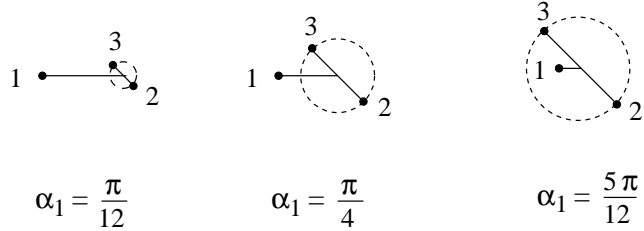}}
\medskip
\caption{
Three-body configurations with the same hyperradius $R$ 
but different hyperangles: $\alpha_1 = {\pi \over 12}$, ${\pi \over 4}$,
and ${5\pi \over 12}$.  
Any configuration with atoms 2 and 3 at the endpoints of a diameter of the 
dashed circle will have the same values of $R$ and $\alpha$ 
as the one shown.}
\label{fig:hypercoord}
\end{figure}

In order to define hyperspherical coordinates, 
we first introduce Jacobi coordinates.  A set of Jacobi coordinates 
consists of the separation vector ${\bm r}_{ij}$ between a pair of
atoms and the separation vector ${\bm r}_{k,ij}$ of the third atom from the
center-of-mass of the pair. For atoms of equal mass, the Jacobi 
coordinates are
\begin{eqnarray}
{\bm r}_{ij} & = & {\bm r}_i - {\bm r}_j \,, 
\qquad
{\bm r}_{k, ij}  =  {\bm r}_k - \mbox{$1\over2$}({\bm r}_i + {\bm r}_j) \,.
\label{jacobi}
\end{eqnarray}
In Fig.~\ref{fig:Jacobi}, we illustrate one of the three possible sets 
of Jacobi coordinates.  The {\it hyperradius} $R$ is the root-mean-square 
separation of the three atoms:
\begin{eqnarray}
R^2 &=& \mbox{$1 \over 3$} \left( r_{12}^2 + r_{23}^2 + r_{31}^2 \right) 
= \mbox{$1 \over 2$} r_{ij}^2 + \mbox{$2 \over 3$} r_{k,ij}^2 \,.
\end{eqnarray}
The hyperradius is small only if all three atoms are close together. 
It is large if any single atom is far from the other two. 
The {\it Delves hyperangle} \cite{Delves60} $\alpha_k$ is defined by
\begin{eqnarray}
\alpha_k=\arctan \left( {\sqrt{3} r_{ij} \over 2 r_{k,ij}} \right) \,,
\label{alphak}
\end{eqnarray}
where $(i,j,k)$ is a permutation of $(1,2,3)$. The range of the hyperangle
$\alpha_k$ is from 0 to ${1\over2}\pi$.  It is near 0 when atom $k$ 
is far from atoms $i$ and $j$, and it is near ${1\over2}\pi$ 
when atom $k$ is near the center of mass of atoms $i$ and $j$.
The magnitudes of the separation vectors can be 
expressed as
\begin{eqnarray}
r_{ij} & = & \sqrt{2} R \sin \alpha_k \,, 
\qquad
r_{k, ij}  =  \sqrt{3/2} \, R \cos \alpha_k \,.
\end{eqnarray}
Examples of 3-body configurations with the same hyperradius $R$ 
but different hyperangles are illustrated in Fig.~\ref{fig:hypercoord}.
The definition in Eq.~(\ref{alphak}) expresses the hyperangle $\alpha_k$
in terms of the Jacobi coordinates ${\bm r}_{ij}$ and 
${\bm r}_{k,ij}$ defined in Eqs.~(\ref{jacobi}).
The other two hyperangles can also be expressed as functions of those
Jacobi coordinates:
\begin{eqnarray}
\sin^2 \alpha_i  &=&  \mbox{$1\over 4$} \sin^2 \alpha_k 
+ \mbox{$3 \over 4$} \cos^2 \alpha_k 
+ \mbox{$1 \over 2$} \sqrt{3} \sin \alpha_k \cos \alpha_k \,
        \hat {\bm r}_{ij} \cdot \hat {\bm r}_{k,ij} \,,
\label{sin2alpha}
\end{eqnarray}
where $(i,j,k)$ is a permutation of $(1,2,3)$.
For fixed $\alpha_k$, the range of $\alpha_i$ is 
\begin{eqnarray}
\left| \mbox{$1 \over 3$} \pi - \alpha_k \right| < \alpha_i 
        <  \mbox{$1 \over 2$} \pi - \left| \mbox{$1 \over 6$} \pi
- \alpha_k \right| \,.
\end{eqnarray}
The consistency with Eq.~(\ref{sin2alpha}) can be verified by
applying the function $\sin^2$ to each term.
Using Eq.~(\ref{sin2alpha}), we can also derive the simple identity
\begin{eqnarray}
\sin^2 \alpha_1 + \sin^2 \alpha_2 +  \sin^2 \alpha_3 =  \mbox{$3\over 2$} \,.
\label{sin2alpha-sum}
\end{eqnarray}
The volume element for the Jacobi coordinates can be written
\begin{eqnarray}
d^3 r_{ij} d^3 r_{k,ij} =  {3 \sqrt{3} \over 4} R^5 dR
\sin^2(2 \alpha_k) d\alpha_k d \Omega_{ij} d \Omega_{k,ij} \,,
\end{eqnarray}
where $d \Omega_{ij}$ and $d \Omega_{k,ij}$ are the differential solid angles
for the unit vectors $\hat {\bm r}_{ij}$ and $\hat {\bm r}_{k,ij}$.

The Schr{\"o}dinger equation for the stationary wave function
$\Psi({\bm r}_1, {\bm r}_2, {\bm r}_3)$ of three atoms with mass $m$
interacting through a potential $V$ is
\begin{eqnarray}
\left(-{\hbar^2 \over 2m} \sum_{i=1}^3 \nabla_i^2 + V ({\bm r}_1, {\bm r}_2,
{\bm r}_3)\right) \Psi = E \Psi \,.
\label{Seq:1}
\end{eqnarray}
If the interaction potential $V$ is translation invariant,
it depends only on 6 independent coordinates. 
The wave function $\Psi$ in the center-of-mass frame also depends on 
only 6 independent coordinates.  A convenient choice consists of 
the hyperradius $R$, one of the hyperangles $\alpha_k$, and the
unit vectors $\hat {\bm r}_{ij}$ and $\hat {\bm r}_{k,ij}$.
We will refer to the 5 dimensionless variables 
($\alpha_k,\hat {\bm r}_{ij},\hat {\bm r}_{k,ij}$) as hyperangular
variables and denote them collectively by $\Omega$.
When expressed in terms of hyperspherical coordinates, the
Schr{\"o}dinger equation for the wave function in the center-of-mass 
frame reduces to
\begin{eqnarray}
\left ( T_R + T_{\alpha_k} + {\Lambda_{k,ij}^2 \over 2mR^2} + V (R, \Omega)
\right ) \Psi = E \Psi \,,
\label{Seq:3}
\end{eqnarray}
where $T_R$ is the hyperradial kinetic energy operator,
\begin{subequations}
\begin{eqnarray}
T_R & = & - {\hbar^2 \over 2m}
\left[ {\partial^2 \ \over \partial R^2}
     + {5 \over R} {\partial \ \over \partial R} \right]
\\& = & {\hbar^2 \over 2m} R^{-5/2}
\left [ - {\partial^2 \ \over \partial R^2} 
     + {15 \over 4 R^2} \right] R^{5/2} \,,
\end{eqnarray}
\end{subequations}
$T_{\alpha_k}$ is the kinetic energy operator associated with the hyperangle
$\alpha_k$,
\begin{subequations}
\begin{eqnarray}
T_\alpha & = & - {\hbar^2 \over 2m R^2}
\left [ {\partial^2 \ \over \partial \alpha^2}
    +4 \cot (2 \alpha) {\partial \ \over \partial \alpha} \right]
\\
       & = & {\hbar^2 \over 2mR^2} {1 \over \sin (2 \alpha)}
\left [ - {\partial^2 \ \over \partial \alpha^2} 
         - 4 \right] \sin (2 \alpha) \,,
\end{eqnarray}
\end{subequations}
and $\Lambda_{k,ij}^2$ is a generalized angular momentum operator:
\begin{eqnarray}
\Lambda_{k,ij}^2 = {{\bm L}_{ij}^2 \over \sin^2 \alpha_k} 
                + {{\bm L}_{k,ij}^2 \over \cos^2 \alpha_k} \,.
\label{gamo}
\end{eqnarray}
The operators ${\bm L}_{ij}$ and ${\bm L}_{k,ij}$ are the conventional
angular momentum operators associated with the vectors
${\bm r}_{ij}$ and ${\bm r}_{k,ij}$, respectively. 

A convenient way to solve the Schr{\"o}dinger equation 
in hyperspherical coordinates is to use the 
{\it adiabatic hyperspherical representation}. 
For each value of $R$, the wave function $\Psi(R, \Omega)$ 
is expanded in terms of a complete set of
hyperangular functions $\Phi_n (R,\Omega)$:
\begin{eqnarray}
\Psi(R, \Omega) = R^{-5/2} \sum_n f_n (R) \Phi_n(R, \Omega) \,.
\end{eqnarray}
The functions $\Phi_n (R, \Omega)$ are solutions to a differential eigenvalue
equation in the hyperangular variables:
\begin{eqnarray}
\left[ T_{\alpha_k} + \Lambda_{k,ij}^2/(2m R^2) + V(R, \Omega) \right]
\Phi_n (R, \Omega) 
= V_n (R) \Phi_n (R, \Omega) \,.
\label{Phi-eigen}
\end{eqnarray}
The hyperradius $R$ is treated as a parameter and the eigenvalue $V_n(R)$ is a
function of that parameter. It can be interpreted as an effective potential for
the channel associated with the hyperangular function $\Phi_n$.  The
orthonormality condition on the hyperangular functions can be written
\begin{eqnarray}
\int d \Omega \, \Phi_n (R, \Omega)^* \Phi_m (R, \Omega) = \delta_{nm} \,,
\label{ortho}
\end{eqnarray}
where the hyperangular integral is
\begin{eqnarray}
\int d \Omega = \int_0^{{1\over2}\pi} d \alpha_k \sin^2(2 \alpha_k) 
\int d \Omega_{ij} \int d \Omega_{k,ij}.
\label{Omega}
\end{eqnarray}
Upon using the orthonormality condition to project onto $\Phi_n$, the
Schr{\"o}dinger equation (\ref{Seq:3}) reduces to a coupled set of
eigenvalue equations for the hyperradial functions $F_n(R)$:
\begin{eqnarray}
\left[ {\hbar^2 \over 2 m} 
\left( - {\partial^2 \ \over \partial R^2} + {15\over 4 R^2} \right) 
        + V_n(R) \right] f_n(R) &&
\nonumber 
\\ 
+ \sum_m \left[ 2 U_{nm}(R) {\partial \ \over \partial R}  
   + W_{nm}(R) \right] f_m(R)
    &=& E f_n (R) \,, 
\label{radeq:S}
\end{eqnarray}
where $U_{nm}(R)$ and $W_{nm}(R)$ are coupling potentials defined by
\begin{subequations}
\begin{eqnarray}
U_{nm}(R) &=&  -{\hbar^2 \over 2 m} \int d \Omega \,
\Phi_n (R, \Omega)^* {\partial \ \over \partial R} \Phi_m (R, \Omega) \,,
\\
W_{nm}(R) &=&  -{\hbar^2 \over 2 m} \int d \Omega \,
\Phi_n (R, \Omega)^* {\partial^2 \ \over \partial R^2} \Phi_m (R, \Omega) \,.
\end{eqnarray}
\end{subequations}

The off-diagonal coupling potentials generally fall 
off more rapidly at large distances than
the hyperspherical potentials by a factor of $1/R^2$ \cite{NFJG01}.  
In the low-energy limit, the 
off-diagonal terms in Eq.~(\ref{radeq:S}) are therefore small 
compared to the diagonal terms.  If the off-diagonal terms are 
neglected, the set of eigenvalue equations decouple:
\begin{eqnarray}
\left[ {\hbar^2 \over 2 m} 
\left( - {\partial^2 \ \over \partial R^2} + {15\over 4 R^2} \right) 
        + V_n(R) + 2 U_{nn}(R) {\partial \ \over \partial R}  
   + W_{nn}(R) \right] f_n(R)
\nonumber
\\
    = E f_n (R) \,. 
\label{radeq:aha}
\end{eqnarray}
This approximation is called the 
{\it adiabatic hyperspherical approximation}.
It was first introduced by Macek in 1968 \cite{Macek68}.
In many cases, the diagonal coupling terms are also small compared to
the hyperspherical potential $V_n(R)$.  The eigenvalue equations 
then reduce to radial Schr{\"o}dinger equations
in the hyperspherical potentials.
One advantage to keeping the diagonal coupling potentials is that 
the adiabatic hyperspherical approximation is then a variational 
approximation \cite{SW-79}.


\subsection{Low-energy Faddeev equation}

One disadvantage of the 3-body Schr{\"o}dinger equation is that 
it does not take advantage of simplifications associated with 
configurations consisting of a 2-body cluster that is well-separated
from the third atom. 
The {\it Faddeev equations } are an equivalent set of equations 
that exploit these simplifications.
We will use the Faddeev equations together with
a restriction to total angular momentum zero and 
some further approximations that can be justified at low energy 
to reduce the 3-body problem to a set of coupled integro-differential 
equations in one variable, the hyperradius $R$.
We  follow closely the treatment of Fedorov and Jensen 
in Ref.~\cite{FJ-93}.

We make the simplifying assumption that
the potential $V$ can be expressed as the sum of three 2-body potentials, 
each of which depends only on the separation $r_{ij}$ of a pair of atoms:%
\footnote{
One could also include an intrinsically 3-body potential 
at the cost of a small complication in the formalism.
See, e.g., Ref.~\cite{Gloeckle}.}
\begin{eqnarray}
V({\bm r}_1, {\bm r}_2,{\bm r}_3)
=V ( r_{12})+V ( r_{23}) +V ( r_{31}) \,.
\label{pairpot}
\end{eqnarray}
We use the same symbol $V$ for the total potential 
and the 2-body potential, distinguishing them by the context 
and by the number of arguments.
The assumption in Eq.~(\ref{pairpot}) is milder than it appears, 
because even if the potential includes 
intrinsically 3-body terms at short distances, 
their universal effects at low-energy can be
reproduced by a sum of pair-wise terms. 

The {\it Faddeev equations} are a set of equations that generate solutions 
to the 3-body Schr\"odinger equation of the form \cite{Faddeev}
\begin{eqnarray}
\Psi ({\bm r}_1, {\bm r}_2, {\bm r}_3) 
&=& \psi^{(1)}({\bm r}_{23},{\bm r}_{1,23}) 
+ \psi^{(2)}({\bm r}_{31},{\bm r}_{2,31}) 
+ \psi^{(3)}({\bm r}_{12},{\bm r}_{3,12}) \,.
\label{Psi-Faddev}
\end{eqnarray}
The Faddeev equations are
\begin{eqnarray}
\left( T_R + T_{\alpha_1} + \frac{\Lambda_{1,23}^2}{2mR^2} \right) \psi ^{(1)}
+ V(r_{23}) \left( \psi^{(1)} + \psi^{(2)} + \psi^{(3)} \right) 
&=& E \psi ^{(1)} \,,
\label{Faddeev}
\end{eqnarray}
together with the two equations obtained by cyclicly permuting the subscripts 
and superscripts $(1,2,3)$.  If $\psi^{(1)}$, $\psi^{(2)}$, and $\psi^{(3)}$
are solutions to this set of equations, then their sum
is a solution to the Schr\"odinger equation (\ref{Seq:3}).
If we set $\psi^{(2)} = \psi^{(3)} = 0$ and take $\psi^{(1)}$ to be a 
function of ${\bm r}_{23}$ only, the Faddeev equation (\ref{Faddeev})
reduces to the 2-body equation for atoms 2 and 3.  Thus the Faddeev 
wave function $\psi^{(1)}$ can naturally take into account 
the correlations between atoms 2 and 3 at large hyperradius $R$
when they are both far away from atom 1.
Note that nontrivial solutions 
$\psi^{(1)}$, $\psi^{(2)}$, and $\psi^{(3)}$ of the Faddeev equations 
can give the trivial solution $\psi^{(1)} + \psi^{(2)} + \psi^{(3)} = 0$ 
to the Schr\"odinger equation.  
Such solutions of the Faddeev equations are called {\it spurious solutions}.

We restrict our attention to states with 
total angular momentum quantum number $L=0$.
For a discussion of higher angular momenta, 
see Refs.~\cite{NFJG01,GasMa02}.
We also make an additional simplifying assumption about the form 
of the wave function.  
The Faddeev wave function $\psi^{(1)}({\bm r}_{23},{\bm r}_{1,23})$
can be decomposed into spherical harmonics for the unit vectors
$\hat {\bm r}_{23}$ and $\hat {\bm r}_{1,23}$:
\begin{eqnarray}
\psi^{(1)}({\bm r}_{23},{\bm r}_{1,23})
= \sum_{l_x,m_x} \sum_{l_y,m_y} f^{(1)}_{l_x m_x,l_y m_y}(R,\alpha_1)
Y_{l_x m_x}(\hat {\bm r}_{23}) Y_{l_y m_y}(\hat {\bm r}_{1,23}) \,, 
\label{psi-Ylm}
\end{eqnarray}
where $l_x, m_x$ and $l_y,m_y$ are the quantum numbers associated with orbital 
angular momentum in the $23$ and $1,23$ subsystems, respectively.
Our simplifying assumption is that the expansion in Eq.~(\ref{psi-Ylm})
is dominated by the $l_x = l_y = 0$ term. Thus we neglect 
any orbital angular momentum of the subsystems $ij$ or $k,ij$.  
This is not an essential assumption, but it greatly simplifies 
the formalism by avoiding sums over the angular momentum quantum numbers
$l_x,m_x$ and $l_y,m_y$.
For a general treatment of the problem including subsystem 
angular momentum, see Ref.~\cite{NFJG01}.
The simplifying assumption of neglecting subsystem angular momentum
is motivated by the general suppression of 
higher orbital angular momentum at low energies.
Ignoring subsystem angular momentum  
is a better approximation for the Faddeev equations than for the
Schr\"odinger equation.  In the Faddeev equations,
the coupling between different angular momenta for the subsystems 
enters only at second order in the 2-body potential $V$ \cite{NFJG01},
while it enters at first order for the Schr\"odinger equation.

Given the simplifying assumption of neglecting subsystem angular momentum, 
the Schr\"odinger wave function in Eq.~(\ref{Psi-Faddev})
for three identical particles reduces to 
\begin{eqnarray}
\Psi ({\bm r}_1, {\bm r}_2, {\bm r}_3) 
= \psi(R, \alpha_1) + \psi(R,\alpha_2) + \psi(R, \alpha_3) \,.
\label{psi3}
\end{eqnarray}
The Faddeev equations for the pairwise potential in Eq.~(\ref{pairpot})
then reduce to a particularly simple set of Faddeev equations:
\begin{eqnarray}
\left(T_R + T_{\alpha_1} - E \right)\psi (R, \alpha_1) 
&+& V (\sqrt{2}R \sin \alpha_1)
\nonumber
\\
&& \hspace{-4cm} 
\times 
\left[\psi (R, \alpha_1) + \psi (R, \alpha_2) + \psi (R, \alpha_3) 
\right] =  0 \,,
\label{Faddeev-5}
\end{eqnarray}
together with the two equations obtained by cyclicly permuting 
$\alpha_1$, $\alpha_2$, and $\alpha_3$.  We can reduce these
three equations to a single equation by exploiting the fact
that the averages of $\psi(R, \alpha_2)$ and $\psi (R, \alpha_3)$ 
over the angular variables $\hat {\bm r}_{23}$ 
and $\hat {\bm r}_{1, 23}$ can be expressed as an
integral operator acting on $\psi (R, \alpha_1)$:
\begin{eqnarray}
\Big\langle \psi (R, \alpha_2) \Big\rangle_{\hat {\bm r}_{23},
\hat {\bm r}_{1, 23}} 
&=& \Big\langle \psi (R, \alpha_3) \Big\rangle_{\hat {\bm r}_{23},
\hat {\bm r}_{1, 23}}
\nonumber \\ && 
= {2 \over \sqrt 3} 
\int^{{1\over2}\pi - |{1\over6}\pi - \alpha_1|}_{|{1\over3}\pi - \alpha_1 |}
  {\sin (2 \alpha^\prime) \over \sin (2 \alpha_1)} \psi (R, \alpha^\prime)
  d \alpha' \,.
\end{eqnarray}
The resulting integro-differential equation for $\psi (R, \alpha)$ is 
\begin{eqnarray}
\left (T_R + T_\alpha - E \right) \psi (R, \alpha)  &=& 
- V (\sqrt{2} R \sin \alpha)  
\nonumber
\\
&& \hspace{-4cm}
\times \bigg[ \psi (R, \alpha)+ {4 \over \sqrt 3} 
\int^{{1\over2}\pi - | {1\over6} \pi - \alpha |}_{| {1\over3}\pi  - \alpha |}
   {\sin (2 \alpha') \over \sin (2 \alpha)} \psi (R, \alpha') d \alpha' 
\bigg] \,.
\label{lowenergy-Faddeev}
\end{eqnarray}
We will refer to this equation as the {\it low-energy Faddeev equation}.

A convenient way to solve this equation is to use a
{\it hyperspherical expansion}.
For each value of $R$, the wave function $\psi (R,\alpha)$
is expanded in a complete set of functions $\phi_n(R, \alpha)$ of the
hyperangle $\alpha$:
\begin{eqnarray}
\psi(R, \alpha) = {1 \over R^{5/2} \sin(2 \alpha)} 
                        \sum_n f_n (R) \phi_n(R, \alpha) \,.
\label{Faddeev-exp}
\end{eqnarray}
The divergence of the prefactor $1/\sin(2 \alpha)$
at the endpoints $\alpha = 0$ and ${1\over2}\pi$
imposes boundary conditions that $\phi_n(R, \alpha)$
must vanish at the endpoints.
The functions $\phi_n(R, \alpha)$ are solutions to an
integro-differential eigenvalue equation in the single variable
$\alpha$:
\begin{eqnarray}
\left[ - {\partial^2 \ \over \partial \alpha^2 } - \lambda_n (R)\right ] 
\phi_n(R, \alpha)  &=& -  {2m R^2 \over \hbar^2} V (\sqrt{2}R \sin \alpha)
\nonumber 
\\
&& \hspace{-4.5cm}
\times
\left[ \phi_n (R, \alpha) + {4 \over \sqrt{3} }
\int_{|{1\over 3} \pi- \alpha|}^{{1\over 2}\pi - |{1\over 6}\pi - \alpha|}
\, \phi_n (R, \alpha') d \alpha' \right] \,.
\hspace{0.9cm}
\label{Faddeev-angular}
\end{eqnarray}
The hyperradius $R$ is treated as a parameter and the eigenvalue 
$\lambda_n(R)$ is a function of that parameter.  
The eigenvalues $\lambda_n(R)$ in Eq.~(\ref{Faddeev-angular}) 
define channel potentials for the hyperradial variable:
\begin{eqnarray}
\label{Vn}
V_n(R) = [\lambda_n (R) -4] {\hbar^2 \over 2m R^2} \,.
\label{Vch}
\end{eqnarray}
Note that since the operator on the right side of Eq.~(\ref{Faddeev-angular})
is not hermitian, hyperangular functions $\phi_m(R,\alpha)$
and $\phi_n(R,\alpha)$
with distinct eigenvalues $\lambda_m(R)$ and  $\lambda_n(R)$
need not be orthogonal functions of $\alpha$.
Their inner products define a matrix $G_{nm}(R)$ that depends on $R$:
\begin{eqnarray}
G_{nm}(R) = \int_0^{{1\over 2}\pi} d \alpha \, \phi_n^*(R,\alpha) 
\phi_m(R,\alpha).
\end{eqnarray}
By inserting the expansion in Eq.~(\ref{Faddeev-exp}) into
the low-energy Faddeev equation (\ref{lowenergy-Faddeev}),
projecting onto $\phi_n^*(R,\alpha)$, and then multiplying 
by the inverse of the matrix $G_{nm}(R)$,
we obtain a coupled set of eigenvalue equations 
for the hyperradial wave functions $f_n(R)$:
\begin{eqnarray}
\left[ {\hbar^2 \over 2 m} 
\left( - {\partial^2 \ \over \partial R^2} + {15\over 4 R^2} \right) 
        + V_n(R) \right] f_n(R) &&
\nonumber 
\\ 
+ \sum_m \left[ 2 P_{nm}(R) {\partial \ \over \partial R}  
   + Q_{nm}(R) \right] f_m(R)
    &=& E f_n (R) \,, 
\label{radeq:F}
\end{eqnarray}
where the coupling potentials $P_{mn}(R)$ and $Q_{mn}(R)$ are
defined by
\begin{subequations}
\begin{eqnarray}
P_{nm}(R) &=&  - {\hbar^2 \over 2 m} \sum_k G^{-1}_{nk}(R)
\!\! \int_0^{{1\over2}\pi} \! d \alpha \, 
\phi_k^* (R, \alpha) {\partial \ \over \partial R} \phi_m (R, \alpha) \,,
\\
Q_{nm}(R) &=&  -{\hbar^2 \over 2 m} \sum_k G^{-1}_{nk}(R) 
\!\! \int_0^{{1\over2}\pi} \! d \alpha \, 
\phi_k ^*(R, \alpha) {\partial^2 \ \over \partial R^2} \phi_m (R, \alpha) \,.
\end{eqnarray}
\label{couplingpot}
\end{subequations}

The set of radial equations (\ref{radeq:F}) looks similar to the
set of radial equations (\ref{radeq:S}) for the adiabatic 
hyperspherical representation of the 3-body Schr{\"o}dinger equation.
The difference is that the channel potentials $V_n(R)$ in 
Eq.~(\ref{radeq:F}) are obtained by solving integro-differential 
eigenvalue equations in only one variable, while the channel 
potentials $V_n(R)$ in Eq.~(\ref{radeq:S}) are obtained by solving 
differential eigenvalue equations in five hyperangular
variables.  The reduction in the number of variables came from our
simplifying assumption of neglecting subsystem angular momenta.

If all the eigenvalues $\lambda_n(R)$ were independent 
of $R$, then the hyperangular functions $\phi_n(R)$ obtained by solving 
Eq.~(\ref{Faddeev-angular}) would be independent of $R$ and the coupling 
potentials $P_{mn}(R)$ and $Q_{mn}(R)$ defined by Eqs.~(\ref{couplingpot}) 
would vanish. Thus in regions of $R$
in which all the eigenvalues $\lambda_n(R)$ vary 
sufficiently slowly with $R$, the coupling potentials 
$P_{mn}(R)$ and $Q_{mn}(R)$ can be neglected.
The adiabatic hyperspherical approximation \cite{Macek68} consists 
of neglecting the off-diagonal coupling terms in Eq.~(\ref{radeq:F}),
in which case the eigenvalue equations decouple.
If the diagonal coupling terms in Eq.~(\ref{radeq:F})
are also neglected, the equations reduce to
radial Schr{\"o}dinger equations
for each of the hyperspherical potentials:
\begin{eqnarray}
\left[ {\hbar^2 \over 2 m} 
\left( - {\partial^2 \ \over \partial R^2} + {15\over 4 R^2} \right) 
        + V_n(R)\right] f_n(R)
\approx E f_n (R) \,.
\label{aha:F}
\end{eqnarray}

If the 2-body potential $V(r)$ vanishes,
the integro-differential eigenvalue equation (\ref{Faddeev-angular})
reduces to a simple differential eigenvalue equation that is easy to solve.
The eigenvalues are independent of $R$:
\begin{eqnarray}
\lambda_n(R) = 4 (n+1)^2, \qquad n=0,1,2,\ldots  \,.
\label{lambda-free}
\end{eqnarray}
The corresponding eigenfunctions are 
\begin{eqnarray}
\phi_n(R,\alpha) = \sin[2(n+1) \alpha] \,.
\end{eqnarray}
The corresponding hyperspherical potentials are
\begin{eqnarray}
V_n(R) = 4n(n+2) {\hbar^2 \over 2 m R^2} \,.
\end{eqnarray}

If the 2-body potential $V(r)$ is short-ranged, there are two regions in which 
the integro-differential eigenvalue equation (\ref{Faddeev-angular})
for the angular function $\Phi_n (R, \alpha)$ can be solved analytically.
One region is $R \sin \alpha$  large enough
that the $V(R \sin \alpha)$ term is small compared to $\lambda_n(R)$.  
In this case, we can
neglect the $R^2 V$ term and the equation reduces to
\begin{eqnarray}
\left[ - {\partial^2 \ \over \partial \alpha^2} - \lambda_n(R) \right]
\phi_n^{\rm (hi)}(R,\alpha) \approx 0 \,.
\end{eqnarray}
The solution that vanishes at the upper endpoint $\alpha = {\pi\over 2}$ is
\begin{eqnarray}
\phi_n^{\rm (hi)} (R, \alpha) \approx \sin \left[ \lambda_n^{1/2}(R)
( \mbox{$\pi \over 2$} - \alpha) \right] \,.
\label {Phi-high}
\end{eqnarray}

The other region for which Eq.~(\ref{Faddeev-angular})
can be solved analytically is $R \sin \alpha$  small enough 
that $\lambda_n(R)$ is small compared to the $V(R \sin \alpha)$ term.
In this case, we can neglect $\lambda_n$.  In the region $\alpha \ll 1$,
the equation reduces to
\begin{eqnarray}
 \left[ - {\partial^2 \ \over \partial \alpha^2}
+ {2m R^2 \over \hbar^2} V(\sqrt{2} R \alpha) \right] 
  \phi_n^{\rm (lo)}(R, \alpha)
\nonumber  \\ \quad
 \approx - {2m R^2 \over \hbar^2} V(\sqrt{2} R \alpha) 
{8 \alpha \over \sqrt{3}} \phi_n^{\rm (hi)} (R, \mbox{$\pi \over 3$}) \,.
\label{Faddeev-10}
\end{eqnarray}
The general solution to this inhomogenous equation is the sum of
a particular solution and the general solution to the homogeneous
equation.  A particular solution is
\begin{eqnarray}
\phi_n^{\rm (lo)}(R,\alpha) = - {8 \alpha \over \sqrt{3}}
\phi_n^{\rm (hi)} (R, \mbox{$\pi \over 3$}) \,.
\end{eqnarray}
The homogeneous equation can be expressed in the form
\begin{eqnarray}
\left[ - {\hbar^2 \over 2mR^2} {\partial^2 \ \over \partial \alpha^2}
+ V(\sqrt{2} R \alpha) \right] \phi_n^{\rm (lo)}(R, \alpha)
\approx 0 \,.
\end{eqnarray}
This is identical to the radial Schr{\"o}dinger equation 
for a pair of particles with zero energy interacting 
through the 2-body potential $V(r)$, where $r = \sqrt{2} R \alpha$. 
If we denote the zero-energy solution by $\psi_0(r)$, 
the most general solution to Eq.~(\ref{Faddeev-10}) is 
\begin{eqnarray}
\phi_n^{\rm(lo)} (R, \alpha) \approx 
A(R)\psi_0 (\sqrt{2} R \alpha) - {8 \alpha \over \sqrt{3}}
\phi_n^{\rm(hi)} (R, \mbox{$\pi\over3$}) \,,
\label{Phi-low}
\end{eqnarray}
where $A(R)$ is an arbitrary function of $R$.


\subsection {Hyperspherical potentials}

Thus far, we have made no assumptions about the potential 
$V({\bm r}_1,{\bm r}_2,{\bm r}_3)$ except that
it is a sum of three pair potentials as in Eq.~(\ref{pairpot}).
We now apply the low-energy Faddeev equation 
in Eq.~(\ref{lowenergy-Faddeev})
to the problem of a 2-body potential with 
large scattering length $|a| \gg \ell$.  
Zhen and Macek have used a variational approach
to obtain approximate equations for the channel eigenvalues \cite{ZM88}.
The integro-differential eigenvalue equation can also be reduced
to an exact transcendental equation for $\lambda_n(R)$.  
We follow closely the derivation of 
Fedorov and Jensen in Ref.~\cite{FJ-93}.

The hyperangular functions $\Phi_n (R, \alpha)$ have the approximate solutions 
given in Eqs.~(\ref{Phi-high}) and (\ref{Phi-low})
in appropriate regions of $R$ and $\alpha$.
In the case of large scattering length, the high-$\alpha$ solution
in Eq.~(\ref{Phi-high}) holds for $R \sin \alpha \gg \ell$.
The low-$\alpha$ solution in Eq.~(\ref{Phi-low}) holds for 
$\alpha \ll 1$ and $R \alpha \gg \ell$. 
For $R \alpha \gg \ell$, the zero-energy solution 
$\psi_0 (r)$ that appears in Eq.~(\ref{Phi-low}) is simply the limit as 
$k \to 0$ of the atom-atom scattering solution given in Eq.~(\ref{psi-AAsin}):
\begin{eqnarray}
\psi_0 (r) = r-a \,,
\label{psi0}
\end{eqnarray}
where $a$ is the scattering length.
Upon inserting $\psi_0(\sqrt{2} \alpha R)$ into Eq.~(\ref {Phi-low}),
we see that the extrapolation of the high-$\alpha$ solution 
in Eq.~(\ref{Phi-high}) to the region $\alpha \ll 1$ is compatible with
the low-$\alpha$ solution in Eq.~(\ref{Phi-low}).
Matching their values at $\alpha = 0$, we
determine the unknown function of $R$ that appears in Eq.~(\ref{Phi-low}):
\begin{eqnarray}
A(R) = - \sin \left[ \lambda^{1/2}(R) \mbox{${\pi \over 2}$} \right] 
\frac{1}{a} \,. 
\label{ABmatch}
\end{eqnarray}
Equating the derivatives with respect to $\alpha$ at
$\alpha = 0$, we obtain the matching equation that determines 
the channel eigenvalues $\lambda_n(R)$\footnote{
In Ref.~\cite{FJ-93}, Fedorov and Jensen used the opposite sign convention 
for the scattering length.  They also made an error that resulted in
the omission of the factor of
$\sqrt{2}$ on the right side of Eq.~(\ref{cheigen}).}:
\begin{eqnarray}
\cos \left ( \lambda^{1/2} \mbox {$\pi \over 2$} \right)
- {8 \over \sqrt {3}} \lambda^{-1/2} 
        \sin \left ( \lambda^{1/2} \mbox {$\pi \over 6$} \right )
= \sqrt{2} \lambda^{-1/2} \sin \left ( \lambda^{1/2} \mbox {$\pi \over 2$} 
  \right ) {R \over a} \,.
\label{cheigen}
\end{eqnarray}
This equation was first derived by Efimov \cite{Efimov71}. 
There are infinitely many solutions 
for the channel eigenvalues $\lambda_n(R)$ at each value of $R$. 
The corresponding hyperangular wave functions are
\begin{eqnarray}
\phi_n (R, \alpha) = 
\sin \big[ \lambda^{1/2}_n (R) (\mbox {$\pi \over 2$}- \alpha) \big] \,.
\label{phin-sol}
\end{eqnarray}
This solution does not satisfy the boundary condition
$\phi_n (R, \alpha) \to 0$ as $\alpha \to 0$, 
because the zero-energy solution in Eq.~(\ref{psi0}) is accurate only for 
$R \alpha \gg \ell$. Thus the solution in Eq.~(\ref{phin-sol}) is not accurate 
in the region $R \alpha \sim \ell$.  It is, however, accurate enough
to calculate quantities that are sufficiently insensitive to 
short distances.

The consistency equation (\ref{cheigen}) 
for the channel eigenvalues has the constant solution
$\lambda(R)= 16$, but it is unphysical.
The hyperangular wave function in Eq.~(\ref{phin-sol}) is
$\phi(R, \alpha) = \sin(4\alpha)$.  The resulting 
Faddeev wave function in Eq.~(\ref{Faddeev-exp})
is proportional to $ \cos(2 \alpha)$.
This is a spurious solution to the low-energy Faddeev equation, 
because the corresponding Schr\"odinger
wave function given by Eq.~(\ref{psi3}) is the trivial solution 
$\Psi = 0$.  This follows from the identity
\begin{eqnarray}
\cos(2 \alpha) = - {4 \over \sqrt{3}}  
\int_{|{1\over3}\pi- \alpha|}^{{1\over2}\pi - |{1\over6}\pi - \alpha|}
{\sin(2 \alpha') \over \sin(2 \alpha)}
\, \cos(2\alpha') d \alpha' \,.
\end{eqnarray}

\begin{figure}[htb]
\bigskip
\centerline{\includegraphics*[width=8.5cm,angle=0]{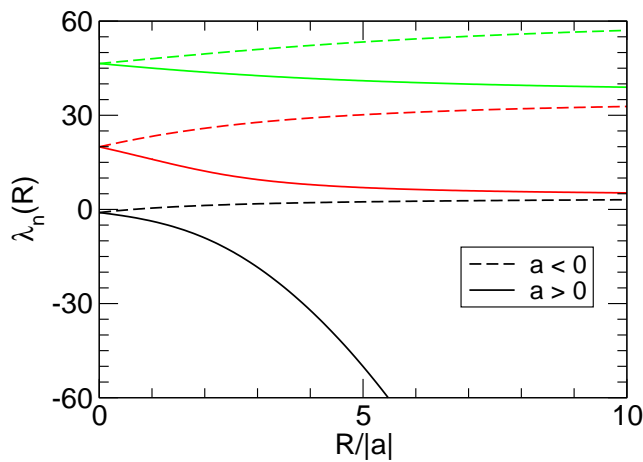}}
\medskip
\caption{The three lowest channel eigenvalues $\lambda_n(R)$
for $a > 0$ (solid lines) and for $a < 0$ (dashed lines).}
\label{fig:cheigen}
\end{figure}

The physical channel eigenvalues $\lambda_n (R)$ 
can be obtained by solving Eq.~(\ref {cheigen}) numerically.  
The lowest three eigenvalues for $a > 0$ and the lowest three
eigenvalues for $a < 0$ are shown as functions of $R/ |a|$ in 
Fig.~\ref{fig:cheigen}.  For $R \ll |a|$, the eigenvalues approach constants
independent of $a$.  The limiting behavior of the lowest eigenvalue 
as $R \to 0$ is
\begin{eqnarray}
\lambda_0 (R) \longrightarrow
- s_0^2 
\left ( 1 + 1.897 { R \over a} \right) \,,
\end{eqnarray}
where $s_0 = 1.00624$ is the solution 
to the transcendental equation (\ref{s0}).
The limiting value of the second lowest eigenvalue as $R \to 0$ is
\begin{eqnarray}
\lambda_1 (R) \longrightarrow 19.94  \,.
\label{eigen0:00}
\end{eqnarray}
The limiting behavior of the lowest eigenvalue as $R \to \infty$ is
\begin{subequations}
\begin{eqnarray}
\lambda_0 (R) & \longrightarrow &
4 \bigg( 1 - { 12 \over \sqrt{2} \,\pi} { |a| \over R} \bigg)
\qquad (a < 0) \,,
\\
& \longrightarrow & - {2 R^2 \over a^2}
\bigg( 1  + { 8 \sqrt{2} \, a \over \sqrt {3} \,R} 
        e^{- \sqrt{2} \, \pi R / 3a} \bigg) 
\qquad (a > 0) \,.
\label{lambda0-largeR}
\end{eqnarray}
\label{lambda0pm-largeR}
\end{subequations}
The limit of the second lowest eigenvalue as $R \to \infty$ is
\begin{subequations}
\begin{eqnarray}
\lambda_1 (R) & \longrightarrow & 36 \qquad (a < 0) \,,
\\
& \longrightarrow & 4 \qquad \hspace{0.2cm} (a > 0) \,.
\label{lamRinfty:1}
\end{eqnarray}
\end{subequations}
For  $n\ge 2$, the limits of the eigenvalues as $R \to \infty$ are
\begin{subequations}
\begin{eqnarray}
\lambda_n (R) & \longrightarrow & 4(n+2)^2 \qquad (a < 0) \,,
\\
             & \longrightarrow & 4(n+1)^2 \qquad (a > 0) \,.
\label{lamRinfty:n}
\end{eqnarray}
\label{lamRinftypm:n}
\end{subequations}
With the exception of $\lambda_0 (R)$ in the case $a >0$,
the limiting values in Eqs.~(\ref{lambda0pm-largeR}--\ref{lamRinftypm:n})
are among those for free particles given in Eq.~(\ref{lambda-free}).

We have simplified the derivation of the matching equation for the channel 
eigenvalues by ignoring the orbital angular momentum of the subsystem 
consisting of the two atoms in a pair or the subsystem 
consisting of a pair and a third atom.
The lowest channel eigenvalue with nonzero subsystem orbital angular 
momentum comes from one unit of angular momentum in the subsystem 
consisting of a pair and a third atom.
This corresponds to the 
$l_x = 0$, $l_y=1$ term in the angular momentum
decomposition of the Faddeev wave function in Eq.~(\ref{psi-Ylm}).
The matching equation for general values of the angular momentum 
quantum number $l_y$ is given in Ref.~\cite{NFJG01}.
The matching equation for $l_y=1$ is \cite{NFJG01}
\begin{eqnarray}
\sin \left ( \lambda^{1/2} \mbox {$\pi \over 2$} \right)
&+& \mbox{$1 \over 3$} \lambda^{1/2} \;
{}_2F_1 \big(\mbox{$1 \over 2$}(3+\lambda^{1/2}),
                \mbox{$1 \over 2$}(3-\lambda^{1/2}),
                \mbox{$5 \over 2$}; \mbox{$1 \over 4$} \big) \,
\nonumber 
\\ 
&=& - {\sqrt{2} \, \lambda^{1/2} \over \lambda - 1} 
\cos \left ( \lambda^{1/2} \mbox {$\pi \over 2$} \right )
{R \over a} \,.
\label{cheigen1b}
\end{eqnarray}
The limiting value of the lowest eigenvalue as $R \to 0$ is
\begin{eqnarray}
\lambda (R) \longrightarrow 8.201  \,.
\label{eigen0:01}
\end{eqnarray}
The limiting behavior of the lowest eigenvalue as $R \to \infty$ is
\begin{subequations}
\begin{eqnarray}
\lambda (R) & \longrightarrow & 25 \qquad\qquad\quad (a < 0) \,,
\\
& \longrightarrow & - {2 R^2 / a^2} \qquad\,   (a > 0) \,.
\end{eqnarray}
\end{subequations}
The lowest eigenvalue for $l_x = 0$, $l_y = 1$ is always lower
than the second lowest eigenvalue for $l_x = l_y =0$.
Thus the neglect of subprocess orbital angular momentum 
is not useful as a quantitative approximation.

\begin{figure}[htb]
\bigskip
\centerline{\includegraphics*[width=8.5cm,angle=0]{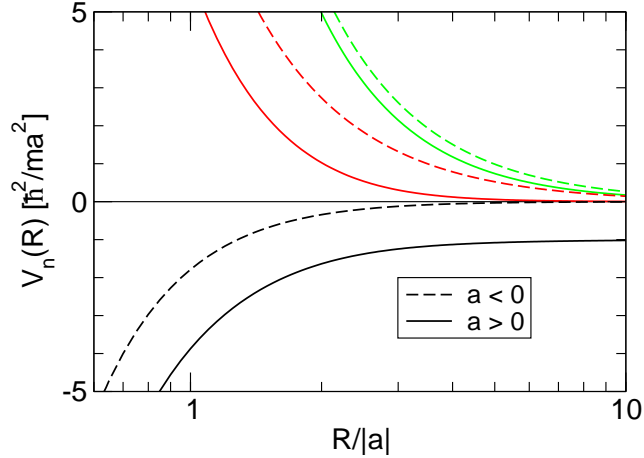}}
\medskip
\caption{The three lowest hyperspherical potentials $V_n(R)$ 
scaled by $\hbar^2/ma^2$
for $a > 0$ (solid lines) and for $a < 0$ (dashed lines).}
\label{fig:hap}
\end{figure}

The hyperspherical potential $V_n(R)$ associated with the eigenvalue 
$\lambda_n(R)$ is given in Eq.~(\ref{Vch}).  
The lowest three hyperspherical potentials for $a > 0$ and for $a < 0$ 
are shown in Fig.~\ref{fig:hap}. 
For either sign of $a$, the lowest hyperspherical potential $V_0(R)$ 
is negative for all $r$ and all the others are positive for all $R$.

As $R \to \infty$, all the hyperspherical potentials asymptote to
the 3-atom threshold $E=0$, with the exception of $V_0(R)$ 
in the case $a > 0$.
We can interpret the asymptotic configurations in the channels 
that asymptote to $E=0$ as 3-atom scattering states.  
For $a > 0$, the lowest hyperspherical potential $V_0(R)$
asymptotes to $-\hbar^2 / ma^2$, which is the dimer binding energy.  
We can interpret the asymptotic configuration in this channel as an 
atom-dimer scattering state.  As $R \to 0$, the hyperspherical potentials 
asymptote to $1/R^2$ potentials with different coefficients:
\begin{eqnarray}
V_n (R) \longrightarrow [ \lambda_n (0) - 4] {\hbar^2 \over 2mR^2},
\hspace{1cm}  R \ll |a| \,.
\label{Vsi}
\end{eqnarray}
The $1/R^2$ potential is attractive for the lowest channel $n = 0$ and 
repulsive for
all the other channels.  We obtain scale-invariant $1/R^2$ potentials, 
because we have taken the scaling limit $\ell \to 0$.  
For finite $\ell$, the hyperspherical potentials have the scale-invariant form 
in Eq.~(\ref{Vsi}) only in the region $\ell \ll R \ll |a|$. 
At short distances $R \sim \ell$, they may be very complicated.

In the adiabatic hyperspherical approximation,
the coupling potentials are ignored and the equations for the 
hyperradial wave functions $f_n(R)$ reduce to Eq.~(\ref{aha:F}).  
This is a good approximation only at limiting values of $R$.  
It is a good approximation in the limit $R \ll |a|$,
because all the eigenvalues $\lambda_n(R)$ approach constant values 
and the hyperangular functions $\phi_0(R,\alpha)$ become 
independent of $R$.  The coupling potentials in Eqs.~(\ref{couplingpot}) 
are therefore suppressed by the derivatives with respect to $R$.
The adiabatic hyperspherical approximation is also a good approximation 
in the limit $R \gg |a|$.
In the case $a<0$, this is again because all the eigenvalues approach 
constant values in the limit.  In the case $a>0$,
this argument does not apply to the coupling potentials 
$P_{n0}(R)$ and $Q_{n0}(R)$, because the lowest eigenvalue $\lambda_0(R)$
diverges to $- \infty$ as $R \to \infty$ as shown in 
Eq.~(\ref{lambda0-largeR}).
The hyperangular wave function $\phi_0(R,\alpha)$ for the lowest channel 
is a hyperbolic function of a real argument.  For $R \gg a$,
it provides an exponential factor 
$\exp[\sqrt{2} R/a ({\pi \over 2} - \alpha)]$ that ensures that the 
wave function has support only in the region $\alpha \sim a/R$.
But for all the other channels, $\lambda_n^{1/2}(R)$ approaches an even 
integer as $R \to \infty$,
according to Eqs.~(\ref{lamRinfty:1}) and (\ref{lamRinfty:n}).
Thus $\phi_n(R,\alpha=0)$ approaches zero as $R \to \infty$.
These zeroes provide suppression of the coupling potentials 
$P_{n0}(R)$ and $Q_{n0}(R)$, which guarantees that the adiabatic 
hyperspherical approximation is also a good approximation 
in the limit $R \gg |a|$.


\subsection {Boundary condition at short distances}
        \label{subsec:bc}

The behavior of the hyperradial wave functions $f_n(R)$ at very small
hyperradius $R \sim \ell$ is determined by the 2-body potential $V(r)$ 
at short distances $r \sim \ell$.  In the scaling limit $\ell \to 0$, 
the information about short distances is lost and it may need to be
reintroduced through boundary conditions on the hyperradial
wave functions.

In the scaling limit, the channel eigenvalues $\lambda_n(R)$ 
are the solutions to Eq.~(\ref{cheigen}).  For $R \ll |a|$, the eigenvalues
approach constant values $\lambda_n(0)$, so the adiabatic hyperspherical
approximation in Eq.~(\ref{aha:F}) is justified.
The hyperradial eigenvalue equation (\ref{aha:F}) then reduces 
in the region $R \ll |a|$ to
\begin{eqnarray}
{\hbar^2 \over 2m} \left [ - { \partial^2 \ \over \partial R^2}
+ {\lambda_n(0) - 1/4 \over R^2} \right ] f_n(R) = Ef_n(R) \,.
\label{eigen-f}
\end{eqnarray}
This looks like the radial Schr{\"o}dinger equation for a particle in 
a $1/R^2$ potential, with the strength of the potential determined by 
$\lambda_n(0)$. 
As illustrated in Fig.~\ref{fig:hap}, in the region $R \ll |a|$,
all the channel potentials with the exception of $V_0(R)$ are repulsive 
$1/R^2$ potentials.  The hyperspherical wave functions $f_n(R)$
for $n \ge 1$ therefore decrease exponentially as $R \to 0$,
and no boundary conditions are required.
In contrast, the potential $V_0(R)$ is an attractive $1/R^2$ potential
for $R \ll |a|$:
\begin{eqnarray}
V_0 (R) \approx - (4 + s_0^2) {\hbar^2 \over 2mR^2}, 
\hspace{1cm} R \ll |a| \,.
\label{V0-si}
\end{eqnarray}
This potential is too singular as $R \to 0$ 
for the hyperradial equation (\ref{eigen-f}) to have well-behaved solutions.
Having taken the scaling limit $\ell \to 0$, 
we have lost information about the boundary condition at $R \to 0$
provided by the 2-body potential $V (r_{ij})$ at short distances.  If the
solution $f (R)$ to the complete problem with the correct boundary
condition at $R \to 0$ was known, 
it could be matched onto the solution of Eq.~(\ref{eigen-f}) simply 
by choosing a hyperradius $R_0$ where Eq.~(\ref{eigen-f}) is accurate and
demanding that the logarithmic derivatives $R_0 f^\prime (R_0) / f (R_0)$ 
match at that point.  Thus the correct treatment of the problem 
at short-distances is equivalent to choosing a matching point $R_0$ 
and specifying the dimensionless number $R_0 f^\prime (R_0) / f (R_0)$.

The matching point $R_0$ can be chosen to lie in the scale-invariant region
$\ell \ll R_0 \ll |a|$.  
If we also choose $R_0 \ll (m|E|/\hbar^2)^{-1/2}$, the energy eigenvalue $E$
in Eq.~(\ref{eigen-f}) can be neglected relative to the channel potential,
and the hyperradial equation reduces to 
\begin{eqnarray}
{\hbar^2 \over 2m} \left[- {\partial^2 \ \over \partial R^2}
- {s_0^2 + 1/4 \over R^2} \right] f_0(R) \approx 0 \,.
\label{eigenR:si}
\end{eqnarray}
This equation has solutions that behave like powers of $R$.
The most general solution is 
\begin{eqnarray}
f_0(R) \approx A R^{1/2+i s_0} + B R^{1/2-i s_0}  
\quad R \ll |a|, 1/\kappa \,.
\label{f0:R}
\end{eqnarray}
where $A$ and $B$ are constants.
The ratio $B/A$ is a complex number that could be determined by
matching to the solution of the problem at short distances $R \sim \ell$.
If $|A|<|B|$,
there is a net flow of probability into the short-distance region.
As will be discussed in detail in Section~\ref{sec:deep},
such a flow of probability is possible if there are deep
(tightly-bound) diatomic molecules.  
In the absence of such deep bound states,
we must have $|A| = |B|$ and the solution in Eq.~(\ref{f0:R}) can be written
\begin{eqnarray}
f_0(R) \approx A R^{1/2} \sin [s_0 \ln (\kappa R) + \alpha] \,,
\label{f0:si}
\end{eqnarray}
where $\alpha$ is a constant and the factor of
$\kappa = (m|E|/\hbar^2)^{1/2}$ in the argument of the logarithm is inserted
to make it dimensionless.  
The phase $\alpha$ is determined by matching to the solution of the problem 
at long distances $R \sim |a|$.  It depends on $a$ and on the
energy $E$.  Since $\alpha$ is dimensionless, it can
depend only on the dimensionless combination $\kappa a$, 
and on the signs of $E$ and $a$.
Computing the logarithmic derivative of the solution
in Eq.~(\ref{f0:si}) and evaluating it at the matching point, we obtain
\begin{eqnarray}
R_0 {f_0^\prime (R_0) \over f_0(R_0)} = 
{ 1 \over 2} + s_0 \cot [ s_0 \ln (\kappa R_0) + \alpha] \,.
\label{separate}
\end{eqnarray}
Inside the argument of the cotangent, the dependence on $R_0$ can be 
separated from the dependence on $E$ and $a$.  
This implies that the wave function at distances $R \gg \ell$ depends only 
on a particular function of the matching point $R_0$ and the logarithmic
derivative $R_0 f'(R_0)/f(R_0)$:
\begin{eqnarray}
\Lambda_0 &=& {1 \over R_0} \exp \left \{ {1 \over s_0} {\rm arccot} \left [ {1
\over s_0} \left( R_0 {f_0'(R_0) \over f_0(R_0)} - {1 \over 2} 
\right ) \right ] \right\} \,.  
\label{Lambda-0}
\end{eqnarray}
The parameter $\Lambda_0$ has dimensions of wave number.
We refer to it as  a scaling-violation parameter, 
because logarithmic scaling violations can affect low-energy observables 
only through their dependence on this parameter.%
\footnote{ 
An equivalent scaling-violation parameter $\Lambda_*$ 
was introduced in Ref.~\cite{BHK99,BHK99b} through the 
renormalization prescription for an effective field theory.}

We can interpret the matching point $R_0$ as a short-distance cutoff.
Configurations with smaller hyperradii $R$ need not be taken into account
explicitly.  Instead their effects on the physics at longer distances are taken
into account through the value of $R_0f'(R_0)/f(R_0)$. All low-energy
observables in the 3-body sector are determined
either by specifying the short-distance cutoff $R_0$ and the
dimensionless number $R_0f'(R_0)/f(R_0)$ or, alternatively, by
specifying the scaling-violation parameter $\Lambda_0$
defined by Eq.~(\ref{Lambda-0}).
This phenomenon of a dimensionless short-distance
parameter and a short-distance cutoff being equivalent to a
dimensionful long-distance parameter is known as {\it
dimensional transmutation}.
It is a familiar feature of quantum chromodynamics, the
quantum field
theory that describes the strong interactions \cite{Gross-Wilczek,Politzer}.
The theory has a single parameter: the
strong coupling constant $\alpha_s$. It can be specified by giving its
value
$\alpha_s(\Lambda)$ when the theory is defined with a large momentum
cutoff $\Lambda$.  When the coupling constant is small, its dependence
on the cutoff can be calculated using perturbation theory. It
satisfies a differential renormalization group equation:
\begin{eqnarray}
\Lambda {\partial \ \over \partial \Lambda} \alpha_s (\Lambda) 
= - {33 - 2n_f \over 6 \pi}
\alpha _s^2 (\Lambda) \,,
\end{eqnarray}
where $n_f$ is the relevant number of quark flavors. This equation
implies that the
parameter $\Lambda_{QCD}$ defined by
\begin{eqnarray}
\Lambda_{\rm QCD} = \Lambda \exp \left[ -{6 \pi} \over (33-2n_f)
\alpha_s (\Lambda) \right]
\end{eqnarray}
is independent of $\Lambda$. The theory can therefore be defined equally
well by
specifying the cutoff $\Lambda$ and the dimensionless parameter
$\alpha_s(\Lambda)$ or
alternatively by specifying the scaling-violation
parameter $\Lambda_{\rm QCD}$
generated by dimensional transmutation. Similarly, the low-energy
theory for the 3-body
problem with large 2-body scattering length can be defined by specifying
$a$, the cutoff
$R_0$, and the dimensionless parameter $R_0f'(R_0)/f(R_0)$ or
alternatively by
specifying $a$ and the scaling-violation parameter $\Lambda_0$
generated by dimensional transmutation.

The logarithmic scaling violations in QCD can be interpreted as the 
result of an {\it anomaly} in the scaling symmetry.  An anomaly in 
this context refers to the violation of a symmetry by quantum effects.  
The classical field equations for QCD have a scaling symmetry 
in the limit in which the masses of all the quarks are set to zero.
But this symmetry is broken by the effects of quantum field 
fluctuations at short distances.  The logarithmic scaling 
violations associated with the Efimov effect can be interpreted 
as the result of an anomaly in a scaling symmetry \cite{Ananos:2002id}.
The scaling symmetry appears if the scaling limit and the resonant 
limit are taken simultaneously.  The symmetry is broken by quantum 
fluctuations at small hyperradii in the 3-body channel.


\subsection{Efimov states in the resonant limit}
        \label{subsec:Efres}

The Efimov effect is the existence of infinitely many 3-body bound states 
with an accumulation point at the 3-atom threshold
in the resonant limit $a \to \pm \infty$.  
A derivation of the Efimov effect within the hyperspherical formalism
was first given by Macek \cite{Macek86}.
We proceed to derive the Efimov effect and deduce some of the 
properties of the Efimov states in the resonant limit. 

In the resonant limit, the adiabatic hyperspherical approximation is 
accurate at all finite values of $R$.  In particular, it is accurate 
for bound states for which hyperradial wave functions $f_n (R)$ fall 
exponentially as $R \to\infty$. 
Since the $n = 0$ channel is the only attractive one, it is the only one 
that supports bound states.  The channel eigenvalue is 
$\lambda_0 (R) = -s_0^2$, so
the Schr\"odinger wave function in the center-of-mass frame reduces to
\begin{eqnarray}
\Psi ({\bm r}_1, {\bm r}_2, {\bm r}_3) = R^{-5/2} f_0 (R) 
\sum_{i = 1}^3 {\sinh \left [ s_0 ({\pi \over 2} - \alpha_i) \right ] 
        \over \sin (2 \alpha_i)} \,.
\end{eqnarray}
Since the channel eigenvalue $\lambda_0 (R) = -s_0^2$ is a constant,
the adiabatic hyperspherical approximation is justified.
The equation for the hyperradial wave function $f_0 (R)$ 
in Eq.~(\ref{eigen-f}) reduces to
\begin{eqnarray}
{\hbar^2 \over 2m} \left [ - { \partial^2 \ \over \partial R^2}
- {s_0^2 + 1/4 \over R^2} \right ] f_0(R) = Ef_0(R) \,.
\end{eqnarray}
The boundary condition at short distances is specified 
by a matching point $R_0$ and the logarithmic derivative
$R_0 f_0^\prime (R_0) / f_0 (R_0)$.  Alternatively, if $R_0$ is in the
scale-invariant region, the boundary condition can be
specified by the 3-body parameter $\Lambda_0$ 
defined by Eq.~(\ref{Lambda-0}).  

If an Efimov state has binding energy $E_T$, a binding wave number 
$\kappa$ can be defined by
\begin{eqnarray}
E_T = {\hbar^2 \kappa^2 \over m} \,.
\end{eqnarray}
The solution to the hyperradial equation (\ref{aha:F}) that decreases
exponentially as $R \to \infty$ is
\begin{eqnarray}
f_0 (R) = R^{1/2} K_{i s_0} (\sqrt{2} \kappa R) \,,
\label{f0-k}
\end{eqnarray}
where $K_{i s_0}(z)$ is a Bessel function with imaginary index.
The boundary condition at short distances determines the discrete 
spectrum of binding energies $E_T^{(n)}$.  In the region $\kappa R \ll 1$, 
the solution in Eq.~(\ref{f0-k}) reduces to
\begin{eqnarray}
f_0 (R) &\longrightarrow&
- \left( \pi / [s_0 \sinh (\pi s_0)] \right )^{1/2} 
R^{1/2} \sin \left[ s_0 \ln (\kappa R) + \alpha_0 \right] \,,
\label{f0-sin}
\end{eqnarray}
where the angle $\alpha_0$ is
\begin{eqnarray}
\alpha_0 = - {1 \over 2} s_0 \ln 2 
- {1 \over 2} \arg {\Gamma (1 + i s_0) \over \Gamma (1 - is_0)} \,.
\end{eqnarray}
Inserting the solution in Eq.~(\ref{f0-sin}) into 
Eq.~(\ref{Lambda-0}), the equation for $\Lambda_0$ reduces to
\begin{eqnarray}
s_0 \ln(\Lambda_0 R_0) = 
{\rm arccot} \left( \cot[ s_0 \ln( \kappa R_0) +\alpha_0 ] \right) \,.
\label{Lambda0-kappa}
\end{eqnarray}
The solutions for $\kappa$ can be written
\begin{eqnarray}
\kappa^{(n)} =  
\left( e^{-\pi/s_0} \right)^{n - n_0} e^{-\alpha_0/s_0} \Lambda_0 \,,
\label{Lambda*-kappa}
\end{eqnarray}
where $n_0$ is an arbitrary integer that arises from the choice 
of the branch of the cotangent in Eq.~(\ref{Lambda-0}).  
The resulting expression 
for the spectrum $E_T^{(n)}$ of the Efimov states 
in the resonant limit can be expressed in the form
\begin{eqnarray}
E_T^{(n)} = \left( e^{-2 \pi/s_0} \right)^{n-n_*}
{\hbar^2 \kappa_*^2 \over m} \,,
\label{B3n-n*}
\end{eqnarray}
where $\kappa_*$ is the binding wave number for the Efimov state labeled by
$n = n_*$.  The spectrum in Eq.~(\ref{B3n-n*}) is geometric, 
with the binding energies of 
successive Efimov states having the ratio $e^{2 \pi / s_0} \approx 515.03$.  
The relation between $\kappa_*$ and $\Lambda_0$ can be expressed 
in a form that does not involve the integers $n_0$ and $n_*$:
\begin{eqnarray}
s_0 \ln(\kappa_*)  &=& 
s_0 \ln (\Lambda_0) - \alpha_0  \mod \pi \,.
\label{kappa-Lam0}
\end{eqnarray}
The relation between these parameters is defined only up to 
multiplicative factors of $e^{\pi / s_0} \approx 22.7$, 
because their definitions involve the arbitrary integers
$n_0$ in Eq.~(\ref{Lambda*-kappa}) and $n_*$ in Eq.~(\ref{B3n-n*}).

\begin{figure}[htb]
\centerline{\includegraphics*[width=8.5cm,angle=0,clip=true]{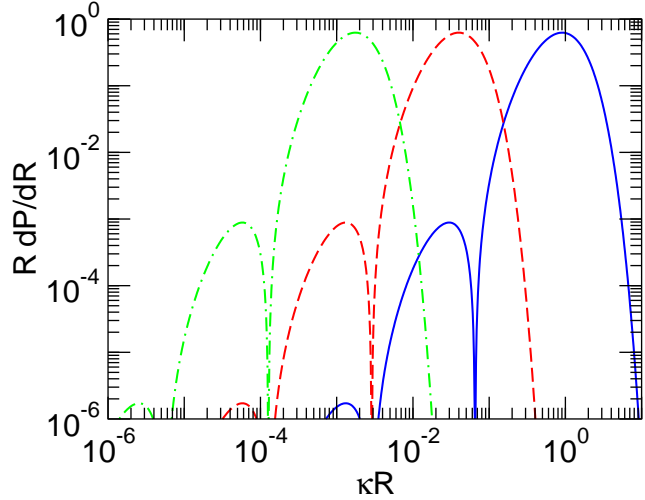}}
\medskip
\caption{
The normalized probability distributions $RdP/dR$  for three successive 
Efimov states in the resonant limit
as functions of $\ln(\kappa R)$, where $\kappa$ is the binding 
wave number of the shallowest of the three states.}
\label{fig:probEf}
\end{figure}

The probability distribution $dP/dR$ for the hyperradius $R$ in an Efimov 
state in the resonant limit is proportional to $f_0(R)^2$, where $f_0(R)$ is 
given in Eq.~(\ref{f0-k}). The normalized probability distributions for three 
successive  Efimov states are shown in Fig.~\ref{fig:probEf}.
A quantitative measure of the size of a 3-body bound state is the expectation 
value of
$R^2$.  For an Efimov state in the resonant limit, this expectation value is
\begin{eqnarray}
\langle R^2 \rangle = 
{\int_0^\infty dR \, R^2 f_0^2 (R) \over \int_0^\infty dR \, f_0^2 (R)} \,.
\end{eqnarray}
Inserting the hyperradial wave function in Eq.~(\ref{f0-k}), we obtain 
\begin{eqnarray}
\langle R^2 \rangle^{(n)} = {2 (1 + s_0^2) \over 3} 
 \left(\kappa^{(n)} \right)^{-2} \,.
\end{eqnarray}
Using the expression for $\kappa$ for the $n^{\rm th}$ Efimov state from
Eq. (\ref{Lambda*-kappa}), the mean-square hyperradius is 
\begin{eqnarray}
\langle R^2 \rangle^{(n)} = \left( e^{2 \pi /s_0} \right)^{n-n_*} 
{2(1 + s_0^2)\over 3} 
\kappa_*^{-2} \,.
\end{eqnarray}
Thus the root-mean-square hyperradius for each successively shallower Efimov
state is larger than the previous one by $e^{\pi / s_0} \approx 22.7$.

The simple geometric spectrum of Efimov states given in Eq.~(\ref{B3n-n*}) 
is obtained only if we take the scaling ($\ell \to 0$) 
and resonant ($a \to \pm \infty$) limits simultaneously. 
If we stay in the resonant limit but do not take the scaling 
limit $\ell \to 0$, the 3-body spectrum will be bounded from
below.  There will be a deepest Efimov state that can be labeled by $n=0$ and
whose binding energy is comparable to the natural ultraviolet cutoff:
\begin{eqnarray}
E_T^{(0)} \sim {\hbar^2 \over m \ell^2} \,.
\end{eqnarray}
There will also be power-law scaling violations that give corrections to the
binding energies.  The leading corrections scale as
$\kappa^{(n)} \ell$, where $\kappa^{(n)}$ is the binding wave number of the
Efimov state, so they go to zero as the binding energy $E_T^{(n)}$ of the
Efimov state goes to 0.  Thus the Efimov spectrum satisfies
\begin{eqnarray}
E_T^{(n)} &\longrightarrow & \left( e^{-2 \pi /s_0}  \right)^{n-n_*} 
{\hbar^2 \kappa_*^2 \over m}  \,,
\qquad {\rm as \ } n \to \infty {\rm \ \ with \ } a = \pm \infty \,.
\label{B3n-resonant}
\end{eqnarray}
There are infinitely many arbitrarily-shallow Efimov states, with an 
accumulation point at zero energy and an asymptotic discrete scaling 
symmetry with discrete scaling factor $e^{\pi /s_0} \approx 22.7$.

The parameter $\kappa_*$ defined by Eq.~(\ref{B3n-resonant}) can be used 
as the 3-body parameter as an alternative to $\Lambda_0$. 
These parameters differ by a multiplicative 
factor that consists of a numerical constant and an arbitrary 
integer power of $e^{\pi/s_0}$.
An advantage of $\kappa_*$ is that it is defined in terms 
of physical observables. It is defined for any system 
with a large scattering length $a$ and with a short-distance tuning 
parameter that can be used to tune the system to the resonant limit 
$a \to \pm \infty$.    One can determine 
$\kappa_*$ from calculations of the Efimov spectrum at a sequence 
of scattering lengths that approach the resonant limit. 
The definition does not depend on any calculational framework.
In contrast, $\Lambda _0$ is defined in Eq.~(\ref{Lambda-0})
in terms of the Schr\"odinger wave function in hyperspherical coordinates.  
It cannot be determined so straightforwardly in some other frameworks, 
such as effective field theory.

The 2-body parameter $a$ can be defined very simply in terms of physical
observables. According to Eq.~(\ref{sig-2}), 
$a^2$ is simply the limiting value of $d\sigma/d\Omega$ as the energy
approaches zero. The sign of $a$ is determined by whether or not there is a
bound state with binding energy given by Eq.~(\ref{N-Efimov}). 
The 3-body parameter 
$\kappa_*$ is defined by a limiting procedure that involves 
tuning $a$ to $\pm \infty$.  However, it can also be determined 
approximately from the measurement of a single physical observable, 
such as the binding
energy $E_T^{(N)}$ of the Efimov state closest to threshold.
One might be tempted to specify the theory by the
physical observables $a$ and $E_T^{(N)}$ instead of $a$ and $\kappa_*$.  
The disadvantage of this choice is that there
are no other 3-body observable that can be expressed in a simple form in terms
of $a$ and $E_T^{(N)}$.  As we shall see in Section~\ref{sec:uni3}, 
the 3-body parameter $\kappa_*$ has the advantage 
that there are some 3-body observables that are given by simple analytic 
expressions in terms of $a$ and $\kappa_*$.


\subsection{Efimov states near the atom-dimer threshold}
        \label{subsec:EfAD}

While there are infinitely many Efmov trimers in the resonant 
limit $a = \pm\infty$, there are only finitely many for any finite 
value of $a$.  Thus for almost all of these states, 
there must be a critical positive value of $a$ at which the 
bound state appears below the atom-dimer threshold and a
critical negative value of $a$ at which it disappears 
through the 3-atom threshold.  Efimov states near the 
atom-dimer threshold can be understood intuitively 
as 2-body systems composed of an atom and a dimer.

We will denote the critical value of $a$ at which the 
Efimov state appears below the atom-dimer threshold by $a_*$.
For $a$ near $a_*$, the existence of an Efimov trimer
close to threshold produces resonant enhancement of 
atom-dimer scattering.  Thus the atom-dimer scattering length 
$a_{AD}$ must diverge as $a \to a_*$. From Efimov's universal 
formula for $a_{AD}$ in Eq.~(\ref{a12-Efimov}), we can see that 
$a_*$ must satisfy 
$s_0 \ln(a_* \kappa_*) + \beta = (n + {1\over2})\pi$, 
where $n$ is an integer, and that the behavior of $a_{AD}$ 
as $a$ approaches $a_*$ from above must be  
\begin{eqnarray}
a_{AD} \approx {b_0 a_* a \over s_0 (a - a_*)} \,.
\end{eqnarray}
Thus if $a-a_*$ is small compared to $a_*$, 
the atom-dimer scattering length is large compared to $a$. 

We can exploit the universality of the 2-body systems 
with large scattering lengths to deduce some properties of the
shallowest Efimov state when it is close to the atom-dimer threshold.
In this case, the 2-body system consists of an atom
of mass $m$ and a dimer of mass $2m$.
We denote the binding energy of the shallowest Efimov state by
$E_T^{(N)}$.  The analog of the universal formula in Eq.~(\ref{B2-uni}) 
is obtained by replacing the reduced mass $m/2$ of the atoms 
by the reduced mass $2m/3$ of the atom and dimer.
Thus the binding energy relative to the 3-atom threshold 
can be approximated by
\begin{eqnarray}
E_T^{(N)}  \approx E_D + {3 \hbar^2 \over 4m a_{AD}^2} \,.
\label{B3approxa}
\end{eqnarray}
The errors in this approximation scale as $(a-a_*)^3$.

We can also use universality to deduce the wave function 
of the Efimov trimer in the limit $a_{AD} \gg a$.  
The Schr\"odinger wave function  can be expressed as the sum of
three Faddeev wave functions as in Eq.~(\ref{Psi-Faddev}). 
In the limit $r_{1,23} \gg r_{23}$, 
the first Faddeev wave function should have the form
\begin{eqnarray}
\psi^{(1)}({\bm r}_{23}, {\bm r}_{1,23}) 
\approx \psi_{AD}(r_{1,23}) \, \psi_D (r_{23}) \,, 
\end{eqnarray}
where $\psi_D(r)$ is the dimer wave function given in Eq.~(\ref{psi-D}) 
and $\psi_{AD}(r)$ is the analogous universal wave function 
for a shallow bound state consisting of two particles 
with large positive scattering length $a_{AD}$:
\begin{eqnarray}
\psi_{AD}(r) = {1 \over r} e^{-r/a_{AD}} \,, 
\label{psi2-AD}
\end{eqnarray}
Expressed in terms of hyperspherical coordinates,
this Faddeev wave function has the form
\begin{eqnarray}
&& \psi(R,\alpha_1) 
\longrightarrow  { 2 \over \sqrt{3} R^2 \sin(2 \alpha_1)} 
\exp \left( - {\sqrt{2} R \sin \alpha_1 \over a} 
- {\sqrt{3} R \cos \alpha_1 \over \sqrt{2} a_{AD}} \right) \,.
\label{Faddeev:AD}
\end{eqnarray}
Most of the support of the probability density $|\Psi|^2$ 
is concentrated in the region in which the hyperradius is very large,
$R \sim a_{AD}$, and one of the three   hyperangles is very small,
$\alpha_i \ll 1$.  The mean-square hyperradius can be calculated 
easily when $a_{AD} \gg a$:
\begin{eqnarray}
\langle R^2 \rangle^{(N)} \approx \mbox{$1 \over 3$} a_{AD}^2 \,. 
\end{eqnarray}
This result can be obtained more easily simply by using the 
universal atom-dimer wave function in Eq.~(\ref{psi2-AD})
and the approximate expression $R^2 \approx {2\over3} r^2$
for the hyperradius.

The dependence of the Faddeev wave function in Eq.~(\ref{Faddeev:AD})
on the hyperangle $\alpha_1$ is compatible 
with the expression for the hyperangular wave function
in Eq.~(\ref{phin-sol}) that was derived from the hyperangular 
eigenvalue equation.   Using the asymptotic expression 
in Eq.~(\ref{lambda0-largeR}) for the lowest channel eigenvalue
$\lambda_0(R)$, the first Faddeev wave function 
in the region $R \gg a$ is predicted to have the form
\begin{eqnarray}
\psi(R,\alpha_1) \approx R^{-5/2} f_0(R) 
{\sinh \left[ ( \mbox{$\pi \over 2$} - \alpha_1) \sqrt{2} R/a \right]
        \over \sin(2 \alpha_1)} \,.
\label{psi1-0}
\end{eqnarray}
In both Eqs.~(\ref{Faddeev:AD}) and (\ref{psi1-0}),
the leading dependence on $\alpha_1$ for $R \gg  a$,
aside from the factor $1/\sin(2 \alpha_1)$, comes from an
exponential factor $\exp(- \sqrt{2} R \alpha_1/a)$.  
We can read off an approximate expression for the hyperradial 
wave function $f_0(R)$ by comparing Eqs.~(\ref{Faddeev:AD}) 
and (\ref{psi1-0}):
\begin{eqnarray}
f_0(R) \approx {4 \over \sqrt{3}} R^{1/2} 
\exp \left( - \mbox{$\pi \over 2$} \sqrt{2} R/a 
        - (\mbox{$3 \over 2$})^{1/2} R/a_{AD} \right) \,.
\end{eqnarray}


%
%

\section{Universality for Three Identical Bosons}
        \label{sec:uni3}

In this section, we describe the universal aspects of the
3-body problem for three identical bosons with large
scattering length in the simple case where the effects of deep 2-body
bound states are negligible. The changes in the universal properties 
due to the effects of deep 2-body
bound states are described in Section~\ref{sec:deep}.


\subsection{Discrete scaling symmetry}
        \label{sec:discrete}

In the 2-body problem in the scaling limit, 
the continuous scaling symmetry in Eq.~(\ref{scaling-1})
is a trivial reflection of the fact that the scattering length $a$ is the only 
scale in the problem. 
There is also a continuous scaling symmetry for the 3-body problem in the
sectors with total angular momentum quantum number $L \geq 1$.
However, in the $L=0$ sector, there are logarithmic scaling violations 
that introduce a second scale.  A convenient choice for the second scale 
is the wave number $\kappa_*$ defined  in Eq.~(\ref{B3-resonant})
by the spectrum of Efimov states in the resonant limit.  
We can of course define a trivial continuous 
scaling symmetry by rescaling $\kappa_*$ as well as $a$.  
Remarkably, however, there is also a nontrivial discrete scaling symmetry 
in which $\kappa_*$ is held fixed while $a$ and kinematic variables, 
such as the energy $E$, are scaled by appropriate integer powers 
of the discrete scaling factor $\lambda_0 = e^{\pi / s_0} \approx 22.7$:
\begin{eqnarray}
\kappa_* & \longrightarrow & \kappa_* \,,
\qquad
a  \longrightarrow  \lambda_0^{m} a \,,
\qquad
E  \longrightarrow  \lambda_0^{-2m} E \,,
\label{dssym}
\end{eqnarray}
where $m$ is any integer.  Under this symmetry, observables
such as binding energies and cross sections,
scale with the integer powers of $\lambda_0$
suggested by dimensional analysis.

The discrete scaling symmetry strongly constrains the dependence 
of the observables on 
the parameters $a$ and $\kappa_*$ and on kinematic variables. 
For example,  the scaling of the atom-dimer cross section is 
$\sigma_{AD} \to \lambda_0^{2m} \sigma_{AD}$.  
The discrete scaling symmetry constrains its dependence 
on $a$, $\kappa_*$, and the energy $E$: 
\begin{eqnarray}
\sigma_{AD} (\lambda_0^{-2m} E; \lambda_0^{m} a, \kappa_*)
= \lambda_0^{2m}  \sigma_{AD} (E; a, \kappa_*) \,,
\label{sigAD:dss}
\end{eqnarray}
for all integers $m$.
At $E=0$, the cross section is simply $\sigma_{AD} = 4 \pi | a_{AD} |^2$,
where $a_{AD}$ is the atom-dimer scattering length.
The constraint in Eq.~(\ref{sigAD:dss}) implies that $a_{AD}$ 
must have the form
\begin{eqnarray}
a_{AD} 
= f \big( 2 s_0 \ln(a  \kappa_*) \big) \, a \,,
\end{eqnarray}
where $f(x)$ is a periodic function with period $2 \pi$.
As another example, the binding energies 
of the Efimov trimers scale as  $E_T^{(n)} \to \lambda_0^{-2m} E_T^{(n)}$.
The constraints  of the discrete scaling symmetry are more intricate
in this case, because it maps each branch of
the Efimov spectrum onto another branch.  
The dependence of the binding energies on $a$ and $\kappa_*$
must satisfy
\begin{eqnarray}
E_T^{(n)}(\lambda_0^{m} a, \kappa_*)  = 
\lambda_0^{-2m} E_T^{(n - m)}(a,\kappa_*) \,.
\end{eqnarray}
This implies that the binding energies for $a>0$ have the form
\begin{eqnarray}
E_T^{(n)}( a, \kappa_*)  = F_n \big(2s_0 \ln(a \kappa_*) \big) 
 {\hbar^2 \kappa_*^2 \over m} \,,
\end{eqnarray}
where the functions $F_n(x)$ satisfy
\begin{eqnarray}
F_n(x + 2 m \pi)  = \left( e^{-2 \pi/s_0} \right)^m F_{n-m}(x).
\end{eqnarray}
The functions $F_n(x)$ must also have smooth limits as $x \to \infty$:
\begin{eqnarray}
F_n(x )   \to  \left( e^{ -2\pi/s_0} \right)^{n-n_*} 
\qquad {\rm as} \;  x \to \infty \,.
\end{eqnarray}

In the 3-body problem, it is again convenient to introduce 
the energy variable $K$ defined by Eq.~(\ref{K-def}).
For a given value of $\kappa_*$,
the possible low-energy 3-body states in the scaling limit 
can be identified with  points in the $(a^{-1}, K)$ plane. 
It is also convenient to introduce the polar coordinates 
$H$ and $\xi$ defined by Eqs.~(\ref{Hxi-def}).
The discrete scaling transformation in Eqs.~(\ref{dssym})
is simply a rescaling of the radial variable with 
$\kappa_*$ and $\xi$ fixed:  $H \to \lambda_0^{-m} H$.

\begin{figure}[htb]
\bigskip
\centerline{\includegraphics*[width=8cm,angle=0]{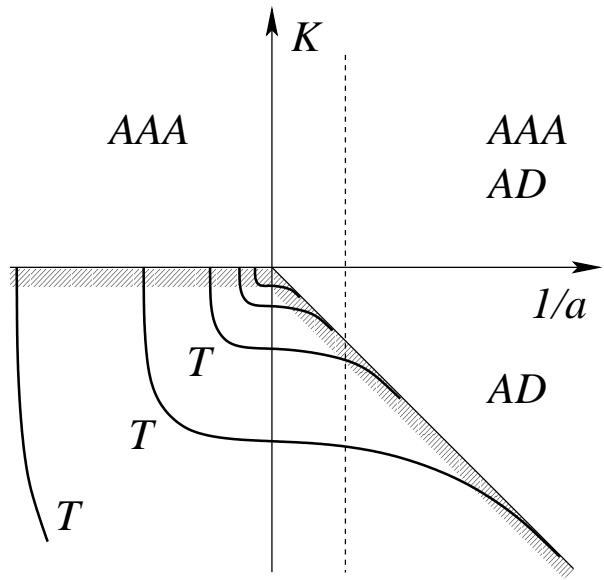}}
\medskip
\caption
{The $a^{-1}$--$K$ plane for the 3-body problem. The allowed regions for
3-atom scattering states and atom-dimer scattering states are
labelled $AAA$ and $AD$, respectively. The heavy lines labeled $T$
are three of the infinitely many branches of Efimov states. 
The cross-hatching indicates the threshold for scattering states.
The axes labelled $1/a$ and $K$ are actually $H^{1/4} \cos\xi$
and $H^{1/4} \sin\xi$.
}
\label{fig:3body}
\end{figure}

The $a^{-1}$--$K$ plane for three identical bosons in the scaling limit 
is shown in Fig.~\ref{fig:3body}.
The possible states are 3-atom scattering states,
atom-dimer scattering states, and Efimov trimers.
The regions in which there are 3-atom scattering states and 
atom-dimer scattering states are labelled $AAA$ and $AD$, respectively.
The threshold for scattering states is indicated by the hatched area.
The Efimov trimers are represented by the heavy
lines below the threshold, some of which are labelled
$T$. There are infinitely many branches of Efimov trimers, 
but only a few are shown.
They intercept the vertical axis at the points
$K = - (e^{-\pi/s_0} )^{n-n_*} \kappa_*$.
Although we have labelled the axes $a^{-1}=H\cos\xi$ and
$K=H\sin\xi$, the curves for the binding energies in Fig.~\ref{fig:3body}
actually correspond to plotting $H^{1/4}\sin\xi$ versus $H^{1/4}\cos\xi$.
This effectively reduces the discrete symmetry
factor 22.7 down to $22.7^{1/4} = 2.2$, allowing a greater range of
$a^{-1}$ and $K$ to be shown in the Figure. 
A given physical system has a specific value of the scattering length, 
and so is represented by a vertical line.
The points on the vertical line above the scattering threshold
represent the continuum of 3-atom and atom-dimer scattering states.
The intersections of the vertical line with the lines labelled $T$
represent Efimov trimers.
Changing $a$ corresponds to sweeping the vertical dashed line 
in Fig.~\ref{fig:3body} horizontally across the page. 
The resonant limit corresponds to tuning the vertical line to the $K$ axis.

\begin{figure}[htb]
\bigskip
\centerline{\includegraphics*[width=8cm,angle=0]{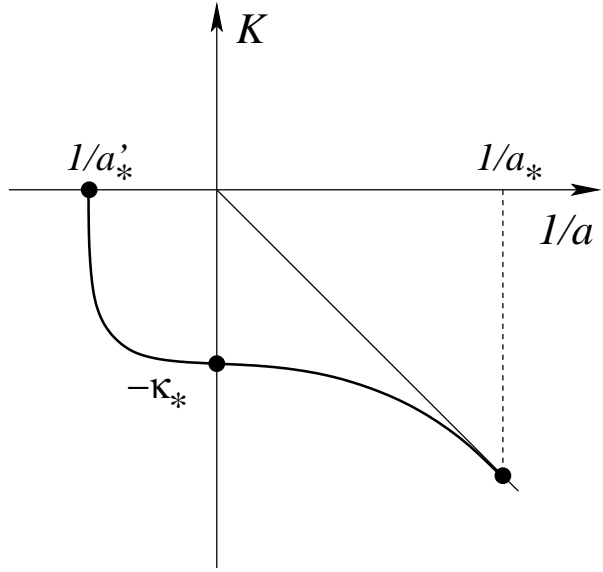}}
\medskip
\caption
{The energy variable $K$ for the branch of Efimov trimers 
labelled by $n=n_*$ as a function of $1/a$. 
In the resonant limit, the binding wave number is $\kappa_*$.
The branch disappears through the atom-dimer threshold at
$a=a_*$ and through the 3-atom threshold at $a = a_*'$.
}
\label{fig:efistar}
\end{figure}

Changing $1/a$ continuously from a large positive value
to a large negative value corresponds to the vertical dashed line in
Fig.~\ref{fig:3body} sweeping from right to left across the 
$1/a = 0$ axis. New Efimov trimers continue to appear at the 
atom-dimer threshold at positive critical values of $1/a$ 
that differ by multiples of $e^{\pi/s_0} \approx 22.7$ 
until there are infinitely many at $1/a = 0$.
For $a<0$, as $1/a$ increases in magnitude, the Efimov trimers
disappear one by one through the 3-atom threshold
at negative critical values of $1/a$ that differ by multiples 
of $e^{\pi/s_0}$.  We will focus on the specific branch of Efimov trimers 
labelled by the integer $n=n_*$, 
which is illustrated in Fig.~\ref{fig:efistar}.
At some positive critical value $a = a_*$, this branch
of Efimov trimers appears at the atom-dimer threshold:
$E_T^{(n_*)}=E_D$.  As $1/a$ decreases, its binding energy 
$E_T^{(n_*)}-E_D$ relative to the atom-dimer threshold increases 
but its binding energy $E_T^{(n_*)}$ relative to the 3-atom threshold
decreases monotonically. As $1/a \to 0$, the binding energy approaches 
a well-defined limit: $E_T^{(n_*)} \to \hbar^2 \kappa_*^2/m$.
As $1/a$ continues to decrease through negative values, 
$E_T^{(n_*)}$ continues to decrease monotonically.
Finally, at some negative critical value $a=a_*'$,
it disappears through the 3-atom threshold.

The most obvious consequence of changing the parameter $\kappa_*$ 
by a multiplicative factor $\lambda$ is to multiply each branch 
of the Efimov trimers
in Fig.~\ref{fig:3body} by a factor $\lambda$ without changing their shapes.  
If the factor $\lambda$ is increased to $e^{\pi/s_0} \approx 22.7$, 
each branch of the Efimov trimers is mapped onto the next branch.  
Not only is the Efimov spectrum identical for $\lambda = 1$ 
and $\lambda = \lambda_0$, but all other 3-body observables are as well.  
Thus in the scaling limit the distinct 3-body systems can be labeled by $a$, 
which can range from $- \infty$ to $+ \infty$, and by $\kappa_*$, 
which is positive and ranges over an interval whose endpoints
differ by a multiplicative factor of $\lambda_0 \approx 22.7$.


\subsection{Efimov's radial law}

In addition to the discrete scaling symmetry, Efimov derived some powerful
constraints on 3-body
observables that he called the {\it radial law} \cite{Efimov79}. 
When expressed in terms
of the polar variables $H$ and $\xi$ defined by Eqs.~(\ref{Hxi-def}), 
an observable is proportional to $H^p$, where the
power $p$ is determined by dimensional analysis
and the coefficient depends on the dimensionless variables 
$H/ \kappa_*$ and $\xi$.
Efimov's radial law strongly constrains the
dependence of the observables on $H/ \kappa_*$ and allows the
calculation of 3-body observables to be reduced to
the calculation of a few universal functions of $\xi$.

The range of the hyperradius $R$ includes two important length scales:
the short-distance scale $\ell$
and the long-distance scale $|a|$.
There are therefore 4 important regions of $R$, and
it is useful to give names to each of these regions:
\begin{itemize}
\item the {\it short-distance region} $R \sim |l|$,
\item the {\it scale-invariant region} $\ell \ll R \ll |a|$,
\item the {\it long-distance region} $R \sim |a|$,
\item the {\it asymptotic region} $R \gg |a|$.
\end{itemize}
The only states whose wave functions can have significant support
in the asymptotic region are
3-atom scattering states and atom-dimer scattering states.
The amplitudes for the incoming scattering states to evolve into outgoing
scattering states are given by the S-matrix.
Efimov's radial law strongly constrains the dependence of the S-matrix on
the radial variable $H$.  It follows from the
approximate scale-invariance of the 3-body problem at length scales $R$
in the scale-invariant region together with the conservation of 
probability in both the short-distance and long-distance regions. 

The simplest states whose wave functions have significant support
in the asymptotic region are atom-dimer scattering states. 
In the center-of-mass frame, such a state can be labelled by the 
wave number ${\bm k}$ of the dimer.  If we project onto a state 
with orbital angular momentum quantum number $L=0$,
the asymptotic form of the  
Faddeev wave function for energy $E = -E_D + 3\hbar^2 k^2/4m$
is the product of the dimer wave function 
$\psi_D(r_{23})= \psi_D(\sqrt{2}R \sin \alpha)$, 
which is given in Eq.~(\ref{psi-D}),
and a scattering wave function in the variable 
$r_{1,23} = \sqrt{3/2} R \cos \alpha$:
\begin{eqnarray}
\psi_{AD}(R,\alpha)  
& \longrightarrow & 
{e^{-\sqrt{2} R \sin \alpha/a}\over R^2 \sin(2 \alpha)}
\left( C e^{i\sqrt{3/2}\, kR\cos \alpha} 
        + D e^{-i\sqrt{3/2}\, kR\cos \alpha} \right) \,,
\label{psi-AD}
\end{eqnarray}
where $C$ and $D$ are arbitrary coefficients.  
The terms with the coefficients $C$ and $D$ represent 
outgoing and incoming atom-dimer scattering states, respectively.

The other states whose wave functions have significant support
in the asymptotic region are 3-atom scattering states. 
Such a state can be labeled by the wave numbers ${\bm k}_1$, ${\bm k}_2$, 
and ${\bm k}_3$ of the three atoms.  In the center-of-mass frame, 
the wave numbers satisfy ${\bm k}_1 + {\bm k}_2 + {\bm k}_3 = 0$ 
and the energy is $E = \hbar^2(k_1^2 + k_2^2 + k_3^2) / 2m$.  
If we further constrain the 3-atom state to have a very low energy $E$, 
total orbital-angular-momentum quantum number $L=0$, 
and no subsystem angular momentum, 
the asymptotic wave function becomes relatively simple.
We can identify a very low-energy 3-atom scattering state with 
a state in the lowest hyperspherical potential 
in Fig.~\ref{fig:hap} that asymptotes to 0 as $R \to \infty$.
This is the lowest potential $V_0(R)$ if $a<0$
and the second lowest potential $V_1(R)$ if $a>0$.  
In either case, the eigenvalue $\lambda_n(R)$ asymptotes to 4.
The corresponding hyperangular wave function is 
$\phi(R,\alpha) = \sin (2 \alpha)$.  
Thus the asymptotic Faddeev wave function is independent of $\alpha$.  
The solution to the hyperradial equation in Eq.~(\ref{aha:F}) with $V_n =0$ 
can be expressed in terms of Bessel functions.  
The asymptotic form of the Faddeev wave function for a 3-atom scattering 
state with total energy $E = \hbar^2 \kappa^2/2m$ is
\begin{eqnarray}
\psi_{AAA}(R,\alpha) \longrightarrow 
{1 \over R^2} \left[ F J_2(\kappa R) + G J_{-2}(\kappa R) \right] \,,
\label{psi-AAA}
\end{eqnarray}
where $F$ and $G$ are arbitrary coefficients.   
The terms with the coefficients $F$ and $G$ represent 
outgoing and incoming 3-atom scattering states, respectively.

The evolution of an incoming scattering state to an outgoing scattering state
is described by a $2\times2$ symmetric unitary matrix  with entries
$S_{AD,AD}$, $S_{AD,AAA}=S_{AAA,AD}$, and $S_{AAA,AAA}$.
These S-matrix elements are nontrivial only in those regions 
of the $a^{-1}-K$ plane where
the corresponding asymptotic states are kinematically allowed.
In the quadrant $0 < \xi < {1\over2}\pi$, both $AAA$ and $AD$ are allowed.
In the quadrant ${1\over2}\pi < \xi < \pi$, only $AAA$ is allowed, 
so $S_{AAA,AAA}$ is the only nontrivial S-matrix element.
In the wedge $-{1\over4}\pi < \xi < 0$, only $AD$ is allowed, 
so $S_{AD,AD}$ is the only nontrivial S-matrix element.
Finally, in the wedge $-\pi < \xi < -{1\over4}\pi$, 
neither $AAA$ nor $AD$ is allowed so the S-matrix is completely trivial.

An essential ingredient in the derivation of Efimov's radial law is the
approximate scale invariance of the 3-body system for hyperradius $R$
in the scale-invariant region $\ell \ll R \ll |a|$.
In this region, the equation for the hyperradial wave function
reduces to Eq.~(\ref{eigenR:si}).  If $R$ is sufficiently deep into the
scale-invariant region, the energy eigenvalue can be neglected.
The most general solution in Eq.~(\ref{f0:R}) then has the simple form
\begin{eqnarray}
f (R) = R^{1/2} \left[A e^{is_0 \ln (H R)} + B e^{-is_0 \ln (H R)}
\right] \,,
\label{f-general}
\end{eqnarray}
where $H$ is an arbitrary variable with dimensions of wave number.
The Faddeev wave function in Eq.~(\ref{psi1-0}) is therefore 
\begin{eqnarray}
\psi_{\rm hw}(R,\alpha) &=& 
{\sinh \left[ ( \mbox{$\pi \over 2$} - \alpha_1) R/a \right]
        \over R^2 \sin(2 \alpha_1)}
\left[ A e^{is_0 \ln (H R)} + B e^{-is_0 \ln (H R)} \right] \,,
\label{psi-hw}
\end{eqnarray}
where $A$ and $B$ are arbitrary coefficients.  
The terms with the coefficients $A$ and $B$ represent 
an outgoing hyperradial wave and an incoming hyperradial wave, 
respectively.

In the short-distance region $R \sim \ell$, the  
wave function becomes very complicated. However, as shown by Efimov,
we can get surprisingly strong constraints on the S-matrix
just by using  the conservation of probability. 
Efimov assumed implicitly that the 2-body problem had no deep 
bound states with binding energies of order $\hbar^2/m \ell^2$ or larger. 
Thus the 2-body potential supports no bound states 
at all if $a<0$ and only the shallow bound state with
binding energy $E_D=\hbar^2/ma^2$ if $a>0$. If there were any deep
2-body bound states, some of the probability in an incoming 
hyperradial wave could flow in to short
distances and then emerge through scattering states of an atom and a
deep 2-body bound state.  Given the assumption that there
are no deep 2-body bound states, the probability in the incoming 
hyperradial wave must be totally
reflected at short distances. Thus the amplitudes
$A$ and $B$  of the incoming and
outgoing hyperradial waves in Eq.~(\ref{f-general}) must be equal
in magnitude, so they must satisfy 
\begin{eqnarray}
A = - e^{2 i\theta_*} B
\label{A/B}
\end{eqnarray}
for some angle $\theta_*$.  
The hyperradial wave function must therefore have the form
\begin{eqnarray}
f (R) \approx \sqrt{H R} \sin \left[ s_0 \ln (cH R) + \theta_* \right] \,,
\label{f-sin}
\end{eqnarray}
where $c$ is an arbitrary constant and
the angle $\theta_*$ is determined by the boundary condition on 
$f(R)$ at short distances. The angle $\theta_*$ can equally well be 
specified by giving the value of the
logarithmic derivative $R_0 f' (R_0)/ f (R_0)$ at a point $R_0$ in the
scale-invariant region. It can be expressed as 
\begin{eqnarray}
\theta_* &=& - s_0 \ln (cH R_0) 
+ \, {\rm arccot} \left[ { 1 \over s_0} \left(
R_0 {f' (R_0) \over f(R_0)} - {1 \over 2}\right) \right] \,.
\end{eqnarray}
All the dependence on $H$ is in the logarithmic term. We can write this
expression in the more compact form
\begin{eqnarray}
\theta_* = - s_0 \ln (cH / \Lambda_0) \,,
\label{theta-star}
\end{eqnarray}
where $\Lambda_0$ is the 3-body parameter 
introduced in Eq.~(\ref{B3-resonant}).

In the long-distance region $R\sim |a|$, the solution for the wave function 
again becomes very complicated. However, no matter how complicated the
wave function, it must conserve probability.
Thus if we identify the asymptotic states at distances $R \ll |a|$ 
and $R \gg |a|$, the evolution of the wave function through 
the long-distance region $R \sim |a|$ can be described by a unitary matrix.
We denote the asymptotic states with probability flowing into the 
long-distance region $R \sim |a|$ by kets $|i \; {\rm in} \rangle$, 
$i=1, 2, 3$, and the asymptotic states with probability
flowing out of that region by kets $|i \; {\rm out}\rangle$, $i = 1, 2, 3$.  
The probability can flow into this region either from the 
scale-invariant region $R \ll |a|$, where the states are hyperradial waves
with Faddeev wave function given by Eq.~(\ref{psi-hw}),
or from the asymptotic region $R \gg |a|$, where the states are 3-atom 
or atom-dimer scattering states, with asymptotic Faddeev wave functions 
given by Eqs.~(\ref{psi-AD}) or (\ref{psi-AAA}), respectively.  
We will denote the states associated with hyperradial waves by 
$| 1 \; {\rm in}\rangle$ and $| 1 \; {\rm out} \rangle$, 
the asymptotic atom-dimer scattering state by $| 2 \; {\rm in} \rangle$ 
and $| 2 \; {\rm out} \rangle$, and the 3-atom scattering states by 
$| 3 \; {\rm in} \rangle$ and $| 3 \; {\rm out} \rangle$.  
It is convenient to normalize these states so that their probability
fluxes are all the same.  The states associated with the
Faddeev wave functions in Eqs.~(\ref{psi-hw}), 
(\ref{psi-AD}), and (\ref{psi-AAA}) can be expressed as
\begin{subequations}
\begin{eqnarray}
| {\rm hw} \rangle & = & A |1 \; {\rm in} \rangle  + B | 
  1 \; {\rm out} \rangle \,,
\\
| AD \rangle      & = & C |2 \; {\rm out} \rangle + D | 
  2 \; {\rm in}  \rangle \,,
\\
| AAA \rangle     & = & F |3 {\; \rm out} \rangle + G | 
  3 \; {\rm in}  \rangle \,.
\end{eqnarray}
\end{subequations}
Note that the outgoing hyperradial wave is an incoming asymptotic state
$| 1  \; {\rm in} \rangle$ as far as the 
long-distance region $R \sim |a|$ is concerned.
The amplitudes for the incoming asymptotic states to evolve 
through the long-distance region into the outgoing asymptotic states 
is described by a unitary $3\times 3$ matrix $s$:
\begin{eqnarray}
s_{ij} = \langle i \; {\rm out} | \hat U | j \; {\rm in} \rangle \,,
\end{eqnarray}
where $\hat U$ is the evolution operator that evolves a wave function
through the long-distance region $R \sim |a|$ over an
arbitrarily large time interval.  Time-reversal invariance implies that
$s$ is a symmetric matrix.

The matrix $s$ depends 
on the interaction potential only through the scattering length $a$.
It is also a function of the energy variable $K$ defined in Eq.~(\ref{K-def}).
Because the matrix  $s_{ij}$ is dimensionless, it can depend on the
variables $K$ and $a$
only through the dimensionless combination $Ka$ or equivalently the
angular variable $\xi$ defined in Eq.~(\ref{Hxi-def}).
In general, the symmetric unitary $3\times3$ matrix $s_{ij}$ is 
determined by 6 real-valued functions of $\xi$,
but it has a much simpler form in some angular regions.
In the region ${1\over2} \pi < \xi < \pi$, there are no $AD$ 
scattering states so the only nontrivial entries of  $s_{ij}$ are 
for indices 1 and 3.  Thus $s_{ij}$ reduces to a symmetric unitary
$2\times 2$ matrix that is described by three real-valued 
functions of $\xi$. In the region $-{1\over4}\pi < \xi < 0$, 
there are no $AAA$ 
scattering states so the only nontrivial entries of  $s_{ij}$ are 
for indices 1 and 2.  Thus $s_{ij}$ again reduces to a 
$2\times 2$ matrix and  is described by only three real-valued 
functions of $\xi$. In the region $-\pi < \xi < -{1\over4}\pi$ 
where both $AAA$ and $AD$ scattering states are forbidden, 
$s_{ij}$ reduces to a $1\times 1$ matrix which is described by a 
single real-valued function of $\xi$.

We now proceed to write down Efimov's radial law
for the S-matrix for low-energy atom-dimer and 3-atom scattering.
We consider only the lowest hyperspherical channels, so the
S-matrix at a given value of the energy $E > 0$ is a $2 \times 2$ matrix.  
The S-matrix elements can be expressed in terms of the elements of $s$ 
as follows:
\begin{subequations}
\begin{eqnarray}
S_{AD,AD} & = & s_{22} 
+ s_{21} {1 \over 1- e^{2i \theta_*} s_{11}} e^{2i \theta_*} s_{12} \,,
\label{RL:AD}
\\
S_{AD,AAA} & = & s_{23} 
+ s_{21} {1 \over 1- e^{2i \theta_*} s_{11}} e^{2i \theta_*} s_{13} \,,
\label{RL:ADAAA}
\\
S_{AAA,AAA} & = & s_{33} 
+ s_{31} {1 \over 1- e^{2i \theta_*} s_{11}} e^{2i \theta_*} s_{13} \,.
\hspace{0.6cm}
\label{RL:AAA}
\end{eqnarray}
\label{RadLaw}
\end{subequations}
In each case, the first term on the right side is due to reflection from the
long-distance region.  If we expand the second term as a power series 
in $s_{11}$, we can
identify the $n^{\rm th}$ term as the contribution from transmission
through the long-distance region and reflection from the short-distance region
followed by $n$ reflections from the long-distance region and from the 
short-distance region before the final
transmission through the long-distance region to the asymptotic region. 
Note that in the regions of $\xi$ for which either $AAA$ or $AD$ 
asymptotic states are forbidden, the form of $s_{ij}$ implies correctly that 
$S_{AD,AAA} = 0$.  Efimov's radial law provides 
very strong constraints on the dependence of the S-matrix elements
on the radial variable $H$, which
enters only through the angle $\theta_*$ given in Eq.~(\ref{theta-star}).
It also constrains the dependence on the angular variable $\xi$,
which is the argument of the symmetric unitary matrix  $s_{ij}$.  
It implies that the calculation of the S-matrix elements for all 
energies and all values of the parameters $a$ and $\kappa_*$
can be reduced to the calculation of a few universal functions of $\xi$.


\subsection {Binding energies of Efimov states}
        \label{subsec:B3}

The binding energy for each Efimov state is a function of $a$ and 
$\kappa_*$.  The simplest application of Efimov's radial law is 
to reduce the calculation of the binding energies for all 
the Efimov states to the calculation of a single universal function of $\xi$.

Efimov states have binding energies that are less than the natural 
ultraviolet cutoff $\hbar^2/m\ell^2$.  The only adiabatic hyperspherical
potential that is attractive in the region $R \gg \ell$ is the lowest one. 
This potential has a scale-invariant region
where the general solution is the sum of an outgoing hyperradial wave and an
incoming hyperradial wave as in Eq.~(\ref{f-general}).
Bound states occur at energies for which the waves reflected from the scaling
region $R \sim |a|$ come into resonance with the waves reflected from the
short-distance region $R \sim \ell$.  The resonance condition is simply
\begin{eqnarray}
\exp(2 i \theta_*) s_{11} = 1 \,.
\label{resonance}
\end{eqnarray}
Note that this is precisely the condition for the vanishing of the 
denominators in the three radial laws in Eqs.~(\ref{RadLaw}).

In the region $-\pi < \xi < - {1\over4}\pi$, the asymptotic
states $AAA$ and $AD$ are kinematically forbidden. 
Thus we can set $s_{22} = s_{33} =1$ and $s_{12} = s_{13} = s_{23} = 0$.  
The only nontrivial entry of the matrix is $s_{11}$. 
The unitarity of the matrix implies 
\begin{eqnarray}
s_{11} = \exp (i \Delta (\xi)) \,,
\end{eqnarray}
where $\Delta/2$ is the phase shift of a hyperradial wave 
that is reflected from the long-distance region.
Combining this with the resonance condition in Eq.~(\ref{resonance}),
we get
\begin{eqnarray}
2\theta_* +  \Delta (\xi)=0 \mod 2\pi \,.
\label{theta-Delta}
\end{eqnarray}
Using the expression for $\theta_*$ in Eq.~(\ref{theta-star}) and the
definitions for $H$ and $\xi$ in Eq.~(\ref{Hxi-def}), 
we obtain Efimov's equation for the binding energies:
\begin{eqnarray}
E_T  + {\hbar^2 \over m a^2} =
\left(e^{-2 \pi/ s_0} \right)^{n-n_*}
\exp \left[ \Delta ( \xi )/s_0 \right] \frac{\hbar^2\kappa_*^2}{m} \,,
\label{B3-Efimov}
\end{eqnarray}
where the angle $\xi$ is defined by
\begin{eqnarray}
\label{xin-def}
\tan \xi = - (mE_T/\hbar^2)^{1/2} \, a  \,.
\end{eqnarray}
Note that the left side of Eq.~(\ref{B3-Efimov}) is 
proportional to the radial variable $H^2$.
For each integer $n$, there is a solution $E_T^{(n)}$ to Eq.~(\ref{B3-Efimov})
that is continuous over a range of $1/a$ that includes 0.
Thus  the binding energy behaves smoothly 
as the scattering length $a$ passes through $\pm \infty$.
We have absorbed the constant $c$ in Eq.~(\ref{theta-star}) 
into the function $\Delta(\xi)$ so that it satisfies
$\Delta(-{1\over2} \pi) = 0$.
Once the universal function $\Delta(\xi)$ has been calculated, 
the binding energies for all the Efimov states for any values of
$a$ and $\kappa_*$ can be obtained by solving Eq.~(\ref{B3-Efimov}).
The equation is the same for different Efimov states except for the
factor of $(e^{-2 \pi/s_0})^n$ on the right side. 
In the resonant limit $a \rightarrow \pm \infty$,
$\xi \to - {1\over2} \pi$ and $\Delta(\xi) \to 0$,
so Eq.~(\ref{B3-Efimov}) reduces to Eq.~(\ref{B3n-resonant}).
It therefore predicts that the Efimov states have a geometric spectrum 
in the resonant limit.

\begin{figure}[htb]
\centerline{\includegraphics*[width=8cm,angle=0,clip=true]{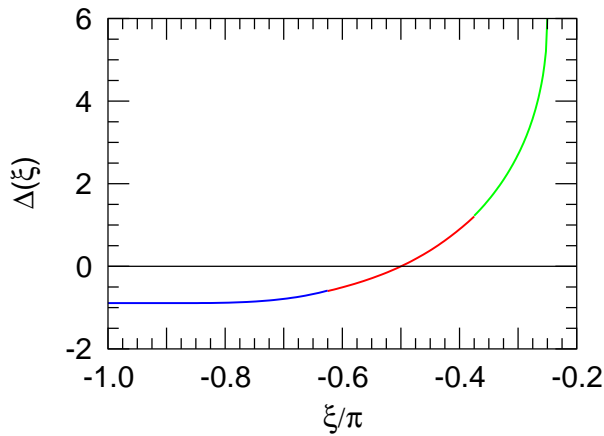}}
\medskip
\caption{The universal function $\Delta(\xi)$ for $-\pi <\xi<-{1\over4}\pi$.}
\label{fig:Delta}
\end{figure}
 
There have been many calculations of the binding energies 
of the Efimov states by solving the Schr{\"o}dinger equation for various
2-body potentials. 
In principle, the function $\Delta(\xi)$ could be 
mapped out by carrying out such calculations for many different 
sequences of potentials that are approaching the scaling limit.
The function $\Delta(\xi)$ can be calculated more directly
using the renormalized zero-range model \cite{FTDA99}
or using effective field theory \cite{BHK99,BHK99b}, because these methods 
allow the range to be set to zero so that one can carry out 
calculations directly in the scaling limit.
In Ref.~\cite{BHK-02a}, the effective field theory of Ref.~\cite{BHK99,BHK99b}
was used to calculate the binding energy $E_T$ of the first few
Efimov trimers numerically as a function of $a$ 
and a 3-body parameter $\Lambda_*$ that 
is defined by the renormalization prescription for an
effective field theory.  The parameter $\Lambda_*$ differs from 
$\kappa_*$ by a multiplicative factor that consists of a 
numerical constant and an arbitrary integer power of $e^{\pi/s_0}$.
The relation can be expressed as 
\begin{eqnarray}
s_0 \ln(\Lambda_*)  \approx s_0 \ln (2.61 \, \kappa_*) 
\mod  \pi \,.
\label{Lamstar-kappa}
\end{eqnarray}
Calculations of the trimer binding energies can be used to determine
$\Delta(\xi)$ over the
entire range of $\xi$.  Simple parametrizations were constructed
that give good approximations for the function $\Delta  (\xi)$ 
in three subsets of the interval $-\pi < \xi < - {1 \over 4} \pi$:
\begin{subequations}
\begin{eqnarray}
\label{expol1}
\xi \in {\textstyle [-{3\pi \over 8},-{\pi \over 4}]:}
&& \Delta=6.04 -9.63x + 3.10x^2  \,, \;
\\[2pt]
\label{expol2}
\xi \in {\textstyle [-{5\pi \over 8},-{3\pi \over 8}]:}
&& \Delta= 2.12y +1.97y^2 + 1.17y^3 \,,
\hspace{0.6cm}
\\[2pt]
\label{expol3}
\xi \in {\textstyle [-\pi,-{5\pi \over 8}]:}
&& \Delta=-0.89+0.28z+0.25z^2 \,,\;
\end{eqnarray}
\label{expol}
\end{subequations}
where the expansion parameters are
\begin{subequations}
\begin{eqnarray}
x&=&(-\mbox{$1\over4$}\pi-\xi)^{1/2} \,,
\label{x-def}
\\
y&=&\mbox{$1\over2$}\pi+\xi \,,
\label{y-def}
\\
z&=&(\pi+\xi)^2 \exp[-1/(\pi+\xi)^2] \,.
\label{z-def}
\end{eqnarray}
\end{subequations}
The function $\Delta(\xi)$ is shown in Fig.~\ref{fig:Delta}.
The parametrizations deviate from the numerical results
for $\Delta(\xi)$ by less than 0.013.  
The discontinuities at $\xi=-{3 \over 8}\pi$ and $\xi=-{5 \over 8}\pi$
are less than 0.016. The expansion variable $x$ defined in 
Eq.~(\ref{x-def}) has a square-root singularity at $\xi=-{1 \over 4}\pi$,
which corresponds to the atom-dimer threshold.
Near this threshold, the 3-body bound state reduces to a 2-body
bound state consisting of an atom and dimer, 
and the square-root singularity follows
from the known analytic behavior of the 2-body problem.  
The expansion variable $z$ defined in Eq.~(\ref{z-def}) 
has an essential singularity at $\xi = - \pi$,
which corresponds to the 3-atom threshold.  
An essential singularity seems to be necessary to reproduce 
the numerical results in this region of $\xi$, but the precise form 
of the essential singularity in Eq.~(\ref{z-def}) is simply empirical.
If an analytic understanding of the form of the 3-atom threshold 
were available, it could be used to construct a better parametrization.

Efimov's universal function $\Delta(\xi)$ has recently been
calculated with a precision of about 12 digits for $a>0$, 
which corresponds to the range 
$-{1\over2} \pi< \xi < -{1\over4} \pi$ \cite{Mohr03}.
The calculation of Ref.~\cite{Mohr03} provides a check on the accuracy 
of the parameterizations in Eqs.~(\ref{expol}). 
The errors in the parameterization of $\Delta (\xi)$
increase slowly from less than 0.002 near $\xi = -{1\over2}\pi$
to about 0.004 at $\xi = -(1.007){1\over4}\pi$, and then increase to 
about 0.012 at $\xi = -{1 \over4} \pi$. 
The precise result for the value of $\Delta(\xi)$ at the point
that corresponds to the atom-dimer threshold is \cite{Mohr-pc}
\begin{eqnarray}
\Delta(-\mbox{$1\over4$} \pi) 
= 6.02730678199 \,.
\label{dDelta}
\end{eqnarray}
The corresponding result from the parameterization 
in Eq.~(\ref{expol1}) is 6.04, which is too large by 0.2\%.  

The results for $\Delta(\xi)$ can be used to determine 
the critical values of the scattering length at which 
Efimov trimers disappear through the atom-dimer threshold
or the 3-atom threshold.  For the branch of Efimov trimers 
labelled by $n=n_*$, whose binding energy
in the resonant limit is $E_T^{(n_*)}  =  \hbar^2 \kappa_*^2/m$,
the critical values $a_*$ and $a_*'$ are illustrated in 
Fig.~\ref{fig:efistar}. The precise result for 
$\Delta(-{1\over4} \pi)$ in Eq.~(\ref{dDelta}) can be used to 
determine the positive critical values of $a$
for which there is an Efimov trimer at the atom-dimer threshold. 
For the branch of Efimov trimers labelled by $n=n_*$, 
the critical value at which $E_T^{(n_*)} = E_D$ is
\begin{eqnarray}
a_*  = 0.0707645086901 \, \kappa_*^{-1} \,.
\label{B3=B2}
\end{eqnarray}
The other critical values are $( e^{\pi/s_0} )^n a_*$, 
where $n$ is an integer.  The value $\Delta(-\pi) \approx -0.89$ 
given in Eq.~(\ref{expol3}) can be used to determine the 
negative critical values of $a$
for which there is an Efimov trimer at the 3-atom threshold.
For the branch of Efimov trimers
labelled by $n=n_*$, the critical value at which $E_T^{(n_*)} = 0$ is
\begin{eqnarray}
a_*' = - 1.56(5) \,\kappa_*^{-1} \,.
\label{B3=0}
\end{eqnarray}
The other critical values are $( e^{\pi/s_0} )^n a_*'$, 
where $n$ is an integer.
The error estimate in Eq.~(\ref{B3=0}) is obtained using an independent
determination of $a_*'$ from the poles in the 3-particle 
elastic scattering amplitude.
In contrast to $a_*$ in Eq.~(\ref{B3=B2}), only a few of digits of precision 
are currently available for $a_*'$.
Comparing Eqs.~(\ref{B3=B2}) and (\ref{B3=0}), we see that 
$a_*'  \approx - 22.0 \;  a_*$.

There is a common misconception in the literature 
that Efimov states must have binding energies that differ 
by multiplicative factors of 515.03.  However, this
ratio applies only in the resonant limit $ a \to \pm \infty$.  
The ratio $E_T^{(n-1)}/E_T^{(n)}$ of the binding energies
of adjacent Efimov trimers can be much smaller than 515 
if $a > 0$ and much larger than 515 if $a < 0$.
The smallest ratios occur at the critical values $( e^{\pi/s_0} )^n a_*$,
where $a_*$ is given in Eq.~(\ref{B3=B2}). 
The accurate results of Ref.~\cite{Mohr03} for the 
binding energies $E_T$ of the first few Efimov states in units of 
$E_D = \hbar^2/ma^2$ are 
\begin{subequations}
\begin{eqnarray}
E_T^{(N)} \hspace{0.3cm} &=& E_D \,,
\\
E_T^{(N-1)} &=& 6.75029015026 \, E_D \,,
\\
E_T^{(N-2)} &=& 1406.13039320 \, E_D \,.
\end{eqnarray}
\end{subequations}
Thus, if $a>0$,  the ratio $E_T^{(N-1)}/E_T^{(N)}$ of the binding energies
for the two shallowest Efimov trimers
can range from about 6.75 to about 208.
The largest ratios occur at the critical values 
$( e^{\pi/s_0} )^n a_*'$, where  $a_*'$ is given in Eq.~(\ref{B3=0}). 
The binding energies $E_T$ of the first few Efimov states 
can be obtained by solving Eq.~(\ref{B3-Efimov}) using the parameterization
in Eq.~(\ref{expol1}) \cite{BHK-02a}:
\begin{subequations}
\begin{eqnarray}
E_T^{(N)} \hspace{0.3cm} &=& 0 \,,
\\
E_T^{(N-1)} &=& 1.09 \times 10^3 \; \hbar^2/ma^2 \,,
\\
E_T^{(N-2)} &=& 5.97 \times 10^5 \; \hbar^2/ma^2 \,.
\end{eqnarray}
\end{subequations}
Thus, if $a<0$,
the ratio $E_T^{(N-1)}/E_T^{(N)}$ of the binding energies 
for the two shallowest Efimov states can range from about 
550 to $\infty$.

For any nonzero value of $a$, 
the binding energy equation (\ref{B3-Efimov}) has a shallowest solution
$E_T^{(N)}$ and infinitely many deeper solutions $E_T^{(n)}$
corresponding to all integers $n < N$.
Thus it predicts that there are infinitely many Efimov states. 
However, this prediction is an artifact of the scaling limit $\ell \to 0$. 
For a system with natural low-energy length scale $\ell$, 
the only states of physical relevance are those whose binding
energies are less than the natural ultraviolet cutoff of order 
$\hbar^2 /m \ell ^2$.  The deeper bound states predicted by 
Eq.~(\ref{B3-Efimov}) are artifacts of the scaling limit.

\begin{figure}[htb]
\bigskip
\centerline{\includegraphics*[width=8.cm,angle=0]{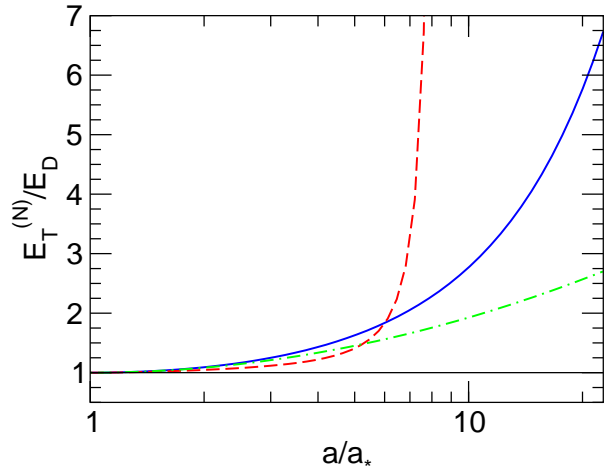}}
\medskip
\caption{
Comparison of the analytic approximations 
in Eqs.~(\ref{B3approxa}) (dashed line) and (\ref{B3approxb}) 
(dash-dotted line)
to the numerical solution for the binding energy of the shallowest Efimov state
(solid line) over the range $a_* < a < e^{\pi/s_0} a_*$.}
\label{fig:B3approx}
\end{figure}

It is useful to have analytic approximations for the Efimov binding energies
for limited ranges of  the scattering length.
One such range is the region of $a$ just above the value $a_*$ 
at which there is an Efimov state at the atom-dimer threshold.
A simple approximation for the binding energy $E_T^{(N)}$ of the 
shallowest Efimov state in this region can be obtained by exploiting 
the fact that the atom-dimer scattering length $a_{AD}$ diverges at $a = a_*$.
The universal result for the binding energy in the limit
$a_{AD} \gg a$ is given in Eq.~(\ref{B3approxa}).
The errors in this approximation scale as $(a-a_*)^3$.
Efimov's universal formula for $a_{AD}$ is given in Eq.~(\ref{a12-Efimov}).
The values of the real numerical constants $b_0$, $b_1$, and $\beta$ 
are given in Section~\ref{sec:atom-dimer} in Eq.~(\ref{a12-explicit}).
One can derive an alternative approximation that also has errors of
order $(a-a_*)^3$ by inserting Efimov's universal formula
for $a_{AD}$ into Eq.~(\ref{a12-Efimov}) 
and expanding in powers of $\ln(a/a_*)$:
\begin{eqnarray}
E_T^{(N)} \approx E_D \big[ 1 + 0.164  \ln^2(a/a_*) \big] \,.
\label{B3approxb}
\end{eqnarray}
In Fig.~\ref{fig:B3approx}, we compare the analytic approximations 
Eqs.~(\ref{B3approxa}) and (\ref{B3approxb}) to the numerical solution 
for the binding energy of the shallowest Efimov state
over the range $a_* < a < e^{\pi/s_0} a_*$.
The large-$a_{AD}$ approximation in Eq. (\ref{B3approxa}) 
has pathological behavior as $a$ approaches at $8.62 a_*$,
because $a_{AD}$ vanishes at that point.
The approximation in Eq. (\ref{B3approxb}) has comparable accuracy 
as $a \to a_*$, and it is better behaved at larger $a$.

Another region in which an analytic approximation for the 
binding energies of the Efimov states can be given is
the resonant region $a \to \pm \infty$.  The binding energies 
in the resonant limit are given in Eq.~(\ref{B3n-n*}).
The leading correction for finite $a$ can be obtained by expanding 
$\Delta(\xi)$ in Eq.~(\ref{B3-Efimov}) to first order around 
$\xi = - {1\over2} \pi$:
\begin{eqnarray}
E_T^{(n)} &=& \left( e^{-2 \pi/s_0} \right)^{n-n_*}
{\hbar^2 \kappa_*^2 \over m}
\left( 1 +  2.11 {(e^{\pi/s_0})^{n-n_*} \over \kappa_* a} \right) \,,
\label{B3approxc}
\end{eqnarray}
where we have used $\Delta'(-{1\over2} \pi) \approx 2.12$
from the parameterization in Eq.~(\ref{expol2}).

It would also be valuable to have an analytic approximation for the 
binding energy of the shallowest Efimov states near the negative 
value $a_*'$ where the Efimov state disappears into the 3-atom continuum.
This would require understanding the analytic behavior 
of the function $\Delta(\xi)$ near the point $\xi = - \pi$.


\subsection{Atom-dimer elastic scattering}
        \label{sec:atom-dimer}

The simplest scattering states in the 3-atom sector are 
atom-dimer scattering states, which we denote by $AD$.
As indicated in Fig.~\ref{fig:3body}, these states are possible only 
if $a>0$. The total energy for an atom and dimer 
in the center-of-mass frame with wave vectors $\pm{\bm k}$ is
\begin{eqnarray}
E = - E_D + {3 \hbar^2 k^2 \over 4m},
\label{E-AD}
\end{eqnarray}
where $E_D=\hbar^2/ma^2$ is the atom-dimer binding energy.
The threshold for atom-dimer scattering is $E= -E_D$ or $k=0$. 
The scattering is elastic up to the dimer-breakup
threshold $E=0$ or $k=2/ (\sqrt{3}a)$.
Above that threshold, the inelastic channel $AD \to AAA$ opens up.

The expression for the differential cross section 
for atom-dimer scattering in terms of the
elastic scattering amplitude $f^{AD}_k (\theta)$ is
\begin{eqnarray}
\frac{d\sigma_{AD}}{d\Omega}=\left| f^{AD}_k (\theta)\right|^2 \,.
\label{dsigma-AD}
\end{eqnarray}
The elastic cross section $\sigma_{AD}(E)$ is obtained by integrating 
over the solid angle of $4\pi$.
The partial wave expansion for the elastic scattering amplitude has the
form
\begin{eqnarray}
f^{AD}_k (\theta) = \sum_{L=0}^\infty {2L+1 \over k \cot \delta^{AD}_L (k) -ik}
 P_L(\cos \theta) \,.
\label{fk-AD}
\end{eqnarray}
The phase shifts $\delta^{AD}_L(k)$ are real-valued 
below the dimer-breakup threshold, 
but they become complex-valued above that threshold.
The expression for the elastic cross section obtained by integrating over
the scattering angle is
\begin{eqnarray}
\sigma_{AD}(E)  = \frac{4\pi}{k^2} \sum_{L=0}^\infty (2 L+1) 
e^{-2 {\rm Im} \delta^{AD}_L(k)} 
\left| \sin \delta^{AD}_L(k) \right|^2  \,.
\label{sigtot-AD}
\end{eqnarray}
Near the atom-dimer threshold, the $L=0$ term
in the partial-wave expansion can be expanded in powers of $k$. The
expansion is conventionally expressed in the form
\begin{eqnarray}
k \cot \delta^{AD}_0(k) = - 1/a_{AD} + \mbox{$1 \over 2$} r_{s,AD} k^2 +
\ldots \,,
\label{rAD-def}
\end{eqnarray}
which defines the atom-dimer scattering length $a_{AD}$ and effective
range $r_{s,AD}$. The optical theorem relates the total cross section,
which is the sum of the elastic cross section in Eq.~(\ref{sigtot-AD})
and the inelastic cross section, to the $\theta \to 0$ limit of the 
elastic scattering amplitude in Eq.~(\ref{fk-AD}):
\begin{eqnarray}
\sigma_{AD}^{\rm (total)}(E)  = \frac{4\pi}{k} \, {\rm Im} \, 
f^{AD}_k(\theta = 0) \,.
\label{optical-AD}
\end{eqnarray}

Since the phase shifts $\delta^{AD}_L(k)$ are dimensionless, they can
depend only on the dimensionless combinations $ka$ and $a \kappa_*$.  
For $L \ge 1$, the phase shifts are insensitive to 3-body interactions at
short distances, and they are therefore universal functions of $ka$ only.  
The $L = 0$ phase shift is sensitive to 3-body interactions at
short distances.  The universal expression for $\delta^{AD}_0(k)$ 
therefore depends on $a \kappa_*$ as well as on $ka$, 
although the dependence on $a \kappa_*$ is strongly constrained by
Efimov's radial law in Eq.~(\ref{RL:AD}). 
For S-wave atom-dimer scattering states, 
the S-matrix element $S_{AD,AD}$ can be expressed 
in terms of the phase shift $\delta^{AD}_0(k)$:
\begin{eqnarray}
S_{AD,AD} = e^{2i \delta^{AD}_0 (k)} \,.
\end{eqnarray}
Below the dimer-breakup threshold, the phase shift is real-valued.
In this region, the $2 \times 2$ submatrix with entries 
$s_{11}$, $s_{12} = s_{21}$, and $s_{22}$ is a unitary matrix.
Using the unitarity of this submatrix, 
we can eliminate $s_{22}$ from the radial law in Eq.~(\ref{RL:AD}) 
and express it in the form given by Efimov \cite{Efimov79}:
\begin{eqnarray}
e^{2i \delta^{AD}_0(k)} = {s_{12} \over s_{12}^*} 
{e^{2i \theta_*} - s_{11}^* \over 1- s_{11} e^{2i \theta_*}} \,.
\label{exp-delta}
\end{eqnarray}

Efimov used his radial law to derive analytically the 
dependence of the atom-dimer scattering length $a_{AD}$
on $a$ and $\kappa_*$.
He showed that $a_{AD}$ must have the form in Eq.~(\ref{a12-Efimov})
where $b_0$, $b_1$, and $\beta$ are universal
constants \cite{Efimov79}.  
This expression diverges at the critical values of $a$ 
given by Eq.~(\ref{B3=B2}) for which there is an Efimov trimer 
at the atom-dimer threshold. The expression in Eq.~(\ref{a12-Efimov}) 
is also consistent with the discrete scaling symmetry
in Eq.~(\ref{dssym}),
because the dimensionless combination $a_{AD}/a$ 
is a periodic function of $\ln(a)$ with period $\pi/s_0$.
We proceed to derive Efimov's formula for $a_{AD}$.
We need the behavior of the matrix $s$ as $k \to 0$.  At $k=0$, 
we must have $s_{12}=0$ and we can set $s_{22} = 1$ by 
a choice of the overall phase of the matrix $s$.  
By the unitarity of $s$, $s_{11}$ must be a pure phase at $k=0$:  
$s_{11} = -e^{2 i\beta^\prime}$ for some angle $\beta^\prime$.  
For small $k$, the most general form 
for the nontrivial entries of $s$ allowed by unitarity is
\begin{subequations}
\begin{eqnarray}
s_{11} &=& 
- e^{2 i \beta^\prime} [1 - 2(b_0 + i b_2) ak + \ldots] \,,
\label{s11-AD}
\\
s_{12} &=& 
(4 b_0 ak)^{1/2} e^{i \beta^\prime} [1 
- (2 b_3 + i b_1 + i b_2) ak + \ldots] \,,
\label{s12-AD}
\\
s_{22} &=& 1 - 2(b_0 + i b_1) ak + \ldots \,,
\label{s22-AD}
\end{eqnarray}
\label{s-AD}
\end{subequations}
where $b_0$, $b_1$, $b_2$, $b_3$, and $\beta'$ are real constants and 
$b_0 > 0$. Inserting these expressions for $s_{ij}$ into 
the radial law in Eq.~(\ref{RL:AD}),
we find that the S-matrix element for atom-dimer scattering 
near the atom-dimer threshold is
\begin{eqnarray}
S_{AD,AD}  \longrightarrow  
1 -  2 i \left[ b_1 - b_0 \tan(\theta_* + \beta') \right] \; a k \,,
\qquad {\rm as \ } E \to - E_D \,.
\end{eqnarray}
Identifying this expression with $1 - 2 i a_{AD} k$, 
using the expression for $\theta_*$  in Eq.~(\ref{theta-star}),
and setting $H=\sqrt{2}/a$, 
we obtain Efimov's expression in Eq.~(\ref{a12-Efimov}) with 
$\beta = \beta^\prime - s_0 \ln (\sqrt 2 \, c \kappa_*/\Lambda_0)$.
The real numerical constants $b_1$ and $b_0$ in Eq.~(\ref{a12-Efimov}) 
were first calculated by Simenog and Sinitchenko \cite{Sim81}.
They were also calculated by Bedaque, Hammer, and van Kolck
using an effective field theory \cite{BHK99,BHK99b}.
A more accurate determination of these constants 
is given in Ref.~\cite{BH02}.
The explicit expression for the atom-dimer scattering length is
\begin{eqnarray}
a_{AD} = \big( 1.46-2.15 \tan [s_0 \ln (a \Lambda_*) +0.09] \big) \; a \,,
\label{a12-explicit}
\end{eqnarray}
where $\Lambda_*$ is a 3-body  parameter
that arises naturally in the renormalization of the effective field theory.
It differs from the parameter $\kappa_*$ defined by the Efimov spectrum 
in the resonant limit by a multiplicative factor that is known to only a few
digits of accuracy and is given in Eq.~(\ref{Lamstar-kappa}).
Since most of the phases of the log-periodic functions that appear in  
3-body observables
in the scaling limit have been calculated using effective field theory, 
we will express them in terms of $\Lambda_*$.  The corresponding results
in terms of  $\kappa_*$ can be obtained by making the substitution
\begin{eqnarray}
s_0 \ln(a\Lambda_*)  = s_0 \ln(a\kappa_*) + 0.97 \mod \pi \,.
\label{Lambdastar-kappastar}
\end{eqnarray}

Efimov's radial law can also be used to deduce the functional form of the
effective range $r_{s,AD}$. 
The expansions in Eq.~(\ref{s-AD})
must be extended to third order in $a k$.
After inserting these expansions into Eq.~(\ref{exp-delta}),
the expression for $k \cot \delta^{AD}_0(k)$ can be expanded in powers of $k$
and compared to Eq.~(\ref{rAD-def}).
After taking into account the constraints from the unitarity of $s$,
there is still a linear term in $k$ as well as a quadratic term.
The coefficients of these terms depend on $b_0$, $b_1$, $b_2$, $b_3$,
$\beta'$, and 4 additional constants. However, $k \cot \delta^{AD}_0(k)$ must 
be an analytic
function of $k^2$, so the coefficients of the odd powers of $k$ must vanish.
This implies, for example, that $b_2 = 0$ and $b_3 = b_0/2$.
After taking into account the constraints of analyticity,
we obtain an expression for $r_{s,AD}$ that involves $\beta'$, $b_0$, $b_1$,
and three additional constants.
Simenog et al.~have calculated $a_{AD}^2 r_{s,AD}$ by solving
the Faddeev equation for an interaction with a small but finite range
\cite{Sim81,Sim86}.
Hammer and Mehen have calculated $r_{s,AD}/a$ 
as a function of $a \Lambda_*$ \cite{HM-01a}
using the effective field theory of Ref.~\cite{BHK99,BHK99b}.
Both calculations agree qualitatively. 
A more accurate expression can be obtained by
fitting the calculated atom-dimer scattering phase shifts  
up to the dimer-breakup threshold (see below) 
and deriving $r_{s,AD}$ from this fit.
The resulting expression for $r_{s,AD}$ is \cite{BH02}
\begin{eqnarray}
r_{s,AD} &=& \big( 1.30 - 1.64 \tan[s_0 \ln(a\Lambda_*)+ 1.07 ] 
\nonumber  \\ && \quad
+ 0.53 \tan^2[s_0 \ln(a\Lambda_*)+ 1.07 ] \big) \; a  \,.
\label{r12}
\end{eqnarray}

\begin{figure}[htb]
\bigskip
\centerline{\includegraphics*[width=8.5cm,angle=0]{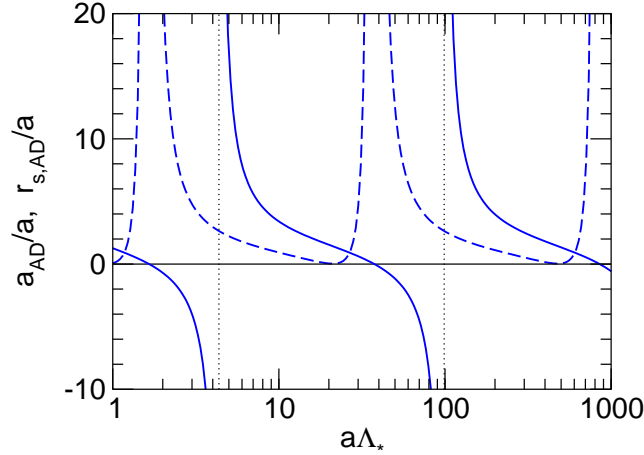}}
\medskip
\caption
{The atom-dimer scattering length $a_{AD}$ (solid line)
 and the effective range $r_{s,AD}$ (dashed line)
 as functions of $a\Lambda_*$.  The vertical dotted lines indicate 
 the location of the poles in $a_{AD}$.}
\label{fig:a12r12}
\end{figure}

In Fig.~\ref{fig:a12r12}, we plot the atom-dimer scattering length $a_{AD}$ 
and effective range $r_{s,AD}$ as functions of $a\Lambda_*$ from 1 to 1000,
which corresponds to a little more than two periods in $\ln(a)$. 
The scattering length $a_{AD}$ diverges at the critical values $a_*^{(n)}$
of the scattering length given in Eq.~(\ref{B3=B2}) 
for which there is an Efimov trimer at the atom-dimer threshold.
The atom-dimer effective range $r_{s,AD}$ behaves smoothly at these points.
The atom-dimer scattering length $a_{AD}$ also has zeroes 
at $a= (e^{\pi/s_0})^n 1.65/\Lambda_*$. The effective range 
$r_{s,AD}$ diverges at this points, but $r_{s,AD}a_{AD}^2$ behaves smoothly.
The atom-dimer effective range achieves its minimum value at 
the points $a=(e^{\pi/s_0})^n 0.94/\Lambda_*$.
Its value at the minimum is consistent with zero 
to within the numerical accuracy.  If it really is zero, it would
indicate that the coefficient of $a$ in Eq.~(\ref{r12}) is the square 
of an expression that is linear in $\tan[s_0 \ln(a\Lambda_*)+ 1.07 ]$.
This would require an additional constraint on the expansion coefficients 
of the matrix $s$ in Eq.~(\ref{s-AD})
beyond those that follow from unitarity and analyticity.
If we assume that there is such a constraint, our best fit
to the expression for  $r_{s,AD}$ is
\begin{eqnarray}
r_{s,AD} =
\big( 1.13 - 0.73 \tan[s_0 \ln(a\Lambda_*)+ 1.07 ] \big)^2 a \,.
\label{r12-square}
\end{eqnarray}

Macek, Ovchinnikov, and Gasaneo \cite{MOG05} have derived an analytic 
expression for the atom-dimer S-wave phase shift $\delta_0^{AD}(k)$
at the dimer-breakup threshold $k=k_0$, where $k_0 a = 2/\sqrt{3}$:
\begin{eqnarray}
\delta_0^{AD}(k_0) = s_0 \ln(a \kappa_*) - \delta_0 + \delta_\infty
+ \arctan { \sin[2s_0 \ln(a \kappa_*) - 2\delta_0]
\over e^{2 \pi s_0} - \cos[2s_0 \ln(a \kappa_*) -2\delta_0]} \,,
\nonumber
\\
\label{deltaAD-analytic}
\end{eqnarray}
where $\delta_0$ and $\delta_\infty$ are real numerical constants.%
\footnote{For later convenience, we have chosen a phase $\delta_0$
that differs from that in Ref.~\cite{MOG05} by ${1\over2}\pi$.}
Using Eq.~(\ref{sigtot-AD}), we can get an analytic expression
for the S-wave contribution to the atom-dimer elastic scattering cross section 
at the dimer-breakup threshold.  Because
$e^{2 \pi s_0} \approx 557$ is large, the phase shift in 
Eq.~(\ref{deltaAD-analytic}) is well approximated by the much 
simpler expression
\begin{eqnarray}
\delta_0^{AD}(k_0) \approx s_0 \ln(a \kappa_*) - \delta_0 + \delta_\infty \,.
\label{deltaAD-approx}
\end{eqnarray}
The corresponding approximation for the $L=0$ term in the
atom-dimer cross section at the dimer-breakup threshold is 
\begin{eqnarray}
\sigma_{AD}^{(L=0)} (E = 0) 
\approx 3 \pi \sin^2[ s_0 \ln(a \kappa_*) - \delta_0 + \delta_\infty ] 
\, a^2 \,.
\label{sigmaAD-approx}
\end{eqnarray}
 
For a complete parameterization of the S-wave phase shift $\delta^{AD}_0 (k)$ 
in the region $ka<2 /\sqrt 3$ below the dimer-breakup threshold, 
we start from the expression
for $\cot \delta^{AD}_0 (k)$ in Eq.~(\ref{exp-delta}).  
Multiplying it by $ka$ and using trigonometric identities,
it can be written in the form 
\begin{eqnarray}
&& ka\,\cot \delta^{AD}_0 (k) = c_1(ka) 
+ c_2(ka) \cot [s_0 \ln (a \Lambda_*) + \phi(ka)] \,.
\label{kacotdel}
\end{eqnarray}
The simple approximation in Eq.~(\ref{deltaAD-approx}) 
for the S-wave phase shift at the dimer-breakup threshold 
requires the functions $c_1(ka)$ and $c_2(ka)$ 
to have specific values at $k_0 a = 2/\sqrt{3}$:
\begin{eqnarray}
c_1(k_0a) \approx 0 \,,
\qquad
c_2(k_0a) \approx 2/\sqrt 3 \,.
\label{c1-c2}
\end{eqnarray}
The functions $c_1(ka)$, $c_2(ka)$, and $\phi(ka)$ have been determined 
over the entire range $0<ka< 2 /\sqrt 3$ \cite{BH02} by calculating the 
phase shifts $\delta^{AD}_0(k)$ numerically using the effective field theory
of Ref.~\cite{BHK99,BHK99b}. 
The results can be parametrized as \cite{BH02}
\begin{subequations}
\begin{eqnarray}
c_1(ka)&=& -0.22 + 0.39 \, k^2 a^2 -0.17 \, k^4 a^4 \,,
\\
c_2(ka)&=& 0.32 + 0.82 \, k^2 a^2 -0.14 \, k^4 a^4 \,,
\\
\phi(ka)&=& 2.64 -0.83 \,k^2 a^2 + 0.23 \, k^4 a^4 \,.
\label{kcot-par:phi}
\end{eqnarray}
\label{kcot-par}
\end{subequations}
The approximate constraints in Eqs.~(\ref{c1-c2}) have been incorporated 
into these parameterizations.

We now discuss the qualitative behavior of the expression for 
$k\cot \delta^{AD}_0(k)$ in Eq.~(\ref{kacotdel}). 
As $a \Lambda_*$ varies with $\xi$ fixed,
the cotangent ranges from $- \infty$ to $+ \infty$. Thus for fixed
$ka$, $k\cot \delta^{AD}_0(k)$ ranges over all possible value from 
$-\infty$ to $+ \infty$. 
A divergence in $k\cot \delta^{AD}_0(k)$ implies a zero
in the S-wave contribution to the differential cross section. At those values
of $k$, the cross section is dominated by the higher partial waves.  
The S-wave contribution vanishes at the atom-dimer threshold $k = 0$
if $a \Lambda_* = (e^{\pi/s_0})^n \times 1.65$ 
and it vanishes at the dimer-breakup threshold $ka = 2/ \sqrt 3$ 
if $a \Lambda_* = (e^{\pi/s_0})^n \times 0.15$. 
A zero in $k\cot \delta^{AD}_0(k)$ implies that the S-wave contribution to the 
differential cross section $d\sigma/d \Omega$ saturates its unitary bound 
of $1/k^2$. The unitarity bound is saturated in the limit $k \to 0$ 
if $a\Lambda_* = (e^{\pi/s_0})^n \times 4.34$, 
which corresponds to the critical values where there is an Efimov state 
at the atom-dimer threshold.
The unitarity bound is saturated at the dimer-breakup threshold 
$ka = 2/\sqrt 3$ if $a \Lambda_* = (e^{\pi/s_0})^n \times 15.7$.

\begin{figure}[htb]
\bigskip
\centerline{\includegraphics*[width=8.5cm,angle=0]{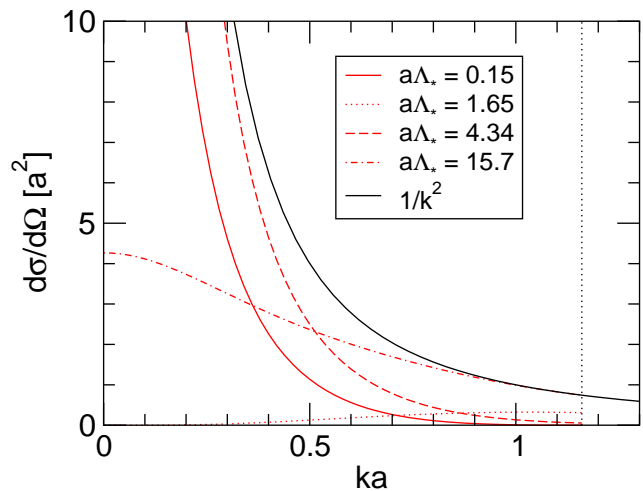}}
\medskip
\caption
{The S-wave contribution to the differential cross section for atom-dimer
scattering in units of $a^2$ as a function of $ka$ for $a\Lambda_*=
0.15,1.65,4.34$, and 15.7. The black solid line is the unitarity bound
$1/k^2$. The vertical dotted line is the dimer-breakup threshold.}
\label{fig:dsigdomS}
\end{figure}

In Fig.~\ref{fig:dsigdomS}, we plot the S-wave contribution to the
differential cross section for atom-dimer scattering as a function
of $ka$ for the four special values 
$a\Lambda_* = 0.15$, 1.65, 4.34, and 15.7 described above. 
The heavy solid line is the unitarity bound $1/k^2$. 
Note that the shape of the differential cross section 
varies dramatically with $a\Lambda_*$.

Using their result in Eq.~(\ref{deltaAD-analytic}) for the 
atom-dimer S-wave phase shift at the 
dimer-breakup threshold, Macek, Ovchinnikov, and Gasaneo \cite{MOG05} 
have deduced the $3 \times 3$ matrix $s$ whose entries appear in 
Efimov's radial laws in  Eq.~(\ref{RadLaw}).
At $E=0$, the nonzero entries of the matrix are
\begin{subequations}
\begin{eqnarray}
s_{11} & = & e^{-2 \pi s_0} e^{-2 i \delta_0}  \,,
\label{s11-3A}
\\
s_{12} & = & \sqrt{1 - e^{-4 \pi s_0}} \, e^{i (\delta_\infty - \delta_0)} \,,
\\
s_{22} & = & -e^{-2 \pi s_0} e^{2 i \delta_\infty}  \,,
\\
s_{33} & = & 1  \,.
\end{eqnarray}
\label{s-E0}
\end{subequations}
We have set $s_{33} = 1$ by the
choice of an overall phase  in the matrix $s$.
Since $e^{2 \pi s_0} \approx 557$ is large,
the diagonal entries $s_{11}$ and $s_{22}$ are very small.
This implies that the lowest hyperspherical potential $V_0(R)$ 
is almost {\it reflectionless} at $E=0$.  
If it were exactly reflectionless, an incoming atom-dimer scattering state 
with $E=0$ would be almost completely transmitted 
through the long-distance region $R \sim a$ 
into an incoming hyperradial wave in the scale-invariant region.
Similarly, an outgoing hyperradial wave  
would be almost completely transmitted through the long-distance region
into an atom-dimer scattering state.


\subsection{Three-body recombination}
        \label{sec:3BR}

Three-body recombination is a process in which three atoms collide to form
a diatomic molecule and an atom. The energy released 
by the binding energy of the molecule goes into the
kinetic energies of the molecule and the recoiling atom.
If the scattering length $a$ is negative, 
the molecule can only be a deep (tightly-bound) diatomic molecule 
with binding energy of order $\hbar^2/m \ell^2$ or
larger. However, if $a$ is positive and unnaturally large ($a \gg \ell$),
the molecule can be the shallow dimer with binding
energy $E_D = \hbar^2/ma^2$. Three-body recombination into deep 
molecules will be discussed in Section~\ref{sec:deep}.
In this section, we assume there are no deep molecules.
We therefore assume $a>0$ and focus on 3-body
recombination into the shallow dimer. 

The 3-body recombination rate depends on the momenta of
the three incoming atoms. If their momenta are sufficiently small, 
the dependence on the momenta can be neglected,
and the recombination rate reduces to a constant. The {\it recombination
event rate constant} $\alpha$ is defined such that the number of
recombination events per time and per volume in a gas of cold
atoms with number density $n_A$ is $\alpha n^3_A$. 
The resulting rate of decrease in the number of atoms 
with energies small compared to $\hbar^2/ma^2$ 
depends on whether the three atoms are in a gas or a
Bose-Einstein condensate:
\begin{subequations}
\begin{eqnarray}
{d \ \over dt} n_A
& = & - 3 \alpha n^3_A \hspace{1cm} {\rm (gas)} \,,
\label{dnA-gas}
\\
& = & - \mbox{$1 \over 2$} \alpha n_A^3 \hspace{1cm} {\rm (BEC)} \,.
\label{dnA-BEC}
\end{eqnarray}
\label{dnA-3br}
\end{subequations}
In Eq.~(\ref{dnA-gas}), the factor of 3 accounts for the three 
low-energy atoms lost per recombination
event. In a Bose-Einstein condensate, the three atoms must
all be in the same quantum state, so the coefficient of $n_A^3$ in 
Eq.~(\ref{dnA-gas}) must be multiplied by 1/3! to
account for the symmetrization of the
wave functions of the three identical particles \cite{KSS-85}. 
The decrease in the 3-body recombination rate by a factor of 6 
when a cold gas condenses into a Bose-Einstein condensate 
was first observed in experiments on $^{85}$Rb atoms \cite{BGMHCW}.

We now restrict our attention to the case of large scattering length and
3-body recombination into the shallow dimer. The condition that the
recombination rate be independent of the momenta of the three atoms is that
their energies are small compared to $\hbar^2/ma^2$. In this case, we
might as well set them all equal to zero. Energy and momentum conservation
then imply that the atom and dimer will emerge with wave numbers 
$k_f=2/(\sqrt{3}a)$.

\begin{figure}[htb]
\centerline{\includegraphics*[width=8.5cm,angle=0,clip=true]{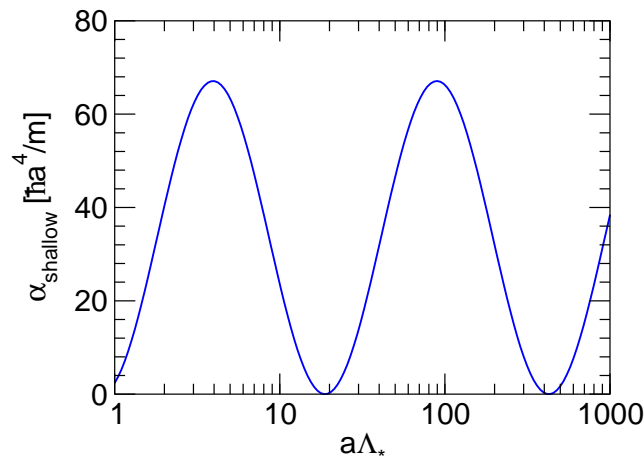}}
\medskip
\caption{The 3-body recombination rate constant
$\alpha_{\rm shallow}$ in
units of $\hbar a^4/m$ as a function of $a\Lambda_*$.}
\label{fig:alpha}
\end{figure}

We denote the contribution to the rate constant $\alpha$ 
from 3-body recombination into the shallow dimer by $\alpha_{\rm shallow}$.
By dimensional analysis, it is proportional to $\hbar a^4/m$ 
with a coefficient that depends only on $a\kappa_*$.
Petrov has derived a remarkable analytic expression for  
$\alpha_{\rm shallow}$ \cite{Petrov-octs}:
\begin{eqnarray}
\alpha_{\rm shallow} = 
{128 \pi^2 (4 \pi - 3 \sqrt{3}) \over
\sinh^2(\pi s_0) + \cosh^2(\pi s_0) \tan^2 [s_0 \ln (a\kappa _*) + \gamma]} \,
{\hbar a^4 \over m} \,,
\label{alpha-analytic}
\end{eqnarray}
where $\gamma$ is a real numerical constant.
The maximum value of the coefficient of $\hbar a^4/m$ is 
\begin{eqnarray}
C_{\rm max} = 
{128 \pi^2 (4 \pi - 3 \sqrt{3}) \over\sinh^2(\pi s_0)} \,.
\label{C-max}
\end{eqnarray}
Its numerical value is $C_{\rm max} = 67.1177$.
We can exploit the fact that $e^{2 \pi s_0} \approx 557$ is large
to simplify the expression in Eq.~(\ref{alpha-analytic}).
It can be approximated with an error of less than 1\% by
\begin{eqnarray}
\alpha_{\rm shallow} \approx 
C_{\rm max} \sin^2 [s_0 \ln (a \kappa _*) + \gamma] \,
{\hbar a^4 \over m} \,.
\label{alpha-sh0}
\end{eqnarray}
This approximate functional form of the rate constant was first deduced
by Nielsen and Macek \cite{NM-99} 
and by Esry, Greene and Burke  
\cite{EGB-99}.\footnote{In the formula for $\alpha_{\rm shallow}$ 
in Ref.~\cite{EGB-99}, $s_0 \approx 1.00624$ is replaced by 1.}
The constants $C$ and $\gamma$ were first calculated accurately
by Bedaque, Braaten and Hammer \cite{BBH-00} using the effective field theory 
of Ref.~\cite{BHK99,BHK99b}.  A more accurate determination of the
phase $\gamma$ was given in Ref.~\cite{BH02}. The resulting expression is
\begin{eqnarray}
\alpha_{\rm shallow} \approx 67.1 \sin^2 [s_0 \ln (a\Lambda _*) + 0.19] \,
{\hbar a^4 \over m} \,.
\label{alpha-sh}
\end{eqnarray}
The relation between $\Lambda_*$ and $\kappa_*$ is given in 
Eq.~(\ref{Lambdastar-kappastar}).
The coefficient of $\hbar a^4 / m$ is shown as a function of
$\ln (a\Lambda_*)$ in Fig.~\ref{fig:alpha}.
The most remarkable feature of the analytic expression in 
Eq.~(\ref{alpha-analytic}) and the approximate expression 
in Eq.~(\ref{alpha-sh}) is that the coefficient of $\hbar a^4/m$ 
oscillates between 0 and 67.1 as a function of $a$.
In particular, $\alpha_{\rm shallow}$ has zeroes at values of $a$ 
that differ by multiplicative factors of $e^{\pi/s_0} \approx 22.7$.
Using Eq.~(\ref{Lambdastar-kappastar}), the locations of the zeroes can be 
expressed as
\begin{eqnarray}
a =  \left( e^{\pi/s_0} \right)^n 0.32 \, \kappa_*^{-1} \,,
\label{alpha-zeroes}
\end{eqnarray}
where $n$ is an integer. The maxima of $\alpha/a^4$ occur near the 
values of the scattering length for which there is an Efimov trimer 
at the atom-dimer threshold: $a=( e^{\pi/s_0} )^n a_*$,
where $n$ is an integer and $a_*$ is given in Eq.~(\ref{B3=B2}).

The oscillations in the coefficient of $\hbar a^4/m$
in the recombination rate constant arise from interference effects
that were first derived within the hyperspherical framework
by Nielsen and Macek and by Esry, Greene, and Burke \cite{NM-99,EGB-99}. 
The 3-body recombination process involves the transition 
from an incoming 3-atom scattering state 
on the second lowest hyperspherical potential $V_1(R)$ 
in Fig.~\ref{fig:hap}
to an outgoing atom-dimer scattering state on the lowest 
hyperspherical  potential $V_0(R)$. 
These two potentials, which correspond to the lowest two solid lines in 
Fig.~\ref{fig:hap}, asymptote to $E=0$ and $E=-E_D$, respectively.
The 3-body recombination process involves a nonadiabatic transition 
between these two adiabatic potentials that takes place in the 
long-distance region $R \sim a$.
The process begins with an incoming 3-atom scattering state
approaching the long-distance region.
The state that emerges from the long-distance 
region is a superposition of an outgoing atom-dimer scattering state 
and an incoming hyperradial wave.
The incoming hyperradial wave flows to short-distances, where 
it is completely reflected into an outgoing hyperradial wave. 
It is then almost completely
transmitted through the $R \sim a$ region into an atom-dimer 
scattering state, because the lowest hyperspherical potential 
is almost reflectionless at $E=0$.
The resulting atom-dimer scattering state can interfere
with the atom-dimer scattering state
that emerges directly from the nonadiabatic transition,
and this interference gives rise to the oscillations of the 
coefficient of $\hbar a^4/m$ in Eq.~(\ref{alpha-analytic}) 
as a function of $\ln(a)$.
The zeroes in the recombination rate constant indicate that the 
interference is totally destructive.
Such exact zeroes are a well-known phenomenon in 
the Landau-Zener-Stueckelberg problem of the nonadiabatic transition
between two adiabatic energy levels with an avoided crossing.

Nielsen and Macek \cite{NM-99} obtained their result  
for $\alpha_{\rm shallow}$ by applying {\it hidden crossing theory}. 
The adiabatic hyperspherical potentials  $V_n(R)$ in Eq.~(\ref{Vch}) 
involve the functions $\lambda_n(R)$ that are solutions 
to Eq.~(\ref{cheigen}). If this equation is used to define
the functions $\lambda_n(R)$ for complex values of $R$, 
the channel potentials  $V_n(R)$ become sheets in the complex $R$ plane 
that are connected at square-root branch points. 
For example, if $a>0$ one can go continuously 
from the lowest adiabatic potential $V_0(R)$ 
to the second lowest one $V_1(R)$ by following a path
that goes around a square-root branch point at $R = (2.59 + 2.97i)a$.
In the analysis of Nielsen and Macek, the two interfering amplitudes
that contribute to $\alpha_{\rm shallow}$ correspond to two WKB
integration contours.  Both contours begin with real-valued $R$ 
in the asymptotic region of the $n=1$ adiabatic potential, 
go continuously around the branch point 
in the complex $R$ plane to the $n=0$ adiabatic potential, 
and eventually end with real-valued $R$ 
in the asymptotic region of the $n=0$ adiabatic potential. 
However, after having made the transition to the $n=0$ adiabatic potential,
the first contour goes immediately out to asymptotic $R$, 
while the second contour first goes to the short-distance region of $R$ 
and then returns through the region $R \sim a$ to the asymptotic $R$.  
Along the path to the short-distance region and back, 
the second contour picks up an additional WKB phase.
Thus the two amplitudes have the same magnitude 
and differ only by a phase that is determined by the short-distance
region of the lowest adiabatic potential.
 
Esry, Greene, and Burke \cite{EGB-99}
obtained their result for $\alpha_{\rm shallow}$
by solving the 3-body Schr{\"o}dinger equation in hyperspherical coordinates. 
They attributed the $\sin^2$ factor in Eq.~(\ref{alpha-sh0}) to St\"uckelberg
oscillations associated with a broad avoided crossing between the $n=0$
and $n=1$ adiabatic potentials for $R$ around $3a$.

The $a^4$ scaling behavior of $\alpha_{\rm shallow}$ was first 
obtained in Ref.~\cite{FRS96}.  However, the coefficient of $a^4$ 
was claimed to be a constant independent of $a$:
$\alpha_{\rm shallow} = 3.9 \, \hbar a^4/m$.
Several independent groups using completely different methods
have shown that the coefficient is actually a log-periodic function of $a$.  
We conclude that there was an error in the analysis of Ref.~\cite{FRS96}.

The power-law scaling behavior $\alpha_{\rm shallow} \sim a^4$
has been verified in experiments with ${}^{133}$Cs atoms \cite{WHMNG03}.
The logarithmic scaling violations associated with the 
log-periodic dependence on $a$ of the coefficient in 
Eq.~(\ref{alpha-analytic}) 
have not yet been observed in experiments.

Since the zeroes in $\alpha_{\rm shallow}$ are so remarkable, it is worth
enumerating some of the effects that will tend to fill in the zeroes, 
turning them into local minima of $\alpha_{\rm shallow}/a^4$. 
First, all universal predictions for the
2-body and 3-body sectors hold only up to corrections suppressed by
powers of $\ell/a$. Thus the zeroes in $\alpha_{\rm shallow}$ really mean 
that the coefficient of $\hbar a^4/m$ goes to zero like $\ell /a$ as 
$\ell \to 0$. The zeroes in $\alpha_{\rm shallow}$ are also exact 
only at threshold. If the recombining atoms have wave numbers of order $k$, 
the zeroes indicate that $\alpha_{\rm shallow}/a^4$ goes to zero like $ka$ 
as $k \rightarrow 0$. Thus
thermal effects that give nonzero momentum to the atoms could tend to
fill in the zeroes.  Finally, if the 2-body potential supports deep
diatomic molecules, their effects will tend to fill in the zeros of 
$\alpha_{\rm shallow}$ as described in Section~\ref{sec:deep-shallow}.  
Furthermore, 3-body recombination into those deep molecules
gives an additive contribution $\alpha_{\rm deep}$ to the
rate constant that will be determined in Section~\ref{sec:deep-3BR}.
The range of validity of the universal expression in 
Eq.~(\ref{alpha-analytic}) has been studied by D'Incao et al.~\cite{dInc04}.
They showed that it is a good approximation only 
for collision energy $E$ in the threshold region $E \lsim E_D$.  
However, their conclusion that 
``universal behavior is limited to the threshold region'' 
is a misinterpretation.
Universality predicts that 3-body observables are determined by $a$ 
and $\kappa_*$ only not only in the threshold region
but at all energies satisfying $E \ll \hbar^2/(m r_s^2)$.
Thus far, the predictions of universality for the recombination rate 
have been calculated only at the threshold $E=0$. 
They have not yet been calculated as a function of $E$.

We now proceed to show how the analytic dependence of the 3-body 
recombination rate constant $\alpha_{\rm shallow}$ on $a$ and $\kappa_*$
in Eq.~(\ref{alpha-analytic}) can be derived 
from Efimov's radial law.  The radial law for the S-matrix element 
$S_{AD,AAA}$ is given in Eq.~(\ref{RL:ADAAA}). 
The recombination rate constant $\alpha$ is determined by the behavior 
of $s_{11}$, $s_{12}$, $s_{13}$, and $s_{23}$ just above 
the dimer-breakup threshold.  
The values of $s_{11}$ and $s_{12}$ at the threshold 
$E=0$ are given in Eq.~(\ref{s-E0}). 
The entries $s_{13}$ and $s_{23}$ vanish at $E=0$.
The leading dependence of these entries on the energy 
$E = \hbar^2 K^2 /m$ can be deduced from threshold laws
for 3-body reaction rates \cite{Delves60}.
They require $s_{13}$ and $s_{23}$ to scale like $K^2$:
\begin{subequations}
\begin{eqnarray}
s_{13} & = & - c_1 e^{-i \gamma_0} a^2 K^2 [ 1 + \ldots] \,,
\label{s13-3A}
\\
s_{23} & = & - c_2 e^{+i \gamma_1} a^2 K^2 [ 1 + \ldots] \,,
\label{s23-3A}
\end{eqnarray}
\label{s-3A}
\end{subequations}
where $c_1$, $c_2$, $\gamma_0$, and $\gamma_1$ are real numerical constants
and $c_1$ and $c_2$ are positive.  The entries of $s$ have expansions 
in powers of $K$.  The unitarity of $s$ imposes constraints on the
numerical constants in these expansions.
For example, it implies $c_2 = [\tanh(\pi s_0)]^{1/2} c_1$. 
Expanding the S-matrix element for 3-body recombination 
in Eq.~(\ref{RL:ADAAA}) to leading order in $K$, we find
\begin{eqnarray}
S_{AD,AAA} & \longrightarrow & 
- c_2 e^{i \gamma_1} a^2 K^2 
{ 1 + e^{2i(\theta_* - \delta_0)} 
\over  1 - e^{-2 \pi s_0} e^{2i(\theta_* - \delta_0)} } \,,
\qquad {\rm as \ } E \to 0^+ \,.
\label{SADAAA:th}
\end{eqnarray}
The 3-body recombination rate constant $\alpha$ is proportional to 
$|S_{AD,AAA} |^2$.  Squaring the expression in Eq.~(\ref{SADAAA:th})
and setting $\theta_* = s_0 \ln ( a \kappa_*)$, we find that the 
dependence on $a$ and $\kappa_*$ agrees with the analytic expression for 
$\alpha_{\rm shallow}$ in Eq.~(\ref{alpha-analytic}).


\subsection {Three-atom elastic scattering}
        \label{sec:3atomscat}

The physical region for 3-atom scattering states is $E>0$, where $E$ is
the total energy of the three atoms. We expect nontrivial 3-body effects
to be most dramatic in the threshold region $E \to 0$, so we will only
consider this limit. 
In the standard plane-wave basis, the T-matrix element 
for 3-body elastic scattering diverges as $E \to 0$ \cite{AR70}. 
The most singular term comes from two successive 2-body scatterings
that involve all three particles, and it is proportional to $a^2/E$.
There are also singular terms proportional to $a^3/\sqrt{E}$ 
and $a^4\ln E$.  To obtain the 3-body elastic scattering rate,
the effects of 2-body elastic collisions, with the third particle 
infinitely far away, must be subtracted \cite{Newton}.

The T-matrix element ${\mathcal T}$ for 3-atom elastic scattering 
depends on the three wave vectors of the incoming atoms 
and the three wave vectors of the outgoing atoms.
The contribution to ${\mathcal T}$ from total orbital angular 
momentum quantum number $L=0$ depends only on the total energy
$E = \hbar^2 K^2/m$ of the three particles.  
Its limiting behavior as $E \to 0$ is
\begin{eqnarray}
{\mathcal T}^{(L=0)} &\longrightarrow & \left( {A \over K^2 a^2} 
+ {B \over K a}
+ C \ln(K |a|) + D_\pm \right) {\hbar a^4 \over m}\,, 
\qquad {\rm as \ }E \to 0 \,.
\label{T3to3-L=0}
\end{eqnarray}
The coefficients $A$, $B$, and $C$ are numerical constants.
The sensitivity to Efimov physics resides only in the 
coefficient $D_\pm$, which depends on the sign $\pm$ of $a$
and is a function of $a \kappa_*$.
The coefficient $D_-$ for the case $a < 0$ is real-valued.
The coefficient $D_+$ for the case $a > 0$ is complex-valued.
Its imaginary part is related to the 3-body recombination rate 
constant $\alpha_{\rm shallow}$ given in Eq.~(\ref{alpha-analytic}) 
by the unitarity condition
\begin{eqnarray}
{\rm Im} \; {\mathcal T} = 3 \, \alpha_{\rm shallow} \,.
\label{ImT-alpha}
\end{eqnarray}

The numerical constants $A$, $B$, and $C$ 
can be calculated from the terms of orders $a^2$, $a^3$, 
and $a^4$, respectively, in the perturbative expansion 
of ${\mathcal T}$, which are given explicitly in Section~\ref{sec:EFT3}.
The coefficients $A$ and $B$ have not been calculated.
The coefficient $C$ is known analytically \cite{BN-99}:
\begin{eqnarray}
C = 384 \pi \big( 4 \pi - 3 \sqrt{3} \,\big) \,.
\label{C:analytic}
\end{eqnarray}
The coefficient $D_\pm$ in Eq.~(\ref{T3to3-L=0}) 
can only be obtained by a nonperturbative calculation.
One might expect that $D_\pm$ could be obtained from the order-$a^4$ 
term in the perturbative expansion of ${\mathcal T}$,
but the perturbative contribution to $D_\pm$ is ultraviolet divergent.  
The perturbative approximation can be expressed in the form
\begin{eqnarray}
D_\pm \approx D^{\rm (pert)}_\pm - C \ln(|a| \Lambda) \,,
\label{D-pert}
\end{eqnarray}
where $\Lambda$ is the ultraviolent cutoff and $D^{\rm (pert)}_\pm$ 
is a numerical constant that depends on how the ultraviolet cutoff 
is implemented.  The difference between the coefficient $D_\pm$ 
and its perturbative approximation can be determined 
by a nonperturbative 3-body calculation.
The coefficient $D_\pm$ can be expressed as the sum 
of the perturbative and nonperturbative contributions:
\begin{eqnarray}
D_\pm = \left[ D^{\rm (pert)}_\pm - C \ln(|a| \Lambda) \right]
+ \left[ D^{\rm (nonpert)}_\pm + C \ln(|a| \Lambda) \right] \,,
\label{D-nonpert}
\end{eqnarray}
where $D^{\rm (nonpert)}_\pm$ depends on $|a| \kappa_*$. 
The dependence on the ultraviolet cutoff cancels in the sum 
of the two contributions in Eq.~(\ref{D-nonpert}).  

The nonperturbative contributions to the coefficients $D_\pm$ 
were calculated in Ref.~\cite{BHM-01}
for a specific regularization of the perturbative T-matrix element:
{\it dimensional regularization} and {\it minimal subtraction}.
In dimensional regularization, momentum integrals
are analytically continued from 3 dimensions to $D$ dimensions,
in which case logarithmic ultraviolet divergences appear as 
poles in $D-3$.
In minimal subtraction, the ultraviolet divergences are removed
simply by subtracting the poles in $D-3$. 
In the case $a<0$, the functional form of the dependence of $D_-$
on $|a| \kappa_*$ was deduced by Efimov \cite{Efimov79}.
In Ref.~\cite{BHM-01}, the calculated result for $D^{\rm (nonpert)}_-$
was fit to that functional form, with the result
\begin{equation}
D^{\rm (nonpert)}_- = 
C \big( 1.23 + 3.16\cot [s_0 \ln(|a| \Lambda_*) -1.38] \big) \,,
\label{d-nonpert}
\end{equation}
where $C$ is given in Eq.~(\ref{C:analytic}).
The coefficient $D^{\rm (nonpert)}_-$
diverges at those negative values of the scattering length for which 
there is an Efimov trimer at the 3-atom threshold: 
$a=( e^{\pi/s_0} )^n a_*'$, where $n$ is an integer and
$a_*'$ is given in Eq.~(\ref{B3=0}).

The nonperturbative contribution to the complex-valued coefficient
$D_+$ was also calculated in Ref.~\cite{BHM-01}.
The calculated result for $D^{\rm (nonpert)}_+$ was found empirically 
to be an oscillatory function of $\ln(a \Lambda_*)$.  
Its real and imaginary parts were fit by the expressions
\begin{subequations}
\begin{eqnarray}
{\rm Re} \, D^{\rm (nonpert)}_+ &\approx& 
C \big( 1.22 + 0.021 \sin^2 [s_0 \ln (a \Lambda_*) - 0.6] \big) \,,
\label{Red-nonpert}
\\
{\rm Im} \, D^{\rm (nonpert)}_+ &\approx& 
C \big( 0.022 \sin^2 [s_0 \ln (a \Lambda_*) + 0.19] \big) \,.
\label{Imd-nonpert}
\end{eqnarray}
\label{ReImd}
\end{subequations}
The imaginary part in Eq.~(\ref{Imd-nonpert}) is consistent 
with the unitarity constraint in Eq.~(\ref{ImT-alpha})
if we use the approximate expression for $\alpha$ in 
Eq.~(\ref{alpha-sh}).
Note that the oscillatory term of the real part in 
Eq.~(\ref{Red-nonpert}) has approximately the same amplitude 
as the imaginary part but a different phase.

We proceed to derive the dependence of the coefficients $D_\pm$ 
in Eq.~(\ref{T3to3-L=0}) on $a \kappa_*$ using Efimov's radial law.
The radial law for the $L=0$ contribution to the S-matrix element 
for 3-atom elastic scattering is given in Eq.~(\ref{RL:AAA}).
It involves the entries $s_{11}$, $s_{13}$ and $s_{33}$ 
of the $3\times3$ matrix $s$. We first consider the case $a<0$.
We need the behavior of the entries of $s$ as a function of the 
energy $E = \hbar^2 K^2/m$ near the threshold.  At $K=0$, 
we must have $s_{13}=0$ and we can set $s_{33} = 1$ by 
a choice of the overall phase of the matrix $s$.  
By the unitarity of $s$, $s_{11}$ must be a pure phase at $K=0$:  
$s_{11} = -e^{2 i\delta}$ for some angle $\delta$. 
All the entries have expansions in powers of $aK$.  
The coefficients are constrained by the unitarity of the 
$3 \times3$ matrix $s$.  The expansion of the off-diagonal 
entry $s_{13}$ must begin at order $a^2K^2$ for the 
dependence of $S_{AAA,AAA}$ on $\kappa_*$ 
to enter only at order $a^4 K^4$.  There is an additional 
constraint  on the coefficients from the fact that the 
expansion of $S_{AAA,AAA}-1$ begins at order $a^2 K^2$. 
The resulting expansions for the entries of $s$ that contribute 
to $S_{AAA,AAA}$ are
\begin{subequations}
\begin{eqnarray}
s_{11} &=& - e^{2 i \delta} 
[1 - i d_1 a K -(\mbox{$1\over2$}d_2^2 + i d_6) a^2 K^2 + \ldots] \,,
\label{s11-AAA}
\\
s_{13} &=& d_0 e^{i \delta} a^2 K^2 
[1 - (d_5 + i \mbox{$1\over2$}d_1) a K +  \ldots] \,,
\label{s12-AAA}
\\
s_{33} &=& 1 - i d_2 a^2 K^2 - i d_3 a^3 K^3 
-(\mbox{$1\over2$}d_0^2 + \mbox{$1\over2$}d_2^2 + i d_4) a^4 K^4 + \ldots \,,
\label{s22-AAA}
\end{eqnarray}
\label{s-AAA}
\end{subequations}
where $d_0$, $d_1$, $d_2$, $d_3$, $d_4$, $d_5$, $d_6$, 
and $\delta$ are real constants and 
$d_0 > 0$.  Inserting the expressions for $s_{ij}$ in 
Eqs.~(\ref{s-AAA}) into the radial law in Eq.~(\ref{RL:AAA}), we obtain 
\begin{eqnarray}
S_{AAA,AAA} &\longrightarrow&  1 - i d_2 a^2 K^2 - i d_3 a^3 K^3 
\nonumber
\\
&& - \big[\mbox{$1\over2$}d_2^2 + i d_4 
	- \mbox{$1\over2$}d_0^2 \cot (\theta_* + \delta) \big] \; a^4 K^4\,, 
\qquad {\rm as \ } E \to 0 \,.
\label{D:a<0}
\end{eqnarray}
The elements of the S-matrix and the T-matrix are related
by $S = 1 + i T$.  Note that the coefficient of the $a^4 K^4$
term has the same functional dependence on $a \kappa_*$ 
as the empirical coefficient $D^{\rm (nonpert)}_-$ in
Eq.~(\ref{d-nonpert}), which is the nonperturbative part 
of the coefficient of $a^4$
in the T-matrix element in Eq.~(\ref{T3to3-L=0}).

We next consider the case $a>0$.
The radial law for $S_{AAA,AAA}$ in Eq.~(\ref{RL:AAA})
involves the entries $s_{11}$, $s_{13}$ and $s_{33}$ 
of the $3\times3$ matrix $s$.
The values of $s_{11}$, $s_{12}$, $s_{22}$, and $s_{33}$ 
at $E=0$ are given in Eqs.~(\ref{s-E0}).
The leading terms in the entries $s_{13}$ and $s_{23}$  
are of order $a^2K^2$ and are given in Eq.~(\ref{s13-3A}).
The entries all have expansions in powers of $K$.  
The coefficients are constrained by the
unitarity of the $3 \times3$ matrix $s$.  There is an additional 
constraint from the fact that the expansion of $S_{AAA,AAA}-1$
begins at order $a^2 K^2$. That expansion includes a term of order
$a^3K^3$ with a constant coefficient and then a term of order $a^4K^4$
whose coefficient $iD$ depends on $a\kappa_*$. The coefficient $D$
has the form
\begin{eqnarray}
D =  c_3  + i c_1^2
- i c_1^2 {e^{2 i (\theta_* -\delta_0)}   \over
1 - e^{-2 \pi s_0} e^{2 i (\theta_* -\delta_0)}} \,,
\label{SAAAAAA-4}
\end{eqnarray}
where $c_1$ is the numerical constant in Eq.~(\ref{s13-3A})
and $c_3$ is another numerical constant. 
The relation between the S-matrix and the T-matrix 
is $S = 1 + i T$. Thus the coefficient $D_+$ of the $a^4$ term
in the T-matrix element in Eq.~(\ref{T3to3-L=0}) must have the 
same dependence on $a \kappa_*$ as in Eq.~(\ref{SAAAAAA-4}).
We can exploit the fact that 
$e^{2 \pi s_0} \approx 557$ is large
to simplify the coefficient in Eq.~(\ref{SAAAAAA-4}).
It can be approximated with an error of less than 1\% by
\begin{subequations}
\begin{eqnarray}
D &\approx& c_3 -i c_1^2 (e^{2 i (\theta_* -\delta_0)} - 1 ) 
\\
&=& c_3 - c_1^2
+ 2 c_1^2 \sin^2(\theta_* -\delta_0 + \mbox{$1\over4$} \pi) 
+ 2 i c_1^2 \sin^2(\theta_* -\delta_0 )  \,.
\label{TAAAAAA-4}
\end{eqnarray}
\end{subequations}
The dependence of this approximate expression on $a \kappa_*$ 
is compatible with that of the empirically determined 
coefficient $D^{\rm (nonpert)}_+$ given by Eqs.~(\ref{ReImd}).
In particular, the amplitudes of the oscillatory terms in 
Eqs.~(\ref{Red-nonpert}) and (\ref{Imd-nonpert}) are equal to 
within the numerical accuracy and the phase difference between 
the oscillations is close to the value ${1 \over 4} \pi \approx 0.79$ 
predicted by Eq.~(\ref{TAAAAAA-4}). More precise results for the 
coefficients in Eq.~(\ref{ReImd}) could have been obtained if the 
constraints from Efimov's radial law had been imposed on the fit.

For a dilute homogeneous Bose gas composed of particles with 
a positive scattering length $a$ and number density $n$,
the leading term in the low-density expansion of the energy per particle 
is given in Eq.~(\ref{Eovern}).  The low-density expansion
is an expansion in powers of the diluteness variable $(na^3)^{1/2}$.
Despite the fractional powers of $a$,
this is a perturbative expansion in powers of $a$.
The dimensionless expansion parameter is the ratio $a/\xi$ of the 
scattering length and the coherence length $\xi = (16 \pi n a)^{-1/2}$.
The first few terms in the low-density expansion have the form
\begin{eqnarray}
{{\mathcal E} \over n} &=& {2 \pi \hbar^2 \over m} a n 
\bigg( 1 + {128 \over 15 \sqrt{\pi}} (na^3)^{1/2} 
+ {8(4 \pi - 3 \sqrt{3}) \over 3} [ \ln(na^3) + 2 d] na^3  \bigg)
\, .
\nonumber\\
\label{Eovern:lde}
\end{eqnarray}
The $(na^3)^{1/2}$ correction was first calculated by Lee and Yang 
in 1957 for the case of particles interacting through a hard-sphere 
potential of radius $a$ \cite{LY57}.  This correction is universal:  it
applies equally well to any potential with a positive scattering 
length $a$.  The $na^3 \ln(na^3)$ correction,
which was first calculated in 1959 \cite{Wu59,HP59,Sawada59},
is also universal in the sense that it depends only on $a$.
The $na^3$ correction is not universal, according to the definition
traditionally used in the theory of the homogeneous Bose gas,
because it depends on few-body parameters other than $a$.
Specifically, it depends on the coefficient $D_+$ in the low-energy 
expansion of the T-matrix element for 3-body 
elastic scattering given in Eq.~(\ref{T3to3-L=0}) \cite{BN-99}.
In the special case of a large scattering length, this correction
is universal by our definition, which we argue is more appropriate 
for this strongly interacting problem.  
The coefficient of $na^3$ in Eq.~(\ref{Eovern:lde}) has a well-defined 
scaling limit as the range of the interaction is tuned to zero.  
It is a log-periodic function of $a \kappa_*$ \cite{BHM-01}:
\begin{eqnarray}
d = D_+^{\rm (nonpert)}/C + 2.36 \,,
\end{eqnarray}
where $D_+^{\rm (nonpert)}$ is given in Eqs.~(\ref{ReImd}).  The imaginary part 
of the coefficient reflects the loss of mean-field energy due to
3-body recombination into the shallow dimer.


\subsection {Helium atoms}
        \label{sec:he4atoms}

Helium atoms provide a beautiful illustration of universality 
in the 3-body system \cite{BH02}.
The binding energies of the $^4$He
trimers have been calculated accurately for a number of different model
potentials for the interaction between two $^4$He atoms.  
For the purposes of illustration, we will use the TTY potential \cite{TTY95}.  
The scattering length for the TTY potential is $a = 188.99 \, a_0$.  
This is much larger than its effective range $r_s = 13.85 \, a_0$, 
which is comparable to the van der Waals
length scale $\ell_{\rm vdw} = 10.2 \, a_0$.  
The TTY potential supports a single 2-body bound state, 
the $^4$He dimer whose binding energy is $E_2= 1.30962$ mK.  
The conversion factor to the atomic energy unit is given 
in Eq.~(\ref{convert}).
The $^4$He dimer was first observed in 1992 \cite{LMKGG}.
The TTY  potential has exactly two 3-body bound states:  
the ground-state trimer, which we label $n = 0$, 
and the excited trimer, which we label $n = 1$. 
There have been several accurate calculations of the binding energies
$E_3^{(0)}$ and $E_3^{(1)}$ for the TTY potential 
\cite{NFJ98,RY00,MSSK01}.  The results agree to within 
0.5\% for both $E_3^{(0)}$ and $E_3^{(1)}$.  
The results of Ref.~\cite{MSSK01} are $E_3^{(0)} = 125.8$ mK
and $E_3^{(1)} = 2.28$ mK. 
The ground state trimer was first observed in 1994 \cite{STo96}.
The excited state has not yet been observed.

Lim, Duffy, and Damert proposed in 1977 that the excited state 
of the $^4$He trimer is an Efimov state \cite{LDD77}.
This interpretation is almost universally accepted.
Some researchers have proposed that the ground state trimer 
is also an Efimov state \cite{BHK99,BHK99b,FTDA99}.
This raises an obvious question: 
what is the definition of an Efimov state? 
The most commonly used definition 
is based on rescaling the depth of the 2-body potential:
$V({\bf r})  \longrightarrow \lambda V({\bf r})$.
A trimer is defined to be an Efimov state if its binding 
energy as a function of the scaling parameter $\lambda$
has the qualitative behavior illustrated in Fig.~\ref{fig:efistar}.
As $\lambda$ is decreased below 1, the trimer eventually 
disappears through the 3-atom threshold.
As $\lambda$ is increased above 1, the trimer eventually 
disappears through the atom-dimer threshold.  
Calculations of the trimer binding energies \cite{Esry96a}
using a modern helium potential show that the excited trimer 
satisfies this definition of an Efimov state
but the ground state trimer does not.  
The excited trimer disappears through the 3-atom threshold
when $\lambda$ is decreased to about $0.97$, 
and it disappears through the atom-dimer threshold 
when $\lambda$ is increased to about 1.1.
The ground state trimer disappears through the 3-atom threshold
when $\lambda$ is about $0.9$. 
However, as $\lambda$ is increased above 1, its binding energy
relative to the atom-dimer threshold continues to increase. 
Thus is does not qualify as an Efimov state by the definition 
given above.

The traditional definition of an Efimov state described above is
not natural from the point of view of universality.
The essence of universality concerns the behavior of a system 
when the scattering length becomes increasingly large.
The focus of the traditional definition is on the endponts 
of the binding energy curve in Fig.~\ref{fig:efistar},
which concerns the behavior of the system 
as the scattering length decreases in magnitude.
The problem is that the rescaling of the potential can move 
the system outside the large-scattering-length region $|a| \gg r_s$
before the trimer reaches the endpoint of the binding energy curve.
We therefore propose a definition of an Efimov state
that is more natural from the universality perspective.
A trimer is defined to be an Efimov state if a deformation 
that tunes the scattering length to $\pm \infty$ moves its 
binding energy along the universal curve illustrated in 
Fig.~\ref{fig:efistar}.  The focus of this definition is on the 
resonant limit where the binding energy crosses the $1/a = 0$ axis.
In particular, the binding energy at this point should be larger 
than that of the next shallowest trimer by about a factor of 515.
In the case of helium, the resonant limit can be reached by rescaling
the 2-body potential by a factor $\lambda \approx 0.97$ \cite{Esry96a}.
At this point, the binding energy of the ground state trimer 
is larger than that of the excited trimer by about a factor of 570.
The closeness of this ratio to the asymptotic value 515
supports the hypothesis that the properties of the ground state trimer 
are largely determined by universality.  Further evidence 
in support of this hypothesis will be presented below.

In order to apply the universal predictions for low-energy 3-body
observables to the case of $^4$He atoms,
we need a 2-body input and a 3-body input  
to determine the parameters $a$ and $\kappa_*$.  
The scattering length $a = 188.99 \, a_0$ itself 
can be taken as the 2-body input.
An alternative 2-body input is the dimer binding energy $E_2$. 
A scattering length $a_D$ can be determined by identifying 
$E_2$ with the universal binding energy of the
shallow dimer: $E_2 = \hbar^2/ma_D^2$.
The result is $a_D = 181.79 \, a_0$.
The 3.8\% difference between $a$ and $a_D$ is a measure of
how close the system is to the scaling limit.
To minimize errors associated with the
deviations of the system from the scaling limit, 
it is best to take the shallowest 3-body
binding energy available as the input for determining $\kappa_*$.  
In the case of $^4$He atoms, this is the binding energy 
$E_3^{(1)}$ of the excited trimer.

We proceed to consider the universal predictions for the 
trimer binding energies. Having identified $E_3^{(1)}$ 
with the universal trimer binding energy $E_T^{(1)}$, 
we can use Efimov's binding energy equation (\ref{B3-Efimov}) 
with $n_* = 1$ to calculate $\kappa_*$ 
up to multiplicative factors of $e^{\pi/s_0} \approx 22.7$
\cite{BH02}.  The result is 
$\kappa_* = 0.00215 \,a_0^{-1}$ or $\kappa_* = 0.00232 \,a_0^{-1}$, 
depending on whether $E_2=1.31$ mK or $a=189.0 \; a_0$ 
is used as the 2-body input.
The intuitive interpretation of $\kappa_*$ is that if a parameter 
in the short-distance potential is adjusted  to tune $a$ to $+\infty$, 
the binding energy $E_3^{(1)}$ 
should approach a limiting value of approximately $\hbar^2 \kappa_*^2/m$,
which is 0.201 mK or 0.233 mK depending on the 2-body input.

\begin{table}[htb]
\begin{tabular}{l|cc|cccc}
potential & $a$ & $a_D$ & $E_3^{(1)}$ & $E_3^{(0)}$ & $E_3^{(-1)}$    \\ 
\hline 
TTY & 100.0 &  96.2 &     2.28    &    125.8   &         --         \\
    &       & input &    input    &    129.1   & $5.38 \times 10^4$ \\
    & input &       &    input    &    146.4   & $6.23 \times 10^4$ \\
\hline
\end{tabular}
\vspace*{0.3cm}
\caption{
Binding energies $E_3^{(n)}$ of the $^4$He trimers
for the TTY potential (row 1)
compared to the universality predictions using as the inputs
either $E_2$ and $E_3^{(1)}$  (row 2) or $a$ and $E_3^{(1)}$  (row 3). 
Energies are given in mK and lengths are given in \AA.
The trimer binding energies for the TTY potential  
are from Ref.~\cite{MSSK01}. (Note that $\hbar^2/m=12.1194$ K\AA$^2$
for $^4$He atoms.)}
\label{tab:He1}
\end{table}

Once $\kappa_*$ has been calculated, we can solve Eq.~(\ref{B3-Efimov}) 
for the binding energies of the deeper Efimov states.  
The prediction for the next two binding energies are shown in 
Table~\ref{tab:He1}.  The prediction for $E_3^{(0)}$
differs from the binding energy of the ground-state trimer by 2.6\% or 16.4\%,
depending on whether $E_2$ or $a$ is taken as the 2-body input.  
The expected error is 
comparable to the maximum of $\ell_{\rm vdW} / a = 5.4$\%  
and $(E_3^{(0)} / E_{\rm vdW} )^{1/2} \approx 50$\%,
where $E_{\rm vdW} = \hbar^2 / m \ell^{~2}_{\rm vdW} \approx 420$~mK
is the natural ultraviolet cutoff for $^4$He atoms.  
The errors are much smaller than expected, 
suggesting that the scaling limit is more robust than one might 
naively expect.

Efimov's equation (\ref{B3-Efimov}) also predicts infinitely many 
deeper 3-body bound states.  
The prediction for the next deepest state is given in Table~\ref{tab:He1}:
$E_3^{(-1)} \approx 5 \times 10^4$ mK.
This is more than two orders of magnitude larger than the natural
ultraviolet cutoff for $^4$He atoms, which is about 420~mK.
We conclude that this state and all the deeper bound states are artifacts 
of the scaling limit. 

Using the above values of $\kappa_*$ for the TTY potential,
we can immediately predict the atom-dimer scattering length $a_{AD}$.
If $a$ is used as the 2-body input, we find $a_{AD} \approx 0.94\, a$,
corresponding to $a_{AD}\approx 178\, a_0$.
If $E_2$ is used as the 2-body input,
we find $a_{AD} \approx 1.19\, a_D$, corresponding to  
$a_{AD}\approx 216\, a_0$.
These values are in reasonable agreement with the
calculation of Ref.~\cite{MSSK01}, which gave $a_{AD} = 248(10) \, a_0$.  
Since $r_s/ a = 7.3$\% for the TTY potential, 
much of the remaining discrepancy can perhaps be attributed 
to effective-range corrections.

\begin{table}[htb]
\begin{tabular}{l|cc|cccc}
potential   &   $a$  & $a_D$ & $E_3^{(1)}$ & $E_3^{(0)}$ & $a_{AD}$
& $\alpha$ \\ 
\hline 
HFDHE2      & 124.6 & 120.8 &  1.67 & 116.7 &   --   &  --   \\
            &       & input & input & 118.5 &  87.9  & 3.79  \\
            & input &       & input & 129.1 &  65.8  & 5.95  \\ 
\hline
HFD-B       &  88.5 &  84.8 &  2.74 & 132.5 & 135(5) &  --   \\
            &       & input & input & 137.5 & 120.2  & 0.064 \\
            & input &       & input & 159.7 & 100.4  & 0.37  \\ 
\hline
LM2M2       & 100.2 &  96.4 &  2.28 & 125.9 & 131(5) &  --   \\
            &       & input & input & 130.3 & 113.1  & 0.45  \\
            & input &       & input & 147.4 & 92.8   & 1.16  \\ 
\hline
TTY         & 100.0 &  96.2 &  2.28 & 125.8 & 131(5) &  --   \\
            &       & input & input & 129.1 & 114.5  & 0.41  \\
            & input &       & input & 146.4 & 94.0   & 1.11  \\
\hline
HFD-B3-FCI1 &  91.0 &  87.0 &  2.62 & 131.3 &   --   & 0.12 \\
            &       & input & input & 133.8 & 123.0  & 0.090 \\
            & input &       & input & 156.1 & 101.5  & 0.48  \\
\hline
\end{tabular}
\vspace*{0.3cm}
\caption{Three-body results for various model potentials
for $^4$He compared to the universality predictions using
either $a_D$ and $E_3^{(1)}$ or $a$ and $E_3^{(1)}$ as the inputs. 
All energies are given in mK, all lengths are given in \AA, 
and $\alpha$ is given in 10$^{-27}$ cm$^6$/s.
The 3-body results for the HFDHE2, HFD-B, LM2M2, and TTY potentials  
are from Ref.~\cite{MSSK01}, while the 3-body results for the HFD-B3-FCI1 
potential are from Refs.~\cite{SEGB02,Esryprivate}.
(Note that $\hbar^2/m=12.1194$ K\AA$^2$ for $^4$He atoms.)
}
\label{tab:He2}
\end{table}

We can also predict the 3-body recombination rate constant 
for $^4$He atoms interacting through the TTY potential.  
The prediction for $\alpha_{\rm shallow}$ is $2.9 \, \hbar a_D^4/m$
or $6.9 \, \hbar a^4/m$, depending on whether the dimer binding energy
$E_2$ or the scattering length $a$ is used as the 2-body input.  
In either case, the coefficient of $\hbar a^4/m$ is much smaller than
the maximum possible value 67.1.  Thus $^4$He atoms are fortuitously 
close to a combination of $a$ and $\kappa_*$ for which 
$\alpha_{\rm shallow}$ is zero.  The 3-body recombination rate constant
has not yet been calculated for the TTY potential, 
so the prediction of universality cannot be tested.

Similar comparisons can be made for other modern $^4$He potentials.
In Table \ref{tab:He2}, we have collected the available 
3-body results and universality predictions
for the HFDHE2 \cite{HFDHE2}, HFD-B \cite{HFD-B}, LM2M2 \cite{AS91}, 
TTY \cite{TTY95}, and HFD-B3-FCI1 \cite{AJM95}
potentials. The 3-body results for the HFDHE2, HFD-B, LM2M2, and 
TTY potentials  are from Ref.~\cite{MSSK01}, 
while the 3-body results for the HFD-B3-FCI1 
potential are from Refs.~\cite{SEGB02,Esryprivate}.
The universality predictions are given for the case
where $a_D$ and $E_3^{(1)}$ are used as the inputs and for the case where
$a$ and $E_3^{(1)}$ are used as the inputs. 
Where a comparison can be made, we find reasonably good agreement 
between the universality predictions and the direct calculations.


\subsection {Universal scaling curves}
        \label{sec:curves}

The logarithmic scaling violations associated with the Efimov effect 
imply that low-energy observables in the scaling limit depend not only 
on the scattering length $a$, but also on the 3-body parameter $\kappa_*$.  
A dimensionless combination of observables 
is called a {\it scaling variable}.  Examples are ratios of 
binding energies, such as $E_T^{(N)} / E_D$, or
ratios of scattering lengths, such as $a_{AD} / a$.  Since they are
dimensionless, 3-body scaling variables must be a function of the
dimensionless combination $a \kappa_*$ only.  By eliminating $\kappa_*$, 
we can express one scaling variable as a function of another.  
Equivalently, in the plane defined by two scaling variables, 
the variation of $\kappa_*$ generates a curve.  
Such a curve is called a {\it universal scaling curve}.

\begin{figure}[htb]
\centerline{\includegraphics*[width=8.5cm,angle=0,clip=true]{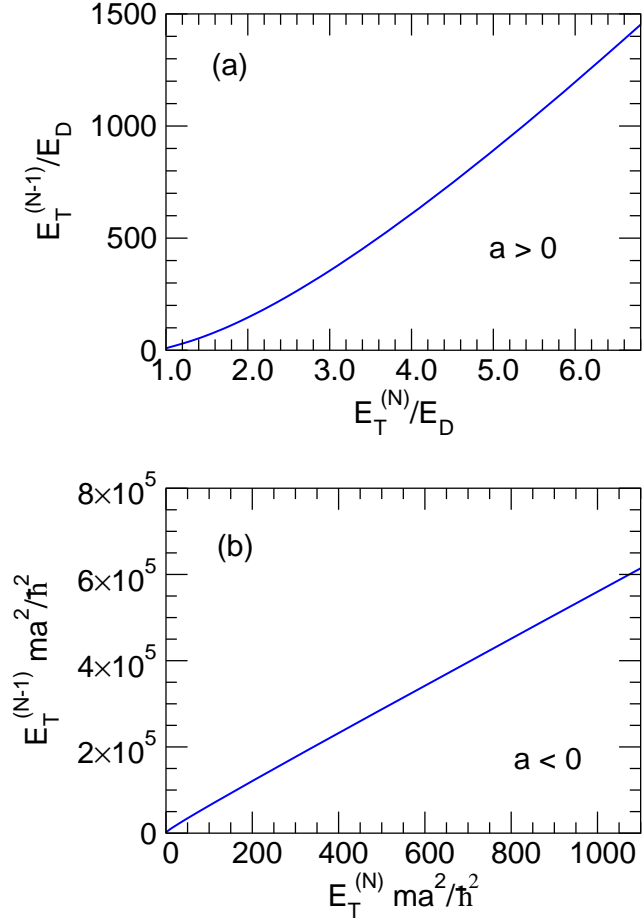}}
\medskip
\caption{
Universal scaling curves for $E_T^{(N-1)}ma^2/\hbar^2$ 
vs.~$E_T^{(N)}ma^2/\hbar^2$ for  (a) $a>0$ and (b) $a<0$.
$E_T^{(N)}$ and $E_T^{(N-1)}$ are the binding energies
of the shallowest and second shallowest Efimov states.
}
\label{fig:B30vsB31}
\end{figure}

The universal scaling curve relating the binding energies $E_T^{(n+1)}$ 
and $E_T^{(n)}$ of two successive Efimov states was calculated in
Ref.~\cite{FTDA99} using the renormalized zero-range model~\cite{AFG95,AF95}. 
The scaling variables in Ref.~\cite{FTDA99} were 
$[( E_T^{(n+1)} - E_a) /  E_T^{(n)}  ]^{1/2}$ 
and $[E_T^{(n)} /E_a]^{- 1/2}$, where $E_a = \hbar^2/ma^2$.  
The scaling curve was calculated over the entire range of  $E_T^{(n+1)}$, 
which includes the resonant limit where $a \to \pm \infty$. 
A more useful pair of scaling
variables that contains much of the same information is  
$E_T^{(N)}  / E_a$ and  $E_T^{(N-1)} /E_a$, 
where  $E_T^{(N)}$ and $E_T^{(N-1)}$ 
are the binding energies of the two shallowest Efimov states. 
The universal scaling curves for $a > 0$ and $a < 0$ were calculated in
Ref.~\cite{BH02} using the effective field theory of 
Ref.~\cite{BHK99,BHK99b}.  
They are shown in Figs.~\ref{fig:B30vsB31}(a) and \ref{fig:B30vsB31}(b).  
For $a > 0$, the ranges of $E_T^{(N)} /E_a$ and  
$E_T^{(N-1)}/E_a$ are 1 to 6.75 and 6.75 to 1406, respectively.  
For $a < 0$, their ranges are 0 to $1.1\times 10^3$ and $1.1 \times 10^3$ 
to $6.0 \times 10^5$, respectively.  If $a$ and $E_T^{(N)}$ are known, 
these universal scaling curves can be used to predict the binding energy 
$E_T^{(N-1)}$ of the second shallowest Efimov state.

\begin{figure}[htb]
\centerline{\includegraphics*[width=8.8cm,angle=0,clip=true]{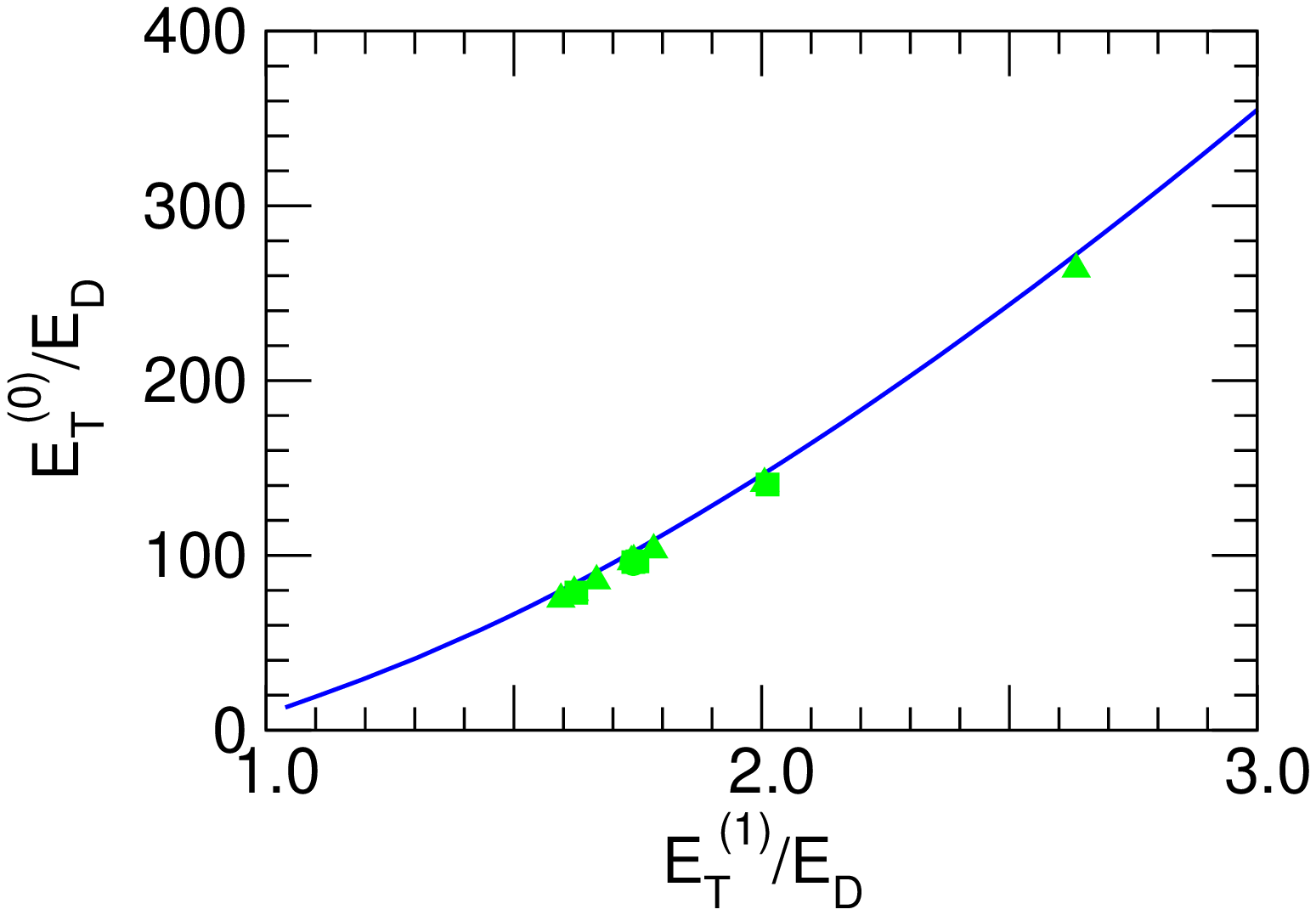}}
\medskip
\caption{
The universal scaling curve in Fig.~\ref{fig:B30vsB31}(a)
in the region relevant for $^4$He atoms.
The data points are fully-converged calculations for various $^4$He
potentials. 
}
\label{fig:B30vsB31he4}
\end{figure}

The availability of accurate numerical calculations of the 3-body binding
energies for various potential models for $^4$He atoms allows for a dramatic
illustration of the universal scaling curves.  The first fully-converged
calculation for a $^4$He potential was carried out by Cornelius and
Gl\"ockle in 1986 \cite{CG86}.  Fully-converged calculations for more
modern $^4$He potentials have recently become available 
\cite{NFJ98,RY00,MSSK01}.  
The potentials for which fully-converged calculations are available 
have scattering lengths $a$ that range widely 
from 167.2 $a_0$ to 235.6 $a_0$. 
If we use the dimer binding energy $E_D$ and the excited trimer binding energy
$E_T^{(1)}$ as inputs, $\Lambda_*$ ranges from 0.922 $a_D^{-1}$ to 1.258
$a_D^{-1}$, where $a_D=(mE_D/\hbar^2)^{1/2}$. 
Thus $\ln \Lambda_*$ ranges over 10\% of its
complete period of $\pi/s_0$, which is enough to trace out 
a significant fraction of the universal scaling curve.  
In Fig.~\ref{fig:B30vsB31he4}, 
the scaling variables $E_T^{(1)}/E_D$ and $E_T^{(0)}/E_D$ 
from the fully converged calculations are shown along with the 
appropriate part of the universal scaling curve from
Fig.~\ref{fig:B30vsB31}(a).  
The numerical results all lie very close to the universal scaling curve.  
Thus the correlation between $E_T^{(1)}$ and $E_T^{(0)}$ is very close 
to that predicted by the scaling limit.  
This is somewhat surprising, because the ground-state binding energy
$E_T^{(0)}$, which ranges from 117 mK to 133 mK depending on the potential, 
is not much smaller than the 
natural ultraviolet cutoff $\hbar^2/m \ell^2_{\rm vdW} = 420$ mK.  
This suggests that the scaling limit is surprisingly robust.

The numerical results in Fig.~\ref{fig:B30vsB31he4} 
all lie systematically below the universal scaling curve.  
This can be explained by the fact that the effective
ranges $r_s$ for the potentials for which fully converged calculations are
available all lie in the narrow range between 13.75 $a_0$ and 13.98 $a_0$.
Much of the discrepancies between the calculated results and the
universal scaling curve can perhaps be explained by the leading power-law
scaling violations, which are first-order in $r_s / a$.

\begin{figure}[htb]
\centerline{\includegraphics*[width=8.5cm,angle=0,clip=true]{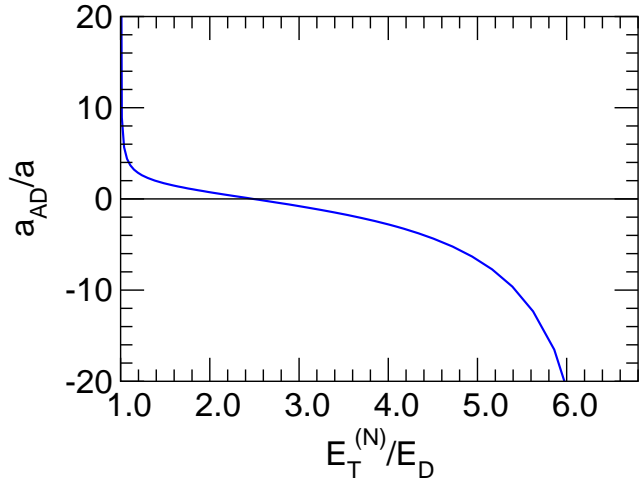}}
\medskip
\caption{
Universal scaling curve for $a_{AD}/a$ vs.~$E_T^{(N)}/E_D$,
where $E_T^{(N)}$ is the binding energy of the shallowest Efimov state.}
\label{fig:AadvsB31}
\end{figure}

The universal scaling curve for 
the scaling variables $E_T^{(1)} / E_D$ and $a_{AD} / a$ 
was calculated in Ref.~\cite{BH02} using the effective field theory 
of Ref.~\cite{BHK99,BHK99b} and is shown in Fig.~\ref{fig:AadvsB31}.  
The ratio $a_{AD} / a$ diverges to $+ \infty$ at
$E_T^{(1)}/E_D = 1$ and to $- \infty$ at $E_T^{(1)}/E_D = 6.75$. 
An approximate expression for this scaling curve in the 
resonant region $a\to +\infty$ can be obtained by using 
Eqs.~(\ref{B3approxc}, \ref{a12-explicit}, \ref{Lambdastar-kappastar}) 
and eliminating $\kappa_* a$.

Several other universal scaling curves have been calculated using the
renormalized zero-range model \cite{AFG95,AF95}.  One is the scaling curve
relating $E_{T'}^{(N+1)}/E_D$ and $E_{T'}^{(N)}/E_D$, 
where $E_{T'}^{(n)}$ is the energy of the virtual state that 
appears when the $n^{\rm th}$ Efimov state disappears
through the atom-dimer threshold \cite{YFDT02}.  The scaling curves relating 
$E_T^{(n)} m a^2/\hbar^2$ and $\langle R^2 \rangle^{(n)} E_T^{(n)}m/\hbar^2$, 
where $\langle R^2 \rangle^{(n)}$ is the mean-square hyperradius
of the $n^{\rm th}$ Efimov state, 
have been calculated for $n =N$ and $N-1$ \cite{YMTF03}.  
Finally the scaling curve relating $\alpha m /\hbar a^4$ and 
$E_D/ E_T^{(N)}$, where $\alpha$ is the 3-body recombination
rate constant, has also been calculated \cite{YFTD03}.


%
%

\section{Effects of Deep Two-Body Bound States}
        \label{sec:deep}

In Section~\ref{sec:uni3}, we assumed implicitly that there are no
deep (tightly-bound) diatomic molecules. In this section, we deduce the
effects of deep molecules on the universal aspects of the 3-body
problem.


\subsection {Extension of Efimov's radial law}
        \label{sec:deep-RL}

In Efimov's derivation of his radial law, he assumed implicitly that the
effects of deep 2-body bound states were negligible. This
assumption implies that all the probability
in an incoming hyperradial wave is reflected back from the short-distance
region $R \sim \ell$ into an outgoing hyperradial wave. 
The resulting expression for the hyperradial wave function in the
scale-invariant region $\ell \ll R \ll |a|$ is Eq.~(\ref{f-sin}).
If there are deep 2-body bound states, this assumption
is not true. Some of the probability in the incoming hyperradial
wave that flows into the short-distance region emerges in the form of
scattering states that consist of an atom and a deep diatomic
molecule with large kinetic energy but small total energy.
We will refer to these states as 
{\it high-energy atom-molecule scattering states}.

The 2-body potentials for the alkali atoms other than hydrogen have many
2-body bound states. If it was necessary to take into account each of
the bound states explicitly, the problem would be hopelessly difficult.
Fortunately the cumulative effect of all the deep 2-body bound states on
low-energy 3-body observables can be taken into account by a simple
extension of Efimov theory.  This extension introduces one additional
low-energy parameter: an inelasticity parameter $\eta_*$
that determines the widths of the Efimov states. 
In the scaling limit, the low-energy 3-body
observables are completely determined by $a$, $\kappa_*$, and $\eta_*$.

The reason the cumulative effects of the deep 2-body bound states can be
described by a single number $\eta_*$ is that all pathways from a low-energy
3-body state with $|E| \ll \hbar^2/m\ell^2$ to a high-energy
atom-molecule scattering state must flow through the lowest 
hyperspherical potential, which in the scale-invariant region has the
form given in Eq.~(\ref{V0-si}).
The reason for this is that in order to reach a high-energy atom-molecule
scattering state, the system must pass through an intermediate 
configuration in which all three atoms are simultaneously close together
with a hyperradius $R$ of order $\ell$ or smaller. It is obvious
that the two atoms that form the bound state must approach to within a
distance of order $\ell$, since the size of the bound state is of
order $\ell$ or smaller. However, the third atom must also approach the
pair to within a distance of order $\ell$. Let $E_{\rm deep}$ be the
binding energy of the 2-body bound state, with 
$E_{\rm deep} \gsim \hbar^2/m\ell^2$.
Energy and momentum conservation then require that the molecule
and the recoiling atom emerge with equal and opposite momenta 
$(4mE_{\rm deep} /3)^{1/2}$, which is of order $\hbar/\ell$
or greater. The third
atom and the pair can deliver the necessary momentum kicks only if they
approach to within short distances of order $\ell$ or smaller. 
Thus any path from a
low-energy 3-body state to a high-energy atom-molecule
scattering state must pass through a configuration with
small hyperradius $R$ of order $\ell$. Such small values of
$R$ are accessible to a low-energy 3-body state
only through the lowest hyperspherical potential.

Efimov's radial law was based on combining the analytic solution to the
hyperradial equation in the scale-invariant region $\ell \ll R \ll |a|$  
in Eq.~(\ref{f-general}) with conservation of probability in the 
short-distance 
region $R \sim \ell$ and in the long-distance region $R \sim |a|$.
If there are deep diatomic molecules, the only aspect that must be 
treated differently is the short-distance region.
Efimov assumed that a hyperradial wave
that flows to short distances is totally reflected back to the scale-invariant 
region.  The amplitude $A$ of the outgoing wave in Eq.~(\ref{f-general})
then differs from the amplitude $B$ of the incoming wave
by a phase as in Eq.~(\ref{A/B}).  If there are deep
molecules, some of the probability in a hyperradial wave
that flows to short distances emerges in the form 
of atom-molecule scattering states.
The fraction of the probability that is reflected back to long distances
through the lowest adiabatic hyperspherical potential is less than 1.
We will denote this fraction by $e^{-4 \eta_*}$ 
and refer to $\eta_*$ as the inelasticity parameter.  
The corresponding boundary condition on the amplitudes 
of the hyperradial waves in Eq.~(\ref{f-general}) is  
\begin{eqnarray}
A = - e^{-2 \eta_* + 2 i \theta_*} B.
\end{eqnarray}

We can now write down the extensions of Efimov's radial laws 
in Eqs.~(\ref{RadLaw})
to the case in which there are deep 2-body bound states.
All that is required is to replace the phase factor $e^{2 i \theta_*}$
associated with reflection from the short-distance region
by the factor $e^{-2 \eta_* + 2 i \theta_*}$:
\begin{subequations}
\begin{eqnarray}
S_{AD,AD} & = & s_{22} 
+  {s_{21} e^{-2 \eta_* + 2i \theta_*} s_{12} 
        \over 1 - e^{-2 \eta_* + 2i \theta_*} s_{11}} \,,
\label{RLdeep:AD}
\\
S_{AD,AAA} & = & s_{23} 
+  {s_{21} e^{-2 \eta_* + 2i \theta_*} s_{13} 
        \over 1 - e^{-2 \eta_* + 2i \theta_*} s_{11}} \,,
\label{RLdeep:ADAAA}
\\
S_{AAA,AAA} & = & s_{33} 
+  {s_{31} e^{-2 \eta_* + 2i \theta_*} s_{13} 
        \over 1 - e^{-2 \eta_* + 2i \theta_*} s_{11}} \,.
\label{RLdeep:AAA}
\end{eqnarray}
\label{RLdeep}
\end{subequations}
All dependence on the radial variable $H$ is contained in the angle 
$\theta_*$, which is still given by Eq.~(\ref{theta-star}).
The symmetric unitary $3\times 3$ matrix $s$
is the same universal function of the angular variable $\xi$ as before. 
The only difference in the radial law is that the S-matrix elements 
now depend also on the inelasticity parameter $\eta_*$.
The remarkable conclusion is that if the universal expressions 
for the S-matrix elements are known in the case $\eta_*=0$,
all the effects of deep 2-body bound states in the scaling limit
can be deduced by the simple substitution 
$\theta_* \to \theta_* + i \eta_*$.

The radial laws can also be generalized to the S-matrix elements 
for transitions from low-energy 3-body scattering states to 
high-energy scattering states
consisting of an atom and a deep diatomic molecule
with large kinetic energy but small total energy.
The transitions from low-energy scattering states to these high-energy 
atom-molecule scattering states involve the wave function in the 
short-distance region $R \sim \ell$.  This wave function may be very
complicated, but it must conserve probability.  We can therefore treat this
region in the same way Efimov treated the scale-invariant region 
$R \sim a$.  If we identify the appropriate asymptotic states, 
the evolution of the wave function between those states will be 
described by a unitary matrix $t$.  
In the scale-invariant region, the asymptotic states
as far as the short-distance region is concerned 
are the outgoing and incoming hyperradial waves 
represented by the first and  second terms on the 
right side of Eq.~(\ref{f-general}).
We denote them by $| 1  \; {\rm out} \rangle$
and $| 1  \; {\rm in} \rangle$, respectively.
Note that the outgoing hyperradial wave is an outgoing asymptotic state
$| 1  \; {\rm out} \rangle$ as far as the short-distance region 
$R \sim \ell$ is concerned,
while it is an incoming asymptotic state $| 1  \; {\rm in} \rangle$
as far as the long-distance region $R \sim |a|$ is concerned.
In the asymptotic region $R \gg |a|$, the asymptotic states 
whose probability can flow directly into or out of the short-distance region
are incoming or outgoing high-energy atom-molecule scattering states.
We denote them by $| X  \; {\rm in} \rangle$
and $| X  \; {\rm out} \rangle$, where $X$ ranges over all the
high-energy atom-molecule scattering states.
The amplitudes for the incoming asymptotic states 
to evolve into the outgoing asymptotic states 
is described by a unitary matrix $t$:
\begin{eqnarray}
t_{ij} = \langle i \; {\rm out} | \hat U | j \; {\rm in} \rangle \,,
\end{eqnarray}
where $\hat U$ is the evolution operator that evolves a wave function
through the short-distance region over an
arbitrarily long time interval.

We already know one element of the
unitary matrix $t$ that describes the evolution of the wave function between
these asymptotic states.  The entry of the matrix that gives the
amplitude of the incoming hyperradial wave $| 1  \; {\rm in} \rangle$
to be reflected into the outgoing hyperradial wave 
$| 1  \; {\rm out} \rangle$ is
\begin{eqnarray}
t_{11} =\exp(- 2 \eta_* + 2i \theta_*) \,.
\end{eqnarray}
The unitarity of the matrix $t$ then determines that the total probability 
for an incoming hyperradial wave to emerge as 
a high-energy atom-molecule scattering state:
\begin{eqnarray}
\sum_X | t_{X1} |^2 = 1 - e^{-4 \eta_*} \,.
\label{sumsq-tX1}
\end{eqnarray}
We can now write down the radial laws for the S-matrix elements 
for the transitions from low-energy scattering states $AAA$ and $AD$
to a high-energy atom-molecule scattering state $X$:
\begin{subequations}
\begin{eqnarray}
S_{X,AD} &=& t_{X1} s_{12} 
+ t_{X1}  {s_{11} e^{- 2 \eta_* + 2i \theta_*} s_{12}
                \over 1-e^{- 2 \eta_* + 2i \theta_*} s_{11}} \,,
\label{RL:XAD}
\\
S_{X,AAA} &=& t_{X1} s_{13} 
+ t_{X1} {s_{11}  e^{- 2\eta_* + 2i \theta_*} s_{13}
                \over 1 - e^{- 2\eta_* + 2i \theta_*} s_{11}} \,.
\hspace{0.6cm}
\label{RL:XAAA}
\end{eqnarray}
\label{RL:XAAAD}
\end{subequations}
The first terms in Eqs.~(\ref{RL:XAD}) and (\ref{RL:XAAA})
are the contributions from transmission 
through the long-distance region followed by transmission 
through the short-distance region to the asymptotic state $X$.  
The second terms include the contributions from arbitrarily many 
reflections of hyperradial waves from the short-distance region 
with amplitude $e^{2i \theta_* - 2 \eta_*}$ 
and from the long-distance region with amplitude $s_{11}$.

The S-matrix elements in Eq.~(\ref{RL:XAAAD}) have a factor $t_{X1}$ 
that depends strongly  on the short-distance behavior 
of the interaction potential.
However, the corresponding rates summed over all 
high-energy atom-molecule scattering states $X$ 
are much less sensitive to short distances.
Squaring the S-matrix elements, summing over the high-energy states $X$, 
and using the unitarity relation in Eq.~(\ref{sumsq-tX1}), we obtain
\begin{subequations}
\begin{eqnarray}
\sum_X | S_{X,AD} |^2 = 
{(1 - e^{-4 \eta_*}) |s_{12}|^2 
        \over | 1 - e^{-2 \eta_* + 2i \theta_*} s_{11} |^2} \,,
\label{sumsq:SXAD}
\\
\sum_X | S_{X,AAA} |^2 = 
{(1 - e^{-4 \eta_*}) |s_{13}|^2 
        \over |1 - e^{-2 \eta_* + 2i \theta_* } s_{11} |^2} \,.
\label{sumsq:SXAAA}
\end{eqnarray}
\end{subequations}
These are the radial laws for the inclusive transitions
from low-energy scattering states into states that include deep molecules.
These inclusive rates are sensitive to short-distances
only through the parameters $a$, $\kappa_*$, and $\eta_*$.


\subsection {Widths of Efimov states}
        \label{sec:deep-widths}

One obvious consequence of the existence of deep diatomic 
molecules is that the Efimov states are no longer sharp states. 
They have widths, because they can decay into an atom and a 
deep molecule.  Thus, the Efimov states are really just resonances 
in the scattering of an atom and a deep molecule.  

The binding energy $E_T$ and width $\Gamma_T$ of an Efimov resonance 
can be obtained as a complex eigenvalue $E=-(E_T+i\Gamma_T/2)$
of the 3-body Schr{\"o}dinger equation.  
If the width is small compared to the binding energy, 
the line shape of the resonance can be approximated by
a Breit-Wigner resonance centered at the energy $-E_T$ 
and with full width at half maximum $\Gamma_T$.
The cross section for the scattering of an atom and a deep molecule 
with total energy $E$ near $-E_T$ is
\begin{eqnarray}
\sigma(E) \approx 
{\Gamma_T^2/4 \over (E+E_T)^2 + \Gamma_T^2/4} \; 
\sigma_{\rm max} \,.
\end{eqnarray}

In the absence of deep molecules, 
the binding energies of Efimov states satisfy Eq.~(\ref{B3-Efimov}), where
$\Delta(\xi)/2$ is the phase shift of a hyperradial wave that is
reflected from the long-distance region $R\sim |a|$.  To obtain the
corresponding equation in the case of deep bound states, we need only
make the substitution $ \theta_* \to \theta_* + i \eta_*$ in
Eq.~(\ref{theta-Delta}):
\begin{eqnarray}
2(\theta_* + i\eta_*) + \Delta(\xi) = 0 \mod 2 \pi.
\end{eqnarray}
This can be satisfied only if we allow complex values of $\xi$
in the argument of $\Delta$. Using the expression
for $\theta_*$  in Eq.~(\ref{theta-star}) and inserting the definition 
of $H$ in Eq.~(\ref{Hxi-def}), we obtain the equation
\begin{eqnarray}
E_T+{i\over 2}\Gamma_T + {\hbar^2 \over ma^2} 
&=& 
\left( e^{-2\pi /s_0} \right)^{n-n_*} 
\exp \left[ {\Delta(\xi)+ 2 i \eta_* \over s_0} \right] 
{\hbar^2 \kappa_*^2 \over m} \,,
\label{B3-Efimov-width}
\end{eqnarray}
where the complex-valued angle $\xi$ is defined by
\begin{eqnarray}
\tan \xi = - \big( m(E_T + i \Gamma_T/2)/\hbar^2 \big)^{1/2} \, a \,.
\end{eqnarray}
To solve this equation for $E_T$ and $\Gamma_T$, we need the analytic
continuation of
$\Delta(\xi)$ to complex values of $\xi$.
The parametrizations for $\Delta(\xi)$ in Eqs.~(\ref{expol})
should be accurate for complex values of $\xi$ with sufficiently small
imaginary parts, except near $\xi=-\pi$ where
the empirical expansion parameter $z$ defined in Eq.~(\ref{z-def})
has  an essential singularity.
If the analytic continuation of $\Delta(\xi)$ were known,
the binding energy and width of one Efimov state
could be used to determine $\kappa_*$ and $\eta_*$.
The remaining Efimov states and their widths could then be
calculated by solving Eq.~(\ref{B3-Efimov-width}).

If the inelasticity $\eta_*$ parameter is extremely small, the right
side of Eq.~(\ref{B3-Efimov-width}) can be expanded to first order in $\eta_*$.
The resulting expression for the width is 
\begin{eqnarray}
\Gamma_T \approx {4 \eta_*\over s_0} 
    \left( E_T +{\hbar^2\over ma^2} \right) \,.
\end{eqnarray}
For the shallowest Efimov states, the order of magnitude of the width is simply
$\eta_* \hbar^2 /ma^2$.  The widths of the deeper Efimov states are
proportional to their binding energies, which behave asymptotically like
Eq.~(\ref{B3-resonant}).  This geometric decrease in the widths of 
shallower Efimov states has been observed in calculations 
of the elastic scattering of atoms with deeply bound molecules \cite{NSE00}.


\subsection{Atom-dimer elastic scattering}
        \label{sec:deepAD}

The effects of deep diatomic molecules modify the universal 
expressions for low-energy 3-body scattering observables
derived in Section~\ref{sec:uni3}.
The radial laws in Eqs.~(\ref{RLdeep}) for the case 
in which there are deep molecules can be obtained from 
Efimov's radial laws in Eqs.~(\ref{RadLaw}) simply by substituting 
$\theta_* \to \theta_* + i \eta_*$.  Thus if the universal 
expression for a scattering amplitude for the case of no deep molecules 
is expressed as an analytic function of $\ln(\kappa_*)$,
the corresponding universal expression for the case in which there are deep
molecules can be obtained simply by substituting
$\ln(a\kappa_*) \to \ln(a\kappa_*) + i \eta_*/s_0$.

Just above the atom-dimer threshold, the S-wave phase shift 
$\delta_0^{AD}(k)$ for atom-dimer scattering can be approximated 
accurately by keeping the first two terms in the 
expansion in Eq.~(\ref{rAD-def}).
The universal expressions for the atom-dimer scattering length $a_{AD}$ 
and for the effective range $r_{s,AD}$ if there are no deep molecules 
are given in Eqs.~(\ref{a12-explicit}) and (\ref{r12-square}).
The corresponding expressions for the case in which there are deep molecules
can be obtained by replacing
 $s_0 \ln (a \Lambda_*)$ by $s_0 \ln (a \Lambda_*)+ i \eta_*$.
The resulting expressions can be written in the form
\begin{subequations}
\begin{eqnarray}
a_{AD} &=& \big( 1.46
 +2.15 \cot [s_0 \ln (a/a_*) + i \eta_*]
\big) \; a \,, 
\label{aAD-deep}
\\
r_{s,AD}&=& \big( 1.13 
+ 0.73 \cot[s_0 \ln(a/a_*)+ 0.98 + i \eta_*]  \big)^2 \; a \,, 
\label{r12-deep}
\end{eqnarray}
\label{ar-deep}
\end{subequations}
where $a_*$ is a value of the scattering length for which
the peak of an Efimov resonance is at the atom-dimer threshold.
The relation between $a_*$ and $\Lambda_*$ can be obtained by 
combining Eqs.~(\ref{B3=B2}) and (\ref{Lambdastar-kappastar}):
\begin{eqnarray}
s_0 \ln(a/a_*) = s_0 \ln(a \Lambda_*) + 1.66 \mod \pi \,.
\label{s0lnastar}
\end{eqnarray}
The expressions in Eqs.~(\ref{ar-deep}) are complex-valued, 
because there are inelastic channels in which the scattering produces
high-energy atom-molecule scattering states.

Near the atom-dimer threshold, the cross section for elastic 
atom-dimer scattering is dominated by S-wave scattering.
At the threshold $E=-E_D$, the differential cross section 
is simply $\left| a_{AD} \right|^2$.
Inserting the expression for the atom-dimer scattering length
in Eq.~(\ref{aAD-deep}) and multiplying by the $4 \pi$ solid angle, 
the cross section has the form
\begin{eqnarray}
\sigma_{AD}(E = -E_D)  
&=& 84.9 \,
{ \sin^2[s_0 \ln(a/a_*) - 0.97] + \sinh^2 \eta_*
\over  \sin^2[s_0 \ln(a/a_*)] + \sinh^2 \eta_*} \,   a^2 \,.
\end{eqnarray}
This expression has maxima near the 
values $a = (e^{\pi/s_0} )^n a_*$ for which the peak of an Efimov resonance 
is at the atom-dimer threshold.
In the strongly inelastic limit $\eta_* \to \infty$, the 
cross section reduces simply to $84.9 \, a^2$.

\begin{figure}[htb]
\bigskip
\centerline{\includegraphics*[width=8cm,angle=0]{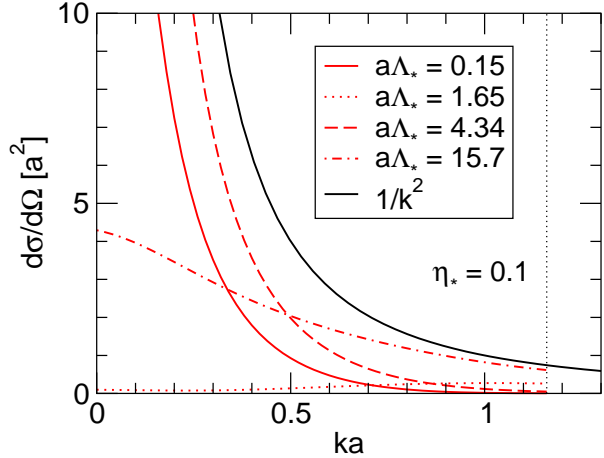}}
\medskip
\caption
{The S-wave contribution to the differential cross section for atom-dimer
scattering in units of $a^2$ as a function of $ka$ for
$a\Lambda_*= 0.15,1.65,4.34,15.7$
and inelasticity parameter $\eta = 0.1$.
The black solid line is the unitarity bound $1/k^2$.
The vertical dotted line is the dimer-breakup threshold.}
\label{fig:dsig-deep}
\end{figure}

We next consider the effect of deep molecules on the cross 
section for atom-dimer scattering below the dimer-breakup threshold.  
The S-wave contribution to the cross section is obtained by inserting 
the S-wave phase shift $\delta_0^{AD}(k)$ into Eq.~(\ref{sigtot-AD}).
The universal expression for the cotangent of the phase shift 
at $ka < 2 /\sqrt 3$
in the case of no deep molecules is given by Eq.~(\ref{kacotdel}).
Explicit parameterizations of the universal
functions $c_1(ka)$, $c_2(ka)$, and $\phi(ka)$ are given in 
Eqs.~(\ref{kcot-par}). 
The phase shift in the case where there are deep molecules
can be obtained by making the substitution
$s_0\ln(a\Lambda_*)\to s_0\ln(a\Lambda_*)+i\eta_*$ in Eq.~(\ref{kacotdel}):
\begin{eqnarray}
ka\,\cot \delta_0^{AD}(k) &=&
c_1(ka) 
+ c_2(ka) \cot [s_0 \ln (a \Lambda_*) + \phi(ka) + i \eta_*] \,.
\label{kacotdel:deep}
\end{eqnarray}
The resulting
differential cross section is shown in Fig.~\ref{fig:dsig-deep} for 
inelasticity parameter $\eta_* = 0.1$ and 
$a\Lambda_*= 0.15,1.65,4.34$, and $15.7$.
Comparing with Fig.~\ref{fig:dsigdomS} which shows the differential
cross section for the same four values of $a \Lambda_*$ but with $\eta_* = 0$, 
we see that the effects of deep 2-body bound states 
fill in the zeroes and prevent the cross section from saturating 
the unitarity bound.


\subsection{Three-body recombination into the shallow dimer}
        \label{sec:deep-shallow}

If there are no deep molecules, the rate constant $\alpha_{\rm shallow}$
for 3-body recombination into the shallow dimer 
has the remarkable form given in Eq.~(\ref{alpha-analytic}), 
which has zeroes at values $a$ that differ by multiples of 
$e^{\pi/s_0} \approx 22.7$.  
One of the effects of deep molecules is to fill in these zeroes.

\begin{figure}[htb]
\bigskip
\centerline{\includegraphics*[width=8cm,angle=0]{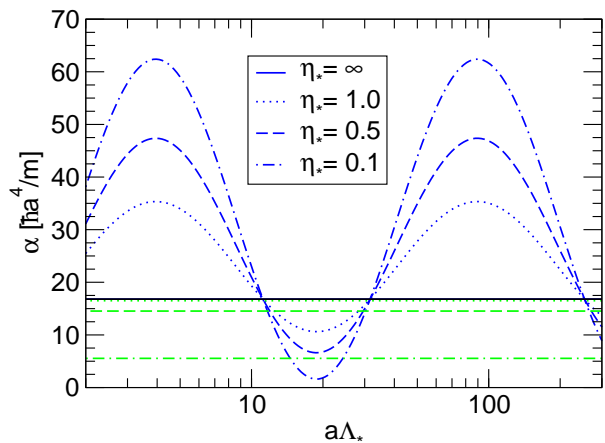}}
\medskip
\caption
{The 3-body recombination rate constants
$\alpha_{\rm shallow}$ (with large-amplitude oscillations) 
and  $\alpha_{\rm deep}$ (with small-amplitude oscillations)
in units of $\hbar a^4/m$ as functions of $a \Lambda_*$ 
for $a >0$ and different values of $\eta_*$.
}
\label{fig:alpha-deep+}
\end{figure}

If there are deep molecules, the radial law for the
S-matrix element for 3-body recombination into the shallow dimer 
is given in Eq.~(\ref{RLdeep:ADAAA}).  For energies 
$E = \hbar^2 \kappa^2 / m$ just above the dimer-breakup threshold, 
the entries of the matrix $s_{ij}$ have the expansions given in 
Eqs.~(\ref{s-3A}).  The limiting behavior of the S-matrix element 
as $E \to 0^+$ is obtained by substituting 
$\theta_* \to \theta_* + i \eta_*$ into Eq.~(\ref{SADAAA:th}):
\begin{eqnarray}
S_{AD,AAA} & \longrightarrow & 
- c_2 e^{i \gamma_1} a^2 K^2 
{ 1 + e^{2i(\theta_* - \delta_0)-2 \eta_*} 
\over  1 - e^{-2 \pi s_0} e^{2i(\theta_* - \delta_0)-2 \eta_*} } \,,
\qquad {\rm as \ } E \to 0^+ \,.
\label{SADAAA:eta}
\end{eqnarray}
The corresponding modification of the analytic expression
for the rate constant $\alpha_{\rm shallow}$  in 
Eq.~(\ref{alpha-analytic})  is
\begin{eqnarray}
\alpha_{\rm shallow} = 
{128 \pi^2 (4 \pi - 3 \sqrt{3})
(\cos^2 [s_0 \ln (a\kappa _*) + \gamma] + \sinh^2\eta_*)
\over
\sinh^2(\pi s_0 + \eta_*) + \sin^2 [s_0 \ln (a\kappa _*) + \gamma] }\,
{\hbar a^4 \over m} \,.
\label{alpha-analytic:eta}
\end{eqnarray}
We can exploit the fact that $e^{2 \pi s_0} \approx 557$ is large
to simplify the expression in Eq.~(\ref{alpha-analytic:eta}).
It can be approximated with an error of less than 1\% by
\begin{eqnarray}
\alpha_{\rm shallow} &\approx& 67.1 \, e^{-2 \eta_*} 
\left( \sin^2[s_0 \ln (a/a_*) + 1.67] + \sinh^2 \eta_* \right)
{\hbar a^4 \over m} .
\label{alpha_sh:deep}
\end{eqnarray}
The relation between $a_*$ and $\Lambda_*$ is given in 
Eq.~(\ref{s0lnastar}). 
The coefficient of $\hbar a^4/m$ is shown as a function of $a \Lambda_*$ 
in Fig.~\ref{fig:alpha-deep+} for several values of $\eta_*$.
As $a$ varies, the coefficient of $\hbar a^4 /m$ oscillates between
about $67.1  e^{-2 \eta_*} \sinh^2 \eta_*$ 
and about $67.1 e^{-2 \eta_*} \cosh^2 \eta_*$. 
Thus one effect of the deep molecules is to eliminate the zeros 
in the rate constant for 3-body recombination into the shallow dimer.
Note that the depth of the minimum is quadratic in $\eta_*$
as $\eta_* \to 0$, so the coefficient of $\hbar a^4/m$ can be very small 
if the inelasticity parameter $\eta_*$ is small.

In the strongly inelastic limit  $\eta_* \to \infty$, 
the coefficient of $\hbar a^4/m$ in Eq.~(\ref{alpha-analytic:eta})
approaches a constant 
16.719 that is extremely close to $1 \over 4$ 
of the maximum coefficient when $\eta_* =0$, which is $C = 67.1177$.
This can be understood from the fact that 
$|s_{32}|^2 / |s_{31}|^2 \to \tanh(\pi s_0)$ as $E \to 0$. 
Since $\tanh(\pi s_0) \approx 0.996$, 
the amplitude for an incoming 3-atom scattering state to be 
reflected from the long-distance region into an outgoing atom-dimer scattering 
state is nearly equal in magnitude to the amplitude for it to be transmitted 
through the long-distance region 
to an incoming hyperspherical wave.  If $\eta_* = \infty$, the incoming 
hyperspherical wave is completely absorbed at short distances and only 
the first amplitude contributes to the recombination rate into the shallow 
dimer. If $\eta_* =0$, the hyperspherical wave is completely reflected at 
short distances and totally transmitted into an outgoing atom-dimer 
scattering state. At special values of $a$, there is constructive 
interference between the two amplitudes and the recombination rate is 
approximately 4 times larger than the contribution from the first amplitude 
alone.


\subsection {Three-body recombination into deep molecules}
        \label{sec:deep-3BR}

The existence of deep diatomic molecules 
opens up additional channels for 3-body recombination.  
If there are no deep molecules,  3-body recombination 
can only produce the shallow dimer if $a>0$ 
and it cannot proceed at all if $a<0$.
If there are deep molecules, they can be produced by 3-body 
recombination regardless of the sign of $a$.
The rate constant $\alpha$ for 3-body recombination 
is defined in Eq.~(\ref{dnA-3br}).
We will denote the inclusive contribution to this rate constant from 
3-body recombination into all the deep molecules by $\alpha_{\rm deep}$.

The radial law for inclusive 3-body recombination into deep molecules 
is given in Eq.~(\ref{sumsq:SXAAA}).
We first consider the case of positive scattering length $a>0$.  
As the energy $E = \hbar^2 K^2 /m$ approaches the dimer-breakup 
threshold, the limiting behaviors of $s_{11}$ and $s_{13}$ are given in 
Eqs.~(\ref{s11-3A}) and (\ref{s13-3A}).  Thus the limiting behavior of
the sum of the squares of S-matrix elements in Eq.~(\ref{sumsq:SXAAA}) 
as $K \to 0$ is 
\begin{eqnarray}
\sum_X | S_{X,AAA} |^2 
\longrightarrow 
{c_1^2 a^4 K^4 (1 - e^{-4 \eta_*} )  \over 
|1 - e^{-2 \pi s_0} e^{2i(\theta_* - \delta_0)-2 \eta_*}|^2} \,,
\qquad {\rm as \ } E \to 0 \,.
\label{sumsq:SXAD0}
\end{eqnarray}
An analytic expression for $\alpha_{\rm deep}$ can be obtained
from Eqs.~(\ref{SADAAA:eta}), (\ref{alpha-analytic:eta}), and
(\ref{sumsq:SXAD0}) by using the facts that 
$\alpha_{\rm deep}/\alpha_{\rm shallow}$ is equal to the ratio
of $\sum_X | S_{X,AAA} |^2$ to $|S_{AD,AAA}|^2$ 
and that $c_2^2/c_1^2 = \tanh(\pi s_0)$:
\begin{eqnarray}
\alpha_{\rm deep} = 
{C_{\rm max} \cosh(\pi s_0) \sinh(\pi s_0) \cosh\eta_* \sinh\eta_*
\over \sinh^2(\pi s_0 + \eta_*) + \sin^2 [s_0 \ln (a\kappa _*) + \gamma] }\,
{\hbar a^4 \over m} \,.
\label{alpha-deep:eta}
\end{eqnarray}
The coefficient of $\hbar a^4/m$ has very weak 
log-periodic dependence on $a \kappa_*$.
We can exploit the fact that $e^{2 \pi s_0} \approx 557$ is large
to simplify the expression in Eq.~(\ref{alpha-deep:eta}).
It can be approximated with an error of less than 1\% by
\begin{eqnarray}
\alpha_{\rm deep} \approx 
16.7 \left( 1 - e^{-4\eta_*} \right) {\hbar a^4 \over m} 
\,, \qquad (a>0) \,.
\label{alpha-deep:a>0}
\end{eqnarray}
The fact that the coefficient of $\hbar a^4/m$
is very nearly constant is a 
consequence of the nearly reflectionless character of the lowest 
adiabatic hyperspherical potential at $E=0$.  
The extremely weak dependence on $a \Lambda_*$ was first observed 
in numerical calculations using an effective field theory
for the case of infinitesimal $\eta_*$ \cite{BH01}. 
The numerical result for the coefficient in Eq.~(\ref{alpha-deep:a>0})
was first derived in Ref.~\cite{Braaten:2003yc}.
The coefficient of $\hbar a^4/m$, which is independent of $a \kappa_*$, 
is shown in Fig.~\ref{fig:alpha-deep+} for several values of $\eta_*$.
	
In the strongly inelastic limit $\eta_* \to \infty$, 
the coefficient of $\hbar a^4/m$ in 
Eq.~(\ref{alpha-deep:eta}) approaches the constant 16.779.
This is extremely close to the coefficient of $\hbar a^4/m$ 
in $\alpha_{\rm shallow}$ in Eq.~(\ref{alpha_sh:deep}) 
in the limit $\eta_* \to \infty$. This can be understood from the fact 
that $|s_{32}| / |s_{31}| \to \tanh(\pi s_0)$ as $E \to 0$.
Since $\tanh(\pi s_0) \approx 0.997$, the amplitude for an incoming 3-atom 
scattering state to be reflected from the long-distance 
region into an outgoing atom-dimer scattering state is almost equal
in magnitude to the amplitude for it to be transmitted through the scaling 
region to an incoming hyperspherical wave.  If $\eta = \infty$, the incoming 
hyperspherical wave is completely absorbed at short distances and 
emerges as high-energy atom-molecule scattering states.
The approximate equality of  $|s_{32}| ^2$ and $|s_{31}|^2$ as $E \to 0$
implies the approximate equality of the 3-body recombination rate 
into the shallow dimer
and the inclusive 3-body recombination rate into deep molecules.

\begin{figure}[htb]
\bigskip
\centerline{\includegraphics*[width=8.5cm,angle=0]{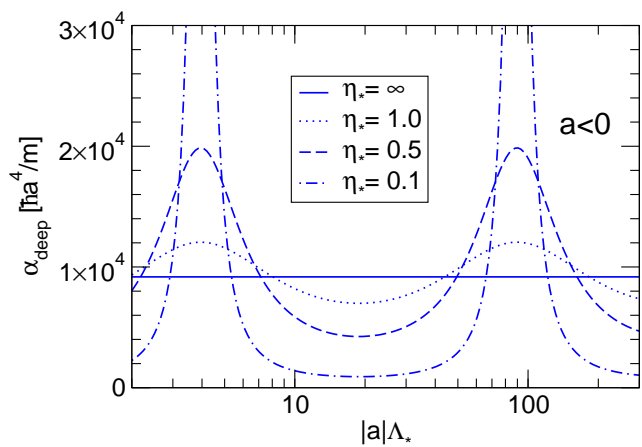}}
\medskip
\caption
{The 3-body recombination rate constant $\alpha_{\rm deep}$ 
in units of $\hbar a^4/m$ as a function of $|a| \Lambda_*$ 
for $a<0$ and different values of $\eta_*$.}
\label{fig:alpha-deep-}
\end{figure}

We now consider the consequences of the radial law 
in Eq.~(\ref{sumsq:SXAAA}) 
for 3-body recombination into deep molecules 
in the case of negative scattering length $a<0$.  
As the energy $E = \hbar^2 K^2 /m$ approaches 
the 3-atom threshold $E=0$, the limiting behavior of the entries 
of the matrix $s$ are given in (\ref{s-AAA}).
Thus the limiting behavior of the sum of the squares 
of the S-matrix elements in Eq.~(\ref{sumsq:SXAAA}) is
\begin{eqnarray}
\sum_X |S_{X,AAA} |^2 & \longrightarrow & 2 d_0 a K
{\sinh(2\eta_*)   \over \sin^2(\theta_* + \gamma) + \sinh^2 \eta_*}\,,  
\qquad {\rm as \ } E \to 0 \,.
\end{eqnarray}
The resulting 3-body recombination constant for $a<0$ is
\begin{eqnarray}
\alpha_{\rm deep} &=& 
{4590 \sinh(2 \eta_*) \over \sin^2 [s_0 \ln (a/a_*')] + 
\sinh^2 \eta_*}
\; {\hbar a^4 \over m}\,, \qquad (a<0) \,,
\label{alpha-deep:a<0}
\end{eqnarray}
where $a_*'$ is a negative value of the scattering length for which
the peak of an Efimov resonance is at the 3-atom threshold.
The relation between $a_*'$ and $\Lambda_*$ can be obtained by 
combining Eqs.~(\ref{B3=0}) and (\ref{Lambdastar-kappastar}):
\begin{eqnarray}
s_0 \ln(a/a_*') = s_0 \ln(|a| \Lambda_*) + 1.72(3) \mod \pi.
\end{eqnarray}
The coefficient of $\hbar a^4/m$ is shown as a function of $a\Lambda_*$ 
in Fig.~\ref{fig:alpha-deep-} for several values of $\eta_*$.
It displays resonant behavior with maxima when the scattering length
has one of the values $(e^{\pi/s_0})^n a_*'$ for which the peak of an 
Efimov resonance is at the 3-atom  threshold. 
The maximum value $9180 \coth \eta_*$ diverges in the limit $\eta_* \to 0$. 
In the  limit $\eta_* \to \infty$, 
the coefficient of $\hbar a^4/m$ in Eq.~(\ref{alpha-deep:a<0}) approaches 
the constant 9180 independent of $a \Lambda_*$.
Thus, the resonant effects associated with Efimov states 
disappear in the limit of strong inelasticity.

The scaling of $\alpha_{\rm deep}$ with $a^4$ was first predicted by
Nielsen and Macek and by Esry, Greene, and Burke \cite{NM-99,EGB-99}.
The existence of a log-periodic sequence of resonances related to 
Efimov states was pointed out by Esry, Greene, and Burke \cite{EGB-99}.
The explicit formula for $\alpha_{\rm deep}$ in Eq.~(\ref{alpha-deep:a<0})
was first derived in Ref.~\cite{Braaten:2003yc}.


\subsection{Dimer relaxation into deep molecules}
        \label{sec:deep-DD}

If there are no deep diatomic molecules, 
atom-dimer scattering is completely elastic below the 
dimer-breakup threshold $ka = 2/\sqrt{3}$.
The existence of deep molecules opens up an inelastic channel 
in which an atom and a shallow dimer with low energy
collide to form an atom and a deep molecule.
The large  binding energy of the molecule is released through the  
large kinetic energies of the recoiling atom and molecule. 
This process is called {\it dimer relaxation}.

The relaxation rate depends on the momenta of the incoming atom and
dimer.  If the momenta are small enough,
the relaxation rate reduces to a constant.
The {\it relaxation event rate constant} $\beta$ is defined so that the
number of relaxation events per time and per volume 
in a gas of very cold atoms with number density $n_A$ 
and very cold dimers with number density $n_D$ is $\beta n_A n_D$.  
The resulting decrease in the number densities is given by
\begin{eqnarray}
{d \ \over d t} n_A & = &  {d \ \over d t} n_D
= - \beta n_A n_D \,.
\label{dn-deact}
\end{eqnarray}
These equations apply equally well if either the
atoms or the dimers or both are in Bose-Einstein condensates.

\begin{figure}[htb]
\centerline{\includegraphics*[width=8.5cm,angle=0,clip=true]{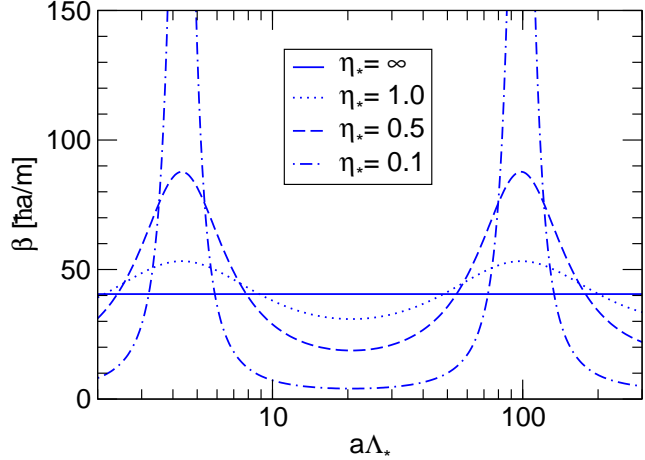}}
\vspace*{0.0cm}
\caption{The dimer relaxation rate constant $\beta$
in units of $\hbar a / m$ for different values of $\eta_*$
as a function of $a \Lambda_*$.}
\label{fig:ddad}
\end{figure}

The radial law for the inclusive dimer relaxation rate is given in 
Eq.~(\ref{sumsq:SXAD}). The total energy of the atom and dimer is expressed
in terms of the  common wave number $k$ of the atom and dimer in 
Eq.~(\ref{E-AD}). Just above the atom-dimer threshold, the nonzero entries 
of the unitary matrix $s$ have the form given in Eq.~(\ref{s-AD}). 
The resulting expression for the sum of the squares of the S-matrix elements 
in Eq.~(\ref{sumsq:SXAD})  reduces  to
\begin{eqnarray}
\sum_X | S_{X,AD} |^2 & \longrightarrow &
2b_0 ak {\sinh (2\eta_*) \over \sin^2(\theta_* + \beta^\prime) 
+ \sinh^2 \eta_*}  \,,
\qquad {\rm as \ } E \to - E_D \,.
\end{eqnarray}
The constants $b_0$ and $\beta^\prime$ are calculated in 
Ref.~\cite{Braaten:2003yc}.  
The resulting expression for the dimer relaxation constant $\beta$ 
defined by Eq.~(\ref{dn-deact}) is given by
\begin{eqnarray}
\beta = 
{20.3 \sinh(2\eta_*) \over \sin^2 [s_0 \ln (a/a_*)] + \sinh^2 \eta_*} 
\; {\hbar a \over m}  \,.
\label{beta:a>0}
\end{eqnarray}
This result was first obtained in Ref.~\cite{Braaten:2003yc}.
The relation between $a_*$ and $\Lambda_*$ is given in 
Eq.~(\ref{s0lnastar}).
The coefficient of $\hbar a/m$ is shown as a function of $a\Lambda_*$ 
in Fig.~\ref{fig:ddad} for several values of $\eta_*$.
It displays resonant behavior with maxima when the scattering length
has one of the values $(e^{\pi/s_0})^n a_*$ for which the peak of an 
Efimov resonance is at the atom-dimer threshold. 
The maximum value $40.6 \coth \eta_*$ diverges in the limit $\eta_* \to 0$. 
In the  limit $\eta_* \to \infty$, 
the coefficient of $\hbar a^4/m$ approaches the constant 40.6
independent of $a \Lambda_*$.
Thus, the resonant effects associated with Efimov states 
disappear  in the limit of strong inelasticity.

The result for the dimer relaxation rate constant in  Eq.~(\ref{beta:a>0}) 
could also have been obtained from the expression for the atom-dimer 
scattering length in Eq.~(\ref{aAD-deep}) using unitarity:
\begin{eqnarray}
\beta = - {6 \pi \hbar \over m} \; {\rm Im} \; a_{AD} \,.
\end{eqnarray}
The optical theorem then implies that the cross section
for inelastic atom-dimer scattering at threshold is
\begin{eqnarray}
\sigma_{AD}^{\rm (inelastic)}(E) \longrightarrow {2 m \over 3 \hbar k} \beta 
\,,\qquad {\rm as\ } E \to - E_D \,.
\end{eqnarray}


%
%

\section{Effective Field Theory}
\label{sec:EFT}

Effective field theory has proved to be a very powerful tool 
for quantitative calculations of the predictions of universality.
In this section, we give an introduction to effective field theory 
and describe how it can be applied to the problem of identical bosons 
with large scattering length in the scaling limit.


\subsection{Effective field theories}
\label{sec:EFT-ep}

Effective theory is a general approach to understanding the 
low-energy behavior of a physical system that has deep roots 
in several areas of physics.  
Some of these roots are described implicitly in Ken Wilson's 
Nobel lecture on the Renormalization Group \cite{Wilson-83}.
Effective field theory is the application of this general method 
to a field theory.  In elementary particle physics, 
the roots of effective field theory have two main branches.  One
branch is concerned with making intuitive sense of the renormalization 
procedure for {\it quantum electrodynamics} (QED) \cite{Lepage-89}.  
The other main branch came from
trying to understand the low-energy behavior of strongly interacting particles
like pions and nucleons. 

In perturbative calculations in QED, intermediate steps are plagued by
{\it ultraviolet divergences} that indicate strong sensitivity to physics at
extremely short distances.  Yet the renormalization procedure allows extremely
accurate predictions of low-energy properties of electrons, positrons, 
and photons in terms of two fundamental parameters:  the
fine structure constant $\alpha$ and the electron mass $m_e$.  
If one tries to introduce any additional parameters into the theory, 
there are ultraviolet divergences that cannot be eliminated 
and the renormalization procedure breaks down.  This would not be a
problem if QED with electrons and photons was a complete theory.  
But in the real world, there are also other heavier charged particles, 
such as muons and pions. 
There are effects from such particles that cannot be absorbed into the
definitions of $\alpha$ and $m_e$.  Adding interaction terms to QED to take
these effects into account destroys the renormalizability of the theory.  

Let us focus specifically on the effects of muons, 
whose mass $m_\mu$ is about 200 times larger than that of the electron.  
The effects of muons can be described with arbitrary accuracy 
by the extension of QED to a quantum field theory that has a muon field 
in addition to the photon and electron fields. 
However, the effects of virtual muons on electrons,
positrons, and photons with energies small compared to $m_\mu$
can also be described with arbitrary accuracy 
by an effective field theory that has only photon and electron fields.
More specifically, the effective field theory approach
involves the construction of a sequence of field theories 
that take into account the effects of muons with
increasing accuracy.  QED is simply the first theory in this sequence.  If the
electrons and photons have momenta of order $p$, the QED predictions for their
scattering amplitudes have errors that are $2^{\rm nd}$ order in $x = p/m_\mu$
and $y=m_e/m_\mu$.  However, the errors can be reduced to $4^{\rm th}$ order in
$x$ and $y$ by using an effective field theory with an additional magnetic
moment interaction.  The additional parameter can be calculated as a function
of $\alpha$, $m_e$, and $m_\mu$.  The errors can be reduced further to $6^{\rm
th}$ order in $x$ and $y$ by adding three additional interaction terms whose
coefficients are calculable as functions $\alpha$, $m_e$, and $m_\mu$. 
Proceeding in this manner, one can take into account the effects of muons 
on electrons, positrons, and photons with momenta small compared to $m_\mu$ 
with arbitrarily high accuracy.

The fundamental quantum field theory that describes 
{\it hadrons}, the particles 
that feel the strong force, is  {\it quantum chromodynamics} (QCD).  
It describes the strong interactions between hadrons in terms of 
gauge interactions between their constituents:
quarks, antiquarks and gluons.  The lightest hadrons are the pions:
$\pi^+$, $\pi^0$, and $\pi^-$. Because the QCD interaction is strong,
the direct calculation of the behavior of pions from QCD is very difficult.
However, effective field theory can provide a
systematically improvable description of the low-energy behavior of 
pions without using any information about QCD other than its
symmetries \cite{Weinberg-79}.  In addition to the space-time symmetries,
QCD has a global symmetry called {\it chiral symmetry}.
The simplest effective field theory for pions is called the
{\it nonlinear sigma model}.  It
has two parameters that can be determined by taking the pion mass
$m_\pi =140$ MeV and the pion decay constant $f_\pi = 93$ MeV as
input. Predictions for the scattering amplitudes of low-energy pions with
momenta of order $p$ have errors that are $4^{\rm th}$ order in 
$x \sim p / (4 \pi f_\pi)$ and $y \sim m_\pi / (4 \pi f_\pi)$. 
However, the errors can be
decreased systematically to $6^{\rm th}$ order in $x$ and $y$ by using an
effective field theory with 10
additional parameters, thus requiring 10 additional
low-energy measurements as input. The error can be decreased even further
to $8^{\rm th}$ order in $x$ and $y$ by adding even more parameters, and so on.
The systematic expansion in $x$ and $y$ generated by this sequence of effective
field theories is called {\it chiral perturbation theory}.\footnote{
See, e.g., Ref.~\cite{DGH92} for a textbook treatment of this
effective field theory.}

Starting from these two main roots, effective field theory has developed
into a universal language for modern elementary particle physics
\cite{Geo93,Kap95,Man96}.
It has two main classes of applications.  One class involves the
systematic development of various low-energy approximations to the
Standard Model of elementary particle physics.  
The other class of applications involves treating the Standard Model 
itself as a low-energy approximation to a more
fundamental theory, such as a unified field theory or string theory.

Effective field theory also has many applications in condensed matter physics
\cite{Shan97,Mari00}.
Examples include the Landau theory of Fermi liquids \cite{Pol92},
phonons \cite{Leutwyler:1996er},
spin waves \cite{Hofmann:1998pp},
the weakly-interacting Fermi gas \cite{Hammer:2000xg},
and the weakly-interacting Bose gas \cite{Andersen03}.


\subsection{Effective theories in quantum mechanics}
\label{sec:EFT-qm}

Most of the applications of effective theories to date have been carried
out within the context of quantum field theory.
However, as pointed out by Lepage \cite{Lepage-97}, the principles of
effective theory apply equally well to problems in quantum mechanics,
such as two particles interacting through a potential $V(r)$.
Suppose we are interested only in the low-energy
observables of the system, where ``low energy" refers to energy
$E$ close to the scattering threshold $E=0$.
The low-energy observables include bound-state energy levels close
to threshold and low-energy scattering cross sections.  Suppose also that 
the potential $V(r)$ is known accurately at long distances $r > r_0$,
but that its short-distance behavior is not known accurately enough 
to calculate the low-energy observables.  
For example, if it is a short-range potential with range smaller 
than $r_0$, then $V(r) = 0$ for $r > r_0$.  
If the particles are real atoms interacting at long
distances through a van der Waals potential, 
then $V(r) \approx -C_6 / r^6$ for $r > r_0$.
Given more and more information about some of the low-energy observables, 
effective theories allow
all other low-energy observables to be calculated with increasingly high
accuracy without having any information about the short-distance
potential.

The basic idea is very simple.  Simply replace $V(r)$ by an
effective potential $V_{\rm eff}(r; c_1)$ that is identical for $r > r_0$ 
and whose form for $r < r_0$
involves an adjustable parameter $c_1$.  For $r < r_0$, the effective
potential need not bear any resemblance to the original potential $V(r)$
as long as it has an adjustable parameter.
Tune the value of this parameter $c_1$ so that
the scattering amplitude at threshold is reproduced exactly.  Then the
Schr{\"o}dinger equation with $V_{\rm eff} (r; c_1)$ will reproduce all
the low-energy observables 
with errors that are linear in $E$.  There is typically some energy scale
$E_0$ at which the errors become roughly 100\%.  We can describe the
errors at energies $|E| < E_0$ as being of order $E/E_0$.
To achieve higher accuracy than order $E/E_0$,
use an effective potential $V_{\rm eff}(r; c_1,c_2)$ with two adjustable
parameters $c_1$ and $c_2$, and tune them to reproduce the scattering
amplitude at threshold and the linear term in its expansion in powers of
the energy $E$. Using this effective potential, all low-energy observables 
involving energies $|E| \ll  E_0$ will
be reproduced with errors of order $(E/E_0)^2$.  If one tunes $N$
parameters, the errors in the S-wave scattering amplitude can be
reduced to order $(E/E_0)^N$, and the errors in other low-energy S-wave
observables will also scale like $(E/E_0)^N$.  Thus a low-energy observable 
can be calculated to increasingly high accuracy by tuning more and more
parameters in the effective potential.  Note that the rate of decrease  
of the error depends on the energy.  The improvement is very rapid
if $|E| \ll E_0$, but there may be no improvement if $|E| \sim E_0$.

It is well-known in the atomic physics community that the determination
of low-energy observables like the scattering length can be improved
by tuning short-distance parameters to fit other low-energy observables.
For example, 
low-energy 2-body observables are known to be extremely 
sensitive to the inner wall of the interatomic potential.
By fine-tuning the inner wall to fit some low-energy 2-body observables,
one can significantly improve the predictions for others.
This method has been used to improve the determination 
of the scattering length for ${}^{23}$Na atoms \cite{TWJJLP96}.
The new insight from effective theory is that tuning more and more
short-distance parameters can give systematically improvable
determinations of low-energy observables with errors that scale as 
increasingly high powers of the energy.

It is easy to prove that by tuning $N$ short-distance parameters,
the errors in S-wave scattering amplitudes can be made to scale like $E^N$.
Let $V_{\rm eff} (r;c)$ be the effective potential that depends on $N$
short-distance tuning parameters $c_1, c_2, . . . , c_N$ that we denote
collectively by $c$. 
Let $u_k(r)/r$ be the radial wave function for S-wave scattering with energy
$E = \hbar^2 k^2/m$ for the true potential $V(r)$,
and let $\delta_0(k)$ be the phase shift for S-wave scattering.
Let $w_k(r;c)/r$ and $\delta_0(k;c)$ be the corresponding quantities 
for the effective potential.
Since $V_{\rm eff}(r; c) = V(r)$ for $r>r_0$,
the Wronskian $u_k(r) w'_k(r;c) - u_k'(r) w_k(r;c)$
must be independent of $r$ in that region.
In the asymptotic region $r \to \infty$, these functions behave like
\begin{subequations}
\begin{eqnarray}
u_k(r) &\longrightarrow&
A \sin[kr + \delta_0(k)] \,,
\\
w_k(r;c) &\longrightarrow&
B \sin[kr + \delta_0(k;c)] \,,
\label{asym}
\end{eqnarray}
\end{subequations}
where $A$ and $B$ are irrelevant constants.
Setting the Wronskian at $r=r_0$ equal to the Wronskian 
of these asymptotic solutions, we have
\begin{eqnarray}
u_k(r_0) w_k'(r_0;c) - u_k'(r_0;c) w_k(r_0) 
&=& k A B \sin[ \delta_0(k) - \delta_0(k;c) ] \,.
\label{Wronskian}
\end{eqnarray}
The left side is an analytic function of $E = \hbar^2 k^2/m$,
because $u_k(r_0)$ and  $w_k(r_0;c)$ are obtained by integrating the
Schr{\"o}dinger equation with parameter $E$ over the finite interval from
$r=0$ to $r=r_0$.  This implies that, up to an overall factor of $k$,
$\delta_0(k)- \delta_0(k;c)$ must be an analytic function of $k^2$.
Note that the phase shifts $\delta_0(k)$ and $ \delta_0(k;c)$ need not 
separately be analytic functions of $k^2$, but the
difference between the phase shifts must be analytic.
Both sides of Eq.~(\ref{Wronskian}) therefore have power series expansion in
$E$ whose coefficients depend on the short-distance parameters
$c_1, \ldots, c_N$.  If those parameters are tuned so that the first $N$
coefficients in the expansion of the left side of Eq.~(\ref{Wronskian})
in powers of $k^2$ vanish, then the differences between the phase shifts
$\delta_0(k)$ and $\delta_0(k;c)$ will be of order $k^{2N-1}$.
This demonstrates that by tuning short-distance parameters 
that affect the
effective potential only in the region $r<r_0$, we can decrease the error
in the phase shifts to higher and higher order in $E$.

The systematic decrease of the errors in the scattering amplitudes 
leads to systematic decrease of the errors in other low-energy observables.
For example, the binding energies $E^{(n)}$ of S-wave bound states
can be determined from the S-wave phase shifts by solving
Eq.~(\ref{BE-eq}).   Thus the sequence of effective potentials 
that give phase shifts with errors that scale as $(E/E_0)^N$
will also give binding energies with errors that scale as $(E^{(n)}/E_0)^N$.

As an illustration of the application of effective theory in quantum mechanics,
we consider a particle in a spherically-symmetric potential $V(r)$ 
that is attractive and proportional to $1/r^2$ 
for $r$ greater than some radius $r_0$:
\begin{subequations}
\begin{eqnarray}
V(r) & =& - \left( \mbox{$1\over 4$} + s_0^2 \right) {\hbar^2 \over 2 m r^2}
\qquad  r > r_0 \,,
\\
& =& V_{\rm short}(r) 
\qquad \hspace{1.5cm}  r < r_0 \,,
\label{Vtrue}
\end{eqnarray}
\end{subequations}
where $s_0$ is a positive parameter.  The coefficient of the 
$1/r^2$ potential is written as ${1\over 4} + s_0^2$ 
because $s_0^2 = 0$ is the critical value above which the potential 
is too singular for the problem to be well-behaved in the limit 
$r_0 \rightarrow 0$.
For example, the spectrum of the Hamiltonian is unbounded 
from below if $s_0^2 > 0$.
We imagine that the short-distance potential $V_{\rm short}(r)$ 
is unknown, but that the energies of bound states can be measured.
The potential $V(r)$ has infinitely many arbitrarily-shallow S-wave 
bound states whose binding energies $E^{(n)}$ have
an accumulation point at the scattering threshold $E=0$.  
As the threshold is approached, the ratio of the binding energies 
of successive states approaches $e^{2\pi/s_0}$.  
The asymptotic spectrum near the threshold therefore has the form
\begin{eqnarray}
E^{(n)}   \longrightarrow 
\left(e^{-2\pi/s_0}\right)^{n-n_*} \hbar^2 \kappa_*^2/m \,,
\qquad {\rm as\ } n \to + \infty \,,
\label{E-low}
\end{eqnarray}
where  $n_*$ is an integer that can be chosen for convenience
and $\kappa_*$ is determined up to a multiplicative factor of 
$e^{\pi/s_0}$ by the short-distance potential.
This geometric spectrum reflects an asymptotic discrete scaling symmetry
in which the distance from the origin is rescaled by the discrete 
scaling factor $e^{\pi/s_0}$.  

The effective theory strategy can be implemented in this problem by 
replacing the potential $V(r)$ by an effective potential 
$V_{\rm eff}(r;\lambda)$ that is identical to $V(r)$ in the region $r>r_0$
but whose behavior in the short-distance region $r < r_0$ depends on a 
tuning parameter 
$\lambda$.  One of the simplest choices for the short-distance potential is a 
spherical delta-shell potential concentrated on a shell with radius  
infinitesimally close to but smaller than $r_0$ \cite{Braaten:2004pg}:
\begin{subequations}
\begin{eqnarray}
V_{\rm eff}(r) & = & - \left( \mbox{$1\over 4$} + s_0^2 \right) 
{\hbar^2 \over 2 m r^2}
\qquad \hspace{0.2cm}  r > r_0,
\\
     & = & -\lambda \frac{\hbar^2}{2mr_0} \delta (r-r_0)
\qquad  r \le r_0 \,.
\label{Vreg}
\end{eqnarray}
\end{subequations}
We will call the dimensionless coefficient $\lambda$ the 
{\it coupling constant}.
Some quantity involving low energies $|E| \ll \hbar^2/mr_0^2$
is selected as a matching quantity, and the
coupling constant is then tuned so that the effective potential 
$V_{\rm eff}(r)$ reproduces the value of the matching quantity for the true 
potential $V(r)$.  
A convenient choice for the matching quantity is the 
bound-state parameter $\kappa_*$ defined by Eq.~(\ref{E-low}). 
The resulting value of the coupling constant depends on $r_0$:
\begin{eqnarray}
\lambda(r_0) = {1\over 2}
- s_0 \cot \left[ s_0 \ln {\kappa_* r_0 \over 2} 
- {\rm arg} \Gamma(1 + i s_0) \right] .
\label{lam-delta}
\end{eqnarray}

\begin{figure}[htb]
\centerline{\includegraphics*[angle=0, width=8cm]{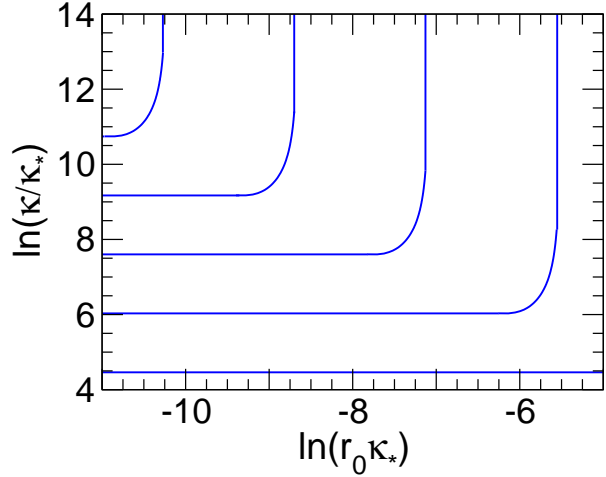}}
\caption{The binding wave numbers $\kappa$ for the deepest bound states 
as a function of $\ln(r_0 \kappa_*)$ for $s_0 = 2$ 
and the delta-shell regularization potential.
}
\label{fig:spectrum-dsh}
\end{figure}

We now consider the bound-state spectrum.  In the effective potential,
the equation for the binding wave number $\kappa$ defined by
$E = - \hbar^2\kappa^2/2m$ is
\begin{eqnarray}
\frac{1}{2} + \kappa r_0 \frac{K'_{is_0} (\kappa r_0)}{K_{is_0} (\kappa r_0)}
- \kappa r_0 \coth (\kappa r_0) = -\lambda (r_0).
\label{bs-delta}
\end{eqnarray}
The spectrum of very shallow bound states has the form (\ref{E-low}).  
The spectrum for the deepest bound states 
is illustrated in Fig.~\ref{fig:spectrum-dsh}.  
At critical values of $r_0$ that differ by multiples of $e^{-\pi/s_0}$,
a new bound state with infinitely large binding energy appears.  
As $r_0$ decreases further, that binding energy rapidly approaches 
its asymptotic value given by (\ref{E-low}).

The binding energies for the true potential $V(r)$ are guaranteed to 
differ from those for the effective potential by errors that scale like 
$E^{(n)}/E_0$, where $E_0= \hbar^2/mr_0^2$ is the energy scale at 
which the effective potential begins to differ significantly from
the true potential.  By tuning a second 
short-distance parameter in the effective potential, 
one could decrease the errors so that they  
scale like $(E^{(n)}/E_0)^2$.

\begin{figure}[htb]
\centerline{\includegraphics*[angle=0, width=8cm]{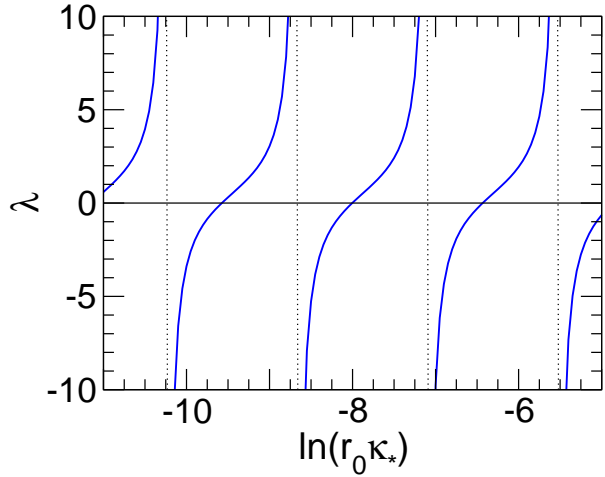}}
\caption{The coupling constant $\lambda(r_0)$ for 
the delta-shell regularization potential 
as a function of ln$(r_0 \kappa_*)$  for $s_0 = 2$.
}
\label{fig:lambda-dsh}
\end{figure}

We can interpret $r_0$ as a short-distance cutoff and $\hbar^2/mr_0^2$ 
as the corresponding ultraviolet energy cutoff.  The tuning of $\lambda(r_0)$ 
can be interpreted as the renormalization of the coupling constant.
As shown in Fig.~\ref{fig:lambda-dsh}, $\lambda(r_0)$ is a 
log-periodic function of $r_0$ with infinite discontinuities.
It jumps discontinuously from $+\infty$ to $-\infty$ as $r_0$ decreases 
through the critical values at which a bound state appears in the spectrum.
The log-periodic behavior of $\lambda(r_0)$ indicates that the renormalization 
is governed by an RG limit cycle. One of the signatures of the RG limit 
cycle is the discrete scaling symmetry of the 
bound state spectrum for the effective potential.
Another simple choice for the effective potential at short distances
is a spherical square-well potential \cite{Beane:2000wh,Bawin:2003dm}.
In this case, the matching condition for the coupling constant $\lambda(r_0)$
has infinitely many solutions.  It can be chosen 
to be a log-periodic function of $r_0$ corresponding to an RG limit cycle, 
but such a choice is not required.
Alternatively, the $1/r^2$ potential can be regularized by a cutoff 
$\Lambda$ in momentum space and renormalized by a momentum-independent
counterterm.  In this case, the counterterm is necessarily a log-periodic 
function of $\Lambda$ corresponding to an RG limit cycle \cite{Hammer:2005sa}.


\subsection{Effective field theories for atoms}

Effective theories can also be used to describe low-energy atoms.
For purposes of illustration, we take the fundamental interaction 
between the atoms to be governed by a 2-body potential $V(r)$.  
The Hamiltonian that describes the $N$-atom system is then
\begin{eqnarray}
\hat H^{(N)}= \sum_{i=1}^N {1 \over 2m} {\bm p} _i^2
+ \sum_{i<j} V(r_{ij}) \,,
\label{H-qm}
\end{eqnarray}
where $r_{ij}=|{\bm r}_{ij}|$ and ${\bm r}_{ij} = {\bm r}_i - {\bm r} _j$. 
There is some natural low-energy length scale $\ell$ associated 
with the potential $V(r)$. 
We are interested only in the low-energy behavior of this
system, where low energy means energy close to the $N$-atom scattering
threshold.  More specifically, we require each atom to have 
kinetic energy small compared to the natural low-energy scale 
$\hbar^2 / m \ell^2$ and we also require each pair of atoms to have 
potential energy small compared to $\hbar^2 / m \ell^2$.

We can describe the low-energy behavior by using an
effective theory.  The simplest possibility
is an effective theory defined by a short-ranged 2-body potential
$V_{\rm eff}(r)$ that depends on a set of short-distance tuning parameters 
$c = (c_1, c_2, \ldots)$:
\begin{eqnarray}
\hat H_{\rm eff}^{(N)}= \sum _{i=1}^N {1 \over 2m} {\bm p}_i^2 +
\sum_{i<j} V_{\rm eff} (r_{ij}) \,.
\label{H-eff}
\end{eqnarray}
One tuning parameter is required to reproduce the scattering length $a$. 
Additional tuning parameters may be required to reproduce
the 2-body scattering amplitude to higher orders in the expansion 
in powers of the energy or to reproduce 3-body or higher $n$-body 
scattering amplitudes to the desired accuracy.

An equivalent formulation of the quantum mechanics of the $N$-atom system
is in terms of a {\it quantum field theory} through the ``second quantization" 
formalism. Instead of coordinate and momentum operators ${\bm r}_i$ 
and ${\bm p}_i$, the theory is formulated in terms of a
quantum field operator $\psi({\bm r})$ that annihilates an atom at the
point ${\bm r}$. If the atoms are bosons, the field operator satisfies
the equal-time commutation relations
\begin{eqnarray}
\big[\psi({\bm r},t), \psi ({\bm r}', t) \big] &=& 0,
\qquad
\big[ \psi({\bm r},t), \psi^\dagger ({\bm r}',t) \big] = \delta^3
({\bm r} - {\bm r}') \,.
\label{cr-qft}
\end{eqnarray}
The time evolution of the quantum field is generated by the Hamiltonian
\begin{eqnarray}
\hat H &=& \int d^3r {\hbar^2 \over 2m} \nabla \psi^\dagger \cdot \nabla \psi
+ {1 \over 2} \int d^3r \int d^3r' \; \psi^\dagger \psi ({\bm r}) 
   V(|{\bm r}-{\bm r}'|) \psi^\dagger \psi ({\bm r}') \,.
\label{H-qft}
\end{eqnarray}
The constraint that there be $N$ particles in the system is implemented
through a number
operator defined by
\begin{eqnarray}
\hat N= \int d^3r \; \psi ^\dagger \psi ({\bm r}) \,.
\label{N-qft}
\end{eqnarray}
A quantum state $|X \rangle$ containing precisely $N$ particles is an
eigenstate of $\hat
N$:
\begin{eqnarray}
\hat N | X \rangle = N | X  \rangle \,.
\label{N-sharp}
\end{eqnarray}
The quantum field theory problem defined by the Hamiltonian in 
Eq.~(\ref{H-qft}),  the commutation relations in Eqs.~(\ref{cr-qft}),
and the constraint in Eq.~(\ref{N-sharp})
is completely equivalent to the $N$-body quantum mechanics problem defined by
the Hamiltonian in Eq.~(\ref{H-qm}) with canonical commutation relations for 
the coordinate and momentum operators. The effective theory defined by the 
effective Hamiltonian in Eq.~(\ref{H-eff}) can also be formulated as a 
quantum field theory by replacing $V(|{\bm r}-{\bm r}'|)$ in Eq.~(\ref{H-qft}) 
with $V_{\rm eff} (|{\bm r}-{\bm r}'|)$.

A class of effective theories that is particularly useful for studying
universal aspects of low-energy physics are those that can be formulated 
as {\it local quantum field theories}.
The Hamiltonian for such a theory can be expressed as the integral of a
Hamiltonian density that depends only on the quantum field $\psi$ 
and its gradients at the same point:
\begin{eqnarray}
\hat H_{\rm eff} = \int d^3r \; {\mathcal H} _{\rm eff} \,.
\end{eqnarray}
There are infinitely many terms that can appear in $H_{\rm eff}$, so we
will write down only a few of them explicitly:\footnote{
The fundamental Hamiltonian in Eq.~(\ref{H-qm}) or (\ref{H-qft}) is invariant 
under Galilean transformations.  This symmetry can be used to constrain the 
terms in the effective Hamiltonian density.}
\begin{eqnarray}
{\mathcal H} _{\rm eff} &=& 
{\hbar^2 \over 2m} \nabla \psi^\dagger \cdot \nabla \psi 
+ \mu \psi^\dagger \psi
+ {g_2 \over 4} (\psi^\dagger \psi)^2
+ {h_2 \over 4} \nabla (\psi^\dagger \psi) \cdot \nabla (\psi^\dagger \psi)
\nonumber \\ && 
+ {g_3 \over 36}(\psi^\dagger \psi)^3 + \ldots \,.
\label{Hdens-eff}
\end{eqnarray}
There are several principles that can be used to reduce 
the number of possible terms in ${\mathcal H} _{\rm eff}$.
If the fundamental Hamiltonian in Eq.~(\ref{H-qft}) has a symmetry,
this symmetry can be imposed on the effective Hamiltonian.
A simple example is the phase symmetry $\psi \to e^{i \alpha} \psi$,
which guarantees conservation of particle number.
It requires that each term in  ${\mathcal H} _{\rm eff}$ have an equal 
number of factors of $\psi$ and $\psi^\dagger$.
We will refer to a term with $n$ factors of both $\psi$ 
and $\psi^\dagger$ as an $n$-body term.
Galilean symmetry imposes particularly powerful 
constraints on ${\mathcal H} _{\rm eff}$.  It forbids any 2-body
terms besides the two terms on the first line of Eq.~(\ref{Hdens-eff}).
The constraints of Galilean symmetry on higher $n$-body terms
are more complicated and will not be given here.
Terms in ${\mathcal H} _{\rm eff}$ that differ by integration by parts are
equivalent, because their difference integrates to a boundary term.
Thus the term $\psi^\dagger \psi \nabla^2 (\psi^\dagger \psi)$
can be omitted, because it is equivalent to the term 
$\nabla (\psi^\dagger \psi) \cdot \nabla (\psi^\dagger \psi)$
in Eq.~(\ref{Hdens-eff}).
Terms with $n$ factors of $\psi$ (and $n$ factors of $\psi^\dagger$)
affect only systems with $n$ or more particles.
Thus if we are considering the 3-body problem, we need only
consider 2-body and 3-body terms in  ${\mathcal H} _{\rm eff}$.
Terms with additional factors of $\nabla$ have effects that
are suppressed by additional powers of the energy $E$.
Thus if we are trying to reproduce the predictions of the 
fundamental Hamiltonian only up to errors that scale like $E^{n+1}$, 
we need only consider terms with up to $2n$ factors of $\nabla$.
The terms shown explicitly in Eq.~(\ref{Hdens-eff}) are sufficient 
to describe 2-body observables up to errors that scale as $E^2$ 
and 3-body observables up to errors that scale as $E$.

The coefficients of the terms in the effective Hamiltonian density in
Eq.~(\ref{Hdens-eff}) are called {\it coupling constants}.
They can be used as tuning parameters  to reproduce low-energy observables. 
The coupling constant $g_2$ for the 2-body contact interaction
can be tuned to reproduce the scattering length. 
The coupling constants $g_2$ and $h_2$ can be tuned simultaneously to
reproduce the scattering length and the effective range associated 
with the 2-body potential.   It may also be necessary to tune the 
coupling constant $g_3$ for the 3-body contact interaction
to reproduce low-energy 3-body scattering amplitudes.
For a theory with short-range interactions,
it is possible to reproduce the low-energy $N$-body scattering amplitudes 
to any desired order in the energy by tuning the coupling constants 
of a local quantum field theory.
This guarantees that the low-energy behavior of an $N$-atom system 
can be described by a local quantum field theory.

One complication of using a local quantum field theory is that it is
ill-defined without an ultraviolet cutoff. We will usually take the 
ultraviolet cutoff to be a cutoff on the wave numbers of atoms 
that can appear in virtual states: $|{\bm k}| < \Lambda$. 
The values of the coefficients will of course depend 
on the cutoff $\Lambda$. With an ultraviolet
cutoff in place, we do not need to be careful about specifying the ordering of
the quantum field operators in the Hamiltonian density (\ref {Hdens-eff}). 
A difference in  operator-ordering can be compensated by a change 
in the coupling constants. In practice, it may be convenient to
use normal-ordered operators, but we will not bother to specify any
operator ordering explicitly.

A local quantum field theory is particularly convenient for describing
the scaling limit of a few-atom system with a large scattering length
$|a| \gg \ell$.
The scaling limit involves taking the range of the interaction to zero, 
but such a limit is built into a local quantum field theory.
One complication is that the large scattering length 
implies strong interactions between the atoms, 
so the field theory must be solved nonperturbatively.
The minimal quantum field theory required to  describe
the 2-atom system in the scaling limit has only a 2-body contact interaction.
Its coupling constant $g_2$ 
can be tuned to give the desired value of the scattering length $a$. 
The minimal quantum field theory required to  describe
the 3-atom system in the scaling limit has also a 3-body contact interaction.
Its coupling constant $g_3$ can be tuned to give the desired value of 
the 3-body parameter $\kappa_*$.  
It would also be possible to reproduce both $a$ and $\kappa_*$ 
by simultaneously tuning $g_2$ and the coupling constant for 
a second 2-body interaction term.  The advantage of using $g_2$ and $g_3$
is that $g_3$ has no effect on the 2-body sector. 
Thus one can first tune $g_2$ 
to get the desired value of $a$ by calculating a 2-body observable, 
and then tune $g_3$ to get the desired 
value of $\kappa_*$ by calculating a 3-body observable.

An important open question is whether additional
tuning parameters would be required to reproduce the low-energy observables 
in the $N$-body sectors, $N=4, 5, 6,\ldots$ to leading order in $\ell / a$. 
This issue will be addressed in Section~\ref{sec:n-body}.


\subsection{Two-body problem}
\label{sec:EFT2}

We will now use our local effective field theory to solve the 2-body
problem. Although the solution is very simple, 
it illustrates many aspects of the solution to the 3-body problem.

\begin{figure}[htb]
\bigskip
\centerline{\includegraphics*[width=8cm,angle=0]{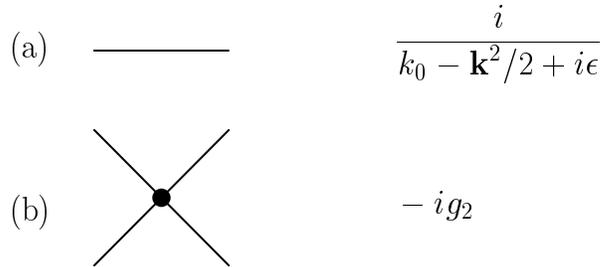}}
\medskip
\caption
{Feynman rules for the Lagrangian in Eq.~(\ref{L-2body}): (a) the propagator
for an atom with energy $k_0$ and momentum ${\bm k}$,
(b) the vertex for the 2-body contact interaction.}
\label{fig:rules2}
\end{figure}

The problem of two identical bosons with large scattering length $a$
in the scaling limit can be described by a local
quantum field theory whose only interaction term is a 2-body contact 
interaction. For practical calculations, it is more convenient to
use the Lagrangian formulation of the effective field theory instead
of the Hamiltonian one from the previous section. The Lagrangian
and Hamiltonian densities are simply related by a Legendre 
transformation. The Lagrangian density is
\begin{eqnarray}
{\mathcal L} = \psi^\dagger \left(i{\partial \ \over \partial t} + {1 \over
2} \nabla^2 \right)\psi - {g_2 \over 4} \left(\psi^\dagger \psi
\right)^2 \,.
\label{L-2body}
\end{eqnarray}
For simplicity of notation, we set $\hbar=1$ and $m=1$ here and in the
remainder of this section.
The coupling constant $g_2$ must be adjusted as a function of the ultraviolet
cutoff $\Lambda$ so that the field theory describes atoms 
with scattering length $a$.

If the effects of the interaction term in Eq.~(\ref{L-2body}) are
calculated as a power series in $g_2$ using perturbation theory, 
the effective field theory describes scattering states of two atoms. 
After renormalization, the scattering amplitude
coincides with the expansion of the universal scattering amplitude 
in Eq.~(\ref{f-2}) in powers of  $k a$.
If the effects of the interaction term are calculated
nonperturbatively, we not only obtain the complete universal expression
for the scattering amplitude in Eq.~(\ref{f-2}) but we find that the
effective field theory also describes bound states with binding energy
given by Eq.~(\ref{B2-uni}). Thus this effective field theory reproduces all
the universal low-energy observables of the 2-body problem with large
scattering length.

All information about the physical observables in the 2-body sector is
encoded in the 4-point Green's function $\langle 0 | {\rm T}( \psi \psi
\psi^\dagger \psi^\dagger) | 0 \rangle$, where $|0 \rangle$ is the vacuum
state, $T$ represents time-ordering, and we have suppressed the time and
space coordinates of the 4 field operators. The physical information is
encoded more succinctly in the truncated connected Green's function in
momentum space which we denote by $i{\mathcal A}$. It is obtained by
subtracting the disconnected terms that have the factored form 
$\langle 0 | {\rm T}(\psi \psi^\dagger) | 0 \rangle
\langle 0 | {\rm T}(\psi \psi^\dagger) | 0 \rangle$,
Fourier transforming in all coordinates, factoring out an overall
energy-momentum conserving delta function, and also factoring out
propagators associated with each of the four external legs. We will refer
to ${\mathcal A}$ as the $2 \to 2$ off-shell amplitude. 

\begin{figure}[htb]
\bigskip
\centerline{\includegraphics*[width=10cm,angle=0]{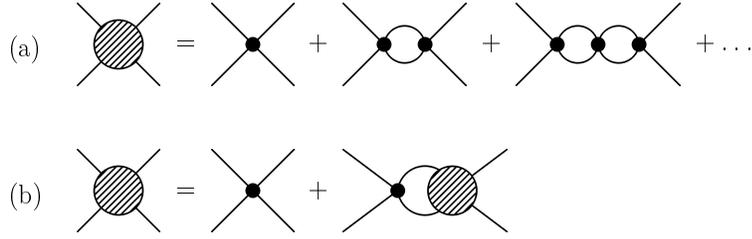}}
\medskip
\caption
{Diagrammatic equations for the $2\to 2$ off-shell amplitude:
(a) the perturbative expansion in $g_2$, 
and (b) the integral equation.}
\label{fig:amp2}
\end{figure}

The amplitude $i \mathcal A$ can be expressed as the sum of connected 
Feynman diagrams constructed out of the propagator in 
Fig.~\ref{fig:rules2}(a) and the vertex in Fig.~\ref{fig:rules2}(b). 
The first three diagrams in the perturbative expansion of $i \mathcal A$ 
in powers of $g_2$ are shown in  Fig.~\ref{fig:amp2}(a).
Energy and momentum are both conserved at every
vertex of the Feynman diagrams.  
For every closed loop, there is an
energy $p_0$ and a momentum ${\bm p}$ that are not determined 
by the external energies and momenta. They must be integrated
over with the measure $d^4p/(2 \pi)^4$.
There are also symmetry factors of
$1/n!$ associated with subdiagrams that are invariant under the
permutation of $n$ internal lines. For example, the second diagram in
Fig.~\ref{fig:amp2}(a) has a symmetry factor of $1/2$ and the third diagram has
a symmetry factor of 1/4. 

In general, the amplitude ${\mathcal A}$ depends on the energies and momenta 
of the four external lines. It is called an {\it off-shell} amplitude, 
because the energy $p_0$ of an external line with momentum ${\bm p}$ need 
not be equal to its
physical value $p^2/2$. The T-matrix element ${\mathcal T}$ for a $2 \to 2$
scattering process is obtained by evaluating ${\mathcal A}$ at the {\it
on-shell point} where $p_0$ is set equal to $p^2/2$ for every external
momentum ${\bm p}$. In the center-of-mass frame, we can take the 2
incoming momenta to be $+{\bm k}$ and $-{\bm k}$ and the two outgoing
momenta to be $+{\bm p}$ and $-{\bm p}$. The amplitude ${\mathcal A}$ then
depends on ${\bm k}$, ${\bm p}$, and the 4 off-shell energies. When the
only interaction is the 2-body contact interaction in
Fig.~\ref{fig:rules2}(b), the amplitude simplifies enormously because it can
depend only on the total momentum and the total off-shell energy $E$. In
the center-of-mass frame, it is a function of $E$ only, so we will
denote it by ${\mathcal A} (E)$. The on-shell point corresponds to setting
$E=2(k^2/2)=2(p^2/2)$, which requires $p=k$. 
The off-shell $2 \rightarrow 2$ amplitude ${\mathcal A}(E)$  encodes all
physical information about the 2-body system at low energies. For
example, the T-matrix element for atoms of momenta 
$\pm {\bm k}$ to scatter into atoms of momenta $\pm {\bm k}'$ 
with $|{\bm k}'| = |{\bm k}| = k$ is
\begin{eqnarray}
{\mathcal T} (k) = {\mathcal A} (E=k^2) \,.
\end{eqnarray}
The conventional scattering amplitude $f_k (\theta)$ for atoms with
momenta $\pm{\bm k}$ to scatter through an angle $\theta$ is
proportional to ${\mathcal A}(E)$  evaluated at the on-shell point:
\begin{eqnarray}
f_k (\theta) = {1 \over 8 \pi} {\mathcal A} (E = k^2) \,.
\label{f-A}
\end{eqnarray}
The limit of the scattering amplitude as $k \to 0$ determines the
scattering length: 
\begin{eqnarray}
a = - {1 \over 8 \pi} {\mathcal A} (0) \,.
\label{a-A}
\end{eqnarray}

The contact interaction in Eq.~(\ref{L-2body}) is ill-defined unless an
ultraviolet cutoff is imposed on the momenta in loop diagrams.
This can be seen by writing down the off-shell amplitude for 2-body
scattering at second order in perturbation theory:
\begin{eqnarray}
{\mathcal A} (E) & \approx & - g_2 - {i \over 2} 
g_2^2 \int {d^3q\over (2\pi)^3}
\int{dq_0 \over 2 \pi} {1 \over q_0 - q^2 /2 + i \epsilon} 
{1 \over E - q_0 - q^2 /2 + i\epsilon} + \ldots \,.
\nonumber\\
\label{A-pert}
\end{eqnarray}
The two terms correspond to the first two diagrams in 
Fig.~\ref{fig:amp2}(a). 
The intermediate lines have momenta $\pm {\bm q}$.
The integral over $q_0$ in Eq.~(\ref{A-pert}) is easily
evaluated using contour integration:
\begin{eqnarray}
{\mathcal A} (E) \approx -g_2 - {1 \over 2} g_2^2 
\int {d^3q \over (2 \pi)^3} {1 \over E - q^2 +i \epsilon} + \ldots \,.
\end{eqnarray}
The integral over ${\bm q}$ diverges. It can be regularized by imposing
an ultraviolet
cutoff $|{\bm q}| < \Lambda$. Taking the limit $\Lambda \gg |E|^{1/2}$,
the amplitude reduces to \footnote{
If the calculation was carried out in a
frame in which the total momentum of the two scattering particles was
nonzero, the simple cutoff $|{\bm q}| < \Lambda$ would give a result that does
not respect Galilean invariance. To obtain a Galilean-invariant result 
requires either using a more sophisticated cutoff or else imposing
the cutoff $|{\bm q}| < \Lambda$ only after an
appropriate shift in the integration variable ${\bm q}$.}
\begin{eqnarray}
{\mathcal A} (E) \approx - g_2 + {g_2^2 \over 4 \pi^2} 
\left(\Lambda - {\pi \over 2} \sqrt {-E -i \epsilon} \right) + \ldots \,.
\label{amp2-2nd}
\end{eqnarray}

The dependence on the ultraviolet cutoff $\Lambda$ can be consistently
eliminated by a perturbative renormalization procedure. A simple choice
is to eliminate the parameter $g_2$ in favor of the scattering length
$a$, which is given by Eq.~(\ref{a-A}):
\begin{eqnarray}
a \approx {g_2 \over 8 \pi} \left( 1- {g_2 \Lambda \over 4\pi^2} + \ldots
\right) \,.
\end{eqnarray}
Inverting this expression to obtain $g_2$ as a function of $a$ we obtain
\begin{eqnarray}
g_2  \approx 8 \pi a \left( 1+ {2 a \Lambda \over \pi}  + \ldots \right) \,,
\end{eqnarray}
where we have truncated at second order in $a$. Inserting the expression
for $g_2$ into Eq.~(\ref{amp2-2nd}) and expanding to second order in $a$,
we obtain the renormalized expression for the amplitude:
\begin{eqnarray}
{\mathcal A} (E) \approx -8 \pi a \left( 1 + a \sqrt{-E - i \epsilon} +
\ldots \right) \,.
\label{A2pert}
\end{eqnarray}
If we evaluate this at the on-shell point $E=k^2$ and insert it into
Eq.~(\ref{f-A}), we find that it reproduces the first two terms in the
expansion of the universal scattering amplitude in Eq.~(\ref{f-2}) 
in powers of $ka$. By calculating ${\mathcal A} (E)$ to higher order 
in perturbation theory, we can reproduce the low-momentum 
expansion of Eq.~(\ref{f-2}) to higher order in $ka$. Thus a perturbative
treatment of the effective field theory reproduces the low-momentum
expansion of the 2-body scattering amplitude.
The perturbative approximation is valid only if the energy satisfies
$E \ll 1/a^2$.

If we are interested in observables involving energy $E \sim 1/a^2$, 
then we must solve the problem nonperturbatively
\cite{vanKolck:1998bw,Kaplan:1998we}.
This is most easily accomplished by realizing that
the Feynman diagrams in Fig.~\ref{fig:amp2}(a) form a geometric series.
Summing the geometric series, the exact expression
for the amplitude is
\begin{eqnarray}
{\mathcal A}(E) = -g_2 \left[ 1 + {g_2 \over 4 \pi^2}
\left( \Lambda - {\pi \over 2} \sqrt {-E -i\epsilon} \right) \right]^{-1} \,.
\label{A-nonpert}
\end{eqnarray}
Alternatively, we can use the fact that summing the diagrams in
Fig.~\ref{fig:amp2}(a) is equivalent to solving the following
integral equation:
\begin{eqnarray}
{\mathcal A}(E) & = & -g_2 - {i \over 2} g_2 \int {d^3q \over (2\pi)^3}
\int{dq_0 \over 2\pi} {1 \over q_0 - q^2/2 + i \epsilon}
{1 \over E - q_0 - q^2/2 + i \epsilon} \, {\mathcal A} (E) \,.
\nonumber
\\
\label{inteq-2}
\end{eqnarray}
The integral equation is expressed diagrammatically in Fig.~\ref{fig:amp2}(b).
Since the function ${\mathcal A}(E)$ is independent of ${\bm q}$ and $q_0$,
it can be pulled
outside of the integral in  Eq.~(\ref{inteq-2}). The integral can be
regularized by imposing an ultraviolet cutoff $\Lambda$.
The integral equation is now trivial to solve and the
solution is given in Eq.~(\ref{A-nonpert}).

The expression for the nonperturbative $2 \to 2$
off-shell amplitude in Eq.~(\ref{A-nonpert}) depends on the
parameter $g_2$ in the Lagrangian and
on the ultraviolet cutoff $\Lambda$. Renormalization can be implemented
by eliminating $g_2$ in favor of a low-energy observable, such as the
scattering length $a$. Using Eq.~(\ref{f-A}), the nonperturbative expression
 for the scattering length is
\begin{eqnarray}
a = {g_2 \over 8 \pi}
\left( 1 + {g_2 \Lambda \over 4 \pi^2} \right)^{-1} \,.
\label{a-g2}
\end{eqnarray}
Solving for $g_2$, we obtain
\begin{eqnarray}
g_2 = 8\pi a \left( 1 - {2 a \Lambda \over \pi} \right)^{-1} \,.
\label{g2-tune}
\end{eqnarray}
Given a fixed ultraviolet cutoff $\Lambda$, this equation prescribes how
the parameter $g_2$ must be tuned in order to give the correct
scattering length $a$.  Note that for $\Lambda \gg 1/|a|$,
the coupling constant $g_2$ is always negative 
regardless of the sign of $a$.
Eliminating $g_2$ in Eq.~(\ref{A-nonpert}) in favor of $a$, 
we find that the nonperturbative off-shell amplitude reduces to
\begin{eqnarray}
{\mathcal A} (E) = {8\pi \over - 1/a + \sqrt {-E -i \epsilon}} \,.
\label{a-npren}
\end{eqnarray}
In this simple case, we find that our renormalization prescription 
eliminates the dependence on $\Lambda$ completely. In general, 
we should expect it to only be suppressed by powers of
$1/(a \Lambda)$ or $E/ \Lambda^2$.  A final step of taking the limit 
$\Lambda \to \infty$ would then be required to obtain results 
that are completely independent of  $\Lambda$.

Evaluating the off-shell amplitude in Eq.~(\ref{a-npren}) 
at the on-shell point $E = k^2$ and inserting
it into the expression for the scattering amplitude in Eq.~(\ref{f-A}), 
we recover the universal expression in Eq.~(\ref{f-2}). 
The differential cross section is therefore given by Eq.~(\ref{sig-2}). 
The nonperturbative off-shell amplitude in Eq.~(\ref{a-npren})
also encodes information about bound states. 
It is an analytic function of the complex energy $E$
except for a branch cut along the positive real axis
and possibly a pole on the negative real axis.
The branch cut is associated with 2-particle scattering states. 
A pole on the negative real axis corresponds to a
bound state. If $a>0$, the amplitude in Eq.~(\ref{a-npren}) has a pole at
$E = -1/ a^2$. This pole indicates that there is a 2-body bound state with
binding energy given by Eq.~(\ref{B2-uni}). If $a<0$, the pole in the off-shell
amplitude is located at $E = e^{3 \pi i} /a^2$, which is on the
second sheet of the complex energy $E$. Such a state is called 
a virtual state.  Therefore there is no 2-body bound state when $a<0$.

The formula for the scattering length in Eq.~(\ref{a-g2}) 
illustrates an important basic principle of
effective theories. Nonanalytic behavior in long-distance observables
generally arises from completely analytic behavior in short-distance
parameters. In this case, the long-distance observable is the scattering
length $a$, and the short-distance parameter is the strength $g_2$ of
the 2-body contact interaction. We should think of the ultraviolet
cutoff $\Lambda$ as some fixed momentum scale that is large compared to
the wave numbers of interest. Particles with wave numbers less than
$\Lambda$ are taken into account explicitly in the field theory.
Particles with wave numbers greater than $\Lambda$ are excluded by this
cutoff. The effects of virtual 2-body states with such wave numbers are
taken into account through the strength $g_2$ of the 2-body contact
interaction. According to Eq.~(\ref{a-g2}), the scattering length diverges
when $g_2$ is tuned to the critical value $-4 \pi^2 / \Lambda$. There is
nothing particularly remarkable about this value as far as
short-distance physics is concerned. The divergence in $a$ arises from
the iteration of quantum fluctuations involving virtual particles with
wave numbers less than $\Lambda$, which generates the term 
$g_2 \Lambda/4 \pi$ in the denominator of Eq.~(\ref{a-g2}). 
The scattering length $a$ is not an analytic function of
$g_2$ at the critical point: a small change in the short-distance
parameter $g_2$ can produce an enormous change in $a$.

The formula in Eq.~(\ref{g2-tune}) illustrates that an arbitrarily large 
coupling
constant is not necessarily pathological in a nonperturbative field theory. 
The expression for $g_2$ diverges as $\Lambda \to \pi / 2 a$.  
However, physical observables are independent of $\Lambda$.  
Thus the effects of the arbitrarily large coupling constant $g_2$ 
must be compensated by equally large effects from
the iteration of quantum fluctuations involving virtual particles.

\begin{figure}[htb]
\bigskip
\centerline{\includegraphics*[width=8.5cm,angle=0]{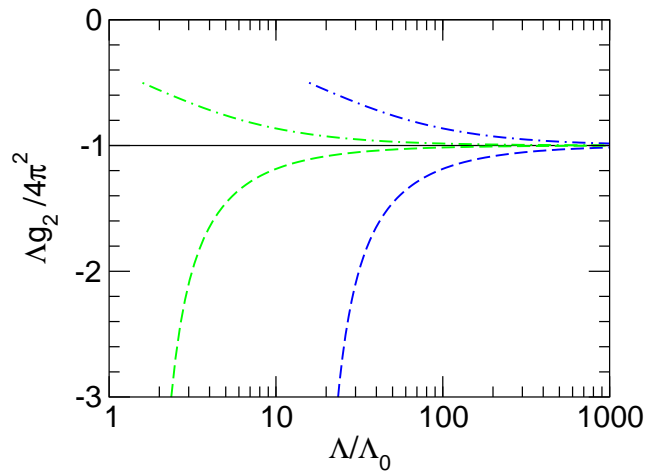}}
\medskip
\caption
{The dimensionless 2-body coupling constant $\hat g_2 = \Lambda g_2/4 \pi^2$
        as a function of the ultraviolet cutoff $\Lambda$ for 
        several values of the scattering
        length $a$.  As $\Lambda \to \infty$, $\Lambda g_2$
        asymptotically approaches an RG fixed point.}
\label{fig:g2RGtra}
\end{figure}

We now discuss the renormalization of this field theory from a 
renormalization group perspective.  We consider a dimensionless 
combination of the coupling constant $g_2$ 
and the ultraviolet cutoff $\Lambda$:
\begin{eqnarray}
\hat g_2(\Lambda)  =   {\Lambda g_2 \over 4 \pi^2} \,.
\label{g2hat}
\end{eqnarray}
Using Eq.~(\ref{g2-tune}),
the dimensionless coupling constant can be written
\begin{eqnarray}
\hat g_2(\Lambda)  = -  \frac {a \Lambda} {a \Lambda - \pi/2} \,.
\label{Lamg2}
\end{eqnarray}
As $\Lambda$ is varied with $a$ fixed, the expression for 
$\hat g_2$ in Eq.~(\ref{g2hat}) maps out a renormalization group 
(RG) trajectory.  The RG trajectories for various values of $a$
are illustrated in Fig.~\ref{fig:g2RGtra}.  
All the points on a given trajectory represent 
the same physical theory with a given scattering length $a$.  
As $\Lambda$ increases, the dimensionless coupling
constant in Eq.~(\ref{Lamg2}) flows towards an ultraviolet fixed point:
\begin{eqnarray}
\hat g_2(\Lambda)  \longrightarrow - 1  
\qquad \mbox{as} \ \Lambda \to \infty \,.
\end{eqnarray}
Using (\ref {a-g2}), we can identify 
the fixed-point theory as the 2-body problem in the resonant limit 
$a \to \pm \infty$. The scaling limit $\ell \to 0$ is implicit 
in our formulation of the problem in terms of a local quantum field theory, 
so the fixed point corresponds to taking the resonant and scaling limits 
simultaneously.  All the RG trajectories in Fig.~\ref{fig:g2RGtra}
are focused towards this fixed point as $\Lambda \to \infty$.  
The focusing of the RG trajectories indicates that as the energy scale becomes
larger and larger compared to $1/ a^2$, 
the system with fixed scattering length $a$
behaves more and more like the resonant limit.  
For a physical system, there is a natural ultraviolet cutoff 
$\Lambda \sim 1/ \ell$ beyond which the physics can
no longer be reproduced by a 2-body contact interaction.  
For $\Lambda > 1 /\ell$, the behavior of the system becomes more 
complicated and it no longer flows toward the fixed point 
corresponding to the resonant and scaling limits.

Further insight into this problem can be achieved by expressing the 
renormalization group flow in terms of a differential equation
for the $\Lambda$-dependence of $\hat g_2$.
By differentiating both sides of Eq.~(\ref{Lamg2}), one can derive 
the differential RG equation \cite{Kaplan:1998we}
\beq
\Lambda\frac{d}{d\Lambda} \hat g_2 = \hat g_2 (1+ \hat g_2) \,.
\label{eq:RGg2}
\eeq
It is clear from this equation that the RG flow of $\hat g_2$  has two 
fixed points: $\hat g_2=-1$ and $\hat g_2=0$. The fixed point
$\hat g_2=-1$ corresponds to the resonant limit $a\to \pm \infty$
discussed above. 
The second fixed point $\hat g_2=0$ corresponds to the noninteracting 
system with $a=0$.   The perturbative expansion for the scattering amplitude 
in Eq.~(\ref{A2pert}) corresponds to an expansion about this fixed point. 
In the theory of the homogenous Bose gas, the expansion 
in powers of the diluteness parameter $(na^3)^{1/2}$ \cite{Andersen03} 
provides another example of an expansion 
around this noninteracting fixed point.

At the two fixed points, the 2-body system is scale invariant.
In general, the continuous scaling symmetry defined by Eqs.~(\ref{scaling-1})
is a mapping of the theory onto another theory with a different 
scattering length.
At the fixed points $a = \pm \infty$ and $a=0$, the continuous scaling symmetry
maps the theory onto itself.   For the $a=0$ fixed point, the scale invariance
is trivial because there are no interactions.  The $a = \pm \infty$ fixed point
has nontrivial scale invariance.  The 2-body system at this fixed point 
is actually invariant under a larger group of conformal symmetry 
transformations \cite{Mehen:1999nd}.


\subsection{Three-body problem}
\label{sec:EFT3}

\begin{figure}[htb]
\bigskip
\centerline{\includegraphics*[width=5cm,angle=0]{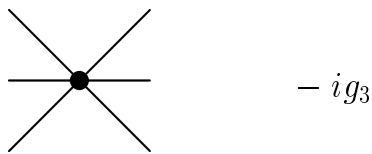}}
\medskip
\caption
{Feynman rule for the 3-body contact interaction.}
\label{fig:vert3}
\end{figure}

We now formulate the 3-body problem in the language of effective field
theory. The problem of three identical bosons with large scattering length
in the scaling limit can be described by
a local quantum field theory whose only interaction terms are 2-body
and 3-body contact interactions. The Lagrangian density is
\begin{eqnarray}
{\mathcal L} &=& \psi^\dagger \left(i{\partial \ \over \partial t} 
+ {1 \over2} \nabla^2 \right)\psi 
- {g_2 \over 4} \left(\psi^\dagger \psi \right)^2
- {g_3 \over 36} \left( \psi^\dagger \psi \right)^3 \,.
\label{L-3body}
\end{eqnarray}
In addition to the Feynman rules in Fig.~\ref{fig:rules2},
there is the additional Feynman rule for the
3-body contact interaction in Fig.~\ref{fig:vert3}.

\begin{figure}[htb]
\bigskip
\centerline{\includegraphics*[width=9cm,angle=0]{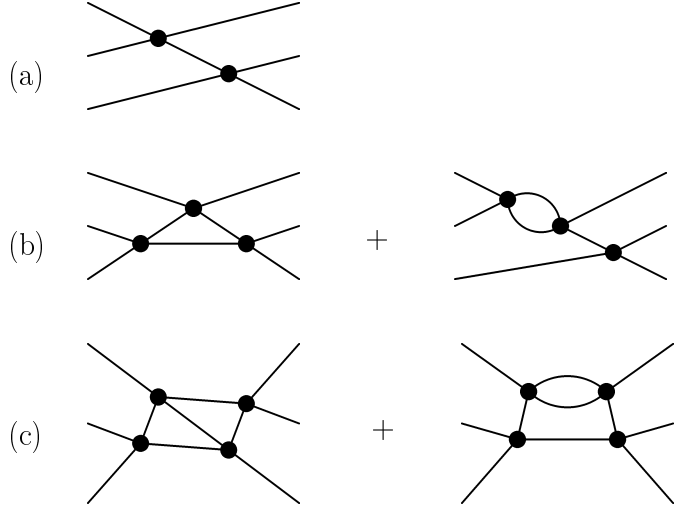}}
\medskip
\caption
{Feynman diagrams for the $3 \to 3$ elastic scattering amplitude:
(a) a tree diagram that contributes to the $E^{-1}$ term, 
(b) one-loop diagrams that contribute to the $E^{-1/2}$ term,
(c) two-loop diagrams that contribute to the $\ln(E)$ term.}
\label{fig:33}
\end{figure}

If the interaction terms in Eq.~(\ref{L-3body}) are treated perturbatively,
the 3-body sector of the quantum field theory describes 3-atom
scattering states. The perturbative expansion in powers of $g_2$ and
$g_3$ coincides after renormalization with an expansion in powers of the
energies of the scattering atoms. The renormalization associated with
$g_2$ can be implemented by making the substitution 
in Eq.~(\ref{g2-tune}) and expanding in powers of $a$. 
If we take all the atoms to have energies proportional to $E$, 
then the T-matrix element can be expanded in powers of $E$.  
The expansion contains singular terms proportional to $E^{-1}$, $E^{- 1/2}$, 
and $\ln (E)$ \cite{AR70}. 
At the leading order $g^2_2$ of the perturbation expansion 
in $g_2$, the $E^{-1}$ term comes from tree diagrams such as the one in
Fig.~\ref{fig:33}(a).  There are higher order contributions to the $E^{-1}$
term from the insertion of the string of one-loop bubble diagrams shown in
Fig.~\ref{fig:amp2} in place of the vertices in Fig.~\ref{fig:33}(a). 
Their effect is simply to renormalize the coupling constant,  
$g_2 \to 8 \pi a$, so that the $E^{-1}$ term is proportional to $a^2$.
There are two classes of contributions 
to the $E^{-1/2}$ term in the T-matrix element for 3-atom elastic scattering.  
At the leading order $g_2^3$, the $E^{- 1/2}$ term comes from the 
1-particle-irreducible 1-loop diagram in Fig.~\ref{fig:33}(b) 
and from the insertion of a 1-loop bubble diagram in
place of either of the vertices in Fig.~\ref{fig:33}(a).  
Again the contributions from higher order diagrams is simply to 
renormalize the coupling constant,
so that the $E^{-1/2}$ term is proportional to $a^3$.
Terms with the logarithmic 
singularity $\ln (E)$ arise first at order $g^4_2$ from the 
1-particle-irreducible 2-loop diagrams with the two topologies
shown in Fig.~\ref{fig:33}(c).  The effects of higher-order diagrams 
in which the vertices in Fig.~\ref{fig:33}(c) are replaced by strings 
of the one-loop bubble diagrams in Fig.~\ref{fig:amp2} is to 
renormalize the coupling constant, so that the coefficient of the 
$\ln (E)$ term is proportional to $a^4$.

The logarithm of $E$ in the T-matrix element arises 
from a scale-invariant region of the 2-loop integral for the diagrams 
in Fig.~\ref{fig:33}(c) that extends from the momentum scale 
$E^{1/2}$ to the ultraviolet cutoff $\Lambda$.  
Thus the logarithmic term is proportional to $a^4 \ln (E/ \Lambda^2)$. 
This logarithmic dependence on the 
ultraviolet cutoff is not removed by renormalization 
of the 2-body coupling constant $g_2$.  The logarithmic dependence 
on $\Lambda$ comes from the region of the loop integrals in which all
the virtual particles in the diagrams in Fig.~\ref{fig:33}(c) have large
momenta of order $\Lambda$.  In coordinate space, this corresponds 
to all the particles having small separations of order $1/\Lambda$.  
The dependence on $\Lambda$ can therefore
be cancelled by the 3-body contact interaction in the Lagrangian 
in Eq.~(\ref{L-3body}) with a coefficient $g_3 (\Lambda)$
that depends on the ultraviolet cutoff $\Lambda$.
The required cutoff-dependence of the 3-body coupling constant  
is described by the differential RG equation \cite{BN-97}
\begin{eqnarray}
\Lambda{d \ \over d \Lambda} g_3 = 384 (4 \pi -3 \sqrt{3}) 
a^4 \,.
\label{g3-rg}
\end{eqnarray}
If the cutoff $\Lambda$ is increased, the
strength of the 3-body contact interaction must be increased in
accordance with Eq.~(\ref{g3-rg}) to cancel the additional short-distance
contributions from four successive 2-body scatterings.

\begin{figure}[htb]
\bigskip
\centerline{\includegraphics*[width=10cm,angle=0]{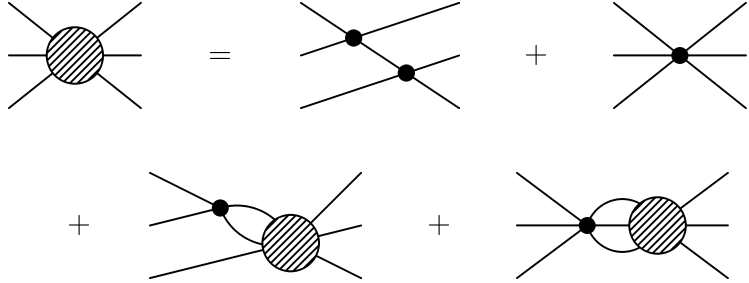}}
\medskip
\caption
{Simple integral equation for the truncated connected 3-body amplitude.  
The first and third diagrams should be summed over the three pairs 
of atoms that interact first.}
\label{fig:integ3}
\end{figure}

In order to calculate observables involving energies $E$ of order
$1/ma^2$, it is necessary to solve the quantum field theory
nonperturbatively. All information about physical observables in the
3-body sector is encoded in the 6-point Green's function $\langle 0|
{\rm T} (\psi \psi \psi \psi^\dagger \psi^\dagger \psi^\dagger) | 0
\rangle$, where we have suppressed the time and space coordinates of the
6 field operators. It is encoded even more succinctly in the truncated
connected 6-point Green's function in momentum space which we will
denote by ${\mathcal A}$.%
\footnote{The 3-atom amplitude ${\mathcal A}$ should not be 
confused with the atom-atom amplitude in Section~\ref{sec:EFT2}, 
which was denoted by the same symbol.}
In the center-of-mass frame, ${\mathcal A}$ is a
function of 4 momentum vectors and 5 off-shell energies. One can in
principle solve the quantum field theory by solving an integral equation
for ${\mathcal A}$. The simplest integral equation for ${\mathcal A}$ is
illustrated in Fig.~\ref{fig:integ3}. It simply states that the Feynman
diagrams that contribute to the amplitude are either tree diagrams or
they involve at least one loop. For those diagrams that involve a loop,
the first interaction is either a 2-body contact interaction or a 3-body
contact interaction. The various possibilities are represented by the
diagrams on the right side of the integral equation in
Fig.~\ref{fig:integ3}. The first two diagrams are tree diagrams and they
constitute the inhomogeneous term in the integral equation. In the
third diagram, the first interaction is a 2-body constant interaction
and it involves a one-loop integral over the amplitude ${\mathcal A}$. In
the fourth diagram, the first interaction is a 3-body contact
interaction and it involves a 2-loop integral over the amplitude 
${\mathcal A}$.

Integral equations in many variables are very difficult to solve
numerically, so the integral equation in Fig.~\ref{fig:integ3} is not
very useful in practice. However, from our understanding of 
universality in the 3-body problem, we can anticipate some of
the features that would appear in the nonperturbative solution. 
In addition to 3-atom scattering states, there must be scattering
states consisting of an atom and a dimer whose binding energy 
is $E_D = 1/a^2$. There must also be a sequence of 3-body bound states
with binding energies that range from order $1/a^2$ to order $\Lambda^2$.
The number of these Efimov states will depend on the ultraviolet cutoff 
$\Lambda$ imposed on the loop momenta, but it should be roughly 
$\ln (|a|\Lambda)/ \pi$ for asymptotically large $\Lambda$. 
These bound states must all emerge dynamically from the
nonperturbative effects encoded in the simple Lagrangian
in Eq.~(\ref{L-3body}).

We conclude this subsection by giving the perturbative expansion for the 
T-matrix element for elastic 3-atom scattering through 4$^{\rm th}$ 
order in the scattering length $a$.
Three atoms with momenta ${\bm k}_1$, ${\bm k}_2$
and ${\bm k}_3$ can scatter into states with momenta
${\bm k}'_1$, ${\bm k}'_2$
and ${\bm k}'_3$ that are allowed by conservation of energy
and momentum. The probability amplitude for $3\to 3$
scattering processes in which all three atoms participate is
given by the connected $T$-matrix element, which we denote
by ${\mathcal T}( {\bm k}_1,{\bm k}_2,{\bm k}_3;
{\bm k}'_1,{\bm k}'_2,{\bm k}'_3 )$. For simplicity, we 
consider only the center-of-mass frame, in which 
${\bm k}_1+{\bm k}_2+{\bm k}_3=0$, and we introduce the shorthand
\begin{eqnarray}
{\mathcal T}(123\to 1'2'3') \equiv
{\mathcal T}({\bm k}_1,{\bm k}_2,{\bm k}_3;
        {\bm k}'_1,{\bm k}'_2,{\bm k}'_3 ) \,.
\end{eqnarray}
The connected $T$-matrix element for $3\to 3$ scattering can be
separated into the terms that involve a single virtual
particle in the intermediate state, which are called 
{\it one-particle-reducible} (1PR), and the remaining terms,
which are called {\em one-particle-irreducible} (1PI):
\begin{eqnarray}
{\mathcal T}( 123\to 1'2'3') &&
= {\mathcal T}^{\rm 1PR}( 123\to 1'2'3') 
+ {\mathcal T}^{\rm 1PI}(123\to 1'2'3') \,,
\label{T3:R+I}
\end{eqnarray}
The 1PR terms can be written down in closed form \cite{BN-99}:
\begin{eqnarray}
{\mathcal T}^{\rm 1PR}(123\to 1'2'3') && 
= \sum_{(123)} \sum_{(1'2'3')}
{ {\mathcal A}(q_{12}^2/4) {\mathcal A}(q_{1'2'}^2/4)
 \over {\bm k}_1 \cdot {\bm k}_2 - ({\bm k}_1+{\bm k}_2)\cdot {\bm k}'_3
        + k_3^{'2} - i \epsilon}  ,
\hspace{0.9cm}
\label{T3:1PR}
\end{eqnarray}
where $q_{12}=|{\bm k}_1 - {\bm k}_2|$, 
$q_{1'2'}=|{\bm k}'_1 - {\bm k}'_2|$, and ${\mathcal A}(E)$ is the 
amplitude for atom-atom scattering given in Eq.~(\ref{a-npren}).
The sums in Eq.~(\ref{T3:1PR}) are over
cyclic permutations of ${\bm k}_1$, ${\bm k}_2$, and 
${\bm k}_3$ and over cyclic permutations of 
${\bm k}'_1$, ${\bm k}'_2$, and ${\bm k}'_3$. 
The summand that is given explicitly in Eq.~(\ref{T3:1PR})
corresponds to the $2\to 2$
scattering of particles 1 and 2 to produce particle $3'$ and
a virtual particle. A subsequent $2\to 2$ scattering of the
virtual particle and particle 3 produces particles $1'$ and
$2'$. Examples of diagrams that contribute to the sum are
the tree diagram in Fig.~\ref{fig:33}(a) and the second 
diagram in Fig.~\ref{fig:33}(b).
The 1PR term in Eq.~(\ref{T3:1PR}) can be expanded as a 
power series in $a$ simply by expanding the amplitudes
${\mathcal A}(E)$ as power series in $a$ using Eq.~(\ref{A2pert}).

The 1PI terms in the T-matrix element cannot be expressed 
in closed form, but we will give explicit expressions for the 
terms of order $a^3$ and $a^4$ in its perturbative expansion.
The term of order $a^3$ comes from one-loop diagrams like the 
first diagram in Fig.~\ref{fig:33}(b). 
It can be expressed as \cite{BN-99}
\begin{eqnarray}
{\mathcal T}^{\rm 1PI}_3( 123\to 1'2'3' ) &&
= - 512 \pi^3 a^3\sum_{(123)} 
\sum_{(1'2'3')} {\mathcal I}(123 \to 1'2'3') \,,
\label{t31pia}
\end{eqnarray}
where the integral ${\mathcal I}$ is a function of the wave vectors 
${\bm k}_i$ and ${\bm k}_i'$:
\begin{eqnarray}
{\mathcal I}(123 \to 1'2'3') &&
\nonumber
\\
&& \hspace{-2.5cm} =
\int{d^3p\over(2\pi)^3}
         {1\over p^2 + {\bm p} \cdot {\bm k}_3 
            + {\bm k}_1 \cdot {\bm k}_2 - i\epsilon}
{1\over  p^2 + {\bm p} \cdot {\bm k}'_3
            + {\bm k}'_1 \cdot {\bm k}'_2-i\epsilon} \, .
\label{int-1loop}
\end{eqnarray}

The terms of order $a^4$ in the perturbative expansion of the 1PI
$T$-matrix element come from adding a one-loop bubble to the 
first one-loop diagram in Fig.~\ref{fig:33}(b)
and from the two-loop diagrams in Fig.~\ref{fig:33}(c). 
The three terms can be expressed in the form \cite{BN-99}
\begin{subequations}
\begin{eqnarray}
{\mathcal T}^{\rm 1PI}_{4a}(123 \to 1'2'3') &&
= 256 i \pi^3 a^4 \sum_{(123)} \sum_{(1'2'3')}
         (q_{12} + q_{1'2'}) {\mathcal I}( 123 \to 1'2'3') \,,
\nonumber 
\\ 
\label{t1PI:4a} 
\\ 
{\mathcal T}^{\rm 1PI}_{4b}(123 \to 1'2'3') &&
 = 8192 \pi^4 a^4 \sum_{(123)} \sum_{(1'2'3')} 
        {\mathcal J}(123 \to 1'2'3') \,,
\label{t1PI:4b} 
\\ 
{\mathcal T}^{\rm 1PI}_{4c}(123 \to 1'2'3') &&
= 4096 \pi^4 a^4 \sum_{(123)} \sum_{(1'2'3')}
        {\mathcal J}'(123 \to 1'2'3')  \,.
\label{t1PI:4c} 
\end{eqnarray}
\end{subequations}
The integrals inside the sums in Eqs.~(\ref{t1PI:4b}) 
and (\ref{t1PI:4c}) are given by
\begin{subequations}
\begin{eqnarray}
{\mathcal J}(123 \to 1'2'3')  &&
= \int{d^3p\over(2\pi)^3} \int{d^3q\over(2\pi)^3} \,
{1\over p^2+q^2+r^2-2mE-i\epsilon}
\nonumber 
\\ 
&& \hspace{-1cm} 
\times {1 \over p^2 + {\bm p} \cdot {\bm k}_3 
        + {\bm k}_1 \cdot {\bm k}_2 -i \epsilon}
 {1 \over q^2 + {\bm q} \cdot {\bm k}'_3 
        + {\bm k}'_1 \cdot {\bm k}'_2 - i\epsilon} \,,
\label{int-2loop:b}
\\ 
{\mathcal J}'(123 \to 1'2'3') && 
= \int{d^3p\over(2\pi)^3} \int{d^3q\over(2\pi)^3}
\left( {1\over p^2+q^2+r^2-2mE-i\epsilon} - {1 \over 2 q^2} \right)
\nonumber 
\\ 
&& \hspace{-1cm}
\times {1 \over p^2 + {\bm p} \cdot {\bm k}_3 
            + {\bm k}_1 \cdot {\bm k}_2-i\epsilon}
{1 \over p^2 + {\bm p} \cdot {\bm k}'_3 
            + {\bm k}'_1 \cdot {\bm k}'_2-i\epsilon} \,,
\label{int-2loop:c}
\end{eqnarray}
\end{subequations}
where $r = |{\bm p} + {\bm q}|$ 
and $E=(k_1^2+k_2^2+k_3^2)/2m$ is the total energy. 
These two-loop integrals are logarithmically ultraviolet 
divergent.  The divergence can be isolated in a term that 
is momentum-independent by subtracting and adding the 
following integrals:
\begin{subequations}
\begin{eqnarray}
{\mathcal J}_{\rm log}(\kappa) 
&=& \int{d^3p\over(2\pi)^3} \int{d^3q\over(2\pi)^3}
{1 \over p^2 + q^2 + r^2 + 2 \kappa^2}
{1 \over (p^2 + \kappa^2) (q^2 + \kappa^2)},
\\ 
{\mathcal J}_{\rm log}'(\kappa) 
&=& \int{d^3p\over(2\pi)^3} \int{d^3q\over(2\pi)^3}
\left( {1 \over p^2 + q^2 + r^2 + 2 \kappa^2} 
- {1 \over 2(q^2 + \kappa^2)} \right)
\nonumber
\\ 
&& \hspace{0.5cm}
\times {1 \over (p^2 + \kappa^2)^2} \,.
\end{eqnarray}
\end{subequations}
Dimensional regularization can be applied to these integrals 
by changing the integrals over 3-dimensional vectors 
to integrals over $D$-dimensional vectors with the following 
prescription for the integration measure:
\begin{eqnarray}
\int {d^3p \over (2 \pi)^3} 
\longrightarrow \Lambda^{3-D} \int {d^Dp \over (2 \pi)^D} \,,
\end{eqnarray}
%
where $\Lambda$, which has dimensions of momentum, 
is called the {\it renormalization scale}. 
The factor of $\Lambda^{3-D}$ ensures that dimensional analysis 
appropriate to 3 dimensions is respected.
The logarithmic ultraviolet divergences then appear as poles in
$D-3$.  The results for these integrals with dimensional 
regularization reduce in the limit $D \to 3$ to
\begin{subequations}
\begin{eqnarray}
{\mathcal J}_{\rm log}(\kappa) 
&=& -{1 \over 96 \pi^3}
         \left(  {1\over D-3} - 0.82735 \right)
         \left( {\kappa \over \Lambda} \right)^{2(D-3)} \,,
\nonumber 
\\ 
\label{polet3b}
\\ 
{\mathcal J}_{\rm log}'(\kappa) 
&=& -{ \sqrt{3} \over 64 \pi^3}
         \left(  {1\over D-3} - 0.39157 \right)
          \left( {\kappa \over \Lambda} \right)^{2(D-3)} \,.
\nonumber 
\\ 
\label{polet3c}
\end{eqnarray}
\end{subequations}
In the minimal subtraction renormalization prescription, 
the poles in $\epsilon$ are subtracted from the Laurent expansion
of the integral in $D-3$.  In the integrals in Eqs.~(\ref{polet3b})
and (\ref{polet3b}), this is equivalent to replacing $1/(D-3)$ by 
$2\ln(\kappa/\Lambda)$ and then setting $D=3$.


\subsection{The diatom field trick}
\label{sec:diatom}

A significant breakthrough 
in applying effective field theory to the 3-body problem with a 
large scattering length was made by Bedaque, Hammer, 
and van Kolck \cite{BHK99,BHK99b}.  They introduced a new effective 
field theory that has been used to calculate many new universal results 
for 3-body observables. 
This effective field theory makes manifest the connnection between
the 3-body problem with a large scattering length 
and renormalization group limit cycles.

Using the effective field theory of Bedaque, Hammer, and van Kolck,
it is very easy to derive the Skorniakov-Ter-Martirosian (STM) equation, 
a simple integral equation for the 3-body problem in the scaling limit.
It is not surprising that many of the universal aspects of the 3-body
problem can be captured in an integral equation simpler than that in
Fig.~\ref{fig:integ3}.  First, the most interesting and subtle aspects of
this problem appear in the sector with total angular momentum quantum
number $L=0$. Second, the integral equation in Fig.~\ref{fig:integ3}
does not exploit the fact that the 2-body problem for this quantum field
theory can be solved analytically. Bedaque, Hammer, and van Kolck found
a way to exploit both of these features.

The STM integral equation is not
an equation for the 6-point Green's function 
$\langle 0| {\rm T}( \psi \psi \psi 
        \psi^\dagger \psi^\dagger \psi^\dagger) | 0 \rangle$ 
discussed in the previous section. 
Instead it is an equation for the 4-point Green's function 
$\langle 0| {\rm T} (d\psi d^\dagger \psi^\dagger) | 0 \rangle$,
where the {\it diatom field} $d$ is a local operator that
annihilates two atoms at a point.  The diatom field is
essentially just the composite quantum field operator $\psi^2$. 
In order to obtain the simplest possible integral equation, it is
useful to construct a new formulation of the quantum field theory that
involves the field $d$ explicitly.\footnote{This trick has been 
used in many other contexts.  A pedagogical treatment in the context
of the $O(N)$ version of $\phi^4$ theory can be found in the Erice 
lectures by Coleman \cite{Coleman}. For early applications in 
nuclear few-body systems, see e.g. Refs.~\cite{Kaplan:1996nv,BK98}.} 
The Lagrangian for the three-boson
system used by Bedaque, Hammer, and van Kolck can be expressed in the form
\begin{eqnarray}
{\mathcal L}_{\rm BHvK} &=& \psi^\dagger \left( i {\partial \ \over \partial
t} + {1 \over 2} \nabla^2 \right) \psi +{g_2 \over 4} d^\dagger d
-{g_2 \over 4} \left( d^\dagger \psi^2 + {\psi^\dagger}^2 d \right)
-{g_3 \over 36} d^\dagger d \psi^\dagger \psi \,.
\nonumber \\
\label{L-BHvK}
\end{eqnarray}
One important feature of this Lagrangian is that there is no direct 
2-body contact interaction term $(\psi^\dagger \psi)^2$.
Another important feature is that there are also no time derivatives 
acting on the diatom field $d$, 
so $d$ is not a dynamical field independent from $\psi$.
Furthermore, the Lagrangian has only linear and quadratic terms in $d$. 
The physics of this quantum field theory
is therefore identical to that of the quantum field theory whose 
Lagrangian is obtained by eliminating $d$ from the 
Lagrangian in Eq.~(\ref{L-BHvK}) by using its equation of motion. 
The equation obtained by varying $d^\dagger$ is
\begin{eqnarray}
d - \psi^2 - (g_3/9 g_2) d\psi^\dagger \psi = 0 \,,
\end{eqnarray}
leading to the solution
\begin{eqnarray}
d = {\psi^2 \over 1 - (g_3/9 g_2) \psi^\dagger \psi} \,.
\end{eqnarray}
The Lagrangian obtained by eliminating $d$ from Eq.~(\ref{L-BHvK}) is
\begin{eqnarray}
{\mathcal L} = \psi^\dagger \left( i{\partial \ \over \partial t} 
+ {1 \over 2} \nabla^2 \right) \psi 
- {g_2 \over 4} {\left( \psi^\dagger \psi \right)^2 
                \over 1 - (g_3/9 g_2) \psi^ \dagger \psi } \,.
\end{eqnarray}
If we expand the interaction term in powers of $\psi^\dagger \psi$,
and truncate after the $(\psi^\dagger \psi)^3$ term, 
we recover the Lagrangian in Eq.~(\ref{L-3body}).  Thus in the 2-body 
and 3-body sectors, the quantum field theory with the Lagrangian 
in Eq.~(\ref{L-BHvK})  is completely equivalent to the quantum field theory 
with the Lagrangian in Eq.~(\ref{L-3body}). 
The Lagrangian in Eq.~(\ref{L-BHvK}) also has $N$-body
contact interaction terms $(\psi^\dagger \psi)^N$ for $N \ge 4$,
but they affect only the $N$-body sectors with $N \ge 4$.

\begin{figure}[htb]
\bigskip
\centerline{\includegraphics*[width=8cm,angle=0]{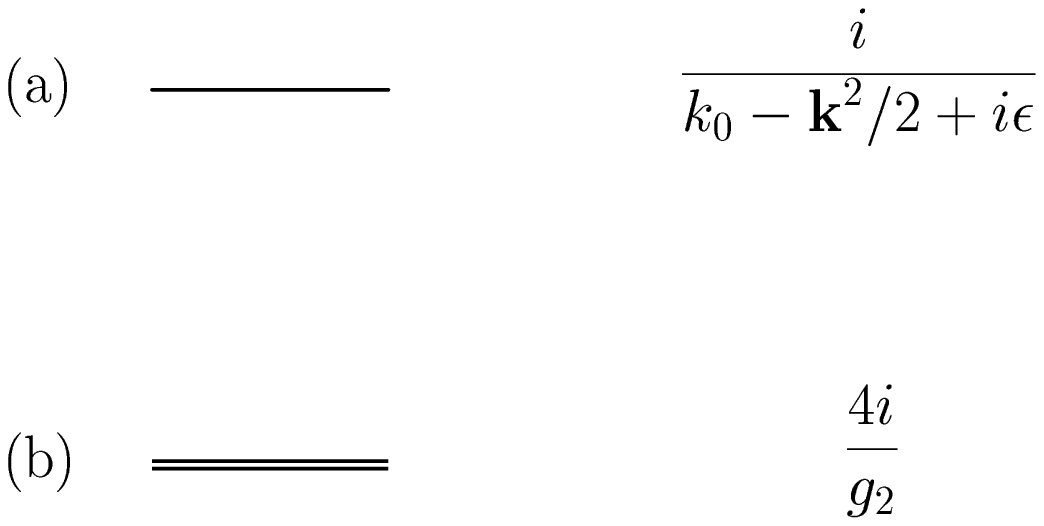}}
\medskip
\caption
{Feynman rules for the Lagrangian in Eq.~(\ref{L-BHvK}): (a) the propagator
for an atom with energy $k_0$ and momentum ${\bm k}$,
(b) the bare propagator for a diatom. Energy and momentum 
are always flowing to the right.}
\label{fig:prop12}
\end{figure}

\begin{figure}[htb]
\bigskip
\centerline{\includegraphics*[width=6cm,angle=0]{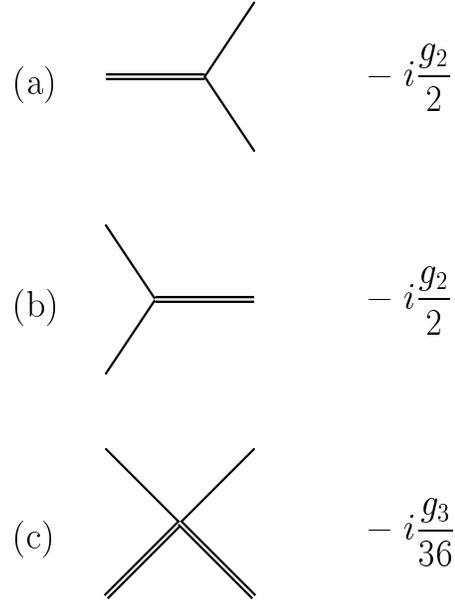}}
\medskip
\caption
{Feynman rules for the Lagrangian in Eq.~(\ref{L-BHvK}):
 interaction vertices. }
\label{fig:vert12}
\end{figure} 

\begin{figure}[htb]
\bigskip
\centerline{\includegraphics*[width=12cm,angle=0]{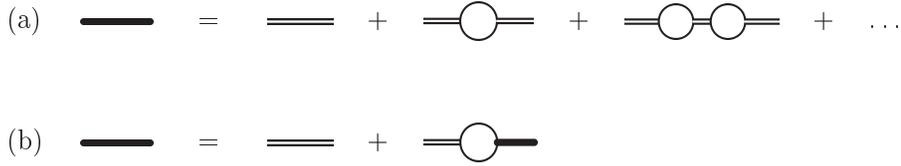}}
\medskip
\caption
{Diagrammatic equations for the complete diatom propagator
$i D(P_0,P)$:
(a) perturbative expansion in powers of $g_2$,
(b) integral equation.}
\label{fig:bubbles}
\end{figure}

The Feynman rules for the Lagrangian  in Eq.~(\ref{L-BHvK}) are shown in
Figs.~\ref{fig:prop12} and \ref{fig:vert12}.
The bare propagator for the diatom field is
simply the constant $4i/g_2$,
which corresponds to no propagation in space or time.
However, there are corrections to the
diatom propagator from the diagrams in Fig.~\ref{fig:bubbles}(a)
which allow the diatom to propagate.
In Feynman diagrams, we represent the complete diatom propagator 
$iD(P_0,P)$ by a thick solid line.
We can calculate the complete diatom
propagator by solving the simple integral equation shown in
Fig.~\ref{fig:bubbles}(b). The loop on the right side is just the integral
in Eq.~(\ref{A-pert}), with $E$ replaced by $P_0-P^2/4$, 
where $P_0$ and ${\bm P}$ are the energy and momentum of the diatom.
The solution for the complete diatom propagator is
\begin{eqnarray}
D(P_0,P) &=& \frac{4}{g_2}
\left[ 1 + {g_2 \over 4 \pi^2} \left( \Lambda - {\pi
\over 2} \sqrt{-P_0 + P^2/4 - i \epsilon} \right) \right]^{-1} ,
\label{diprop}
\end{eqnarray}
where $\Lambda$ is a cutoff on the loop momentum in the bubbles. After
making the substitution given in Eq.~(\ref{g2-tune}), the
expression for the complete diatom propagator is
\begin{eqnarray}
D(P_0, P) =
{32 \pi \over g_2^2} \left[ 1/a - \sqrt{-P_0 + P^2/4 -i \epsilon}
\right]^{-1} .
\label{propdiatom}
\end{eqnarray}
Note that all the dependence on the ultraviolet cutoff is now in the
multiplicative factor $1/g_2^2$.
The complete diatom propagator differs from the off-shell 2-body amplitude
${\mathcal A}$ in Eq.~(\ref{a-npren}) only by a multiplicative constant. For
$a>0$, it has a pole at $P_0 = -1/a^2 + P^2/4$ corresponding to a dimer
of momentum ${\bm P}$ and binding energy $E_D=1/a^2$.
As $P_0$ approaches the dimer pole, the limiting
behavior of the propagator is
\begin{eqnarray}
D(P_0, P) \longrightarrow
{Z_D \over P_0 - (- 1/a^2+ P^2/4) + i \epsilon} \,,
\label{dimer-pole}
\end{eqnarray}
where the residue factor is
\begin{eqnarray}
Z_D= 64\pi/(a g_2^2) \,.
\label{dimer-Z}
\end{eqnarray}
If we regard the composite operator $d$ as a quantum field that
annihilates and creates dimers, then $Z_D$ is the wave function 
renormalization constant for that field.
The renormalized propagator $Z_D^{-1} D(P_0,P)$ is completely 
independent of the ultraviolet cutoff.


\subsection{STM integral equation}

\begin{figure}[htb]
\bigskip
\centerline{\includegraphics*[width=9cm,angle=0]{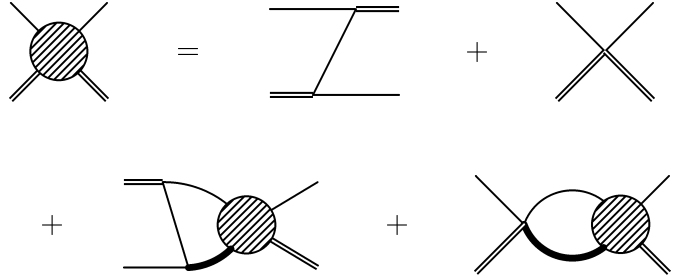}}
\medskip
\caption
{The integral equation for the 3-body amplitude ${\mathcal A}$.}
\label{fig:inteq12}
\end{figure}

The first derivation of the Skorniakov-Ter-Martirosian (STM)
equation using Feynman diagrams was carried out in 
Ref.~\cite{Koma64}.
The derivation is particularly simple in the effective field theory 
introduced by Bedaque, Hammer, and van Kolck.  The STM equation
is an integral equation for the Fourier
transform of the amputated connected part of the Green's function
$\langle 0| {\rm T} (d \psi d^\dagger \psi^\dagger) | 0 \rangle$, 
whichwe will denote by ${\mathcal A}$.%
\footnote{The atom-diatom amplitude ${\mathcal A}$ should not be confused 
with the atom-atom amplitude in Section~\ref{sec:EFT2}
or the 3-atom amplitude in Section~\ref{sec:EFT3}, 
which were denoted by the same symbol.}
The integral equation is shown diagrammatically in Fig.~\ref{fig:inteq12}.
It simply states that the
Feynman diagrams that contribute to ${\mathcal A}$ are either tree diagrams
or they involve at least one loop. For those diagrams that involve a
loop, the first interaction must be one of the vertices in
Fig.~\ref{fig:vert12}. The two tree diagrams on the right side of
Fig.~\ref{fig:inteq12} constitute the inhomogeneous term in the
integral equation. The two loop diagrams involve 1-loop integrals over
the same amplitude ${\mathcal A}$. Note that the thick black lines in
Fig.~\ref{fig:inteq12} represent the complete diatom propagator given in
Eq.~(\ref{propdiatom}).

In the center-of-mass frame, we can take the external momenta of the
atom and diatom to be $-{\bm p}$ and $+{\bm p}$ for the incoming lines
and $-{\bm k}$ and $+{\bm k}$ for the outgoing lines.  We take their
energies to be $E_A$ and $E-E_A$ for the incoming lines and $E_A'$ and
$E-E_A'$ for the outgoing lines. The amplitude ${\mathcal A}$ is then a
function of the momenta ${\bm p}$ and ${\bm k}$ and the energies $E$,
$E_A$ and $E_A'$. The integral equation involves a loop over the
momentum $-{\bm q}$ and energy ${q_0}$ of a virtual atom. Using the
Feynman rules from Figs.~\ref{fig:prop12} and \ref{fig:vert12}, we obtain
\begin{eqnarray}
{\mathcal A} ({\bm p}, {\bm k}; E, E_A, E_A')
&=& - \left[ {g_2^2/4 \over E-E_A-E_A' - ({\bm p} + {\bm k})^2 /2 +i
\epsilon}
+{g_3 \over 36} \right]
\nonumber
 \\
&& \hspace{-3.5cm}
+ {32\pi i\over g_2^2} \int {d q_0 \over 2 \pi} \int{d^3q \over (2 \pi)^3}
\left[ {g_2^2/4 \over E-E_A - q_0 - ({\bm p} + {\bm q})^2/2 + i\epsilon}
        +{g_3 \over 36} \right]
\nonumber
 \\
&&  \hspace{-1cm} \times
{1 \over q_0 - q^2/2 +i \epsilon}\;
\frac{{\mathcal A} ({\bm q}, {\bm k}; E, q_0, E_A')}
{1/a - \sqrt{-(E-q_0) + q^2 /4 -i \epsilon} } \,.
\end{eqnarray}
The integral over $q_0$ can be evaluated by contour integration. This
sets $q_0 = q^2/2$, so the amplitude ${\mathcal A}$ inside the integral has the
incoming atom on-shell. 

We obtain a simpler integral
equation if we also set the energies of both the initial and final atoms in
${\mathcal A}$ on-shell: $E_A = p^2/2$, $E_A' = k^2/2$.
Thus only the diatom lines have energies that are off-shell.\footnote{
This trick of putting one leg on-shell has also been used to simplify 
integral equations for relativistic bound states, 
such as positronium in QED \protect\cite{Gross69}.}
The resulting integral equation is
\begin{eqnarray}
 {\mathcal A} ({\bm p}, {\bm k}; E, p^2/2, k^2/2)
&=& - {g_2^2 \over 4} 
\left[ {1 \over E - (p^2 + {\bm p} \cdot {\bm k} + k^2) +i \epsilon} 
        + {g_3 \over 9 g_2^2} \right]
\nonumber
 \\
&&  \hspace{-1.5cm}
-8 \pi \int{d^3q \over (2 \pi)^3}
\left[ {1 \over E - (p^2 + {\bm p} \cdot {\bm q} + q^2) + i \epsilon}
        +{g_3 \over 9 g_2^2} \right]
\nonumber
 \\
&&  \hspace{1cm} \times
\frac{{\mathcal A} ({\bm q}, {\bm k}; E, q^2/2, k^2/2)}
{- 1/a + \sqrt{- E + 3q^2 /4 -i \epsilon} } \,.
\label{BhvK:general}
\end{eqnarray}
This is an integral equation with three integration variables 
for an amplitude ${\mathcal A}$ that depends explicitly 
on seven independent variables.  There is also an additional implicit variable
provided by an  ultraviolet cutoff $|{\bm q}| < \Lambda$
on the loop momentum.  
If we set $g_3 = 0$ and ignore the ultraviolet cutoff, 
the integral equation in Eq.~(\ref{BhvK:general}) is equivalent to the 
Skorniakov-Ter-Martirosian (STM) equation, 
an integral equation for three particles interacting via
zero-range 2-body forces derived by Skorniakov and Ter-Martirosian 
in 1957 \cite{STM57}.  In the following, we will refer to the 
generalization of the STM equation with 
a nonzero 3-body coupling $g_3$ simply as the {\it STM3 equation}.

It was shown by Danilov that the STM equation has no unique 
solution in the case of identical bosons \cite{Dan61}. He also
pointed out that a unique solution could be obtained if one 
3-body binding energy is fixed. 
Kharchenko was the first to solve the STM equation with a finite 
ultraviolet cutoff that was tuned to fit observed 3-body data.  
Thus the cutoff was treated 
as an additional parameter \cite{Khar73}. Below we will
show that this ad hoc procedure is indeed justified and emerges
naturally when the STM3 equation is renormalized.

If we restrict our attention to the sector of the 3-body problem with 
total orbital angular momentum $L=0$, we can further simplify 
the integral equation.  The projection onto $L=0$ can be accomplished
by averaging the integral equation over the cosine
of the angle between ${\bm p}$ and ${\bm k}$: $x={\bm p}\cdot{\bm k}/
(pk)$. It is also convenient to multiply the amplitude ${\mathcal A}$
by the wave function renormalization factor $Z_D$ given in 
Eq.~(\ref{dimer-Z}).
We will denote the resulting amplitude by ${\mathcal A}_S$:
\begin{eqnarray}
{\mathcal A}_S(p, k; E) \equiv Z_D 
\int_{-1}^1 \! {dx\over 2}\,
{\mathcal A} ({\bm p}, {\bm k}; E, p^2/2, k^2/2) .
\label{A-def}
\end{eqnarray}
Furthermore, it is convenient to express the 3-body coupling constant 
in the form
\begin{eqnarray}
g_3 = - {9 g_2^2 \over \Lambda^2} H(\Lambda) \,.
\label{g3g2}
\end{eqnarray}
Since $H$ is dimensionless, it can only  be a
function of the dimensionless variables $a \Lambda$ and 
$\Lambda/ \kappa_*$, where $\kappa_*$ is the 3-body 
parameter defined by the spectrum of Efimov states in the resonant limit, 
Eq.~(\ref{B3n-resonant}). We will find that $H$ is a function of
$\Lambda/ \kappa_*$ only. 
 
The resulting equation is the STM3 
integral equation in its simplest form:
\begin{eqnarray}
{\mathcal A}_S (p, k; E)  &=& {16 \pi \over a} 
\left[ {1 \over 2pk} \ln \left({p^2 + pk + k^2 -E -i \epsilon \over
p^2 - pk + k^2 - E - i \epsilon}\right) + {H(\Lambda) \over \Lambda^2} \right]
\nonumber
\\
&& \hspace{-1.5cm}
+ {4 \over \pi} \int_0^\Lambda dq \, q^2
\left[{1 \over 2pq} \ln\left( {p^2 +pq + q^2 - E - i \epsilon \over
p^2 - pq + q^2 -E -i \epsilon}\right) + {H(\Lambda) \over \Lambda^2} \right]
\nonumber
\\
&& \hspace{1cm}\times 
\frac{ {\mathcal A}_S (q, k; E)}
{- 1/a + \sqrt{3q^2/4 -E -i \epsilon} }\,.
\label{BHvK}
\end{eqnarray}
Note that the ultraviolet cutoff $\Lambda$ on the
integral over $q$ has been made explicit. 
A change in the endpoint $\Lambda$ of the loop integral
should be compensated by the $\Lambda$-dependence of the function 
$H$ in Eq.~(\ref{BHvK}).
More specifically, $H$ must be tuned as a function of $\Lambda$ 
so that the cutoff dependence of the solution ${\mathcal A}_S (p, k; E)$
of Eq.~(\ref{BHvK}) decreases as a power of $\Lambda$.  This will 
guarantee that ${\mathcal A}_S (p, k; E)$ has a well-behaved limit 
as $\Lambda \to \infty$.  Note that the $H/\Lambda^2$ term in the 
inhomogeneous term of Eq.~(\ref{BHvK}) can be omitted, 
since it goes to zero in the limit $\Lambda \to \infty$.

We will see in Section~\ref{sec:RGlc} that the function $H$ 
in Eq.~(\ref{BHvK}) must have the form
\begin{eqnarray}
H (\Lambda) = {\cos [s_0 \ln (\Lambda/ \Lambda_*) + \arctan s_0]
\over \cos [s_0 \ln (\Lambda/ \Lambda_*) - \arctan s_0]} \,.
\label{H-Lambda}
\end{eqnarray}
This equation defines a 3-body scaling-violation parameter 
$\Lambda_*$ with dimensions of momentum. 
Note that $H$ is a periodic function of 
$\Lambda/\Lambda_*$, so $\Lambda_*$ is defined only up to a 
multiplicative factor of $(e^{\pi/s_0})^n$, where $n$ is an integer.
The relation between $\Lambda_*$ and the 3-body parameter $\kappa_*$ 
defined by the spectrum of Efimov states in the resonant limit, 
which is given in Eq.~(\ref{B3n-resonant}), can be expressed as
\beq
s_0 \ln(\kappa_*) \approx s_0 \ln(0.381\, \Lambda_*)  \mod \pi \,.
\label{eq:kapslams}
\eeq
This relation can be obtained by using the STM3 integral equation 
in Eq.~(\ref{BHvK}) to calculate the binding energy of the Efimov 
trimers in the resonant limit $a = \pm \infty$.


\subsection{Three-body observables}

The solution ${\mathcal A}_S (p, k; E)$ to the STM3
integral equation in Eq.~(\ref{BHvK})
encodes all information about 3-body observables in the sector with total
orbital angular momentum quantum number $L=0$.
In particular, it contains information about the 
binding energies $E_T^{(n)}$ of the Efimov states. 
For a given ultraviolet cutoff $\Lambda$, the amplitude 
${\mathcal A}_S (p, k; E)$ has a finite number of poles in $E$
corresponding to the Efimov trimers whose binding energies
are less than about $\Lambda^2$.  As $\Lambda$ increases,
new poles emerge corresponding to deeper Efimov trimers.
In the limit $\Lambda \to \infty$, the locations of these poles 
approach the energies $-E_T^{(n)}$ of the Efimov trimers.
The residues of the poles of ${\mathcal A}_S (p, k; E)$
factor into functions of $p$ and functions of $k$:
\begin{eqnarray}
{\mathcal A}_S (p, k; E)\longrightarrow 
{ {\mathcal B}^{(n)}(p) {\mathcal B}^{(n)}(k) \over E + E_T^{(n)} } \,,
\qquad {\rm as \ } E \to -E_T^{(n)} .
\end{eqnarray}
Matching the residues of the poles on both sides of Eq.~(\ref{BHvK}),
we obtain the bound-state equation
\begin{eqnarray}
{\mathcal B}^{(n)}(p) &=&  {4 \over \pi} \int_0^\Lambda \! \! dq \, q^2
\left[{1 \over 2pq} \ln {p^2 +pq + q^2 - E - i \epsilon \over
p^2 - pq + q^2 -E -i \epsilon} + {H \over \Lambda^2} \right]
\nonumber
\\
&& \quad \times \left[ - 1/a + \sqrt{3q^2/4 -E
-i \epsilon} \right]^{-1} {\mathcal B}^{(n)}(q) \,.
\label{BHvK-homo}  
\end{eqnarray}
The values of $E$ for which this homogeneous integral equation has 
solutions are the energies $-E_T^{(n)}$ of the Efimov states.
For a finite ultraviolet cutoff $\Lambda$,
the spectrum of $E_T^{(n)}$ is cut off around $\Lambda^2$, 
so the number of Efimov states is
roughly $\ln (|a| \Lambda)/ \pi$. To find deeper Efimov states,
one simply needs to increase the cutoff.  Most of the results on the binding
energies of Efimov states in Section~\ref{subsec:B3} were obtained by
solving the homogeneous integral equation in Eq.~(\ref{BHvK-homo}).

\begin{figure}[htb]
\bigskip
\centerline{\includegraphics*[width=11cm,angle=0]{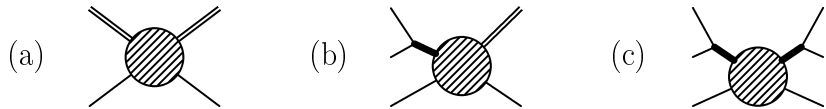}}
\medskip
\caption
{Amplitudes for (a) atom-dimer scattering, (b) 3-body recombination, 
and (c) $3 \to 3$ scattering. 
Diagrams (b) and (c) should be summed over the three pairs of atoms that can 
interact first.  Diagram (c) should also be summed over the three pairs of 
atoms that can interact last.}
\label{fig:3br}
\end{figure}

The S-wave phase shifts for atom-dimer scattering can be determined
from the solution ${\mathcal A}_S (p, k; E)$ to the 
STM3 integral equation in Eq.~(\ref{BHvK}).
The T-matrix element for the elastic scattering of an atom and a dimer 
with momenta $k$ is given by the amplitude ${\mathcal A}$ 
evaluated at the on-shell point $p=k$ and $E= -E_D + 3k^2/4$
and multiplied by a wave function renormalization factor $Z_D^{1/2}$ 
for each dimer in the initial or final state.
It can be represented by the Feynman diagram in 
Fig.~\ref{fig:3br}(a).  The blob represents the amplitude ${\mathcal A}$
or equivalently $Z_D^{-1} {\mathcal A}_S$.
The external double lines correspond 
to asymptotic dimers and are associated with factors $Z_D^{1/2}$.
The S-wave contribution to the T-matrix element is
\begin{eqnarray}
{\mathcal T}_{AD \to AD}^{(L=0)} = 
{\mathcal A}_S (k, k; 3k^2/4-1/a^2) \,.
\end{eqnarray}
Note that the factors of $Z_D$ multiplying ${\mathcal A}_S$ cancel.
The differential cross section for elastic atom-dimer scattering is
\begin{eqnarray}
d \sigma_{AD \to AD}
= {2 \over 3 k} \left| {\mathcal T}_{AD \to AD}(k) \right|^2
{k \over 6 \pi^2} d \Omega \,.
\label{T-AD}
\end{eqnarray}
The flux factor $2/(3 k)$ is the inverse of the relative velocity 
of the atom and the dimer.  The phase space factor 
$k d \Omega/(6 \pi^2)$ takes into account energy and momentum 
conservation and the standard normalization of momentum eigenstates:
\begin{eqnarray}
\int {d^3 p_A \over (2 \pi)^3} {d^3 p_D \over (2 \pi)^3} 
(2 \pi)^4 \delta^3({\bm p}_A + {\bm p}_D)
\delta(\mbox{$1\over2$} p_A^2 + \mbox{$1\over4$}p_D^2 - E) 
&& 
= {1 \over 6 \pi^2} (4 E/3)^{1/2} \int d\Omega \,.
\nonumber \\
\end{eqnarray}
Comparing the expressions for the differential cross section
in Eqs.~(\ref{T-AD}) and (\ref{dsigma-AD}), we see that the
T-matrix element differs from the scattering amplitude $f_k(\theta)$ 
by a factor of $3 \pi$.
Using the expression for the S-wave term 
in the scattering amplitude from Eq.~(\ref{fk-AD}),
the S-wave phase shift is given by
\begin{eqnarray}
{1 \over k\cot\delta^{AD}_0 (k) -ik}
= {1 \over 3 \pi} {\mathcal A}_S (k, k; 3k^2/4-1/a^2) \,.
\label{T12}
\end{eqnarray}
In particular, the atom-dimer scattering length is given by
\begin{eqnarray}
a_{AD}
= - {1 \over 3 \pi} {\mathcal A}_S (0, 0; -1/a^2) \,.
\end{eqnarray}
The results on atom-dimer elastic scattering 
in Section~\ref{sec:atom-dimer} were obtained
from the expression in Eq.~(\ref{T12}).

The threshold 3-body recombination rate can also be obtained 
from the solution ${\mathcal A}_S (p, k; E)$ to the STM3 integral 
equation in Eq.~(\ref{BHvK}). This is possible only at threshold, 
because a 3-atom scattering state becomes pure $L=0$ only
in the limit that the energies of the atoms go to zero. 
The T-matrix element for the recombination process can be represented 
by the Feynman diagram in Fig.~\ref{fig:3br}(b)
summed over the three pairs of atom lines that can attach 
to the diatom line.
The blob represents the amplitude $Z_D^{-1} {\mathcal A}_S$ evaluated 
at the on-shell point $p=0$, $k = 2/(\sqrt{3}\, a)$, and $E=0$.
The solid line represents the diatom propagator $iD(0,0)$
evaluated at zero energy and momentum, which is given by 
Eq.~(\ref{diprop}).  The factor for the atom-diatom vertex is $-ig_2/2$.
The wave function renormalization factor $Z_D^{1/2}$ for the
final-state dimer is given by Eq.~(\ref{dimer-Z}).
In the product of factors multiplying ${\mathcal A}_S$, 
the dependence on $g_2$ and $\Lambda$ can be eliminated in favor 
of the scattering length $a$ given in Eq.~(\ref{a-g2}).
Taking into account a factor of 3 from the three Feynman diagrams,
the T-matrix element is
\begin{eqnarray}
{\mathcal T}_{AAA \to AD} = 6\sqrt{\pi a^3}\, 
{\mathcal A}_S (0, 2/(\sqrt{3}a);0) \,.
\end{eqnarray}
The differential rate $dR$ for the recombination of three atoms
with energies small compared to the dimer binding energy
can be expressed as
\begin{eqnarray}
dR = \left| {\mathcal T}_{AAA \to AD} \right|^2
{k \over 6 \pi^2}\, d \Omega \,,
\label{dR-T}
\end{eqnarray}
where $k = 2/(\sqrt{3} a)$.
The time rate of change of the number density $n_A$
of low-energy atoms is obtained 
by integrating Eq.~(\ref{dR-T}) over the $4 \pi$ solid angle 
and multiplying by the number density 
per volume-cubed of triples in the gas, which is $n_A^3/3!$.
Thus the recombination rate constant $\alpha$ on the right side
of Eq.~(\ref{dnA-gas}) is
\begin{eqnarray}
\alpha = {8 a^2 \over \sqrt{3}} 
\big| {\mathcal A}_S (0, 2/(\sqrt{3}a);0) \big|^2 \,.
\label{alpha-AS}
\end{eqnarray}
This expression was used to calculate the approximate expression
for the 3-body recombination constant given in Eq.~(\ref{alpha-sh}).
The expression for $\alpha$ in Eq.~(\ref{alpha-AS}) gives the 
recombination rate for three atoms with distinct momenta in the limit 
in which they all approach zero.  
If the atoms are all in exactly the same state,
which is the case if they are in a Bose-Einstein condensate, 
the rate must be divided by $3!$ to account for the symmetrization 
of the wave function of the identical particles.

The T-matrix element for 3-atom elastic scattering can be represented 
by a sum of Feynman diagrams like the one shown in Fig.~\ref{fig:3br}(c).
The T-matrix element has not yet been calculated.
However, in Ref.~\cite{BHM-01}, the 4-point Green's function
$\langle 0 | T(\psi d \psi^\dagger d^\dagger) |0\rangle$, 
which is related to the $3\to 3$ scattering amplitude, was
calculated at the three-atom threshold. 
That calculation was used to extract the results for the
nonperturbative constants given in Eqs.~(\ref{d-nonpert})
and (\ref{ReImd}).


\subsection{Renormalization group limit cycle}
\label{sec:RGlc}

The form of the exact renormalized diatom propagator in
Eq.~(\ref{propdiatom}) is consistent with the continuous scaling symmetry
given in Eqs.~(\ref{scaling-1}). 
In the integral equation (\ref{BHvK}), this scaling
symmetry is broken by the ultraviolet cutoff on the integral and by the
3-body terms proportional to $H/ \Lambda^2$. To see that the cutoff and
the 3-body terms are essential, we can try setting $H=0$ and taking
$\Lambda \rightarrow \infty$. The resulting integral equation has exact
scaling symmetry. We should therefore expect its solution 
${\mathcal A}_S (p,k; E)$ to behave asymptotically as $p \rightarrow \infty$ 
like a pure power of $p$. Neglecting the inhomogeneous term,
neglecting $E$ and $1/a^2$ compared to $q^2$,
and setting ${\mathcal A}_S \approx p^{s-1}$, 
the integral equation reduces to \cite{Dan61,MiF62,MiF62b,DaL63}
\begin{eqnarray}
p^{s-1} = {4 \over \sqrt{3} \pi p} \int_0^\infty dq \, q^{s-1} 
\ln {p^2 + pq + q^2 \over p^2 -pq + q^2} \,.
\end{eqnarray}
Making the change of variables
$q = xp$, the dependence on $p$ drops out, and we obtain
\begin{eqnarray}
1 = {4 \over \sqrt{3} \pi} \int_0^\infty dx \, x^{s-1} 
\ln {1 + x + x^2 \over 1 -x + x^2} \,.
\end{eqnarray}
The integral is a Mellin transform that can be evaluated
analytically.  The resulting equation for $s$ is
\begin{eqnarray}
1 = {8 \over \sqrt{3} s} {\sin (\pi s/6) \over \cos (\pi s/2)} \,.
\end{eqnarray}
This is identical to the angular eigenvalue equation (\ref{cheigen})
in the limit $R \gg |a|$
in the adiabatic hyperspherical representation of the 3-body
Schr{\"o}dinger equation. The solutions with the lowest values of $|s|$ are
purely imaginary: $s = \pm i s_0$, where $s_0 \approx 1.00624$. The most
general asymptotic solution therefore has two arbitrary constants:
\begin{eqnarray}
{\mathcal A}_S (p, k; E) \longrightarrow A_+ \, 
p^{-1+is_0} + A_- \, p^{-1-is_0} 
\,,\qquad
{\rm as \ } p \to \infty \,.
\end{eqnarray}
The inhomogeneous term in the integral equation (\ref{BHvK}) will
determine one of the constants. The role of the 3-body term in the
integral equation is to determine the other constant, thus giving the
integral equation a unique solution.

By demanding that the solution of the integral equation (\ref{BHvK}) 
has a well-defined limit as $\Lambda \to \infty$, 
Bedaque, Hammer, and van Kolck deduced the $\Lambda$-dependence 
of $H$ and therefore of $g_3$ \cite{BHK99,BHK99b}. 
The leading dependence on $\Lambda$ on the right side of the
STM3 integral equation in Eq.~(\ref{BHvK}) as $\Lambda \to \infty$ 
is a log-periodic term of order $\Lambda^0$ that comes from the 
region $q \sim \Lambda$.  
There are also contributions of order $1/\Lambda$
from the region $|a|^{-1},k,|E|^{1/2} \ll q \ll \Lambda$,
which have the form
\begin{eqnarray}
{8 \over \pi\sqrt{3}} \int^\Lambda dq \, 
\left( \frac{1}{q^2}+ {H\over\Lambda^2}\right)
(A_+ \, q^{+is_0} + A_- \, q^{-is_0}) \,.
\label{1overLambda}
\end{eqnarray}
The sum of the two terms will decrease even faster as $1/\Lambda^2$ 
if we choose the function $H$ to have the form
\begin{eqnarray}
H(\Lambda) =
{ A_+ \Lambda^{i s_0}/(1-is_0) + A_- \Lambda^{-i s_0}/(1+is_0)
\over A_+ \Lambda^{i s_0}/(1+is_0)   + A_- \Lambda^{-i s_0}/(1-is_0) } .
\label{H-tune}
\end{eqnarray}
The tuning of $H$ 
that makes the term in Eq.~(\ref{1overLambda}) decrease like
$1/\Lambda^2$ also suppresses the contribution from the region
$q \sim \Lambda$ by a power of $1/\Lambda$  so that it goes to 0 
in the limit $\Lambda \to \infty$ \cite{BHK99,BHK99b}.
By choosing $A_\pm = (1 + s_0^2)^{1/2} \Lambda_*^{\mp i s_0}/2$
in Eq.~(\ref{H-tune}),
we obtain the expression for $H$ in Eq.~(\ref{H-Lambda}).

\begin{figure}[htb]
\bigskip
\centerline{\includegraphics*[width=8.5cm,angle=0]{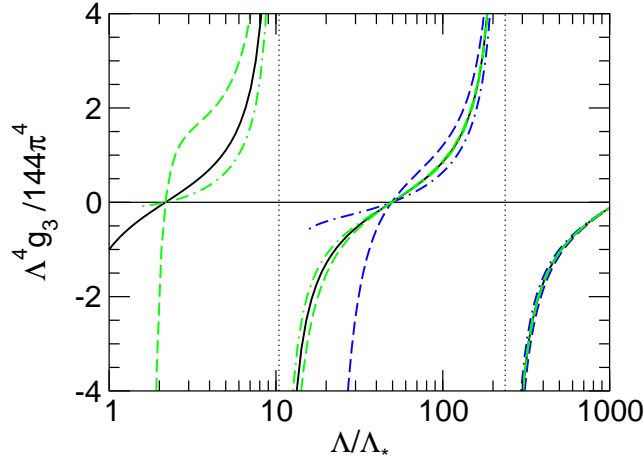}}
\medskip
\caption
{The dimensionless 3-body coupling constant 
        $\hat g_3 = \Lambda^4 g_3/144 \pi^4$
        as a function of the ultraviolet cutoff $\Lambda$ for a fixed
        value of $\Lambda_*$ and several values of the scattering
        length $a$.  As $\Lambda \to \infty$, $\hat g_3$
        asymptotically approaches the RG limit cycle shown as a
        heavy solid line.}
\label{fig:g3RGtra}
\end{figure}

The dimensionless 2-body coupling constant $\hat g_2$ 
is introduced in Eq.~(\ref{Lamg2}). It is convenient to also 
introduce a dimensionless 3-body coupling constant $\hat g_3$ by
\begin{eqnarray}
\hat g_3(\Lambda) = {\Lambda^4 g_3 \over  144 \pi^4 } \,.
\label{g3hat}
\end{eqnarray}
Using Eqs.~(\ref{g3g2}) and (\ref{g2-tune}),
the dimensionless 3-body coupling constant can be written
\begin{eqnarray}
\hat g_3(\Lambda) = - 
\left( {a \Lambda \over a \Lambda - \pi/2} \right)^2 H(\Lambda) \,.
\label{Lam4g3}
\end{eqnarray}
%
As $\Lambda$ is varied with $a$ and $\Lambda_*$ fixed, the expression for 
$\hat g_3$ in Eq.~(\ref{g3hat}) maps out a renormalization group 
(RG) trajectory.  The RG trajectories for a fixed value of $\Lambda_*$
and various values of $a$ are illustrated in Fig.~\ref{fig:g2RGtra}.  
All the points on a given trajectory represent 
the same physical theory with given values of $a$ and $\Lambda_*$.  
As $\Lambda$ increases, the dimensionless coupling
constant in Eq.~(\ref{Lam4g3}) flows towards an ultraviolet limit cycle:
\begin{eqnarray}
\hat g_3(\Lambda) \longrightarrow  -   
{\cos [s_0 \ln (\Lambda/ \Lambda_*) + \arctan s_0]
        \over \cos [s_0 \ln (\Lambda/ \Lambda_*) - \arctan s_0]} \,,
\qquad
{\rm as \ } \Lambda \to \infty \,.
\label{lam4g3:as}
\end{eqnarray}
This limit cycle corresponds to the theory 
in the resonant limit $a = \pm \infty$.
In Fig.~\ref{fig:g3RGtra}, the dimensionless 3-body coupling constant 
$\hat g_3$ is shown as a function of $\Lambda$  for several different 
values of the scattering lengths. The RG trajectories for finite $a$
are rapidly focused to the limit cycle.

Further insight into this problem can be achieved by expressing the 
renormalization group flow in terms of a differential equation
for the $\Lambda$-dependence of $\hat g_3$.
By differentiating both sides of Eq.~(\ref{Lam4g3}), one can derive 
the differential RG equation 
\beq
\Lambda\frac{d}{d\Lambda} \hat g_3  = 
{1 + s_0^2 \over 2} \left(  \hat g_2^2 + {\hat g_3^2 \over  \hat g_2^2} \right)
+ (3 - s_0^2 + 2 \hat g_2) \hat g_3 \,.
\hspace{0.5cm}
\label{eq:RGg3}
\eeq
At the fixed point $\hat g_2 = -1$, the right side of Eq.~(\ref{eq:RGg3})
has no real roots for  $\hat g_3$, so there is no ultraviolet fixed point.
Instead the ultraviolet behavior is governed by the fixed point 
$\hat g_2 = -1$ and the limit cycle in Eq.~(\ref{lam4g3:as}) for $\hat g_3$.

One might expect to be able to derive the perturbative differential
RG equation for $g_3$ in Eq.~(\ref{g3-rg}) from the nonperturbative 
differential RG equation in Eq.~(\ref{eq:RGg3}).  However, the connection 
between these equations is obscured by a fundamental difference 
in renormalization schemes.  The perturbative RG equation in 
Eq.~(\ref{g3-rg}) is derived within a renormalization scheme in which 
power ultraviolet divergences are subtracted from Green's functions, 
so that renormalization of the coupling constants is only required 
to remove residual logarithmic divergences.  For example, in this 
renormalization scheme, all
the loop corrections in the expression for $g_2$ in Eq.~(\ref{g2-tune})
are subtracted away and the expression reduces to $g_2 = 8 \pi a$.
In the nonperturbative renormalization scheme corresponding to the 
differential RG equations for $\hat g_2$ in Eq.~(\ref{eq:RGg2})
and  $\hat g_3$ in Eq.~(\ref{eq:RGg3}), power ultraviolet divergences 
are removed by renormalization of the coupling constants.
The renormalization of $\hat g_3$ also removes from 3-body amplitudes 
cutoff-dependent effects that remain bounded as $\Lambda \to \infty$, 
but have log-periodic dependence on $\Lambda$.

One remarkable feature of the renormalization of $g_3$
is that there are values of the ultraviolet cutoff 
for which the dimensionless coupling constant $\Lambda^4 g_3$ diverges.
The divergences occur as $\Lambda$ approaches the values 
\begin{eqnarray}
\Lambda_n' =  \left( e^{\pi/s_0} \right)^n 
\exp \big( [ \mbox{$1\over 2$} \pi + \arctan s_0]/s_0 \big) \Lambda_* \,,
\label{def-Lambda_n'}
\end{eqnarray}
where $n$ is an integer. At these points, $\Lambda^4 g_3$ increases to
$\infty$, jumps discontinuously to $- \infty$, and then continues to
increase. There is a simple intuitive explanation for this remarkable
behavior. The ultraviolet cutoff $\Lambda$ excludes Efimov states with
binding energies greater than about $\Lambda^2$. As $\Lambda$ increases,
the strength of the 3-body contact interaction must increase in order to
keep low-energy observables invariant. This 3-body contact interaction
takes into account the effects of short-distance 3-body configurations
that are excluded by the cutoff, including the excluded Efimov states.
When $\Lambda$ reaches a critical value, $\Lambda^4 g_3$ 
becomes infinite. At this critical value, a new Efimov state
with binding energy of order $\Lambda^2$ appears in the spectrum
and $\Lambda^4 g_3$ jumps discontinuously to $- \infty$, 
because the 3-body contact interaction no longer needs to take into 
account the virtual effects of that Efimov state.

Another interesting feature of the renormalization of $g_3$
is that there are values of the ultraviolet cutoff 
for which $g_3$ vanishes.
The zeros occur as $\Lambda$ approaches the values 
\begin{eqnarray}
\Lambda_n =  \left( e^{\pi/s_0} \right)^n  
\exp \big( [ \mbox{$1\over 2$} \pi - \arctan s_0]/s_0 \big) \Lambda_* \,.
\label{def-Lambda_n}
\end{eqnarray}
At these discrete values of $\Lambda$, there is no need for a 3-body
contact interaction.  One can therefore simplify the generalized STM integral
equation in Eq.~(\ref{BHvK}) by setting $H=0$ and $\Lambda=\Lambda_n$
\cite{HM-01a}.  
One can take the limit $\Lambda \to \infty$ by increasing the integer $n$.
However, one must choose $n$ large enough that corrections suppressed 
by $1/(a \Lambda_n)$ and $k/\Lambda_n$ are negligible. In practice, 
$n=1$ or 2 is often large enough to get a few digits of accuracy.
This simple trick turns out to be very useful for practical calculations. 
It justifies Kharchenko's ad hoc procedure 
of fitting the cutoff in the STM equation to a 3-body datum and 
then using the same cutoff to predict other data \cite{Khar73}.


\subsection{Effects of deep 2-body bound states}

Effective field theory can also be used to take into account
the effects of deep 2-body bound states.
The coefficient $g_3$ of the 3-body contact interaction in
Eq.~(\ref{L-BHvK}) must be generalized to a complex-valued coupling constant.
The real and imaginary parts of $g_3$ 
can then be tuned simultaneously as functions of the ultraviolet cutoff
to reproduce the correct values for the two 3-body parameters 
$\kappa_*$ and $\eta_*$ that together with $a$ determine 
the low-energy 3-body observables.

In the absence of deep 2-body bound states, a sufficient condition for 
solutions to the STM3 integral equation in Eq.~(\ref{BHvK})
to have a well-behaved limit as $\Lambda \to \infty$ 
is that the 3-body coupling constant $g_3$ must have 
the form given by Eqs.~(\ref{g3g2}) and (\ref{H-Lambda}).  
Since the function $H(\Lambda)$ in 
Eq.~(\ref{H-Lambda}) is an analytic function of $\ln \Lambda_*$, the 
integral equation defines the amplitude ${\mathcal A}_s(q, k; E)$ as 
an analytic function of $\ln \Lambda_*$.  The STM3 integral equation 
continues to have a well-behaved limit as $\Lambda \to \infty$ if 
$\Lambda_*$ in the expression for $H(\Lambda)$ in Eq.~(\ref{H-Lambda})
is replaced by the complex variable $\Lambda_* e^{i \eta_* / s_0}$:
\begin{eqnarray}
H(\Lambda) = 
{\cos [ s_0 \ln (\Lambda/ \Lambda_*) + \arctan s_0 - i \eta_*]
\over \cos [ s_0 \ln (\Lambda/ \Lambda_*) - \arctan s_0 - i \eta_*]} \, .
\label{H-complex}
\end{eqnarray}
If the dependence of the amplitude ${\mathcal A}_s (q,k; E)$ on $\Lambda_*$ 
is known analytically, the effect of the parameter $\eta_*$ can be 
obtained simply by the substitution 
$\ln \Lambda_* \to \ln \Lambda_* + i \eta_*/s_0$.
The bound-state equation in Eq.~(\ref{BHvK-homo}) also defines the energy 
eigenvalues for the Efimov states as analytic functions of 
$\ln \Lambda_*$.  If analytic expressions for the binding energies 
as functions of $\Lambda_*$ were known, the effects of the parameter 
$\eta_*$ could again be determined by simple substitution.  
One limit in which the 
analytic expression is known is the resonant limit $a = \pm \infty$.  
In this case, the Efimov binding energies in the absence of deep 2-body 
bound states satisfy Eq.~(\ref{B3n-resonant}). Since $\kappa_*$ differs 
from $\Lambda_*$ only by a multiplicative constant, the effect of the 
parameter $\eta_*$ can be determined by the substitution 
$\kappa_* \to \kappa_* e^{i \eta_* / s_0}$.  
If there are deep 2-body bound states, the 
resonance energies $E_T^{(n)}$ of the trimers and their widths 
$\Gamma_T^{(n)}$ in the resonance limit must satisfy
\begin{eqnarray}
E_T^{(n)} + \frac i 2 \Gamma_T^{(n)} \longrightarrow && 
\left( e^{-2 \pi/s_0} \right)^{n-n_*} 
e^{2i \eta_* /s_0} {\hbar^2 \kappa^2_* \over m} \,,
\nonumber \\
&& \hspace*{1cm}
{\rm as \ } n \to \infty {\rm \ \ with \ } a = \pm \infty \,.
\end{eqnarray}
The ratios of the widths and the resonance energies approach a 
constant as the 3-atom threshold is approached: 
\begin{eqnarray}
\Gamma_T^{(n)}/E_T^{(n)} & \longrightarrow &
2 \tan (2 \eta_* /s_0)\,,
\quad
{\rm as \ } n \to \infty {\rm \ \ with \ } a = \pm \infty.
\end{eqnarray}
This equation, which involves only physical observables, provides an 
operational definition of the parameter $\eta_*$. 
It can be used to determine $\eta_*$ for 
any system that can be tuned to the resonant limit $a = \pm \infty$.

The effects of deep 2-body bound states are particularly simple 
if the inelasticity parameter $\eta_*$ is infinitesimally small.
For example, the rate constants for 3-body recombination into deep molecules
are given with an accuracuy of better than 1\% 
by Eqs.~(\ref{alpha-deep:a>0}) and (\ref{alpha-deep:a<0}).
Their expansions to first order in $\eta_*$ are
\begin{subequations}
\begin{eqnarray}
\alpha_{\rm deep} & \approx & 67.1 \, \eta_* {\hbar a^4 \over m} \,,
\qquad  \hspace{1.6cm} (a>0) \,,
\label{alpha-deepinf:a>0}
\\
\alpha_{\rm deep} & \approx & 
{9180 \, \eta_* \over \sin^2 [s_0 \ln (a/a_*')]}
\; {\hbar a^4 \over m}\,,
\qquad (a<0) \,. 
\label{alpha-deepinf:a<0}
\end{eqnarray}
\label{alpha-deepinf}
\end{subequations}
Note that the expression for $\alpha_{\rm deep}$
in Eq.~(\ref{alpha-deepinf:a<0}) diverges at values of $a$
that differ from $a_*'$ by multiples of $e^{\pi/s_0}$.
These divergences are cut off in Eq.~(\ref{alpha-deep:a<0})
by summing corrections of higher order in $\eta_*$ to all orders.

\begin{figure}[htb]
\bigskip
\centerline{\includegraphics*[width=9cm,angle=0]{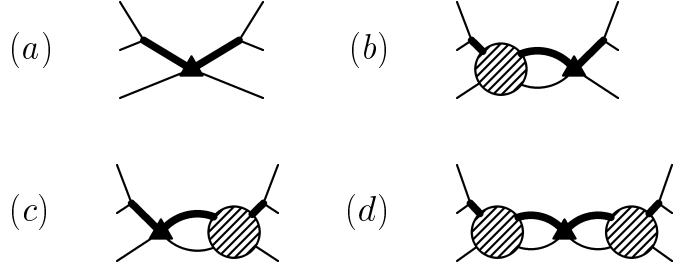}}
\medskip
\caption
{Feynman diagrams with a single insertion of the 
$d^\dagger d \psi^\dagger \psi$ 
vertex proportional to $h_3$ represented by the triangle.
Diagram (d) dominates in the limit $\Lambda\to\infty$.
}
\label{fig:ih3}
\end{figure}

Calculations of the effects of deep 2-body bound states 
have been carried out using effective field theory 
for the case in which the inelasticity parameter $\eta_*$ 
is infinitesimally small.
In this case, the imaginary part of the coupling constant $g_3$ 
is also infinitesimally small and can be treated as a perturbation.  
At leading order in ${\rm Im} \, g_3$, the tuning of ${\rm Re} \, g_3$ 
is unaffected and is given by Eq.~(\ref{g3g2}).
An integral equation for the first-order change in the amplitude 
${\mathcal A}$ can be obtained by substituting 
${\mathcal A} \to {\mathcal A} + \delta {\mathcal A}$ and 
$g_3 \to g_3 + i \, {\rm Im}\, g_3$
into the integral equation in Eq.~(\ref{BhvK:general})
and expanding to first order in $\delta {\mathcal A}$ and ${\rm Im} \, g_3$.
However, it is simpler to express the first-order change 
$\delta {\mathcal A}$ as the sum of all amputated connected 
Feynman diagrams with a single insertion of a 
$d^\dagger d \psi^\dagger \psi$ vertex with the Feynman rule 
${\rm Im} \, g_3/36$.  The resulting set of diagrams is shown in 
Fig.~\ref{fig:ih3}, where the triangle represents the 
$d^\dagger d \psi^\dagger \psi$ 
vertex and the blob represents the amplitude ${\mathcal A}$.
One can easily show that the S-wave amplitude can be expressed 
in the factored form
\begin{eqnarray}
\delta {\mathcal A}_S(p,k;E) &=& -i\, {\rm Im} \, g_3
 {64 \pi^2 a^2 \over g_2^2} 
\big( 1 + {\mathcal C}_S(p;E) \big) 
\big( 1 + {\mathcal C}_S(k;E) \big),
\label{dAS-BS}
\end{eqnarray}
where ${\mathcal C}_S(p;E)$ is given by an integral of ${\mathcal A}_S$
over one of its arguments:
\begin{equation}
{\mathcal C}_S(p;E) = {a \over 4 \pi^2} \int_0^\Lambda 
{ dq \, q^2   {\mathcal A}_S(p,q;E) \over
 - 1/a + \sqrt{3q^2/4 -E -i \epsilon} } \,.
\label{BS-AS}
\end{equation}
Since ${\mathcal A}_S$ scales like $q^{-1}$ as $q \to \infty$, 
the integral is linearly divergent. Thus the 1 in the factors 
$(1 + {\mathcal C}_S)$ in Eq.~(\ref{dAS-BS}) can be neglected as 
$\Lambda \to \infty$.  To obtain a finite limit for 
$\delta {\mathcal A}_S$ as $\Lambda \to \infty$, 
${\rm Im} \, g_3/g_2^2$ must be proportional to $1/\Lambda^2$.  
The function ${\mathcal C}_S(p;E)$ can be obtained by solving the 
integral equation for ${\mathcal A}_S(p,q;E)$ for all values of $q$ 
and inserting the solution into the integral in Eq.~(\ref{BS-AS}).
This method was used in Ref.~\cite{BH01} to calculate the 
rate constants $\alpha_{\rm deep}$ for $a>0$ and $a<0$ 
to first order in ${\rm Im} \, g_3$.
The functional dependence of the results on $a$ and $\Lambda_*$ 
were found to have the forms shown in Eq.~(\ref{alpha-deepinf}).  
For the special cutoff values $\Lambda_n$
at which ${\rm Re}\, g_3=0$ (cf.~Eq.~(\ref{def-Lambda_n})),
the relation between the infinitesimal parameters
${\rm Im} \, g_3$ and $\eta_*$ were found to be%
\footnote{Note that a factor of $1/(32 \pi^2)$ was omitted from the 
expression for $h' = -{\rm Im} \,  g_3/(9 g_2)$ in  Ref.~\cite{BH01}.}
\begin{eqnarray}
{\rm Im} \,  g_3 \approx - 23.3 \, \eta_* \, {g_2^2 / \Lambda_n^2} \,.
\label{Img3-num}
\end{eqnarray}
The dependence on $a$ of the rate constants $\alpha_{\rm deep}$
in Eqs.~(\ref{alpha-deepinf}) was derived using the generalization 
of Efimov's radial laws to the case in which there are 
deep 2-body bound states.  The fact that these results can be 
reproduced using a purely numerical method based on 
effective field theory increases our confidence in the 
generalization of the radial laws.

If the expression for the complex-valued 3-body coupling constant $g_3$
given by Eqs.~(\ref{g3g2}) and (\ref{H-complex}) is expanded 
to first order in $\eta_*$, the result is
\begin{eqnarray}
{\rm Im} \, g_3 = - 
{9  \sin [ 2 \arctan s_0] 
\over \cos^2[ s_0 \ln (\Lambda/ \Lambda_*) - \arctan s_0 ]} \, \eta_* \,
{g_2^2 \over \Lambda^2} \, .
\label{Img3-anal}
\end{eqnarray}
There is a factor 2.6 discrepancy between 
the analytic result for ${\rm Im} \, g_3$
in Eq.~(\ref{Img3-anal}) at $\Lambda=\Lambda_n$
and the numerical result in Eq.~(\ref{Img3-num}).


%
%

\section{Universality in Other Three-body Systems}
        \label{sec:beyond}

In this section, we summarize the universal information that is known 
about 3-body systems with large scattering lengths other than 
three identical bosons in 3 space dimensions.


\subsection{Unequal Scattering Lengths}

There are 3-body systems in which the three particles all have the same mass, 
but the three pairs of particles need not all have the same scattering lengths.
For example, different hyperfine spin states of the same atom have the same 
mass, but the scattering lengths $a_{ij}$ can be different for each pair
$ij$ of hyperfine states.   As another example, different isotopes of a heavy 
atom have nearly the same masses.  Finally, the proton and neutron have 
nearly equal masses.  It is therefore worthwhile to consider the universal 
behavior of systems with equal masses and scattering lengths that are large 
but not all equal.
Note that a multiple fine-tuning of the interactions might be required to make 
more than one scattering length large simultaneously.

One of the most basic questions for a 3-body system is whether the Efimov 
effect occurs in that system.  This question can be answered by determining
the channel eigenvalue $\lambda_0(R)$ for the lowest hyperspherical potential
in the scaling limit.  The Efimov effect occurs if $\lambda_0(R)$ is negative 
at $R = 0$.  If $\lambda_0(0) = - s_0^2$,  the discrete scaling factor
in the Efimov effect is $e^{\pi/s_0}$.  The Efimov effect in general 3-body 
systems was first discussed by Amado and Noble \cite{AN72}
and by Efimov \cite{Efimov72,Efimov73}.  
A summary of their results is as follows. If only one of the 
three scattering lengths is large, the Efimov effect does not occur.
If two of the scattering lengths are large,
the Efimov effect occurs with a discrete scaling factor of about 1986.1
unless two of the three particles are identical fermions,
in which case the Efimov effect does not occur.
If all three scattering lengths are large,
the Efimov effect occurs with a discrete scaling factor of about 22.7.

To derive these results, 
we first consider the case of three distinguishable atoms with equal masses 
and large scattering lengths $a_{12}$, $a_{23}$, and $a_{31}$.  
The other cases can be obtained from this one by taking appropriate limits.
If three atoms are distinguishable, 
the Schr{\"o}dinger wave function need not be symmetric under
interchange of the atoms.  We restrict our attention to total angular momentum
$L=0$ and assume that for each Faddeev component
$\psi^{(i)}({\bm r}_{jk}, {\bm r}_{i, jk})$,
there is no orbital angular momentum associated with either the $jk$ or
$i, jk$ subsystems.  In this case, the Schr{\"o}dinger wave function can be
expressed in the form
\begin{eqnarray}
\Psi ({\bm r}_1, {\bm r}_2, {\bm r}_3) = \psi^{(1)} (R, \alpha_1) 
+ \psi^{(2)}(R, \alpha_2) 
 + \psi^{(3)} (R, \alpha_3) \,.
\label{Psi123}
\end{eqnarray}
If $R \sin \alpha_i$ is large enough that the 2-body potential
$V(R \sin \alpha_i)$ can be neglected, the $i^{\rm th}$ Faddeev
wave function must have the form
\begin{eqnarray}
\psi^{(i)} (R, \alpha_i) \approx F^{(i)} (R)
{\sin [\lambda^{1/2} (R) ({\pi \over 2} - \alpha_i)] \over 
  \sin (2 \alpha_i)} \,.
\label{psiF:i}
\end{eqnarray}
The Faddeev equations can also be solved approximately in the region
$\alpha_i \ll 1$. The matching equations for these solutions are \cite{NFJG01}
\begin{eqnarray}
&&\left[
\cos \left( \lambda^{1/2} \mbox {$\pi \over 2$} \right)
\left( \begin{tabular}{ccc} 
        \ 1 \ & \ 0 \ & \ 0 \ \cr
        \ 0 \ & \ 1 \ & \ 0 \ \cr
        \ 0 \ & \ 0 \ & \ 1 \ \end{tabular} \right) \right. 
- {4 \over \sqrt {3}} \lambda^{-1/2} \sin \left( \lambda^{1/2} 
\mbox {$\pi \over 6$} \right)
\left( \begin{tabular}{ccc} 
        \ 0 \ & \ 1 \ & \ 1 \ \cr
        \ 1 \ & \ 0 \ & \ 1 \ \cr
        \ 1 \ & \ 1 \ & \ 0 \ \end{tabular} \right)
\nonumber
\\[4pt]
&& \quad
\left. - \sqrt{2} \lambda^{-1/2} \sin \left( \lambda^{1/2} 
\mbox {$\pi \over 2$} \right)
\left( \begin{tabular}{ccc}
        $R/a_{23}$ &      0       &      0       \cr
             0       & $R/a_{31}$ &      0       \cr
             0       &      0       & $R/a_{12}$ \end{tabular} \right)
\right]
\times 
\left( \begin{tabular}{c} $F^{(1)}$ \cr $F^{(2)}$ \cr $F^{(3)}$ \end{tabular}
\right)
= 0 \,.
\nonumber\\
\label{match-123}
\end{eqnarray}
The consistency condition for a
nontrivial solution is that the determinant of the $3\times3$ matrix
on the left side of Eq.~(\ref{match-123}) vanishes.
The solutions to this
equation are the possible hyperangular eigenvalues $\lambda_n(R)$.
If atoms 2 and 3 are identical bosons, then $a_{12}=a_{31}$ and
we must impose the constraint $F^{(2)} = F^{(3)}$.
The matching equations then reduce to
\begin{eqnarray}
&&\left[
\cos \left( \lambda^{1/2} \mbox {$\pi \over 2$} \right)
\left( \begin{tabular}{cc} \ 1 \ & \ 0 \ \cr
                        \ 0 \ & \ 1 \ \end{tabular} \right) 
- {4 \over \sqrt {3}} \lambda^{-1/2} \sin \left( \lambda^{1/2} 
  \mbox {$\pi \over 6$} \right)
\left( \begin{tabular}{cc} \ 0 \ & \ 2 \ \cr
                        \ 1 \ & \ 1 \ \end{tabular} \right)\right.
\nonumber
\\[4pt]
&& \quad
\left. - \sqrt{2} \lambda^{-1/2} \sin \left( \lambda^{1/2} 
   \mbox {$\pi \over 2$} \right)
\left( \begin{tabular}{cc}
        $R/a_{23}$ &      0       \cr
             0       & $R/a_{31}$ \end{tabular} \right)
\right]
\times 
\left( \begin{tabular}{c} $F^{(1)}$ \cr $F^{(2)}$ \end{tabular} \right)
= 0 \,.
\label{match-122}
\end{eqnarray}
The consistency condition that determines the eigenvalues $\lambda_n(R)$
is that the determinant of the $2\times2$ matrix
on the left side of Eq.~(\ref{match-122}) vanishes.
If atoms 1, 2, and 3 are all identical bosons, 
then $a_{23}=a_{23}=a_{31}$
and we must impose the constraints $F^{(1)} = F^{(2)} = F^{(3)}$.
In this case, the matching equation reduces to Eq.~(\ref{cheigen}).

We next consider the case of two large scattering lengths $a_{23}$ and
$a_{31}$, with the third scattering length $a_{12}$
having a natural size of order $\ell$.  In the scaling limit,
we can set $a_{12}=0$.  The matching condition for $\psi^{(3)}$
requires $F^{(3)} = 0$. The matching equations for $\psi^{(1)}$
and $\psi^{(2)}$ then reduce to
\begin{eqnarray}
&&\left[
\cos \left( \lambda^{1/2} \mbox {$\pi \over 2$} \right)
\left( \begin{tabular}{cc} 
        \ 1 \ & \ 0 \ \cr
        \ 0 \ & \ 1 \ \end{tabular} \right) 
 - {4 \over \sqrt {3}} \lambda^{-1/2} \sin \left( \lambda^{1/2} 
  \mbox {$\pi \over 6$} \right)
\left( \begin{tabular}{cc} 
        \ 0 \ & \ 1 \ \cr
        \ 1 \ & \ 0 \ \end{tabular} \right)\right.
\nonumber
\\[4pt]
&& \quad
\left. - \sqrt{2} \lambda^{-1/2} \sin \left( \lambda^{1/2}  
   \mbox {$\pi \over 2$} \right)
\left( \begin{tabular}{cc}
        $R/a_{23}$ &      0       \cr
             0       & $R/a_{31}$ \end{tabular} \right)
\right] \times
\left( \begin{tabular}{c} $F^{(1)}$ \cr $F^{(2)}$ \end{tabular} \right)
= 0 \,.
\label{match-120}
\end{eqnarray}
The consistency condition that determines the eigenvalues $\lambda_n(R)$
is that the determinant of the $2\times2$ matrix
on the left side of Eq.~(\ref{match-120}) vanishes.
If atoms 1 and 2 are identical bosons, 
we must impose the constraint $F^{(1)} = F^{(2)}$.
Setting $a_{23}=a_{31}=a$, the matching equation is
\begin{eqnarray}
\cos \left( \lambda^{1/2} \mbox {$\pi \over 2$} \right)
- {4 \over \sqrt {3}} \lambda^{-1/2} 
\sin \left( \lambda^{1/2} \mbox {$\pi \over 6$} \right)
= \sqrt{2} \lambda^{-1/2} \sin \left( \lambda^{1/2} 
  \mbox {$\pi \over 2$} \right) {R \over a} \,.
\label{cheigen-110b}
\end{eqnarray}
If atoms 1 and 2 are identical fermions,
then $a_{12}$ must be exactly 0 and
we must impose the constraint $F^{(1)} = - F^{(2)}$.
Setting $a_{23}=a_{31}=a$, the matching equation is
\begin{eqnarray}
\cos \left( \lambda^{1/2} \mbox {$\pi \over 2$} \right)
+ {4 \over \sqrt {3}} \lambda^{-1/2} \sin \left( \lambda^{1/2} 
  \mbox {$\pi \over 6$} \right)
= \sqrt{2} \lambda^{-1/2} \sin \left( \lambda^{1/2} 
  \mbox {$\pi \over 2$} \right) {R \over a} \,.
\label{cheigen-110f}
\end{eqnarray}
 
Finally, we consider the case of a single large scattering length 
$a_{23} = a$,
with the other two scattering lengths having natural sizes
of order $\ell$. In the scaling limit, we can set $a_{12}=a_{31}=0$.
The matching conditions for $\psi^{(2)}$ and $\psi^{(3)}$
then require $F^{(2)} = F^{(3)} = 0$.
The matching equation for $\psi^{(1)}$ is
\begin{eqnarray}
\cos \left( \lambda^{1/2} \mbox{$\pi \over 2$} \right)
=  \sqrt{2} \lambda^{-1/2} \sin \left( \lambda^{1/2} 
  \mbox{$\pi \over 2$} \right) {R \over a} \,.
\label{cheigen-100}
\end{eqnarray}

We now ask whether the Efimov effect occurs and if so,
what the discrete scaling factor is.
We first consider the case of three distinguishable atoms with large scattering
lengths $a_{12}$, $a_{23}$, and $a_{31}$.
The matching equations are given in Eq.~(\ref{match-123}).
At $R=0$, the consistency condition reduces to
\begin{eqnarray}
&&\left[ \cos \left( \lambda^{1/2} \mbox{$\pi \over 2$} \right)
+ {4 \over \sqrt 3} \lambda^{-1/2} \sin \left( \lambda^{1/2} 
   \mbox{$\pi \over 6$} \right)
\right]^2 
\nonumber \\ && \qquad\times
\left[ \cos \left( \lambda^{1/2} \mbox{$\pi \over 2$} \right)
- {8 \over \sqrt 3} \lambda^{-1/2} \sin \left( \lambda^{1/2} 
  \mbox{$\pi \over 6$} \right)
\right] = 0 \,.
\label{cheigen0-123}
\end{eqnarray}
If atoms 2 and 3 are identical bosons, the matching equations are given in
Eq.~(\ref{match-122}).  At $R=0$, the consistency condition is the same as 
Eq.~(\ref{cheigen0-123}) except that the first factor on the left side 
is not squared.  The possible values of $\lambda(0)$ are identical.
A negative value of $\lambda(0)$ comes only from the vanishing of the 
second factor on the left side of Eq.~(\ref{cheigen0-123}).  This condition
is identical to the matching equation
for three identical bosons in Eq.~(\ref{cheigen}).  
It has the negative solution $\lambda = -s_0^2$,
with $s_0 \approx 1.00624$.  Thus there is an Efimov effect
with discrete scaling factor $e^{\pi / s_0} \approx 22.7$.

We now consider the case of two large scattering lengths $a_{13}$ and
$a_{23}$.  If the three atoms are distinguishable,
the matching equations are given in Eq.~(\ref{match-120}).
At $R=0$, the consistency condition reduces to
\begin{eqnarray}
&&\left[ \cos \left( \lambda^{1/2} \mbox{$\pi \over 2$} \right)
+ {4 \over \sqrt 3} \lambda^{-1/2} \sin \left( \lambda^{1/2} 
   \mbox{$\pi \over 6$} \right)
\right] 
\nonumber \\ &&\qquad \times
\left[ \cos \left( \lambda^{1/2} \mbox{$\pi \over 2$} \right)
- {4 \over \sqrt 3} \lambda^{-1/2} \sin \left( \lambda^{1/2} 
   \mbox{$\pi \over 6$} \right)
\right] = 0 \,.
\label{cheigen0-120}
\end{eqnarray}
If atoms 1 and 2 are identical bosons, the matching equation is given
in Eq.~(\ref{cheigen-110b}). At $R=0$, this implies the vanishing
of the second factor on the left side of Eq.~(\ref{cheigen0-120}).
This equation has a single negative solution $\lambda_0 = - s_0^2$,
with $s_0 \approx 0.4137$.  Thus there is an Efimov effect with
discrete scaling factor $e^{\pi /s_0} \approx 1986.1$.
If atoms 1 and 2 are identical fermions, the consistency equation is given
in Eq.~(\ref{cheigen-110f}). At $R=0$, this implies the vanishing
of the first factor on the left side of Eq.~(\ref{cheigen0-120}).
This equation has no negative solutions, so there is no Efimov effect. 

Finally we consider the case of a single large scattering length 
$a_{23}=a$. The matching equation is given in Eq.~(\ref{cheigen-100}).
At $R = 0$, there are no negative solutions for $\lambda$,
so there is no Efimov effect.


\subsection{Unequal masses}
\label{sec:mass}

In the 2-body sector, the universal results for particles of unequal 
masses are only a little more complicated than those for the equal-mass
case in Sections~\ref{sec:uni2AA} and \ref{sec:uni2D}.
Let the atoms 1 and 2 have a large scattering length $a_{12}$
and unequal masses $m_1$ and $m_2$.  
The 2-body reduced mass is defined by
\begin{eqnarray}
m_{12} &=& {m_1 m_2 \over m_1 + m_2},
\end{eqnarray}
It approaches $m_1$ 
in the limit $m_1\to 0$ and $m_2$ in the limit $m_1 \to \infty$.
In the case of equal masses $m$, it reduces to $m_{12} = m/2$. 
If the momenta of the two atoms are 
$\pm \hbar {\bm k}$, their total kinetic energy is 
$E = \hbar^2 k^2/(2m_{12})$.  The differential cross section for 
elastic scattering is
\begin{eqnarray}
{d \sigma \over d \Omega} = 
{a_{12}^2 \over 1 + a_{12}^2 k^2} \,.
\end{eqnarray}
The cross section is obtained by integrating over the total solid angle 
of $4 \pi$.  If $a_{12}$ is large and positive, the atoms 1 and 2 form 
a shallow dimer with binding energy
\begin{eqnarray}
E_D = 
{\hbar^2 \over 2 m_{12} a_{12}^2} \,.
\end{eqnarray}

In the general 3-body system, the three masses can be unequal 
and any combination of the three scattering lengths can be large.
The Efimov effect in general 3-body 
systems was first discussed by Amado and Noble \cite{AN72}
and by Efimov \cite{Efimov72,Efimov73}.  The special case in which two of the 
three particles have the same mass was also discussed by
Ovchinnikov and Sigal \cite{OS79}. The conditions for the
existence of the Efimov effect and the value of the discrete scaling factor 
depend on the ratios of the masses.
We summarize briefly the results for the extreme cases
in which two masses are equal and the third mass 
is either much larger or much smaller.
In the case of two heavy particles and one light particle, 
the Efimov effect occurs provided the heavy-light scattering length is large.
In the case of one heavy particle and two light particles, the Efimov effect 
occurs only if all three pairs of particles have large scattering lengths.

In the 3-body problem with unequal masses $m_1$, $m_2$, and $m_3$, 
it is convenient to introduce a 3-body reduced mass:
\begin{eqnarray}
m_{123} &=& {m_1 m_2 m_3 \over m_1 m_2 + m_2 m_3 + m_3 m_1} \,.
\end{eqnarray}
It approaches $m_1$ in the limit $m_1\to 0$ 
and $m_{23}$ in the limit $m_1 \to \infty$.
In the case of equal masses $m$, it reduces to $m_{123}=m/3$.  

The hyperspherical coordinates defined for equal-mass particles in
Section~\ref{subsec:hyper} can be generalized to
the case of unequal masses.
A set of Jacobi coordinates consists of the
separation vector ${\bm r}_{ij}$ between a pair of atoms, which is defined in
Eq.~(\ref{jacobi}), and the separation vector of atom $k$ from the
center-of-mass coordinate of atoms $i$ and $j$:%
\begin{eqnarray}
{\bm r}_{k, ij} = {\bm r}_k - {m_i {\bm r}_i + m_j {\bm r}_j 
\over m_i + m_j} \,.
\end{eqnarray}
The hyperradius is the square root of a weighted average 
of the separations of the pairs of atoms:
\begin{eqnarray}
R^2 = {m_1 m_2 r_{12}^2 + m_2 m_3 r_{23}^2 + m_3 m_1 r_{31}^2
        \over m_1 m_2 + m_2 m_3 + m_3 m_1} \,.
\label{Rdef-masses}
\end{eqnarray}
The hyperangles $\alpha_k$ are defined by
\begin{eqnarray}
\tan \alpha_k = 
\left ( {m_{ij}^2 (m_1 + m_2 + m_3) \over m_1 m_2 m_3} \right )^{1/2}
        {r_{ij} \over r_{k, ij}} \,.
\end{eqnarray}
The magnitudes of the separation vectors are
\begin{subequations}
\begin{eqnarray}
r_{ij} &=&
\left( {m_1 m_2 m_3 \over m_{ij} m_{123} (m_1 \! + \! m_2 \! + \!  m_3)} 
\right)^{1/2}  \!\!  R \sin \alpha_k \,,
\hspace{0.9cm}
\\
r_{k, ij} &=&
\left ( {m_{ij} \over m_{123}} \right)^{1/2} R \cos \alpha_k \,.
\end{eqnarray}
\end{subequations}
In Ref.~\cite{NFJG01}, the authors used a definition of the hyperradius $\rho$ 
that depends on an arbitrary mass parameter $\mu$:
\begin{eqnarray}
\rho^2 = {m_1 m_2 + m_2 m_3 + m_3 m_1 \over \mu (m_1 + m_2 + m_3)} R^2 \,.
\end{eqnarray}
Thus our definition in Eq.~(\ref{Rdef-masses})
corresponds to a specific choice for $\mu$.

By introducing hyperspherical coordinates in the Faddeev equation for the 
3-body system, the hyperangular variables can be separated from the 
hyperradius $R$.  The hyperangular eigenvalue 
equations determine the channel eigenvalues $\lambda_n (R)$.  
The channel potentials that appear in the 
coupled set of differential equations for the hyperradial functions
can be written%
\footnote{Our channel eigenvalues $\lambda_n$ corresponds to 
$\lambda_n + 4$ in the notation of Ref.~\cite{NFJG01}.}
\begin{eqnarray}
V_n (R) = - \frac {(\lambda_n (R) - {1\over4}) \hbar^2 (m_1 + m_2 + m_3)}
{2 (m_1 m_2 + m_2 m_3 + m_3 m_1) R^2} \,.
\label{Vn-m1m2m3}
\end{eqnarray}

The question of whether the Efimov effect occurs in a 3-body system
can be answered by determining
the channel eigenvalue $\lambda_0(R)$ for the lowest hyperspherical potential
in the scaling limit.  In the absence of subsystem orbital angular momenta,
the Schr\"odinger wave function can be expressed as the sum of three 
Faddeev wave functions as in Eq.~(\ref{Psi123}).
If $R \sin \alpha_i$ is large enough, the Faddeev wave functions
must have the form given in Eq.~(\ref{psiF:i}).  
The approximate solutions of the Faddeev equations in the region 
$\alpha_i \ll 1$ leads to the matching conditions \cite{NFJG01}
\begin{eqnarray}
&&\left[\cos \left( \lambda^{1/2} \mbox{$\pi \over 2$} \right)
- \left( {m_1 m_2 m_3 \over m_{jk} m_{123} (m_1 +m_2 + m_3)} \right)^{1/2} 
\lambda^{-1/2} \sin \left( \lambda^{1/2} \mbox{$\pi \over 2$} 
\right){R \over a_{jk}}\right]
F^{(i)}
\nonumber
\\[4pt]
&&  \quad
-2 \lambda^{-1/2} \left[
  {\sin \left[ \lambda^{1/2} \left({\pi \over 2} 
- \gamma_{ij} \right) \right]
        \over \sin (2 \gamma_{ij})} F^{(j)}
+ {\sin \left[ \lambda^{1/2} \left({\pi \over 2} 
- \gamma_{ik} \right)      \right]
        \over \sin (2 \gamma_{ik})} F^{(k)}\right]=0 \,,
\nonumber
\\
\label{match-general}
\end{eqnarray}
where $i, j, k$ is a permutation of 1, 2, 3 and the angle $\gamma_{ij}$
satisfies
\begin{eqnarray}
\tan \gamma_{ij} = 
\left( {m_k (m_1 + m_2 + m_3) \over m_i m_j} \right)^{1/2} \,.
\label{gamma-ij}
\end{eqnarray}
This angle is in the range $0 <\gamma_{ij} < {1\over2} \pi$. 
The three angles satisfy 
\begin{eqnarray}
\gamma_{12}  + \gamma_{23} + \gamma_{31} = \pi \,.
\end{eqnarray}
The three equations corresponding to cyclic permutations 
of $i, j, k$  in Eq.~(\ref{match-general}) can be
expressed as a homogeneous matrix equation for the 3-component
vector $(F^{(1)}, F^{(2)}, F^{(3)})$.
The consistency condition that there be a nontrivial solution is that the
determinant of the $3 \times 3$ coefficient matrix be zero.
The $R\to 0$ limit of this equation was first deduced by Efimov
\cite{Efimov72,Efimov73}. The solutions to this equation are the channel
eigenvalues $\lambda_n(R)$.
If particles $i$ and $j$ are identical bosons, then we must impose the
constraint $F^{(i)} = F^{(j)}$.
If particles $i$ and $j$ are identical fermions, then we must impose the
constraint $F^{(i)} = - F^{(j)}$.
In either case, the consistency condition reduces to the vanishing of the
determinant of a $2 \times 2$ matrix.

\begin{figure}[htb]
\medskip
\centerline{\includegraphics*[width=8.5cm,angle=0]{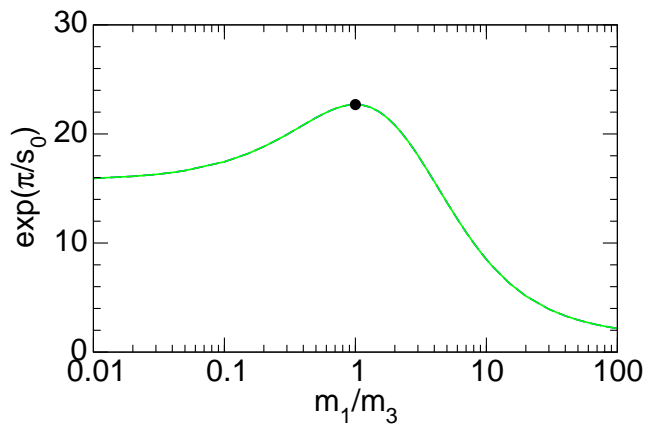}}
\medskip
\caption{
Discrete scaling factor $e^{\pi/s_0}$ 
for two particles of equal mass $m_1 = m_2$ as a function of the 
mass ratio $m_1/m_3$ for the case 
in which all three pairs have large scattering lengths.
The particles 1 and 2 can be either identical bosons or 
distinguishable. The dot indicates the case of three identical bosons.
}
\label{fig:dsf2}
\end{figure}

We first consider the case in which all three pairs have a large scattering 
length. This excludes the possibility of any pair of particles being identical 
fermions. The Efimov effect occurs if the lowest channel eigenvalue 
$\lambda_0(R)$ in the scaling limit has a negative value at $R = 0$.
The consistency condition at $R=0$ does not depend
on the scattering lengths.  For any values of the masses, there is a single
channel eigenvalue with a negative value $\lambda_0(0) = - s_0^2$ at $R=0$.
The Efimov effect therefore occurs with a discrete scaling factor 
$e^{\pi/s_0}$.  
The discrete scaling factor is largest if all three masses are equal.
It has the same value 
$e^{\pi / s_0} \approx 22.7$ as for three identical bosons.  
In the case of two equal-mass particles,
the discrete scaling factor is the same whether
the equal-mass particles are identical bosons or distinguishable.
The discrete scaling factor for $m_1 = m_2$
is shown in Fig.~\ref{fig:dsf2} as a function of the mass ratio 
$m_1/m_3 = m_2/m_3$.
In the limit $m_1 = m_2 \ll m_3$, 
the discrete scaling factor approaches $15.7$.
In the limit $m_1 = m_2 \gg m_3$, it approaches 1.

We now consider the case in which only two pairs have large
scattering lengths.  If $a_{31}$ and $a_{23}$ are large, 
the equation for $F^{(3)}$  in 
Eq.~(\ref{match-general}) should be ignored and $F^{(3)}$ should be set 
to zero in the equations for $F^{(1)}$ and $F^{(2)}$.
If the particles 1 and 2 are distinguishable, 
the resulting consistency condition reduces at $R= 0$ to
\begin{eqnarray}
&&\left[ \cos \left ( \lambda^{1/2} \mbox{$ \pi \over 2$} \right ) 
+ {2\lambda^{-1/2} \sin \left [ \lambda^{1/2} \left ( {\pi \over 2} 
- \gamma_{12} \right ) 
\right ] \over \sin (2 \gamma_{12})} \right]
\nonumber
\\
&&\qquad \times
\left[ \cos \left ( \lambda^{1/2} \mbox{$ \pi \over 2$} \right ) 
- {2\lambda^{-1/2} \sin \left [ \lambda^{1/2} \left ( {\pi \over 2} 
- \gamma_{12} \right ) 
\right ] \over \sin (2 \gamma_{12})} \right]
= 0 \,.
\label{eq:a2dis}
\end{eqnarray}
There is a single negative eigenvalue $\lambda_0 = -s_0^2$
for any values of the masses.  
It comes from the vanishing of the second factor on the left side 
of Eq.~(\ref{eq:a2dis}).  The discrete scaling factor $e^{\pi/s_0}$ 
depends on the masses only through
the angle $\gamma_{12}$ defined in Eq.~(\ref{gamma-ij}).
If particles 1 and 2 are identical bosons, 
the matching condition reduces at $R= 0$ 
to the vanishing of the 
second factor on the left side of Eq.~(\ref{eq:a2dis}).
Thus there is an Efimov effect with the same discrete scaling factor 
as in the case in which the two equal mass particles are distinguishable.
If particles 1 and 2 are identical fermions, the matching equation 
reduces at $R= 0$ to the vanishing of the 
first factor on the left side of Eq.~(\ref{eq:a2dis}).
This equation has no negative solutions for $\lambda$. 
Thus there is no Efimov effect associated with hyperangular channels
with no subsystem angular momentum.  However, as will be discussed in 
Section~\ref{sec:idf}, there is an Efimov effect associated
with hyperangular channels with nonzero subsystem angular momentum
if the mass ratio $m_3/m_1$ exceeds a critical value.

\begin{figure}[htb]
\medskip
\centerline{\includegraphics*[width=8.5cm,angle=0]{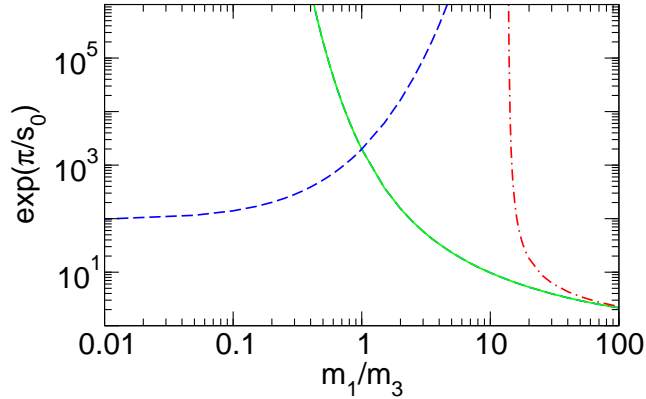}}
\medskip
\caption{
Discrete scaling factor $e^{\pi/s_0}$ for two particles of equal mass 
$m_1 = m_2$ as a function of the mass ratio 
$m_1/m_3$ for the cases in which two pairs have large scattering lengths.
If $a_{23}$ and $a_{31}$ are large, particles 1 and 2 
can be either identical bosons or distinguishable particles (solid line)
or else identical fermions (dash-dotted line).
If $a_{12}$ and $a_{31}$ (or $a_{12}$ and $a_{23}$) are large,
particles 1 and 2 must be distinguishable particles (dashed line).
}
\label{fig:dsf1}
\end{figure}

The discrete scaling factor for the cases in which only two pairs have large
scattering lengths are illustrated in Fig.~\ref{fig:dsf1}.
We consider only the special case in which particles 1 and 2 have 
the same masses $m_1 = m_2$, and we plot the discrete scaling factor 
as a function of the mass ratio $m_1/m_3$.
If the large scattering lengths are $a_{23}$ and $a_{31}$ 
and if particles 1 and 2 are either identical bosons or distinguishable,
the discrete scaling factor is $e^{\pi/s_0}$,
where $\lambda = - s_0^2$ is a negative solution to Eq.~(\ref{eq:a2dis}).
As $m_1/m_3$ increases from 0 to 
1 to $\infty$, $e^{\pi/s_0}$ decreases monotonically from $\infty$ to 1986.1
to 1 as shown in Fig.~\ref{fig:dsf1}.
The case in which particles 1 and 2 are identical fermions, 
for which the discrete scaling factor is also shown in Fig.~\ref{fig:dsf1}, 
is discussed in Section~\ref{sec:idf}.
If the large scattering lengths are either $a_{12}$ and $a_{31}$ 
or $a_{12}$ and $a_{23}$, particles 1 and 2 must be distinguishable.
The equation that determines the discrete scaling factor is 
Eq.~(\ref{eq:a2dis}), with $\gamma_{12}$ replaced by $\gamma_{23}=\gamma_{31}$.
As $m_1/m_3$ increases from 0 to 
1 to $\infty$, $e^{\pi/s_0}$ increases monotonically from 94.36 to 1986.1
to $\infty$ as shown in Fig.~\ref{fig:dsf1}.

In the case of two heavy atoms and a third atom that is much lighter, 
the Efimov effect can be understood intuitively using the 
Born-Oppenheimer approximation \cite{FRS79}.  We take the 
heavy masses to be $m_1 = m_2 = M$ 
and the light mass to be $m_3 = m$ with $m \ll M$.  
The hyperradius $R$ defined in Eq.~(\ref{Rdef-masses}) can be 
identified with the separation $r_{12}$ of the two heavy atoms.  
We take the coordinates of the three particles in the center-of-mass 
frame to be ${\bm r}_1 = + {1 \over 2} {\bm R}$, 
${\bm r}_2 = - {1 \over 2} {\bm R}$, 
and ${\bm r}_3 = {\bm r}$.  We assume that the potential between 
the heavy atoms can be neglected, so that the 3-body potential 
$V({\bm r}_1, {\bm r}_2, {\bm r}_3)$ can be expressed as the sum 
of two pairwise potentials $V(r_{23})$ and $V(r_{31})$.  
In this case, the 3-body Schr\"odinger equation in the center-of-mass 
frame can be reduced to
\begin{eqnarray}
\left[ - \frac {\hbar^2}{M} \nabla^2_R - \frac {\hbar^2}{2m} \nabla^2_r 
        + V(|{\bm r} + \mbox{$1\over2$} {\bm R}|) 
        + V(|{\bm r} - \mbox{$1\over2$} {\bm R}|) \right] \Psi
&& 
= E \, \Psi \,.
\label{Schr3-B0}
\end{eqnarray}
In the Born-Oppenheimer approximation, which becomes exact in the 
limit $M/m \to \infty$, the wave function is expressed 
in the factored form
\begin{eqnarray}
\Psi ({\bm r}, {\bm R}) = \psi ({\bm r}, {\bm R} ) \, \phi ({\bm R}) \,.
\end{eqnarray}
where $\psi ({\bm r},{\bm R})$ can be interpreted as the wave function for 
the light particle in the presence of the two heavy particles with fixed 
positions $\pm {1 \over 2}{\bf R}$.  The 3-body Schr\"odinger equation in 
Eq.~(\ref{Schr3-B0}) can be separated into two coupled equations.  The 
first is the Schr\"odinger equation for $\psi$:
\begin{eqnarray}
\left[ - \frac {\hbar^2}{2m} \nabla^2_r + V(|{\bm r} + \mbox{$1\over2$} 
{\bm R}|) 
+ V(|{\bm r} - \mbox{$1\over2$} {\bm R}|) \right] \psi 
= \epsilon (R) \, \psi \,.
\label{Schr-L}
\end{eqnarray}
The second is the Schr\"odinger equation for the heavy particles in an 
effective potential that is determined by the eigenvalue in Eq.~(\ref{Schr-L}):
\begin{eqnarray}
\left [ - \frac {\hbar^2}{M} \nabla^2_R + \epsilon (R) \right ] \phi = 
  E \, \phi \,.
\label{Schr-H}
\end{eqnarray}

Fonseca, Redish and Stanley applied the Born-Oppenheimer approach 
to the problem of a potential $V(r)$ for which an analytic equation 
could be obtained for the Born-Oppenheimer potential $\epsilon (R)$ 
\cite{FRS79}.
The potential was tuned to give a large positive scattering length $a$ 
in the heavy-light system.  Thus the heavy and light particles form
a shallow bound state with binding energy $E_2 = \hbar^2/(2ma^2)$.  
The Born-Oppenheimer potential can be expressed in the form
\begin{eqnarray}
\epsilon (R) = - \frac {\hbar^2}{2m} \kappa^2 (R) \,.
\end{eqnarray}
In the scaling limit, the equation for $\kappa (R)$ reduces to \cite{FRS79}
\begin{eqnarray}
(\kappa - 1/a) R = e^{- \kappa R} \,.
\label{kappa-BO}
\end{eqnarray}
The solution interpolates between a $1/R^2$ potential for $R \ll a$ and a 
Yukawa potential for $R \gg a$:
\begin{subequations}
\begin{eqnarray}
\epsilon (R) & \longrightarrow & 
- x_0^2 \frac {\hbar^2}{2mR^2} \,,
\qquad \hspace{1.8cm} \mbox{as\ } R \to 0 \,,
\label{epsBO-0}
\\
& \longrightarrow & 
- \frac {\hbar^2}{2ma^2} - \frac {\hbar^2}{maR} e^{-R/a}\,, 
\qquad \mbox{as\ } R \to \infty \,.
\label{epsBO-inf}
\end{eqnarray}
\end{subequations}
In Eq.~(\ref{epsBO-0}), the number $x_0 = 0.567143$ in the prefactor 
of $\hbar^2/2mR^2$ is the solution to the equation $x = e^{-x}$.  
Expressing the asymptotic potential in Eq.~(\ref{epsBO-0}) in 
the form in Eq.~(\ref{V0-si}) with $m$ replaced by $M$, 
we find that $s_0$ is
\begin{eqnarray}
s_0 \approx 0.567143 \, (M/m)^{1/2} \,.
\label{s0-BO}
\end{eqnarray}
In the limit $M/m \to \infty$, $s_0$ approaches $\infty$ and
the discrete scaling factor $e^{\pi / s_0}$ 
approaches 1.  
In Eq.~(\ref{epsBO-inf}), the first term is just the binding energy 
of the shallow dimer.  
The second term is a Yukawa potential that arises from the 
exchange of the light particle between the two heavy particles.

If we apply the hyperspherical formalism to this problem, 
the matching condition given in Eq.~(\ref{match-general})
reduces to 
\begin{eqnarray}
\cos \left( \lambda^{1/2} \mbox{$\pi \over 2$} \right) 
&-& (2m \lambda/M)^{-1/2} 
\sin \left[ \lambda^{1/2} \mbox{$\pi \over 2$} - (2m \lambda/M)^{1/2} \right]
\nonumber
\\
& =& (2m \lambda/M)^{-1/2} 
\sin \left( \lambda^{1/2} \mbox{$\pi \over 2$} \right) {R \over a} \,.
\label{match-BO}
\end{eqnarray}
For negative $\lambda$, the cosines and sines in Eq.~(\ref{match-BO})
become hyperbolic functions with real arguments.  
Keeping only the leading exponentials in the hyperbolic functions,
the matching equation in Eq.~(\ref{match-BO}) reduces to 
Eq.~(\ref{kappa-BO}) for the Born-Oppenheimer potential with 
$\kappa = (2m \lambda/M)^{1/2} R^{-1}$.
The matching equation Eq.~(\ref{match-BO}) is more general, 
because it also applies for negative values of the scattering length.

The widths of Efimov resonances composed of two heavy atoms
and one light atom (or electron) have been calculated by 
Pen'kov \cite{Pen99}.
He considered a model in which the two identical heavy 
bosonic atoms with mass $M$ form a deep S-wave bound state 
with binding energy $E_{\rm deep}$ and the interaction 
between a heavy atom and the light atom of mass $m$ 
is tuned to the resonant limit $a = \pm \infty$.    
The Efimov trimers appear as resonances 
in the scattering of the light atom and the deep diatomic 
molecule composed of the two heavy atoms.  Pen'kov
obtained an analytic expression for the widths $\Gamma_T^{(n)}$
of the Efimov trimers.  In the limit $M \gg m$,
Pen'kov's result approaches \cite{Pen99}
\begin{eqnarray}
{\Gamma_T^{(n)} \over E_T^{(n)}} 
&\longrightarrow & 103.0 \, \exp \left( -1.260 \, \sqrt{M/m} \right)
\sin^2 \left[ \mbox{$1\over2$} s_0 \ln (4E_{\rm deep}/E_T^{(n)})\right]\,,
\nonumber
\\ &&  \qquad
{\rm as \ } n \to \infty {\rm \ \ with \ } a = \pm \infty \,.
\label{Gam-Penkov}
\end{eqnarray}
where $s_0$ is given in Eq.~(\ref{s0-BO}).
Note that in spite of the appearance of the binding energy 
$E_T^{(n)}$ on the right side of Eq.~(\ref{Gam-Penkov}),
the ratio of the width to the binding energy is the same 
for all Efimov states.  This follows from
the fact that the binding energies $E_T^{(n)}$ differ by 
integral powers of the discrete scaling factor $e^{2\pi/s_0}$.
Thus Pen'kov's result is consistent with an exact discrete scaling 
symmetry in the resonant limit.


\subsection{Two identical fermions}
\label{sec:idf}

If two atoms are identical fermions, their S-wave scattering length 
must vanish.  However, the fermions can have a nonzero scattering length
with another atom.  If that scattering length is large, 
3-body systems consisting of two fermions and the third atom have universal
properties.  We will first describe the universal results 
that have been calculated and then discuss the conditions for the Efimov
effect in this system.

We first give some universal results for the special case in which the
identical fermions and the third atom have the same mass $m$.  
We take the third atom to be an orthogonal spin state of the same atom, 
and we label the spin states $\uparrow$ and $\downarrow$.
The large scattering length is $a_{\uparrow \downarrow}=a$. 
If $a>0$, the two spin states form a shallow dimer with binding energy 
$E_D=\hbar^2/(m a^2)$ that we label $D$.
The Efimov effect does not occur in this case, so the universal 
3-body predictions are completely determined by $a$.
The atom-dimer scattering length was first calculated by Skorniakov 
and Ter-Martirosian in 1956 \cite{STM57}:
\begin{eqnarray}
a_{\uparrow D} = a_{\downarrow D} = 1.2 \, a \,.
\label{aAD:fermion}
\end{eqnarray}
The 3-body recombination rate constant has been calculated 
by Petrov \cite{Petrov-03}.  The rate of decrease 
in the number density $n_\uparrow$ of low-energy atoms 
with spin $\uparrow$ from the 3-body recombination processes
$\uparrow \uparrow \downarrow \longrightarrow \uparrow D$ 
and $\uparrow \downarrow \downarrow \longrightarrow \downarrow D$ 
has the form
\begin{eqnarray}
{d \ \over dt} n_\uparrow  =   
- 2 \alpha \langle \epsilon_\uparrow \rangle n_\uparrow^2 n_\downarrow 
- \alpha \langle \epsilon_\downarrow \rangle n_\uparrow n_\downarrow^2 \,,
\label{3br:fermion}
\end{eqnarray}
where $\langle \epsilon_\uparrow \rangle$ and
$\langle \epsilon_\downarrow \rangle$ are the average kinetic energies 
of the atoms in the spin states $\uparrow$ and $\downarrow$, respectively. 
The 3-body recombination event rate constant $\alpha$ defined by 
Eq.~(\ref{3br:fermion}) is \cite{Petrov-03}
\begin{eqnarray}
\alpha = 148 \, a^6/\hbar \,.
\label{alpha:fermion}
\end{eqnarray}

If the fermions with spins $\uparrow$ and $\downarrow$ also form deep 
diatomic molecules, low energy atoms and dimers can be lost from a system 
through atom-dimer collisions via dimer relaxation.  In the limit $a \gg 
\ell$, where $\ell$ is the natural low-energy length scale, the rate 
constant for this process scales like a power of $a$:
\begin{eqnarray}
\beta_{AD} = B \, ( a / \ell)^{-2-2\nu} \hbar a/m \,.
\label{betaAD}
\end{eqnarray}
The coefficient $B$ depends on the details at short distances,
but the exponent of $a$ is universal.
The relaxation process requires all three atoms to approach within 
a distance of order $\ell$.  Since two of these three atoms are identical
fermions, we might expect the prefactor of $\hbar a/m$ in 
Eq.~(\ref{betaAD}) to be suppressed by $(\ell/a)^2$.  However, it 
actually scales like $a^{-2-2\nu}$, 
where $\nu = 1.166$ is an anomalous dimension \cite{PSS03,PSS04}.  
This anomalous scaling behavior 
dramatically suppresses the relaxation rate when $a$ is large.

Some universal results have been calculated explicitly 
for the case of two identical fermions
with mass $m_1=m_2$ and a third particle with mass $m_3$. 
As discussed below, the Efimov effect occurs in this system only 
if the mass ratio $m_1/m_3$ exceeds a critical 
value~\cite{Efimov72,Efimov73}. 
If $m_1/m_3 < 13.6$,
there is no Efimov effect and universal results for 3-body 
observables depend only on $a$ and the masses.
We label the identical fermions $A_1$ and the third particle $A_3$. 
The large scattering length is $a_{13}=a$. 
If $a>0$, the particles $A_1$ and $A_3$
form a shallow dimer labelled $D$ with binding energy 
$\hbar^2/(2 m_{13} a^2)$.
The atom-dimer scattering length $a_{A_1 D}$ has been calculated 
as a function of $m_1/m_3$ by Petrov~\cite{Petrov-03}.
The ratio $a_{A_1 D}/a$ is a monotonically increasing function of
$m_1/m_3$.  
For $m_1 \ll m_3$, the ratio seems to approach 1.
For equal masses $m_1 = m_3$, its value is 1.2, 
in agreement with Eq.~(\ref{aAD:fermion}).
It increases to about 2.3 for $m_1/m_3 = 13.6$.
In Ref.~\cite{Petrov-03}, the result for $a_{A_1 D}/a$
was given as a function of the mass ratio
$m_1/m_3$, even for $m_1/m_3 > 13.6$.
When the mass ratio exceeds this critical value for the Efimov effect,
one would expect $a_{A_1 D}$ to also depend on a 3-body parameter.

Petrov has also calculated the 3-body recombination rate constants 
associated with the process $A_1 A_1 A_3 \to A_1 D$
for mass ratios in the range $0 < m_1/m_3 < 13.6$~\cite{Petrov-03}.
The equation analogous to Eq.~(\ref{3br:fermion})
for the time-derivative of the number density 
$n_1$ of the atoms $A_1$ includes the term 
$-2 \alpha \langle \epsilon_1 \rangle n_1^2 n_3$,
where $\langle \epsilon_1 \rangle$ is the average kinetic energy of the
atoms $A_1$.  The event rate constant $\alpha$ vanishes at the endpoints
of the range $0 < m_1/m_3 < 13.6$ and also at the intermediate value 
$m_1/m_3=8.62$. These zeros are the results of interference effects 
and are analogous to the zeros 
in the 3-body recombination rate constant for
identical bosons given in Eq.~(\ref{alpha-analytic}).
The dimensionless ratio $\alpha_1 \hbar/a^6$ has local maxima 
of about 60 near $m_1/m_3= 4$ and about 3 near $m_1/m_3 =12$.
If $m_1/m_3=1$, its value is 148, as given in Eq.~(\ref{alpha:fermion}).
For $m_1/m_3>13.6$, $\alpha$ presumably depends also on a
3-body parameter.

We proceed to consider the conditions for the Efimov effect 
in the system consisting of two identical fermions of mass $m_1$ 
and a third atom of mass $m_3$.
The Efimov effect occurs if there is a 
hyperangular channel eigenvalue $\lambda(R)$ that in the scaling limit 
is negative as $R \to 0$.  The matching equation for the case in 
which particles 1 and 2 are identical fermions and there is no 
subsystem orbital angular momentum was deduced in Section~\ref{sec:mass}.
The matching equation at $R=0$ is that the first factor
on the left side of Eq.~(\ref{eq:a2dis}) must vanish.  
This equation has no negative solutions for $\lambda(0)$, 
so no Efimov effect arises from this angular momentum channel.  
However, if $m_1/m_3$ is sufficiently large, 
there is a lower eigenvalue in a
channel with one unit of angular momentum
in the subsystem consisting of a pair and a third atom.
This corresponds to the 
$l_x = 0$, $l_y=1$ term in the angular momentum
decomposition of the Faddeev wave function in Eq.~(\ref{psi-Ylm}).
The matching equation for the hyperangular eigenvalue for this 
component of the wave function is \cite{NFJG01}
\begin{eqnarray}
\sin \left ( \lambda^{1/2} \mbox {$\pi \over 2$} \right)
&-& {1 \over 3} \lambda^{1/2} \, \cos (\gamma_{12})\:
{}_2F_1 \big( \mbox{$1 \over 2$}(3+\lambda^{1/2}),
                \mbox{$1 \over 2$}(3-\lambda^{1/2}),
                \mbox{$5 \over 2$}; \cos^2 \gamma_{12} \big) 
\nonumber 
\\ 
& = & - \sqrt{2} {\lambda^{1/2} \over \lambda - 1} 
\cos \left ( \lambda^{1/2} \mbox {$\pi \over 2$} \right )
{R \over a} \,,
\label{cheigen1f}
\end{eqnarray}
where $\gamma_{12}$ is given by Eq.~(\ref{gamma-ij}).
The lowest eigenvalue has a negative value $\lambda(0) = - s_0^2$
at $R=0$ if $m_1/m_3$ exceeds the critical value 13.607.
The discrete scaling factor is shown as a function of $m_1/m_3$ 
in Fig.~\ref{fig:dsf1}. As $m_1/m_3$ increases from the critical 
value to $\infty$, $e^{\pi/s_0}$ decreases monotonically from 
$\infty$ to 1.


\subsection{Particles with a spin symmetry}

We now turn to the case of particles with 
distinct states that are related by a symmetry. 
We will refer to these states as spin states, although they could
equally well be states associated with some internal
symmetry such as isospin. A general
treatment of this case was given by Bulgac and Efimov in 
Ref.~\cite{BE75}. It is more complex than the previously
considered cases for several reasons:
\begin{enumerate}
\item
There can be more than one 
spin configuration leading to a given total spin
of the three-particle system under consideration.
\item
The level spectrum does not in general show the same simple
regularities as in the spinless case. A typical spectrum looks like a 
superposition of several, strongly interacting  spectra for the 
spinless case.
\item
The magnitude of the attraction between the particles depends
on a number of factors:
the particle masses and spins, 
the total spin of the state considered,
the number of channels with large scattering lengths and their spins, 
and the strength of the coupling between the spin configurations of 
different particle pairs.
\end{enumerate}
In the following, we will illustrate some of these new features in
more detail. For a more complete treatment, we refer the reader to
Ref.~\cite{BE75}. 

For a state of total spin $\sigma$, 
the Schr\"odinger wave function can be decomposed in the form
\begin{eqnarray}
\Psi_\sigma ({\bm r}_1, {\bm r}_2, {\bm r}_3) &=& 
\sum_{\sigma_{23}} \psi^{(1)}_{\sigma_{23}} (R, \alpha_1) 
\chi_{\sigma,\sigma_{23}}
+\sum_{\sigma_{31}} \psi^{(2)}_{\sigma_{31}}(R, \alpha_2) 
\chi_{\sigma,\sigma_{31}} 
\nonumber\\&&\qquad
+ \sum_{\sigma_{12}} \psi^{(3)}_{\sigma_{12}} (R, \alpha_3)
\chi_{\sigma,\sigma_{12}} \,,
\label{Psi123fer}
\end{eqnarray}
%
which is a generalization of
the decomposition in Eq.~(\ref{Psi123}). In each of the terms,
the sum is over the spin quantum number $\sigma_{ij}$
of the pair $ij$. Only channels with
large scattering length will contribute to Eq.~(\ref{Psi123fer}).
The symbol $\chi_{\sigma,\sigma_{ij}}$ denotes the spin function for
the total spin $\sigma$ and the spin $\sigma_{ij}$ for the pair $ij$. 
The number of terms in the sum is equal to the number of ways 
the total spin $\sigma$
can be obtained by first coupling the spins of the pair $ij$ to the spin
$\sigma_{ij}$ and then coupling $\sigma_{ij}$ and the spin of the third
particle $\sigma_k$ to the total spin $\sigma$.
We assume that the orbital angular momenta are all zero.

Using Eq.~(\ref{Psi123fer}), one can obtain a matching equation 
analogous to Eq.~(\ref{match-122}).
A new feature of the matching equation is that there can be several negative
hyperangular eigenvalues $\lambda$. This is related to the fact
that, in general, there is no unique way of obtaining a given total 
spin of the three particles from coupling the single particle spins.
These different spin configurations are mixed by the interaction, and as 
a result some configurations may lead to attraction while others 
lead to repulsion. As a consequence, the number of negative eigenvalues can 
not be larger than the number of ways the total spin can be obtained by
coupling the individual spins of the particles. 

An example that will be discussed in the next subsection is 
the triton channel in the case of three nucleons.
There are two ways to obtain the total isospin ${1 \over 2}$ 
and spin ${1 \over 2}$ of the triton from nucleons which also 
have isospin ${1 \over 2}$ and spin ${1 \over 2}$.
The consistency condition at $R=0$ becomes
\begin{eqnarray}
&&\left[ \cos \left( \lambda^{1/2} \mbox{$\pi \over 2$} \right)
+ {4 \over \sqrt 3} \lambda^{-1/2} \sin \left( \lambda^{1/2} 
\mbox{$\pi \over 6$} \right)
\right] 
\nonumber \\ && \qquad
\times
\left[ \cos \left( \lambda^{1/2} \mbox{$\pi \over 2$} \right)
- {8 \over \sqrt 3} \lambda^{-1/2} \sin \left( \lambda^{1/2} 
\mbox{$\pi \over 6$} \right)
\right] = 0 \,.
\label{cheigen_trit}
\end{eqnarray}
%
Only the vanishing of the second factor on the right side
can lead to a negative solution for the eigenvalue $\lambda$.
This condition is identical to the matching equation
for three identical bosons in Eq.~(\ref{cheigen}).  
It has the negative solution $\lambda = -s_0^2$,
with $s_0 \approx 1.00624$.  Thus there is an Efimov effect
with discrete scaling factor $e^{\pi / s_0} \approx 22.7$.
This is a consequence of the Pauli principle which relates the 
symmetry in the spin-isospin and coordinate-space wave 
functions of the triton.
A negative eigenvalue corresponding to attraction is
possible only in the channel yielding a totally symmetric 
coordinate-space wave function and a totally antisymmetric 
spin-isospin wave function. As a consequence, the triton system 
has the same discrete scaling factor as identical bosons.

If there is more than one negative eigenvalue, 
the Efimov spectrum ceases to be geometric in the resonant limit in 
which all large scattering lengths are tuned to $\pm \infty$.
For $n$ negative eigenvalues, one can think of this as arising
from superimposing $n$ independent geometric spectra. 
If one allows mixing from the coupling between different spin
configurations, the levels interact. Close levels repel each other,
but the total number of levels does not change. 
For explicit examples of configurations with more than one negative 
eigenvalue, we refer the reader to Ref.~\cite{BE75}.


\subsection{Dimensions other than 3}

The hyperspherical expansion can be generalized to a continuous number of
spacial dimensions $d$.
A completely general discussion of the dependence of the Efimov effect
on $d$ is given in Ref.~\cite{NFJG01}.  We will simplify the discussion by
considering only the case of identical bosons.
The hyperangular eigenvalue equation
has a more complicated form than the equation for $d=3$ 
in Eq.~(\ref{Faddeev-angular}). One can choose conventions for the 
eigenvalues $\lambda_n(R)$ so that 
the hyperradial equation in the adiabatic hyperspherical approximation 
has the same form as in 3 dimensions:%
\footnote{Our channel eigenvalue $\lambda_n$ corresponds to 
$\lambda_n +(d-1)^2$ in the notation of Ref.~\cite{NFJG01}.}
\begin{eqnarray}
{\hbar^2 \over 2m}
\left( - {\partial^2  \over \partial R^2} + {\lambda_n(R) -
{1\over4} \over R^2} \right) f_n (R) \approx E f_n (R) \,.
\label{aha:d}
\end{eqnarray}
In the neighborhood of $d=3$, the consistency equation that determines
the eigenvalue $\lambda_n(R)$ reduces in the limit $R \to 0$ to
\begin{eqnarray}
&&\cos \left( \lambda^{1/2} \mbox{$\pi \over 2$} \right)
\nonumber
\\
&&+ 2 \, \sin \left( d \mbox{$\pi \over 2$} \right)
\:{}_2F_1 \big( \mbox{$1\over2$}(d-1+\lambda^{1/2}),
                 \mbox{$1\over2$}(d-1-\lambda^{1/2});
                 \mbox{$1\over2$}d;\mbox{$1\over4$} \big)
= 0 \,.
\end{eqnarray}
In the scale-invariant region
where the energy $E$ can be ignored
and $\lambda_n(R)$ can be approximated by $\lambda_n(0)$,
the radial equation in Eq.~(\ref{aha:d})
has the power-law solutions $f_n (R) = R^p$, where the power $p$ satisfies
\begin{eqnarray}
p (p-1) = \lambda_n (0) - \mbox{$1\over4$} \,.
\end{eqnarray}
The Efimov effect occurs if $p$ is imaginary,
which requires $\lambda_n(0) < 0$.
The Efimov effect occurs only for a narrow range of dimensions: 
\begin{eqnarray}
2.30 < d < 3.76 \,.
\end{eqnarray}
The only integer dimension for which the Efimov effect occurs is $d = 3$.
In particular, the Efimov cannot occur in $d=2$ dimensions \cite{NJF99}.  

Since there is no Efimov effect in 2 dimensions,
the universal predictions are completely determined by the
masses and scattering lengths of the particles.  As an illustration, we
describe the spectrum of shallow bound states for the case of identical bosons
with mass $m$ and large scattering length $a$.  
We denote the binding energies of $N$-body bound states by $E_N$.
Shallow few-body bound states exist only if $a > 0$. 
There are various conventions 
for the effective-range expansion in 2 dimensions.
We follow the conventions of Ref.~\cite{Verhaar84}
in which the scattering length $a$ and the effective range $r_s$
are defined by
\beq
\mbox{$1\over2$} \pi \cot\delta_0(k)=\gamma+\ln\left( 
      \mbox{$1\over2$}ka \right)
   + \mbox{$1\over4$} r_s^2 k^2 + {\mathcal O}(k^4) \,,
\eeq
where $\gamma \simeq 0.577216$ is Euler's constant.
The binding energy of the shallow dimer 
in the scaling limit is given by
\begin{eqnarray}
E_2 = 4 e^{-2 \gamma} {\hbar^2 \over ma^2} \,.
\end{eqnarray}
The leading correction is second order in $r_s/a$.
In the scaling limit, there are two shallow 3-body bound states
with binding energies \cite{BruTjo79,NJF99,Hammer:2004as}
\begin{subequations}
\begin{eqnarray}
E_3^{(1)} & = & 1.2704091(1) \, E_2 \,,
\\
E_3^{(0)} & = & 16.522688(1) \, E_2 \,.
\label{eq:3bdy2D}
\end{eqnarray}
\end{subequations}
The 4-body system in two dimensions also has exactly two
bound states with binding energies \cite{Platter:2004ns}
\begin{subequations}
\begin{eqnarray}
E_4^{(1)} & = & 25.5(1) \, E_2 \,,
\\
E_4^{(0)} & = & 197.3(1) \, E_2 \,.
\label{eq:4bdy2D}
\end{eqnarray}
\end{subequations}

In the case of weakly interacting bosons in two dimensions, 
one can derive a number of interesting properties of shallow 
$N$-body bound states in the scaling limit \cite{Hammer:2004as}. 
For each $N$, we will refer to the deepest of the shallow bound states 
as the $N$-boson droplet.  The properties of $N$-boson droplets 
with $N$ large compared to 1 are universal and quite remarkable. 
The size $R_N$ of the $N$-boson droplet satisfies
\begin{equation}
  \label{RNratio}
  \frac{R_{N+1}}{R_N} \approx 0.3417,\qquad N\gg 1 \,.
\end{equation}
The size decreases geometrically with $N$: adding an
additional boson into an existing $N$-boson droplet reduces the size of the
droplet by almost a factor of three.  Correspondingly, the binding energy
$E_N$ of the $N$-boson droplet increases geometrically with $N$:
\begin{equation}
  \label{BNratio}
  \frac{E_{N+1}}{E_N} \approx 8.567, \qquad N\gg 1 \,.
\end{equation}
Thus the separation energy for one particle 
is approximately 88\% of the
total binding energy.  This is in contrast to most other physical
systems, where the ratio of the single-particle separation energy 
to the total binding energy decreases to zero as the number of particles 
increases. The numbers $E_3^{(0)}/E_2=16.5$ and 
$E_4^{(0)}/E_3^{(0)}=11.9$ obtained from the exact 3-body and
4-body results in Eqs.~(\ref{eq:3bdy2D}) and (\ref{eq:4bdy2D}) appear to
be converging toward the universal prediction for large $N$
in Eq.~(\ref{BNratio}).

The origin of the peculiar behavior of $N$-boson droplets for large $N$ 
lies in the asymptotic freedom of bosons 
in two dimensions with a zero-range potential.
As the separation $r$ of two bosons decreases, their interaction 
strength decreases asymptotically to zero as $1/\ln(r)$.
For a potential with a small but nonzero effective range $r_s$, 
Eqs.~(\ref{RNratio}) and (\ref{BNratio}) 
are valid for $N$ large compared to 1 but small compared to
a critical value given by
\begin{equation}
N_{\rm crit} \approx 0.931 \ln\frac{R_2}{r_s}
  + {\mathcal O}(N^0) \,,
\end{equation}
where $R_2$ is the size of the dimer.  For $N \sim N_{\rm crit}$,
the size $R_N$ of the droplet is comparable to the range
of the potential and universality is lost.  If the ratio of $R_2$ to $r_s$
is exponentially large,  then $N_{\rm crit}$ is much larger than one.


\subsection{Few-nucleon problem}

The concept of universality in few-body systems with large scattering length
was originally developed in nuclear physics.
It is therefore worthwhile to describe the nuclear physics 
context in some detail.

Since the nuclear force is short range, 
nucleon-nucleon interactions at low energy should be dominated by S-waves.
As discussed in Section~\ref{PartNuclargeA},
isospin symmetry implies that there are two independent 
S-wave nucleon-nucleon scattering channels.
The scattering lengths $a_s$ and $a_t$ for these two channels are
both relatively large compared to the
natural low-energy length scale $\ell=\hbar/m_\pi c$.
The effective-range expansion for S-wave phase shifts in Eq.~(\ref{kcot})
is an expansion in powers of the energy.  {\it Effective-range theory},
which goes back to Schwinger, Blatt, and Bethe  
\cite{Schwinger47,Blatt48,Bethe49}, is defined by the truncation 
of this expansion after the effective-range term as in  
Eq.~(\ref{kcotdelta:rs}).  Effective-range theory gives a remarkably good 
description of the 2-nucleon system.  It reproduces the S-wave $np$  
phase shifts to better than 5\% out to a momentum of about 150 MeV/$c$
in the center-of-mass frame.  
It reproduces the binding energy of the deuteron to an accuracy of 0.2\%.

The {\it zero-range model} is defined by the more severe truncation
of the effective-range expansion after the scattering length term,
as in Eq.~(\ref{kcot-scaling}).  It is equivalent to approximating 
the nuclear forces by zero-range potentials whose depths are tuned 
to reproduce the 2-body scattering lengths $a_s$ and $a_t$.
Considering its simplicity,
the zero-range model gives a surprisingly good description of the 
2-nucleon system.  It reproduces the S-wave $np$  
phase shifts to better than 20\% out to center-of-mass momenta 
of about 40 MeV. It gives the binding energy of the deuteron 
with an error of 37\%.  This modest success of the zero-range model
motivates an approach in which the effective range and other coefficients 
in the low-energy expansion of the phase shifts are treated 
as perturbations. 
If the first-order corrections in the effective ranges are included in the 
S-wave phase shifts, they are accurate to better than 6\% out 
to kinetic energies as high as 70 MeV.
If the first-order corrections in the effective ranges are included
in the cotangents of the S-wave phase shifts, it is equivalent to 
effective-range theory, which is much more accurate, as described above. 
The error in the deuteron binding energy decreases to 16\%
and then to 8\% upon including the first-order and 
second-order corrections in Eq.~(\ref{B2-exp}). 
Thus this approach works reasonably well in the 2-nucleon system.

Pioneering work in applying the zero-range model to the 3-nucleon 
system was carried out by Skorniakov and Ter-Martirosian in 1957
\cite{STM57} and by Danilov and Lebedev in 1963 \cite{DaL63}.
In 1981, Efimov proposed a new approach to the low-energy 
few-nucleon problem in nuclear physics that was based on 
perturbation theory around the zero-range model
\cite{Efimov81}.
Efimov described it as a ``qualitative approach,'' 
but it can more accurately be described as ``semi-quantitative,'' 
because its goal was the understanding 
of few-body observables at about the 10\% level. 
The traditional nuclear physics community, on the other hand, 
was accustomed to reproducing the observed binding energies
of the light nuclei to several digits of accuracy
using nuclear forces described by phenomenological potentials with a
large number of adjustable parameters.

Remarkably, Efimov's program also works reasonably
well in the 3-nucleon system at momenta small compared to $m_\pi$. 
The only 3-nucleon bound states are the {\it triton} and the $^3$He nucleus,
which are $pnn$ and $ppn$ bound states, respectively.
These nuclei have no excited states.
The $ppn$ system is complicated by the Coulomb interaction between the two
protons, so we focus on the $pnn$ system.
The Efimov effect makes it necessary to impose a boundary condition 
on the wave function at short distances \cite{Efimov81}.  
The boundary condition can be fixed by using either 
the spin-doublet neutron-deuteron scattering length 
or the triton binding energy as input.  
If the deuteron binding energy is used as the 2-body input
and if the boundary condition is fixed by using the spin-doublet 
neutron-deuteron scattering length as input,
the triton binding energy is predicted with an accuracy of 6\%.
The accuracy of the predictions
can be further improved by taking into account the effective range 
as a first-order perturbation \cite{Efimov91}.
Thus the triton can be identified
as an Efimov state associated with the deuteron being a $pn$ bound state
with large scattering length \cite{Efimov81}.

Efimov's program was implemented within an effective field theory 
framework by Bedaque, Hammer, and van Kolck \cite{BK98,BHK98,BHK00}.
In Ref.~\cite{BHK00}, they found that the renormalization of the 
effective field theory requires a $SU(4)$-symmetric 3-body
interaction with an ultraviolet limit cycle. $SU(4)$-symmetry
was introduced by Wigner in 1937 as generalization of the 
$SU(2)\times SU(2)$ spin-isospin symmetry, allowing for a mixing of 
spin and isopin degrees of freedom in symmetry transformations
\cite{Wigner37}. It is satisfied to a high degree in the energy 
spectra of atomic nuclei. Exact Wigner symmetry requires the 
2-body scattering lengths $a_s$ and $a_t$ to be equal. However,
if the two body scattering lengths are large, it is a very good
approximation even if they are different
since the symmetry-breaking terms are proportional
to the inverse scattering lengths.\footnote{See 
Ref.~\cite{Mehen:1999qs} for a discussion of
Wigner's $SU(4)$ symmetry in the two-nucleon system.}
The 3-body force introduced by
Bedaque, Hammer, and van Kolck depends on a parameter $\Lambda_*$ 
that is determined through a renormalization condition
that plays the same role as Efimov's boundary condition.
In addition to the triton,
the effective field theory of Ref.~\cite{BHK00} also
predicts infinitely many deeper bound states in the triton channel,
with the ratio of the binding energies 
of successive states approaching the universal constant 
$\lambda_0^2 \approx 515$, but they are artifacts of the scaling limit.
The effective field theory is ideally suited to calculating 
corrections to the universal results in the scaling limit. 
The leading corrections come from the effective
range and are discussed in subsection \ref{subsec:er}. 
The effective field theory also allows for a straightforward
treatment of electroweak interactions of the nucleons 
(see Refs.~\cite{Beane:2000fx,Bedaque:2002mn} and references therein).

\begin{figure}[htb]
\bigskip
\centerline{\includegraphics*[width=8.5cm,angle=0]{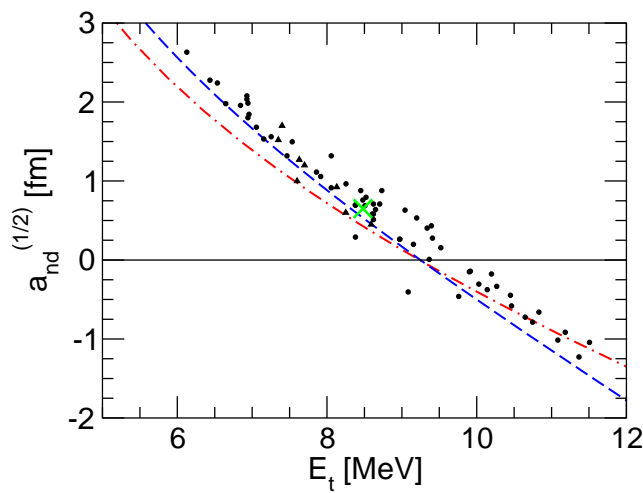}}
\medskip
\caption{
The Phillips line at leading order (dash-dotted line) and next-to-leading
order (dashed line) from the EFT calculation of Ref.~\cite{BGHR03}. 
The dots correspond to the
predictions for the triton binding energy and doublet scattering length in
different models with the same 2-body scattering lengths and effective
ranges as inputs. The cross is the experimental result.
(Figure taken from Ref.~\cite{BGHR03}.)
}
\label{fig:nucphil}
\end{figure}

A peculiar universal feature of the three-nucleon system is the Phillips 
line \cite{Phillips68}. If the predictions of different nucleon-nucleon
potentials for the triton binding energy $E_t$ and the spin-doublet
neutron-deuteron scattering length  $a_{nd}^{(1/2)}$ are plotted
against each other, they fall close to a straight line.\footnote{A similar 
correlation exists between the binding energy of the $^3$He nucleus and the 
spin-doublet proton-deuteron scattering length.}
This correlation between $E_t$ and $a_{nd}^{(1/2)}$ is called the 
Phillips line and cannot be understood in conventional potential
models. The Phillips line is shown in Fig.~\ref{fig:nucphil}. 
The dots correspond to the potential model predictions 
while the cross marks the experimental value. 

The Phillips line can easily be understood within Efimov's
program \cite{EfT85,EfT88a,EfT88b} and its effective field theory 
implementation \cite{BHK00}. 
If corrections to the {\it scaling limit} 
are neglected, all low-energy
3-body observables depend only on the singlet and triplet
S-wave scattering lengths $a_s$ and $a_t$ and the 3-body parameter
$\Lambda_*$. Since the nucleon-nucleon potentials reproduce the 
nucleon-nucleon phase shifts, they all have the same scattering
lengths. The short distance part of the potentials which is encoded
in the 3-body parameter $\Lambda_*$, however, is
not constrained by the phase shifts and in general is different for 
each potential. The different potential model calculations must therefore
fall close to a line that can be parametrized by the 
parameter $\Lambda_*$. 
Due to the uncertainty from higher order corrections in the 
expansion in $\ell/|a|$, the Phillips line becomes a band of finite
width. For an estimate of the error band for the Phillips line, 
see Ref.~\cite{Griesshammer:2004pe}. 
In Fig.~\ref{fig:nucphil}, we show the Phillips line from 
the effective field theory calculation of Ref.~\cite{BGHR03}
at leading order and including the first-order effective-range correction. 
The leading-order curve already agrees reasonably well with the 
predictions from potential models. Including the linear
effective-range correction improves the agreement with potential models 
and also moves the curve closer to the experimental value.
The Phillips line for the proton-deuteron system where the Coulomb 
interaction is present was discussed in Refs.~\cite{TDA87,AdhiDa88}. 
A similar correlation between the triton charge radius and binding energy 
has been traced back to the variation of the 
parameter $\Lambda_*$ as well \cite{Platter:2005sj}.

The success of Efimov's program for the few-nucleon problem 
can be explained by the fact that QCD is close to the critical 
trajectory for an infrared limit cycle in the 
3-nucleon sector \cite{BH03}.  The parameters of QCD include 
the masses $m_u$ and $m_d$ of the up and down quarks.
The inverse scattering lengths $1/a_s$ and $1/a_t$ for the 2-nucleon system
depend strongly on $m_u + m_d$, with the dependence on 
the mass difference $m_u-m_d$ entering only at second order.
Effective field theories with nucleon and pion fields
have been used to extrapolate $a_s$ and $a_t$ in the variable $m_u+m_d$
\cite{Beane:2001bc,Beane:2002vq,Beane:2002xf,Epelbaum:2002gb}.
The physical observable most sensitive to $m_u + m_d$ is the 
pion mass, whose physical value is $m_\pi = 138$ MeV.
The extrapolation in $m_u+m_d$ can therefore be interpreted as an
extrapolation in $m_\pi$.  These extrapolations in $m_u+m_d$ suggest that 
$1/a_s$ and $1/a_t$ vanish at points $m_{\pi,s}$ and $m_{\pi,t}$ 
that are not too much larger than the physical pion mass.
The quark mass difference $m_u-m_d$ provides an additional tuning parameter
that could be used to make $1/a_s$ and $1/a_t$ vanish at the same
point: $m_{\pi,s} = m_{\pi,t}$.
This is the critical point for an infrared limit cycle.
At this critical point, 
the binding energy of the deuteron is exactly zero and
there is also a bound state at threshold in the spin-singlet iso-triplet 
channel.  The triton has infinitely many excited states
with an accumulation point at the 3-nucleon threshold.  
The ratio of the binding
energies of successive bound states rapidly approaches 
the universal constant $\lambda_0^2 \approx 515.03$.
Thus tuning the quark masses $m_u$ and $m_d$ to the critical point
puts QCD on a critical trajectory for an infrared limit cycle.

The natural formulation of Efimov's program for the nuclear few-body 
problem is in terms of an effective field theory in which nucleons 
interact through contact interactions. 
There are other nuclear physics applications in which higher partial 
waves play a more important role.These systems can also be described 
using effective field theory \cite{Griesshammer:2005ga,Birse:2005pm}. 
Due to the absence of the Efimov effect, 3-body
forces are suppressed in these systems.
For calculations of neutron-deuteron scattering phase shifts in higher
partial waves using effective field theory, 
see Refs.~\cite{Griesshammer:2004pe,Gabbiani:1999yv}. 
The nuclear few-body problem has also been studied using
effective field theories with explicit pion fields. 
For the current status of these calculations,
see Refs.~\cite{Beane:2000fx,Bedaque:2002mn,Meissner04,Phillips:2002da} 
and references therein.


\subsection{Halo nuclei}

\begin{figure}[htb]
\bigskip
\centerline{\includegraphics*[width=4cm,angle=0]{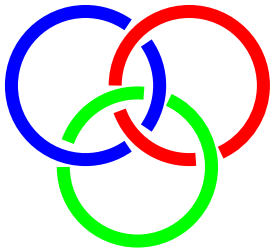}}
\medskip
\caption{Borromean rings.}
\label{fig:borrorings}
\end{figure}

A special class of nuclear systems exhibiting universal behavior
are {\it halo nuclei} \cite{Riisa94,Bertsch95,ZDFJV93,JRFG04}.
A halo nucleus is one that consists of a tightly bound core 
surrounded by one or more loosely bound valence nucleons.
The valence nucleons are characterized by a very low
separation energy compared to those in the core. 
As a consequence, the radius of the halo nucleus is large 
compared to the radius of the core.  A trivial example is the deuteron, 
which can be considered a 2-body halo nucleus. The RMS radius of
the deuteron ${\langle r^2 \rangle}^{1/2}\approx 3$ fm is about  
four times larger than the size of the constituent nucleons. 
Halo nuclei with two valence nucleons are particularly interesting 
examples of 3-body systems.  If none of the 
2-body subsystems are bound, they are called {\it Borromean} halo nuclei. 
This name is derived from the heraldic symbol of the Borromeo family 
of Italy, which consists of three rings interlocked in such way that 
if any one of the rings is removed the other two separate.
Figure~\ref{fig:borrorings} shows an illustration of the Borromean
rings.

The separation of scales in halo nuclei leads to universal properties
that are insensitive to the structure of the core (see, e.g., 
Refs.~\cite{BeE91,EBH97,AFT97,DFHT00}). The most
carefully studied Borromean halo nuclei are $^6$He and $^{11}$Li, 
which have two weakly bound valence neutrons \cite{ZDFJV93,JRFG04}. 
In the case of $^6$He, the core is a $^4$He nucleus, 
which is also known as an $\alpha$ particle.
The two-neutron separation energy for $^6$He is about 1 MeV,
small compared to the binding energy of the $\alpha$ particle
which is about 28 MeV.   The $n \alpha$ system has no bound states, 
because the $^5$He nucleus is unstable.
The $^6$He nucleus is therefore Borromean.
There is, however, a strong P-wave 
resonance in the $J=3/2$ channel of $n \alpha$ scattering,
and this leads to the binding of $^6$He.  Thus $^6$He
can be interpreted as a bound state of an $\alpha$-particle
and two neutrons, both of which are in 
$P_{3/2}$ configurations.\footnote{For an 
effective field theory treatment of  $n\alpha$-scattering, see 
Refs.~\cite{BHK02,BHK03}.}
The structure of $^{11}$Li is somewhat more complicated because a 
larger number of partial waves contribute.

Halo nuclei with two valence nucleons in S-wave states are candidates for 
Efimov states.  Such a state would also be a Borromean halo nucleus
if all three pairs have negative scattering lengths and no deep 2-body 
bound states.
Among the possible candidates for Efimov states are $^{18}$C and $^{20}$C,
which both consist of a core nucleus with spin and parity quantum numbers
$J^P=0^+$ and two valence neutrons.  The nuclei $^{17}$C and $^{19}$C 
are both expected to have $\frac{1}{2}^+$ states near threshold,
implying a shallow neutron-core bound state and therefore a large 
neutron-core scattering length \cite{Federov:1994cf}.

The simplest strange halo nucleus is the hypertriton, a 3-body
bound state of a proton, neutron, and a strange particle called the $\Lambda$. 
The total binding energy is only about 2.4 MeV. 
The hypertriton is not Borromean, because the proton-neutron subsystem 
has a bound state, the deuteron. 
The separation energy for the $\Lambda$,
$E_\Lambda = 0.13 \pm 0.05$ MeV \cite{Ju73,Dav91}, is tiny
compared to the binding energy $E_d = 2.225$ MeV of the 
deuteron.  The hypertriton can therefore also be considered a 2-body 
halo nucleus.
It has been studied in both 2-body and 3-body approaches. See, e.g.
Refs.~\cite{Cobis:1996ru,Fedorov:2001wj,Congl92} and references therein.
The hypertriton has also been studied in the effective
field theory for short-range interactions \cite{Hammer:2001ng}.
The scattering lengths in the ${}^3S_1$ $NN$ channel
and in the ${}^1S_0$ $N \Lambda$ channel
were both assumed to be large.
As in the case of the triton, the renormalization requires a
3-body parameter and involves a limit cycle. 
The discrete scaling factor for this case is only 10.2, 
roughly a factor two smaller than in the triton case.


%
%

\section{Frontiers of Universality}
        \label{sec:frontiers}

In this section, we discuss some problems at the frontiers of 
universality: the $N$-body problem for $N \geq 4$,
higher-order effective-range corrections, and the case of a large
P-wave scattering length.


\subsection{The $N$-body problem for $N \geq 4$}
        \label{sec:n-body}

Amado and Greenwood wrote a paper in 1972 entitled 
``There is no Efimov effect for four or more particles'' \cite{AG72}.
They showed that, for $N \geq 4$, the tuning of the binding energy 
of an $(N-1)$-body bound state to zero cannot produce an 
infinite number of $N$-body bound states with an accumulation point 
at $E=0$.
We will refer to this result as the {\it Amado-Greenwood theorem}.
Note that the Amado-Greenwood theorem does not forbid an 
infinite number of $N$-body bound states with an accumulation point 
at some energy other than zero.

The example of four identical bosons is instructive. 
First consider a negative scattering length at the value $a_*'<0$ 
for which there is an Efimov trimer at the 3-atom threshold. 
The Amado-Greenwood theorem implies that
there cannot be an infinite sequence of tetramers 
with an accumulation point at the threshold $E=0$. 
Next consider a positive scattering length at the value $a_*$ 
for which there is an Efimov trimer at the atom-dimer threshold
and assume there are no deeper dimers or trimers.
Then there is an infinite sequence of 4-body bound states 
with an accumulation point at the atom-dimer threshold $E=-E_D$.
This is a simple consequence of the Efimov effect
and the fact that the atom-dimer 
scattering length $a_{AD}$ diverges at $a=a_*$. 
If $a$ is just a little larger than $a_*$, then $a_{AD} \gg a$ and 
an Efimov trimer is essentially a 2-body bound state 
consisting of a dimer of size $a_*$ and a third atom separated 
by a distance of order $a_{AD}$.
In the atom-atom-dimer system,
two of the three pairs have a resonant interaction with large scattering
length $a_{AD}$.  The atom-atom-dimer system consists of
two identical bosons with mass $m$ and a third particle with mass $2m$.
Thus at the critical point where $a_{AD} \to \pm \infty$,
the Efimov effect produces infinitely many tetramers 
with discrete scaling factor 2.016$\times$10$^5$. 
This value can be read off from Fig.~\ref{fig:dsf1}.

Adhikari and Fonseca used the Born-Oppenheimer approximation
to study the possiblity of an Efimov effect in the 4-body system 
consisting of three identical heavy particles and one light particle
with a large scattering length $a$ between the heavy particles and 
the light particle \cite{AF81}.
They concluded that, in the resonant limit $a \to \infty$,
there can be at most a finite number of 4-body bound states near 
the scattering threshold. 
This result is consistent with the Amado-Greenwood theorem.
 
In the 3-body problem, exact numerical solutions are facilitated by the 
Faddeev equations.
The generalization of the Faddeev equations to the $N$-body problem 
with $N \ge 4$ was given by Yakubovsky \cite{Yakubovsky:1966ue}.
He set up a system of coupled integral equations which are in unique 
correspondence to the $N$-body Schr{\"o}dinger equation and have a kernel which
gets connected after a finite number of iterations. An equivalent 
set of equations was given independently by Grassberger and Sandhas
\cite{GrSa67}. Due to the complexity of these equations, exact 
numerical solutions for $N=4$ have only recently been obtained.

In nuclear physics, exact numerical solutions of the bound
state problem for four nucleons interacting through a potential
are now standard. (See \cite{Nogga:2000uu,Epelbaum:2002ji}
and references therein for more details.) The 4-nucleon scattering 
problem is much more complicated and no exact numerical solution is
available to date. The binding energies of the ground and excited states for
nuclei up to atomic mass 
number $A=10$ have been calculated using quantum Monte Carlo
methods and the no-core shell model
\cite{Pieper:2001mp,Pieper:2002ne,Navratil:2003ef}.

In molecular physics, the only numerically exact $N$-atom calculations 
for $N \ge 4$ that we are aware of are for ground-state binding energies.
There has been a large interest in the 
calculation of the properties of clusters of helium atoms. 
The ground-state binding energies for the $N$-body clusters 
($^4$He)$_N$ up to $N=10$ have been calculated 
using the diffusion Monte Carlo method \cite{Lewe97}.  
Using an approximate numerical method that combines Monte Carlo methods
with the hyperspherical expansion,  Blume and Greene have also calculated
the binding energies of the ground state and excited states 
for ($^4$He)$_N$, as well as the scattering
lengths for elastic $^4$He + ($^4$He)$_{N-1}$ scattering,
up to $N=10$ \cite{BlGr00}.

An important issue in the 4-body system with a large 2-body scattering length
is how many parameters are required to describe the system in the scaling 
limit, that is, up to corrections that decrease like $l/|a|$ as 
$a \to \pm \infty$. In the case of identical bosons,
low-energy 4-body observables necessarily 
depend on the 2-body parameter $a$ and the 3-body parameter $\kappa_*$.
But are any additional 4-body parameters required?
There are theoretical arguments in support of both answers to this question. 
There is a renormalization argument for zero-range  2-body
potentials that indicates that an additional $N$-body parameter is required
to calculate $N$-body binding energies for all $N \ge 3$ \cite{AFG95}. 
On the other hand, a power-counting argument within the effective
field theory framework suggests that no additional parameters
should be necessary to calculate $N$-body observables for $N>3$  
\cite{Lepage-pc}. 
There is some circumstantial evidence in favor of this later possibility
from the 4-body problem in nuclear physics. There is a correlation called the 
``Tjon line'' between the binding energy $E_t$ of the triton
and the binding energy $E_\alpha$ of the $\alpha$ particle \cite{Tjo75}.
Calculations of these binding energies using modern nucleon-nucleon
interaction potentials give results that underestimate both 
binding energies but cluster along a line in the $E_t$-$E_\alpha$ plane.
By adding a 3-nucleon potential whose strength is adjusted to get
the correct value for $E_t$, one also gets an accurate result for 
$E_\alpha$. (See Ref.~\cite{Nogga:2000uu} for some recent calculations
with modern nuclear forces). 

Platter, Hammer, and Mei\ss ner have recently studied the universal
properties of the four-boson system with short-range interactions in an
effective quantum mechanics approach \cite{Platter:2004qn}.
They constructed the effective interaction potential at leading order 
in the large scattering length and computed the 4-body binding energies 
using the Yakubovsky equations. They found that cutoff independence of the 
4-body binding energies does not require the introduction of a 4-body 
force. This suggests that 2-body and 3-body interactions are sufficient 
to renormalize the 4-body system. They have applied their equations 
to $^4$He atoms and calculated the binding energies of the 
ground state and the excited state of the $^4$He tetramer. 
Using the binding energy $E_2$ of the $^4$He  dimer as the 2-body input 
and the binding energy $E_3^{(1)}$ of the excited state of the $^4$He trimer 
as the  3-body input, they found good agreement 
with the results of Blume and Greene \cite{BlGr00}. 

\begin{figure}[htb]
\centerline{\includegraphics*[width=6.8cm,angle=0]{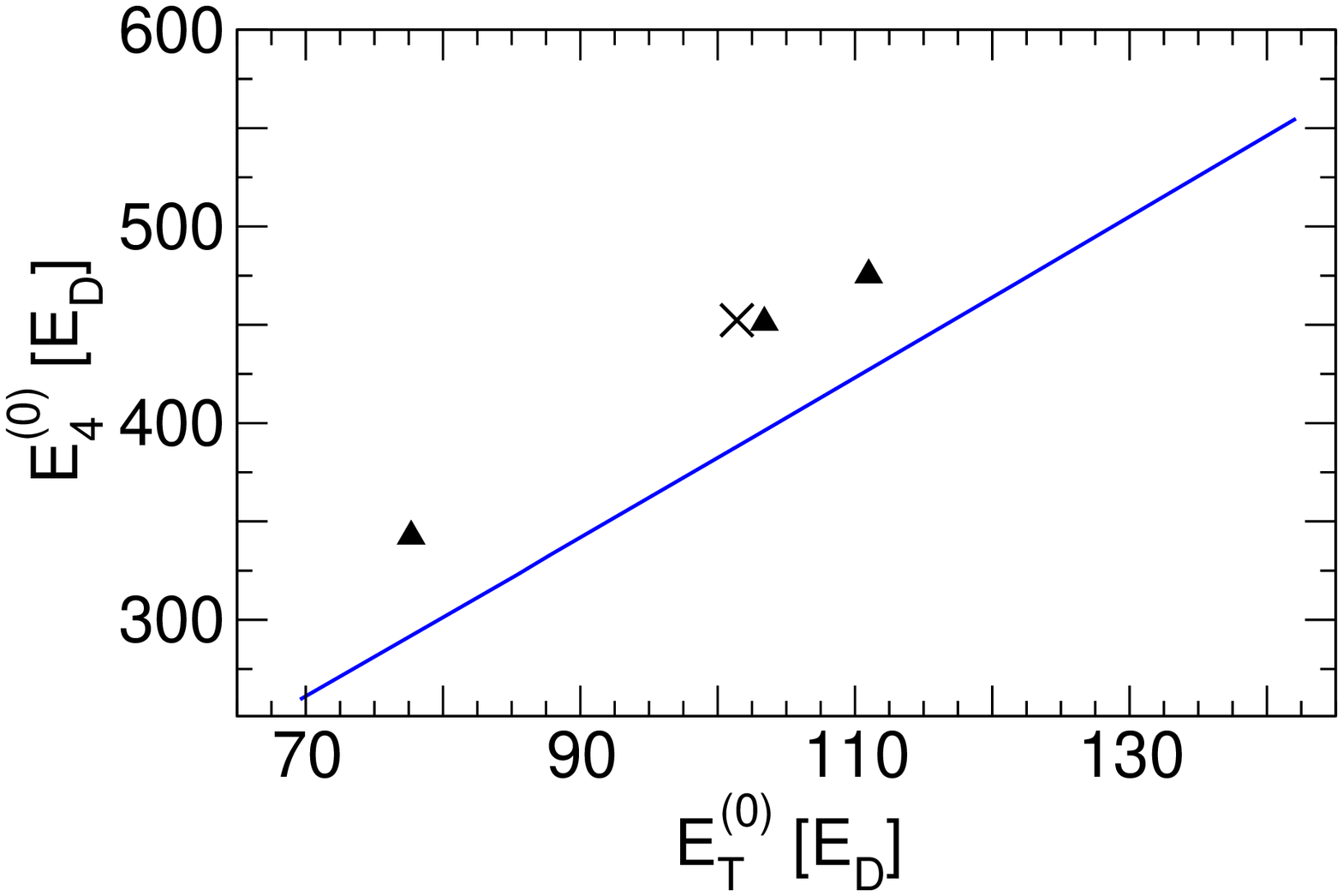}
\:\includegraphics*[width=6.8cm,angle=0]{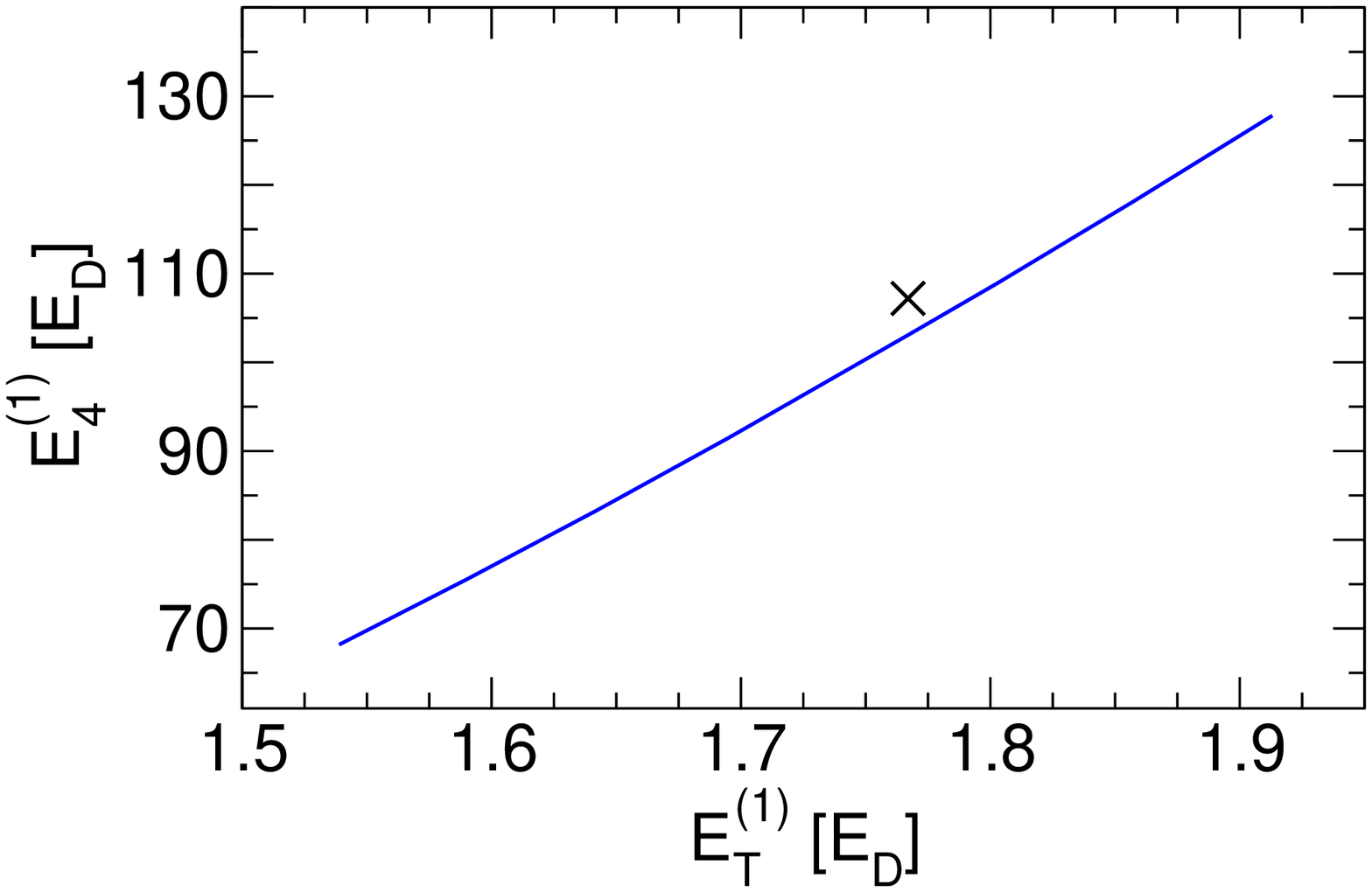}}
\caption{
Universal scaling curves for the binding energies of the 
ground states (left panel) and excited states (right panel)
of the $^4$He trimer and the $^4$He tetramer.  
The crosses are the results  for the LM2M2 potential \cite{BlGr00}.  
The triangles are the results for the TTY, HFD-B, and HFDHE2 potentials 
\cite{Lewe97,Naka83}. 
(Figure taken from Ref.~\cite{Platter:2004qn}.)
}
\label{fig:tjonline}
\end{figure}

The authors of Ref.~\cite{Platter:2004qn}
also observed a correlation between the 
binding energies of the $^4$He tetramer and the $^4$He trimer
similar to the Tjon line in nuclear physics.
We denote the binding energies of 4-body bound states by $E_4^{(n)}$.
The universal scaling curves for the binding energies of the 
ground states of ($^4$He)$_4$ and ($^4$He)$_3$ 
and for the binding energies of the excited states of ($^4$He)$_4$
and ($^4$He)$_3$ are shown in Fig.~\ref{fig:tjonline}.
The calculations of the binding energies for modern $^4$He potentials 
fall close to the universal scaling curves from the effective theory. 
The crosses are the results for the ground state and the excited state 
of the $^4$He tetramer for the LM2M2 potential \cite{BlGr00}. 
For the ground state of the tetramer, calculations with
other $^4$He potentials are available as well. The triangles are the results 
for the TTY, HFD-B, and HFDHE2 potentials  from Refs.~\cite{Lewe97,Naka83}.
The universal scaling curves are very close to linear 
over the range of binding energies relevant to $^4$He atoms.
The universal scaling curves shown in Fig.~\ref{fig:tjonline} are well
approximated by the following linear equations \cite{Platter:2004qn}:
\begin{subequations}
\begin{eqnarray}
\label{eq:tjon1}
E_4^{(0)} &\approx&  4.075 \; E_T^{(0)} - 24.752 \; E_D ,
\\
E_4^{(1)} &\approx&  159.4 \; E_T^{(1)} - 178.0 \; E_D.
\label{eq:tjon4}
\end{eqnarray}
\end{subequations}
The accuracy of Eq.~(\ref{eq:tjon1}) is better than 5\% for 
$69 \leq  E_T^{(0)}/E_D \leq 142$ and the accuracy of
Eq.~(\ref{eq:tjon4}) is better than 2\% for 
$1.52 \leq E_T^{(1)}/E_D \leq 1.92$.
These relations can be used to predict the tetramer ground
and excited state energies to leading order in $\ell/a$
for any potential for which  the dimer binding energy 
and one of the trimer binding energies are known.

The Tjon line for the correlation between the binding energies 
of the ground states of ($^4$He)$_4$ and ($^4$He)$_3$ is evident
from  the results of the four  potentials shown in 
the left panel of Fig.~\ref{fig:tjonline}. 
All four points are systematically above the universal scaling curve
by about the same amount.
If calculations of the binding energy of the excited state of  ($^4$He)$_4$ 
were available for other potentials, they would also fall on a line parallel 
to the universal scaling curve for the excited states. 
For the LM2M2 potential, the results lie above the predictions of the 
universal scaling curves by 
3.5\% for $E_4^{(1)}$ and by 12.1\% for $E_4^{(0)}$.
The leading contribution to the deviations from the universal predictions 
are expected to come from corrections that are first order 
in the effective range $r_s$.
For the shallowest  4-body bound states, the leading corrections to the 
universal predictions for $E_4^{(n)}$ are expected to be proportional  
to $E_4^{(n)} r_s/a$.  The ratio of the effective-range corrections to 
$E_4^{(0)}$ and $E_4^{(1)}$ is then predicted to be 
$E_4^{(0)}/E_4^{(1)}=3.8$.  This is close to the observed ratio
of the deviations from the universal predictions, which is 3.4.
These results provide strong support for the hypothesis that low-energy 4-body 
observables in the scaling limit are completely determined by  $a$ 
and a single 3-body parameter. 

\begin{figure}[htb]
\centerline{\includegraphics*[width=8cm,angle=0]{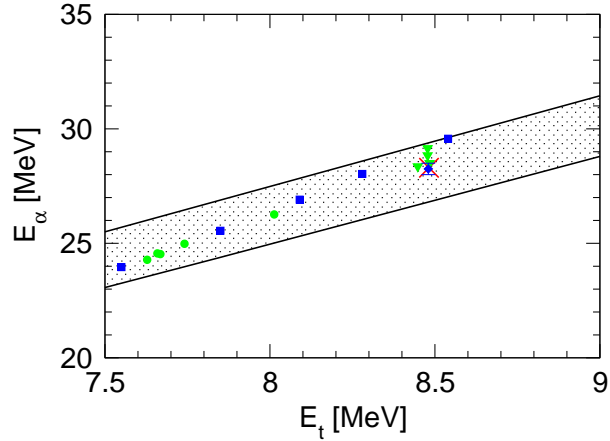}}
\caption{\label{fig:tjonalpha}
The correlation between the binding energies of the 
triton and the $\alpha$-particle. The lower (upper) line
shows the leading order result using  $a_s$ and $E_d$
($a_s$ and $a_t$) as the 2-body inputs. The data points are from 
calculations using various $NN$ potentials with and without 3-body
forces. The cross shows the experimental point.
(Figure taken from Ref.~\cite{Pl_alpha}.)
}
\end{figure}

This work was recently extended to the four-nucleon system in
Ref.~\cite{Pl_alpha}. 
While the four-nucleon system is more complicated due to spin and
isospin degrees of freedom, there is still only one 3-body
parameter that enters at leading order in $\ell/|a|$.
In Fig.~\ref{fig:tjonalpha}, we show the result for the 
nuclear Tjon line with the spin-singlet 
$np$ scattering length $a_s$
and the deuteron binding energy $E_d$ as the 2-body inputs (lower line)
and the result with the $np$  
scattering lengths $a_s$ and $a_t$ as the 2-body inputs (upper line). 
Both lines generate a band that gives a naive estimate of higher 
order corrections in $\ell/|a|$.
We also show some calculations using phenomenological 
potentials \cite{Nogga:2000uu} and a chiral EFT potential with explicit
pions \cite{Epelbaum:2002vt}. The cross shows the experimental point.
All calculations and the experimental point lie within the shaded 
area defined by the two curves.
Using the triton binding energy $E_t=8.48$ MeV as the 3-body input,
one can obtain a prediction for the $\alpha$-particle binding energy
$E_\alpha$. The result is $E_\alpha = 29.5$ MeV (26.9 MeV)
if $a_s$ and $a_t$ ($a_s$ and $E_d$) are used as 2-body input.
This variation is consistent with the expected 30\% accuracy of a leading 
order calculation in $\ell/|a|$. The results agree
with the (Coulomb corrected) experimental value 
$E_\alpha=29.0 \pm 0.1$ MeV to within 10\%.

The universal result for a 4-body observable has also been calculated 
recently for the system consisting of a pair of identical fermions in each 
of two spin states.  We will refer to the spin states as $\uparrow$ and 
$\downarrow$.  If there is a large positive scattering length $a$ between 
the fermions with spins $\uparrow$ and $\downarrow$, they can bind to form a 
shallow dimer $D$.  The dimer is a boson and its binding energy is given by 
the universal formula in Eq.~(\ref{B2-uni}).  In the low-energy limit, the 
scattering of a pair of these dimers is determined by the dimer-dimer 
scattering length $a_{DD}$.  This 4-body observable was recently calculated 
independently by two different groups \cite{PSS03,PSS04,BBF03} with the result
\begin{equation}
a_{DD} = 0.60 ~ a \,.
\end{equation}
In Ref.~\cite{PSS03,PSS04}, the authors used the zero-range approximations 
and the boundary-condition method to derive an integral equation for a 
factor $f ({\bf r}, {\bf R})$ in the Schr\"odinger wave function that 
describes configurations in which one pair of atoms with spins $\uparrow$ 
and $\downarrow$ has separation ${\bf r}$, the other has separation 
${\bf r}^\prime \to 0$, and the two pairs are separated by ${\bf R}$.  
The dimer-dimer scattering length is determined by the behavior of 
$f({\bf r},{\bf R})$ as $|{ \bf R} | \to \infty$.  In Ref.~\cite{BBF03}, 
the authors calculated $a_{DD}$ by solving the Yakubovsky equations for 
dimer-dimer scattering numerically for the system in which fermions with 
spins $\uparrow$ and $\downarrow$ interact through a short-range potential 
$V(r)$ with a large scattering length $a$.

If the fermions with spins $\uparrow$ and $\downarrow$ can also form deep 
diatomic molecules, low energy dimers can be lost from a system through 
dimer-dimer collisions via dimer relaxation.  In the limit $a \gg \ell$, 
where $\ell$ is the natural low-energy length scale, the rate constant 
$\beta_{DD}$ for this process scales like a power of $a$:
\begin{equation}
\beta_{DD} = B \, ( a / \ell)^{-2-2\nu} \hbar a/m \,.
\label{betaDD}
\end{equation}
The coefficient $B$ depends on the details at short distances,
but the exponent of $a$ is universal.
In the dominant relaxation process, both atoms of the relaxing dimer 
and one of the other atoms all approach within 
a distance of order $\ell$.  Since two of these three atoms are identical
fermions, we might expect the prefactor of $\hbar a/m$ in 
Eq.~(\ref{betaAD}) to be suppressed by $(\ell/a)^2$.  However, it 
actually scales like $a^{-2-2\nu}$, where $\nu = 0.773$ is an anomalous 
dimension \cite{PSS03,PSS04}.  This anomalous scaling behavior 
suppresses the relaxation rate when $a$ is large.


\subsection{Effective-range corrections}
\label{subsec:er}

Corrections to the scaling limit in the 3-body 
system can be calculated using effective field theory. 
The most important correction comes from the effective range $r_s$. 
To illustrate the problems involved, we review the 
calculation of the range corrections to S-wave atom-dimer scattering 
for spinless bosons. 
A generalization of the STM3 integral equation in Eq.~(\ref{BHvK})
that includes the effective-range correction to all orders reads
\begin{eqnarray}
&& {\mathcal A}_S (p, k; E) 
=  16 \pi \gamma
\left[ {1 \over 2pk} \ln \left({p^2 + pk + k^2 -E -i \epsilon \over
p^2 - pk + k^2 - E - i \epsilon}\right) + {H(\Lambda) \over \Lambda^2} \right]
\nonumber\\
&&  \quad
+ {4 \over \pi} \int_0^\Lambda dq \, q^2
\left[{1 \over 2pq} \ln \left({p^2 +pq + q^2 - E - i \epsilon \over
p^2 - pq + q^2 -E -i \epsilon}\right) + {H(\Lambda) \over \Lambda^2} \right]
\nonumber
\\
&&  \quad\:\,
\times \left[ -{\gamma} + (\mbox{$3\over4$} q^2 -E -i \epsilon )^{1/2} 
        - \mbox{$1\over2$} r_s (\mbox{$3\over4$}q^2-E-\gamma^2) \right]^{-1} 
{\mathcal A}_S (q, k; E) \,,
\label{BHvK-range}
\end{eqnarray}
where $r_s$ is the effective range 
and $\gamma$ is the position of the pole in the binding momentum 
$(-E)^{1/2}$ of the atom-atom Green's function:
\begin{eqnarray}
\gamma = \left( 1 - \sqrt{1 - 2 r_s/a} \right) {1 \over r_s} \,.
\label{gamma-ars}
\end{eqnarray}
If $a>0$, $\gamma^2$ is the binding energy of the shallow dimer;
if $a<0$, $-\gamma^2$ is the energy of the shallow virtual state.
In the scaling limit $r_s \to 0$, $\gamma$ reduces to $1/a$.
the two quantities differ if the effective range is included.
We have chosen  $\gamma$ and $r_s$ as our 2-body inputs
instead of $a$ and $r_s$, because this choice keeps the location 
of the dimer pole fixed which leads to a better convergence 
of the effective-range expansion \cite{PRS99}. 
The factor $Z_D$ in the definition of the amplitude 
${\mathcal A}_S (p, k; E)$ in Eq.~(\ref{A-def}) introduces a factor 
of $1 - \gamma r_s$ that has been absorbed into ${\mathcal A}_S$, 
so that in Eq.~(\ref{BHvK-range}) the effective range 
appears only in the diatom propagator.
In principle one could obtain the range corrections to all orders 
by solving the integral equation in Eq.~(\ref{BHvK-range}). 
A potential problem comes from the dimer propagator with the 
effective range included. In addition to the pole from the shallow 
dimer, it has also a deep pole. For negative effective range, 
the pole is on the unphysical sheet and it causes no problems.
For positive effective range, the pole is on the physical sheet
and it leads to problems in the solution of the integral equation.

\begin{figure}[htb]
\bigskip
\centerline{\includegraphics*[width=9cm,angle=0]{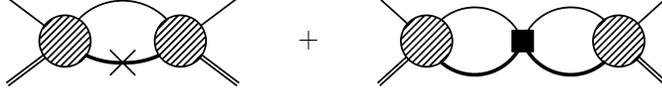}}
\medskip
\caption{Leading order range corrections. Not shown are diagrams that
  vanish as $\Lambda\to\infty$.}
\label{fig:rcorr}
\end{figure}

One possible solution is to calculate the linear range correction
perturbatively \cite{BHK99b,HM-01b}. In this case, the 
renormalization can be carried out analytically.
Writing
\begin{subequations}
\begin{eqnarray}
{\mathcal A}_S (q, k; E) &=& {\mathcal A}_{S,0} (q, k; E)
+ {\mathcal A}_{S,1} (q, k; E)  + \ldots \,
\\
H(\Lambda) &=& H_0(\Lambda)+ H_1(\Lambda)+\ldots \,,
\end{eqnarray}
\end{subequations}
we split the scattering amplitude into a piece of order $(\gamma r_s)^0$,
a piece of order $(\gamma r_s)^1$, and higher order pieces. 
One can then show \cite{HM-01b} that, up to terms that are suppressed
as $\Lambda\to\infty$,  ${\mathcal A}_{S,1} (q, k; E)$ is given by
the diagrams shown in Fig.~\ref{fig:rcorr}.
The diagram on the left-hand side is the contribution of the effective 
range correction. This diagram is logarithmically divergent in the 
ultraviolet. Note also that the range insertion enters only 
on the dimer line. The second diagram is the contribution of
a subleading piece of the contact 3-body force without derivatives.
The 3-body force diagram is required to renormalize 
the ultraviolet divergence of the range correction.
However, its behavior is fully determined by $\Lambda_*$ as we will show
in the following.
Evaluation of the Feynman diagrams in Fig.~\ref{fig:rcorr} leads to
\bqa
{\mathcal A}_{S,1} (k, k; E) &= &
\frac{r_s}{4 \pi^2}\int_0^\Lambda 
     \frac{dq \,q^2 {\mathcal A}_{S,0}^2 (q, k; E)}
     {-\gamma + \sqrt{3q^2/4-E}}
\nonumber\\
&& \qquad
+ \frac{1}{\pi^3 \gamma} \frac{H_1(\Lambda)}{\Lambda^2}
    \left[\int_0^\Lambda\frac{dq \,q^2 {\mathcal A}_{S,0} (q, k; E)}
     {-\gamma + \sqrt{3q^2/4-E}}\right]^2 \,,
\eqa
where we have set $p=k$ for simplicity. The high-$q$ behavior of the
off-shell amplitude ${\mathcal A}_{S,0} (q, k; E)$ is known exactly:
\beq
{\mathcal A}_{S,0} (q, k; E) \longrightarrow
\frac{{\mathcal N}(k,E)}{q} \cos[s_0 \ln(q/\Lambda_*)+\delta]  \,,
\eeq
where $\delta$ is a phase that depends only on $\gamma/\Lambda_*$. 
The important point is that the dependences on $k$ and $E$
and on $q$ factorize for large $q$.
The ultraviolet divergent piece of the two diagrams can therefore be written as
\bqa
\delta {\mathcal A}_{S,1}^{\rm (div)} &=& {\mathcal N}(k,E)^2
  \Bigg( \frac{r_s}{2\sqrt{3}\pi^2}
  \int^\Lambda \frac{dq}{q}  \cos^2[s_0 \ln(q/\Lambda_*)+\delta]
\nonumber
\\
&&  \qquad
+ \frac{4}{3\pi^3\gamma}\frac{H_1(\Lambda)}{\Lambda^2}
       \bigg[\int^\Lambda dq \, \cos[s_0 \ln(q/\Lambda_*)+\delta] 
   \bigg]^2 \Bigg) \,,
\hspace{0.9cm}
\label{AS-div}
\eqa
where the (hidden) lower integration bound is large compared to $k$ but
otherwise arbitrary. Since the energy dependent term ${\mathcal N}(k,E)^2$
factorizes, the linear range correction can be renormalized by adjusting
$H_1(\Lambda)$ so that there is a cancellation of the terms of order
$\Lambda^0$ in Eq.~(\ref{AS-div}).
Thus $H_1(\Lambda)$ is fully determined by the 
effective range $r_s$ and the leading order parameters $\gamma$ and
$\Lambda_*$. The asymptotic phase $\delta$ can be extracted from the 
leading order solution. No new 3-body parameter enters at this order.

Another way to calculate the range corrections is
to expand the dimer propagator in Eq.~(\ref{BHvK-range}) to linear order
in $r_s$ and solve the corresponding integral equation \cite{BGHR03}.
This approach  resums a selected class 
of (small) higher-order effective-range contributions.
The renormalization can no longer be carried out analytically and a small 
cutoff dependence remains. However, this cutoff dependence can be used to 
estimate the errors from higher order corrections by varying the cutoff
over a natural range of values. Furthermore, this approach can be more
easily extended to higher orders. The systematics of higher-order
power corrections and 3-body forces has been discussed in 
Refs.~\cite{BGHR03,Griesshammer:2004pe} and we refer the reader to 
these papers for more details. A general classification of 3-body
forces using renormalization group techniques has recently been given
in Refs.~\cite{Griesshammer:2005ga,Birse:2005pm}.
Afnan and Phillips \cite{Afnan:2003bs} have used a 
subtraction method suggested in Ref.~\cite{HM-01a} to obtain a
renormalized equation that includes the range corrections. They first
solve a subtracted integral equation for the half-off-shell
amplitude ${\mathcal A}_S$ at threshold. Then they use this result to derive 
the full-off-shell amplitude at threshold which determines the 
physical scattering amplitude at all energies. Due to
the subtraction, the 3-body force term drops out of the equations
but one still requires a 3-body datum to fix the subtraction
constant.

To our knowledge, explicit calculations of the range corrections
have to date only been carried out for nuclear systems.
For the spin-quartet S-wave channel (because of the Pauli principle) 
and generally for channels with $L \geq 1$ 
(because of the angular momentum barrier),
no 3-body parameter enters in the first three orders and the 3-body 
calculations are straightforward.
The second-order range corrections
to the scattering length \cite{BK98} and the phase shift \cite{BHK98}
in the spin-quartet S-wave neutron-deuteron channel
have been calculated using effective field theory. 
The range corrections
to the higher partial waves in neutron-deuteron scattering
were calculated in Ref.~\cite{Gabbiani:1999yv}.
The spin-doublet S-wave channel has the same structure as the case
of spinless bosons discussed above. 
A 3-body parameter is required at leading order. 
The first-order range correction 
to the spin-doublet S-wave neutron-deuteron scattering length 
$a_{nd}^{(1/2)}$ is naively infinite and cannot be calculated without 
renormalization. Efimov
and Tkachenko showed that the corrections to the 
triton binding energy contain the same infinity and they derived the linear
range correction to the Phillips line which is manifestly finite
\cite{EfT85,EfT88a,EfT88b}. 
This correction slightly shifts the Phillips line and
moves it closer to the potential model points (cf. Fig.~\ref{fig:nucphil}).
The linear range corrections have also been calculated by 
Efimov \cite{Efimov91,Efimov93}.
The linear range corrections for spin-doublet S-wave neutron-deuteron
scattering at finite energy were first calculated 
perturbatively in Ref.~\cite{HM-01b}. (See Ref.~\cite{Afnan:2003bs} for
a calculation using the subtraction method.) 
In the EFT counting, this corresponds to a next-to-leading order (NLO)
calculation. No new 3-body parameter was required at this order.
The first calculation to next-to-next-to-leading
order (N$^2$LO) was carried out in Ref.~\cite{BGHR03}.
At this order both the quadratic range corrections and a 
second 3-body parameter contribute.
Due to the second 3-body parameter there is no universal Phillips 
line at this order. For the linear range corrections, the methods 
of Refs.~\cite{HM-01b,BGHR03,Afnan:2003bs} agree very well.

A thorough analysis of the power-law corrections near the RG limit cycle
has been carried out for the Glazek-Wilson model \cite{GW03b}, 
a discrete Hamiltonian model described in Section~\ref{sec:RGlimcyc}.
The renormalization group flow was linearized around the limit cycle,
as in Eq.~(\ref{RG-lin}), and the complete set of eigenvectors
of the linearized RG flow was deduced.  The model has no relevant 
operators, a single marginal operator that corresponds to flow 
along the limit cycle, and infinitely many irrelevant operators.
The critical exponents for the irrelevant operators are all integers.
An important general feature of RG limit cycles was established in 
Ref.~\cite{GW03b}:  although the irrelevant operators may vary with the 
phase around the limit cycle, their critical exponents must be 
constants independent of that phase.


\subsection{Large P-wave scattering length}

The phase shifts for higher partial waves have effective-range expansions
analogous to the expansion for the S-wave phase shift in Eq.~(\ref{kcot}).
If the leading term in the effective-range expansion for a higher partial 
wave is unnaturally large, it can also lead to universal low-energy behavior. 
The simplest such case is a large P-wave scattering length. 
The $L=1$ contribution to the atom-atom scattering amplitude
in Eq.~(\ref{pwe}) can be written as
\begin{equation}
f_k^{L=1}(\theta)={3 k^2 \cos\theta \over k^3 \cot\delta_1(k)-ik^3} \,.
\label{fk-Pwave}
\end{equation}
The effective-range expansion for the P-wave phase shift can be written in a
form analogous to that for the S-wave phase shift in (\ref{kcot}):
\begin{equation}
 k^3 \cot\delta_1(k)=-1/a_p + \mbox{$1\over2$} r_p k^2 + \ldots \,,
\label{ere-Pwave}
\end{equation}
which defines the {\it P-wave scattering length} $a_p$ 
and the {\it P-wave effective range} $r_p$.
Dimensional analysis shows that $a_p$ has dimensions of volume
while $r_p$ has dimensions of inverse length. For
simplicity, however, we will still refer to $a_p$ and $r_p$
as a scattering length and an effective range.

Large P-wave scattering lengths are relevant to some halo
nuclei. An example is the $^6$He nucleus.  A P-wave resonance in 
$n \alpha$ scattering, where $n$ is a neutron and $\alpha$ is the 
$^4$He nucleus, plays an important role in the binding of $^6$He. 
The resonance produces a large scattering length in the $P_{3/2}$ 
channel corresponding to
total angular momentum quantum number $3 \over 2$. 

The effects of the large P-wave scattering length have been treated 
using two different scenarios.
In one scenario, the P-wave scattering length $a_p$ was assumed to be 
unnaturally large,
while $r_p$ and the coefficients of higher terms in the effective-range 
expansion in Eq.~(\ref{ere-Pwave}) were assumed to have natural values. 
In this scenario, which requires a single fine-tuning,
the unitarity term $ik^3$ in the denominator in Eq.~(\ref{fk-Pwave}) 
can be neglected at leading order. 
This scenario  has been applied to $n \alpha$ scattering close 
to threshold \cite{BHK03}. 
In the other scenario, both $a_p$ and $r_p$ were 
assumed to be unnaturally large \cite{BHK02}. 
In this scenario, which requires a double fine tuning, 
the unitarity term $ik^3$ in the denominator in Eq.~(\ref{fk-Pwave}) 
generates a rich pole structure in the complex energy plane. 
This scenario has also been applied to $n \alpha$ scattering~\cite{BHK02}. 
Which of these scenarios is most useful for a given system with a large P-wave 
scattering length depends on the scales of the system under consideration.  
For more details, we refer the reader to Refs.~\cite{BHK02,BHK03}.

Suno, Esry, and Greene have studied 3-body recombination in 
a system consisting of three identical fermions with a
large P-wave scattering length \cite{SEG02}.  
The recombination was into deep molecules 
bound by the P-wave potential between 2 identical fermions.
The rate of decrease in the number densities of low-energy fermions 
from the 3-body recombination process  has the form
\begin{eqnarray}
{d \ \over dt} n  =   
- 3 \alpha \langle \epsilon^2 \rangle n^3 \,,
\end{eqnarray}
where $\langle \epsilon^2 \rangle$ is the average of the square of
the kinetic energy of the fermions.   
If the event rate constant $\alpha$ is completely determined by the 
large P-wave scattering length, it should have the scaling behavior 
$m a_p^{8/3}/\hbar^3$.  This scaling behavior was 
observed in some of their numerical calculations.


\subsection{Scattering models}

One might wish to be able to calculate $N$-body observables for 
particles with large scattering lengths from first principles.  
However, for $N \ge 3$, this is prohibitively difficult for any 
physical system.  Even for $^4$He atoms, where the fundamental 
starting point can be taken as the electrodynamics of electrons 
and $^4$He nuclei, the most accurate 3-body calculations proceed 
through the intermediate step of constructing a 
potential model for the interaction between two $^4$He atoms.  
Calculations from first principles in the sector consisting of 
two $^4$He atoms are used as inputs in the construction of the 
potential.  Calculations of 3-body observables are then
carried out by solving the 3-body Schr\"odinger equation for this 
model potential.  For more complicated particles, 
such as alkali atoms which have dozens of electrons, 
even the calculation of 2-body 
observables from first principles is prohibitively difficult.

Since the low-energy behavior of particles with large scattering 
lengths is insensitive to the details of their interactions at 
short distances, the potential provides an inefficient encoding 
of the relevant physics.  The sensitivity to short distances 
enters primarily through the scattering length and other constants 
that describe low-energy scattering.  This motivates a more 
phenomenological approach in which the interactions are 
described by a {\it scattering model}, which can be specified by a 
parameterization of low-energy scattering amplitudes.  
The parameters of the scattering model can be treated as 
phenomenological parameters that can be tuned to reproduce 
the observed low-energy observables of the 2-body system.  
If S-wave interactions dominate, the scattering model is 
conveniently expressed as a parameterization of 
$k \cot \delta_0(k)$, where $\delta_0(k)$ is the S-wave phase shift.

For particles that interact through a short-range potential, 
the effective-range expansion can be used to define a sequence 
of increasingly accurate scattering models.  The first few models 
in the sequence are given by
\begin{subequations}
\begin{eqnarray}
k \cot \delta_0 (k) & = & -1/a \,,
\label{era-0}
\\
                & = & -1/a + \mbox{$1\over2$} r_s k^2 \,,
\label{era-1} 
\\
                & = & -1/a + \mbox{$1\over2$} r_s k^2 
                           - \mbox{$1\over4$}  P_s k^4 \,.
\label{era-2} 
\end{eqnarray}
\end{subequations}
The model in Eq.~(\ref{era-0}) is called the {\it zero-range model}.  
This model is a good starting point for describing the interactions 
between two distinguishable particles or two identical bosons.  
Since all higher coefficients in the effective-range expansion 
have been set to zero, this model is by definition the scaling limit.  
The model in Eq.~(\ref{era-1}) is called {\it effective-range theory}.  
For two distinguishable particles, the leading term in the P-wave 
phase shift in Eq.~(\ref{ere-Pwave}) may be equally important.  
Effective-range theory includes effective-range corrections 
proportional to $r_s/a$.  It also includes corrections that are 
higher order in $r_s /a$ to all orders.  For example, if $a > 0$, 
the binding energy of the shallow dimer is given by 
Eq.~(\ref{B2minus}), and if $r_s > 0$, the model includes a deep 
2-body bound state whose binding energy is given approximately
by Eq.~(\ref{B2plus}).  For particles that interact through a 
short-range potential, the model in Eq.~(\ref{era-2}) provides
an even more accurate description of S-wave interactions at low 
energies.  Unfortunately this is not true for real atoms, 
because the potential at long distance has a van der Waals tail 
that falls off like $1/r^6$.  As a consequence, all the higher 
partial waves give contributions to the scattering amplitude 
proportional to $k^4$.

Each of the models specified by Eqs.~(\ref{era-0})-(\ref{era-2}) 
will exhibit universal low-energy behavior as the scattering 
length $a$ is tuned to $\pm \infty$.  
Low-energy 3-body observables in this limit 
will approach functions of $a$ and the 3-body parameters 
$\kappa_*$ and $\eta_*$.  In the zero-range model, 
there is universal behavior for all $a$.  The parameters 
$\kappa_*$ and $\eta_*$ enter through ambiguities in the 
solutions of integral equations that determine 3-body observables.  
In effective-range theory, universal behavior appears when 
$|a| \gg |r_s |$.  Since $r_s$ is the only dimensionful parameter 
that remains when $a = \pm \infty$, the 3-body parameter 
$\kappa_*$ must have the form $\kappa_* = A/r_s$, 
where $A$ is a numerical constant that is only defined modulo 
multiplicative factors of $e^{\pi/s_0}$.  
As described below, the constant $A$ has been calculated 
for the case $r_s < 0$.   In this case, the model has no deep 
2-body bound states, so $\eta_* = 0$.
If $r_s > 0$, the effective-range theory has a single deep 
2-body bound state, so $\eta_*$ is nonzero.  
It is expected to have the form
$\eta_* = B \kappa_*/ r_s$, where $B$ is a numerical constant.

For an atom near a Feshbach resonance, the scattering length 
$a(B)$ as a function of the magnetic field $B$ can be 
approximated by Eq.~(\ref{Feshbach}).  The effective range 
$r_s(B)$ of the atoms is also a function of the magnetic field.  
As long as $|a(B)|$ is large compared to the natural low-energy 
length scale $\ell_{vdW}$, the few-body system can be 
approximated by the zero-range model defined by (\ref{era-0}) 
with the parameter $a$ replaced by $a(B)$.  A more accurate 
approximation is the effective-range theory defined by 
Eq.~(\ref{era-2}), with the parameters $a$ and $r_s$ replaced 
by $a(B)$ and $r_s(B)$.

An even more accurate description of atoms near a Feshbach resonance 
can be obtained by using a scattering model that describes more 
accurately the physics responsible for the Feshbach resonance.  
If the resonance arises from weak coupling to a closed channel 
in which there is a diatomic molecule extremely close to the 
atom-atom threshold, the interactions can be approximated by 
the {\it resonance model} \cite{KMCWH02}.  This model, 
which has three parameters, can be defined by
\begin{eqnarray}
k \cot \delta_0(k) = 
- \left [\lambda + \frac {g^2}{k^2 - \nu} \right ]^{-1} \,.
\label{kcot:RM}
\end{eqnarray}
The scattering length is
\begin{eqnarray}
a = \lambda - g^2/\nu \,.
\end{eqnarray}
The dependence of the approximation in Eq.~(\ref{Feshbach}) 
on the magnetic field can be reproduced by taking $\lambda$ 
and $g$ to be independent of $B$ and $\nu$ to be linear in 
$B-B_{\rm res}$.  Thus $\lambda$ can be identified with the 
off-resonant scattering length, $g$ is the strength of 
the coupling to the closed channel, and $\nu$ is proportional 
to the detuning energy of the molecule in the closed channel 
with respect to the atom-atom threshold.  The resonance model 
will exhibit universal behavior as $\nu$ is tuned to zero, 
which corresponds to tuning $B$ to $B_{\rm res}$.  
Three-body observables will approach universal functions of $a$, 
the 3-body parameter $\kappa_*$, which is a function of $\lambda$ 
and $g$, and the 3-body parameter $\eta_*$, which is a 
function of the dimensionless combination $\lambda g^2$.

Three-body observables have been calculated in a scattering model 
that is a special case of the resonance model \cite{Petrov04}.  
The model is defined by
\begin{eqnarray}
k \cot \delta_0(k) = - 1/a - R_* k^2 \,,
\end{eqnarray}
where $R_*$ is a positive parameter.  Comparing with 
Eq.~(\ref{kcot:RM}), we see that this is just the resonance model 
with the background scattering length set to 0.  The parameter $g$ 
and $\nu$ are given by $a = -g^2/\nu$ and $R_* = -1/g^2$.  
This model is also identical to the effective-range theory 
defined by Eq.~(\ref{era-1}) with a negative effective range: 
$r_s = -2 R_*$.  The atom-dimer 
scattering length $a_{AD}$ and the 3-body recombination rate constant 
$\alpha$ have been calculated in this model as functions of $a$ and 
$R_*$ \cite{Petrov04}.  For $a \gg R_*$, they have the universal behavior 
given in Eqs.~(\ref{a12-explicit}) and (\ref{alpha-analytic}).  
The 3-body parameter $\kappa_*$ was determined 
in Ref.~\cite{Petrov04} to be
\begin{eqnarray}
s_0 \ln (\kappa_*) = s_0 \ln (2.5 /R_*) \mod \pi \,.
\end{eqnarray}
The cross-over to the universal behavior occurs when $a$ is 
comparable to $R_*$.  The first divergence of the atom-dimer 
scattering length occurs when $a = 0.45 R_*$.  This marks the 
emergence of the first Efimov state below the atom-dimer threshold.  
The second and higher divergences of $a_{AD}$ occur at values 
of $a$ that are well-approximated by the universal predictions: 
$a = (e^{\pi /s_0})^n \, 0.64 \, R_*$, $n = 1, 2, \dots$, 
where $e^{\pi/s_0} \approx 22.7$.  The first zero of the 
recombination rate constant occurs at $a = 3.3 R_*$.  
The second and higher zeroes occur at values of $a$ that are
well approximated by the universal predictions:
$a = (e^{\pi s_0})^n \, 2.9 \, R_*$, $n = 1, 2, \dots\;$.

\section*{Acknowledgments}
We thank J.O.~Andersen, P.F.~Bedaque, V.~Efimov, B.~Esry,
U.~van Kolck, J.~Macek, and K.G.~Wilson for comments on the manuscript.
EB is thankful for the hospitality of  
Nordita, Fermilab, the Aspen Center for Physics,
the Kavli Institute for Theoretical Physics, and the Institute
for Nuclear Theory, where parts of this review were written.
HWH  is thankful for the hospitality of the Ohio State University,
where parts of this review were written.
This research was supported in part by DOE grants DE-FG02-91-ER4069,
DE-FG02-00ER41132, and DE-FG02-05ER15715
and by NSF grant PHY-0098645.


\end{document}